\documentclass[10pt]{article}
\usepackage[left=2cm,top=1cm,right=2cm,bottom=2cm]{geometry}
\usepackage{setspace}
\usepackage{amsmath}
\usepackage{amsthm}
\usepackage{amssymb}
\usepackage{amsbsy}
\numberwithin{equation}{section}
\usepackage{cleveref}
\usepackage[retainorgcmds]{IEEEtrantools}
\usepackage{graphicx}
\usepackage{subcaption}

\author{A.F.A.~Bott$^1$, C.~Graziani$^2$, P.~Tzeferacos$^{1,2}$, T.G.~White$^1$, D.Q.~Lamb$^{2}$, \\G.~Gregori$^1$, A.A.~Schekochihin$^{1,3}$\\
\\
\small $^1$Deparment of Physics, University of Oxford, Parks Road, Oxford OX1 3PU, UK\\
\small $^2$FLASH Center for Computational Science, University of Chicago, 5640 S. Ellis Ave, Chicago, IL 60637, USA\\
\small $^3$Merton College, Merton Street, Oxford OX1 4JD, UK\\}
\title{Proton imaging of stochastic magnetic fields}

\begin{document}


\maketitle

\begin{abstract}
Recent laser-plasma experiments~\cite{F13,H15,T16, T17} 
report the existence of dynamically significant magnetic fields, whose statistical characterisation is essential for a complete understanding of the physical processes these experiments are attempting to investigate. In this paper, we show how a proton imaging diagnostic can be used to determine a range of relevant magnetic field statistics, including the magnetic-energy spectrum. To achieve this goal, we explore the properties of an analytic relation between a stochastic magnetic field and the image-flux distribution created upon imaging that field. This `Kugland image-flux relation' was previously derived~\cite{K12} under simplifying assumptions typically valid in actual proton-imaging set-ups. We conclude that, as in the case of regular electromagnetic fields, features of the beam's final image-flux distribution often display a universal character determined by a single, field-scale dependent parameter -- the contrast parameter $\mu \equiv d_s/\mathcal{M} l_B$ -- which quantifies the relative size of the correlation length $l_B$ of the stochastic field, proton displacements $d_s$ due to magnetic deflections, and the image magnification $\mathcal{M}$. For stochastic magnetic fields, we establish the existence of four contrast regimes -- linear, nonlinear injective, caustic and diffusive -- under which proton-flux images relate to their parent fields in a qualitatively distinct manner. As a consequence, it is demonstrated that in the linear or nonlinear injective regimes, the path-integrated magnetic field experienced by the beam can be extracted uniquely, as can the magnetic-energy spectrum under a further statistical assumption of isotropy. This is no longer the case in the caustic or diffusive regimes. We also discuss complications to the contrast-regime characterisation arising for inhomogeneous, multi-scale stochastic fields, which can encompass many contrast regimes, as well as limitations currently placed by experimental capabilities on one's ability to extract magnetic field statistics. The results presented in this paper are of consequence in providing a comprehensive description of proton images of stochastic magnetic fields, with applications for improved analysis of individual proton-flux images, or for optimising implementation of proton-imaging diagnostics on future laser-plasma experiments. 

\end{abstract}

\section{Introduction}

\begin{figure}[htb]
\centering
\includegraphics[width=0.8\textwidth]{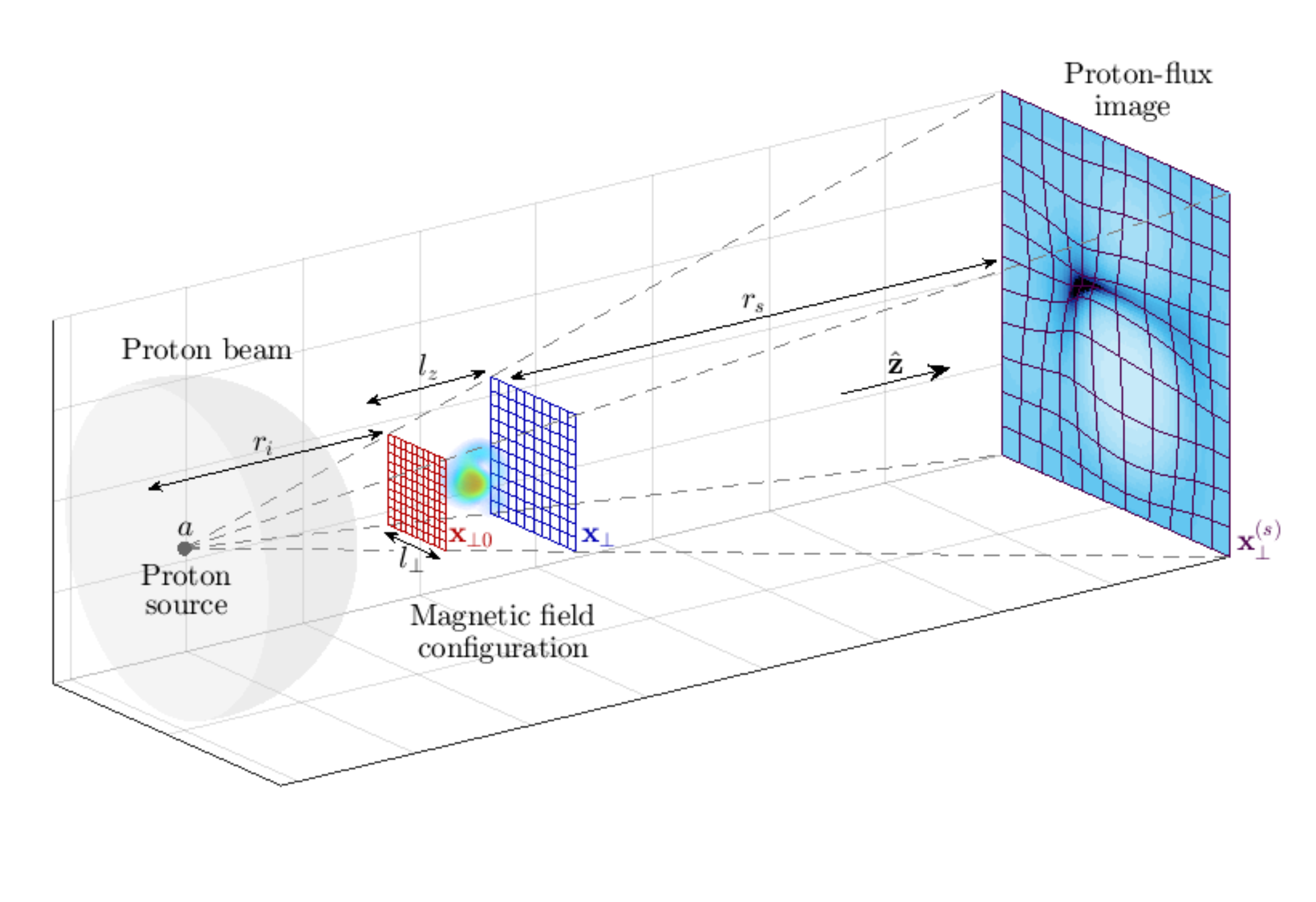}
\caption{\textit{Set-up of proton-imaging diagnostic, with identified theoretical parameters.} A particular magnetic field configuration is imaged by propagating an approximately planar proton beam through the structure. Spatially varying magnetic forces cause non-uniform deflection of protons, resulting in a particular proton-flux image. Here $r_i$ is the distance from the proton source to the magnetic field configuration, $a$ the proton source size, $l_{\bot}$ the perpendicular length scale of the configuration, $l_z$ the parallel scale, $r_s$ the distance from the configuration to the detector, $\mathbf{x}_{\bot0}$ a perpendicular coordinate system for the imaging beam prior to interaction with the magnetic field, $\mathbf{x}_{\bot}$ a perpendicular coordinate system after the interaction, and $\mathbf{x}_{\bot}^{\left(s\right)}$ the image coordinate system. \label{PRsetup}}
\end{figure}

Proton imaging (also known as proton radiography) is an important electromagnetic-field diagnostic used in laser-plasma experiments, with applications spanning laboratory astrophysics and inertial fusion~\cite{M04,M06,L06}.
The diagnostic is implemented by passing an approximately uniform beam of imaging protons through a plasma onto a spatially resolved detector. While inside the plasma, the imaging protons experience Lorentz forces arising from electromagnetic fields, which result in a non-uniform \textit{image-flux distribution}. An idealised picture of a typical imaging set-up is shown in Figure \ref{PRsetup}. 

When implemented successfully, proton imaging can help characterise electromagnetic structures, the measurement of which is crucial for understanding a wide range of plasma dynamics. However, as a two-dimensional diagnostic, a conventional proton imaging set-up is unable to describe completely a full three-dimensional electromagnetic field configuration, instead yielding the two-dimensional image-flux distribution. 
Additionally, it is well known that the morphology and strength of image-flux structures do not correspond directly to the equivalent properties of the electromagnetic field~\cite{K12}. For proton imaging to provide useful measurements of electromagnetic fields, an interpretation of a proton-flux image in terms of the field creating it is therefore required.

Typically there are two ways in which the interpretation of proton-flux images is carried out. The first consists of simulating some artificial field, either by a bespoke electromagnetic-field generation method~\cite{L15}, 
or with a more general plasma-simulation tool, and then introducing a pre-defined proton beam: this is then propagated via a numerical scheme from source to detector. The resulting proton-flux image is compared to experimentally obtained images; similarities between the two are interpreted as evidence for the artificial field being similar to the experimental field~\cite{H15,B02}. 
Such a technique can be enhanced further by running optimisation schemes on proton ray-tracing codes applied to parametrised test fields to find the best possible fit between an electromagnetic-field structure and its associated proton-flux image~\cite{R05}.
This \textit{forward-propagation} technique has been successfully used to describe electromagnetic fields produced in a range of laser-plasma experiments~\cite{R08,B06a,S12}. 

An alternative approach is to perform a general analysis of the evolution of the proton beam analytically, and so derive an analytical expression for the image-flux distribution in terms of the fields creating it~\cite{K12}. The tractability of the relevant arguments is a result of the protons' large speed and low density relative to macroscopic plasma properties: this enables a range of physical processes to be neglected, leaving forces due to electromagnetic fields already present in the plasma solely responsible for alterations to the proton beam's dynamics (this is discussed in Appendix \ref{NegPhysPros}). In addition, a number of assumptions -- such as that of a uniform mono-energetic beam from a point source, paraxality, small proton deflection angles and point projection -- allow for further simplifications. Typically, the image-flux distribution can be related to path-integrated fields~\cite{GTLL16}. 

Both methods have strengths and weaknesses. Forward-propagation techniques require fewer assumptions than analytic theory and, in the case of a simple proton-flux image, finding the electromagnetic fields generating that image is usually tractable. 
However, if analytic theory is valid, it is useful for determining precisely what field statistics are retained in a proton-flux image. It can also lead to practical methods for extracting those statistics directly from an image.

The analytic approach is particularly useful when using proton imaging to investigate stochastic magnetic field configurations. Fields of this sort have recently been studied in a range of laser-plasma experiments. Observations of magnetic field generation via the Weibel instability have been made of late~\cite{F13,H15}, while investigations into the turbulent-dynamo mechanism -- which is thought to generate multi-scale fields in a broad range from system size to resistive scale~\cite{SCTM04,SICMRY07} -- have promising results~\cite{T16,T17}.  
Both experiments are thought to involve stochastic magnetic fields of magnitudes $B_{rms} \sim 0.01 -5 \,\mathrm{MG}$. Given such field strengths and the typical sizes of these experiments, the deflection of imaging protons due to magnetic fields dominates all other processes acting on the imaging proton beam. Therefore, any image-flux structures detected using a proton imaging diagnostic can be associated with magnetic structures. While dominant, magnetic forces experienced by the protons lead to small deflection angles, meaning that the approximations required for the use of analytic theory are often valid. This being the case, we can take advantage of analytic theory not requiring much prior knowledge of magnetic-field statistics (unlike forward-propagation techniques) without risking inaccurate results due to poor assumptions. In this paper, we therefore focus predominantly on constructing an analytic theory of proton imaging for stochastic magnetic fields.

Analytic theory for regular electromagnetic structures is relatively well understood. Kugland \textit{et.}~\textit{al.}~\cite{K12} carried out analyses of various structures based on real-space conservation of proton flux and documented a wide range of features typically observed in proton-flux images. In particular, using simplifying assumptions (stated precisely in Section \ref{Assum}) they derived an analytic relation between path-integrated fields experienced by an imaging beam and the resulting image-flux distribution, which from now on we will refer to as the \emph{Kugland image-flux relation}. They then noted the existence of a dimensionless parameter -- the contrast parameter, $\mu$ -- which effectively characterises image-flux phenomena (for magnetic fields, $\mu$ is defined precisely in Section \ref{Contrast}) . Physically, $\mu$ is a measure of the relative magnitude of proton displacements resulting from electromagnetic forces and the size of electromagnetic structures. For small constrasts, the image flux is related linearly to the path-integrated fields, with the consequence that the latter can in theory be directly reconstructed from proton-flux images~\cite{GTLL16}. 
However, as $\mu$ increases this relationship becomes highly nonlinear; beyond some critical value of $\mu$ some regions of proton flux overlap themselves, and intense structures (\textit{caustics}) appear due to focusing of the proton beam~\cite{K12}. 

A complete discussion of analytic theory of proton imaging for stochastic magnetic fields does not appear to exist. However, a few authors have  considered various  aspects of the general problem. Graziani \emph{et. al.}~\cite{GTLL16} 
developed a linear, small-$\mu$ theory, applied it to multi-scale stochastic magnetic fields, and constructed a method for extracting the \textit{magnetic-energy spectrum} from the image-flux correlation function assuming statistical isotropy and homogeneity of the field. This spectrum can, in turn, be used to calculate the mean magnetic-energy density. There has also been extensive work modelling the diffusive evolution of a beam of charged particles through stochastic magnetic fields. Dolginov and Toptygin~\cite{DT67} 
derived a governing equation for the ensemble-averaged distribution function of non-interacting, unmagnetised test particles using quasi-linear theory; this has been further developed by other authors~\cite{J72,HS67}. 
The result of such theories is typically a proton diffusion tensor, the form of which depends on the properties of both the beam and the field~\cite{P65}. 

In this paper, we aim to determine the magnetic-energy spectrum from a proton image-flux distribution. If known, the magnetic-energy spectrum can in turn be used to deduce both typical magnetic-field strengths and spatial-correlation scales. We wish to address two unknowns: first, the circumstances under which the extraction of the magnetic-energy spectrum is possible; second, how this extraction is done when it is feasible. 

Our approach to answering these questions will be to use the Kugland image-flux relation between the path-integrated magnetic field and the image-flux distribution. We explain in Section \ref{PlasMap} that the path-integrated field is sufficient to characterise the magnetic-energy spectrum uniquely, provided magnetic fluctuations are assumed isotropic and homogeneous. Consequently, if it can be shown that the path-integrated field is extractable from a proton-flux image (or not), then the same is true for extracting the magnetic-energy spectrum for isotropic, stochastic magnetic fields. 

Bearing this last statement in mind, we study the properties of the Kugland image-flux relation for the case of stochastic magnetic fields. Much as with regular electromagnetic structures, the types of image-flux features which manifest for stochastic magnetic fields are determined by $\mu$. More specifically, we outline four regimes of distinctive image-flux phenomena in terms of $\mu$. For each regime, we investigate the possibility of the path-integrated magnetic field being reconstructed from an image-flux distribution, and if this reconstruction is possible, we provide a methodology for its practical implementation.  

The text is organised in the following manner. In Section \ref{InterpProtImag}, we discuss generally the interpretation of proton images resulting from stochastic magnetic fields. More specifically, in Section \ref{StatCharStocMagField} we present the motivation for wishing to determine the magnetic-energy spectrum when using a proton imaging set-up to probe a stochastic magnetic field. Section \ref{PlasMap} introduces the Kugland image-flux relation, along with its correspondence to the magnetic-energy spectrum. By investigating the properties of the Kugland image-flux relation, Section \ref{Contrast} presents a general categorisation of image-flux features arising due to stochastic magnetic fields in terms of $\mu$. In Section \ref{Contrastregimes}, we describe four distinct contrast regimes arising at different values of $\mu$: \textit{linear} (Section \ref{LinRgme}), \textit{nonlinear injective} (Section \ref{NonLinInjRgme}), \textit{caustic} (Section \ref{CauRgme}), and \textit{diffusive} (Section \ref{HighConRgme}). This description is combined with a numerical example illustrating key characteristics of each regime. We also describe possible methods for extracting the path-integrated field (and hence the magnetic-energy spectrum), as well as tests for identifying the likely contrast regime in which a given proton-flux image was formed. In Section \ref{NumReconTest}, we illustrate the success (or failure) or the proposed methods for reconstructing the path-integrated field and magnetic-energy spectrum on the same numerical example used to characterise the contrast regimes.

In Section \ref{TechnicalComp}, we explore various technical issues and complications to the theory presented in Sections \ref{InterpProtImag} and \ref{Contrastregimes}. Section \ref{Assum} outlines the assumptions required for the Kugland image-flux relation to be valid. In Section \ref{TheoCom}, we discuss two theoretical complications to the contrast-regime classification of imaging set-ups applied to stochastic magnetic fields that are significant for multi-scale, or inhomogeneous, anisotropic stochastic fields. Both complications provide examples of situations in which the methods proposed for extracting the magnetic-energy spectrum from a proton-flux image are restricted. Finally, Section \ref{ExpCom} describes three other limitations placed by current experimental capabilities on one's ability to determine the magnetic-energy spectrum using proton imaging.  

Throughout this paper, parameters and coordinate systems are defined as illustrated in Figure \ref{PRsetup}, and a glossary of notation and mathematical conventions adopted in this paper is given in Appendix A. We have in general deferred lengthy proofs of equations to (cross-referenced) appendices. Descriptions of all the numerical algorithms used in this paper are presented in Appendix \ref{NumSim}. A table of contents in given after the appendices.

\section{Interpretation of proton-flux images generated by stochastic magnetic fields} \label{InterpProtImag}

\subsection{Statistical characterisation of stochastic magnetic fields} \label{StatCharStocMagField}

This paper is concerned with recovering statistical properties of stochastic magnetic fields using a proton-imaging diagnostic -- so we begin with a brief outline of how stochastic magnetic fields are described statistically. We consider static magnetic fields of the form
\begin{equation}
\mathbf{B}\!\left(\mathbf{x}\right) = \bar{\mathbf{B}}\!\left(\mathbf{x}\right) +\delta \mathbf{B}\!\left(\mathbf{x}\right) \, , \label{magfielddef}
\end{equation}
where $\bar{\mathbf{B}}$ is some non-stochastic mean field varying on the global scale of the plasma, constrained to be inside a cuboid region, dimensions $l_\bot \times l_\bot \times l_z$ (see Figure \ref{PRsetup}), and $\delta \mathbf{B}$ is a fluctuating field with correlation length $l_B \ll l_z, l_\bot$ ($l_B$ is defined precisely in Appendix \ref{FurtherStatCharMagFieldsCorrDef}). The mean and fluctuating fields are distinguished by introducing a spatial averaging operator $\left<\cdot\right>$, such that $\left<\bar{\mathbf{B}}\right> = \bar{\mathbf{B}}$, and $\left<\delta \mathbf{B}\right> = 0$. This averaging operator can be interpreted either as an average over intermediate scales $l$ such that $l_B \ll l \ll l_z, l_\bot$, or some type of ensemble average. We focus on static fields only because with respect to the fast motion of the protons the evolution of magnetic fields is assumed slow (see Section \ref{Assum} for a discussion of the validity of this assumption).

We will concentrate on one particular magnetic field statistic in this paper: the magnetic-energy spectrum. This is defined by
\begin{equation}
E_B\!\left(k\right) \equiv \int \mathrm{d}\Omega \, k^2\left<\left|\delta \hat{\mathbf{B}}\!\left(\mathbf{k}\right)\right|^2\right> \, , \label{magengspecsec2}
\end{equation}
where $\delta \hat{\mathbf{B}}\!\left(\mathbf{k}\right)$ is the three-dimensional Fourier transform of the fluctuating magnetic field, $\mathbf{k}$ is the wavevector, $k = \left|\mathbf{k}\right|$, and the integral is over solid angles $\Omega$ in $\mathbf{k}$ space. The magnetic-energy spectrum does not in general provide a complete description of a stochastic magnetic field~\cite{A81}. 
However, the magnetic-energy spectrum is a suitable focus for three reasons. Firstly, it can be used to deduce both the fluctuating RMS field strength $B_{rms} \equiv \left<\delta \mathbf{B}^2\right>^{1/2}$ and magnetic field correlation length $l_B$ via relations
\begin{IEEEeqnarray}{rCl}
B_{rms}^2 & = & 8 \pi \int_0^{\infty} \mathrm{d}k \, E_B\!\left(k\right) \, , \label{Brmsspec} \\
l_B & = & \frac{\pi}{2} \frac{\int_0^{\infty} \mathrm{d}k \, E_B\!\left(k\right)/k}{\int_0^{\infty} \mathrm{d}k \, E_B\!\left(k\right)} \, . \label{corrlengthspec}
\end{IEEEeqnarray}
The latter result is derived in Appendix \ref{FurtherStatCharMagFieldsCorrSpec}. Secondly, for the special case of isotropic Gaussian statistics the magnetic-energy spectrum is sufficient to provide a complete statistical characterisation of the stochastic field~\cite{A81}.
Thirdly, even for non-Gaussian stochastic magnetic fields (as physical fields often are), in many situations the magnetic-energy spectrum has a special significance. It describes energy distribution over wavenumber scales, and so is often the focus of theoretical predictions~\cite{G15}.
Throughout the rest of this paper, we therefore explore the question of whether a proton-imaging diagnostic can be used to extract the magnetic-energy spectrum. 

A discussion of the mathematical characterisation of stochastic magnetic fields by an alternative mathematical object -- the magnetic autocorrelation function -- is given in Appendix \ref{FurtherStatCharMagFields}. This is an equivalent quantity to the magnetic-energy spectrum. It is sometimes useful to invoke it instead of the spectrum: the most physically intuitive definition of the magnetic-field correlation length is in terms of magnetic autocorrelation function, and for the derivations of spectral relations \eqref{deffieldspec} and \eqref{linfluxspec3} presented in Appendices \ref{DeflfieldCorr} and \ref{RelFluxRMSLinThy} respectively, the magnetic autocorrelation function enables a clearer characterisation of the precision of approximations made than does the magnetic-energy spectrum. 

\subsection{The plasma-image mapping} \label{PlasMap}

In the Introduction, we discussed the importance of the Kugland image-flux relation between the path-integrated magnetic field and image-flux distribution when seeking to determine the magnetic-energy spectrum; here we outline the form of this relation and its correspondence to the energy spectrum. The Kugland image-flux relation was originally derived from particle conservation in real space~\cite{K12}. In Appendix \ref{KinThyPrad}, we present an alternative derivation from first principles using kinetic theory; any reader who wishes to understand in greater mathematical detail the origin of the Kugland image-flux relation is encouraged to explore this Appendix. 
 
Physically, magnetic forces experienced by the imaging protons lead to global restructuring of the imaging beam. For arbitrary magnetic-field configurations and imaging-beam parameters, characterising this restructuring is extremely complicated, particularly for a stochastic magnetic field. This is made clear by stating the general relationship between the proton beam distribution function and the final proton image (see Appendix \ref{KinThyPradGen}). 
However, under some assumptions that are typically valid for proton-imaging set-ups, the complexity of such a relation is greatly reduced~\cite{K12}. We assume that a number of dimensionless parameters are small, namely the size of the proton source relative to the plasma $a/r_i$, the paraxial parameter
\begin{equation}
\delta \alpha \equiv \frac{l_\bot}{r_i} \ll 1\label{paraxialdef}
\end{equation}
of the proton imaging set-up, the point-projection parameter
\begin{equation}
\delta \beta \equiv \frac{l_z}{r_s} \ll 1 \, ,
\end{equation}
and the magnitude of angular deflections  $\delta \theta$ away from the initial proton trajectories [the magnitude of this for stochastic fields is stated subsequently in Section \ref{Contrastregimes}, equation \eqref{RMSdeflecangle}]. A full discussion of the validity of these assumptions and their significance in simplifying the general proton-imaging problem is given in Section \ref{Assum}.
 
In the limit where the above asymptotic parameters are indeed small, the form of the beam as a two-dimensional near-planar sheet (see Figure \ref{PRsetup} and Appendix \ref{KinThyPradInitCon}) is retained following interaction with the magnetic field.
Furthermore, internal re-distribution of proton flux within the sheet is entirely determined by velocity perturbations acquired inside the plasma. More specifically, it can be shown (\cite{K12}; see also Appendix \ref{KinThyPradSingProt}) that an imaging proton with initial perpendicular position $\mathbf{x}_{\bot0}$ ends up with final perpendicular position on the detector
\begin{equation}
\mathbf{x}_{\bot}^{\left(s\right)}\!\left(\mathbf{x}_{\bot0}\right) \approx
\frac{r_s+r_i}{r_i} \mathbf{x}_{\bot0} + \frac{r_s}{V} \, \mathbf{w}\!\left(\mathbf{x}_{\bot0}\right) \, .\label{divmappingSec2}
\end{equation}
Here $V$ is the initial speed of the proton beam, assumed mono-energetic, and the perpendicular velocity deflection $\mathbf{w}\!\left(\mathbf{x}_{\bot0}\right)$ caused by magnetic forces of an imaging proton with initial perpendicular position $\mathbf{x}_{\bot0}$ is 
\begin{equation}
\mathbf{w}\!\left(\mathbf{x}_{\bot0}\right) \approx \frac{e}{m_p c} \hat{\mathbf{z}} \times \int_0^{l_z} \mathrm{d}s \; \mathbf{B}\!\left(\mathbf{x}\!\left(s\right)\right) \, , \label{pathintfieldSec2}
\end{equation}
where $e$ is the proton charge, $m_p$ the proton mass, and $\mathbf{x}\!\left(s\right)$ the proton trajectory. 
We can view $\mathbf{w}\!\left(\mathbf{x}_{\bot0}\right)$ as a function of the initial perpendicular position. We will call this function the \textit{perpendicular-deflection field} for the remainder of this paper. By conservation of proton flux within the imaging beam, the image-flux distribution $\Psi\!\left(\mathbf{x}_{\bot}^{\left(s\right)}\right)$ is then given by the Kugland image-flux relation~\cite{K12}:
\begin{equation}
\Psi\!\left(\mathbf{x}_{\bot}^{\left(s\right)}\!\left(\mathbf{x}_{\bot0}\right)\right) = \sum_{\mathbf{x}_\bot^{\left(s\right)} = \mathbf{x}_\bot^{\left(s\right)}\!\left(\mathbf{x}_{\bot0}\right)}\frac{\Psi_{0}}{\left|\det{\nabla_{\bot0}\!\left[\mathbf{x}_{\bot}^{\left(s\right)}\!\left(\mathbf{x}_{\bot0}\right)\right]}\right|} \, , \label{screenfluxSec2}
\end{equation}
where $\Psi_{0}$ is the initial flux distribution (assumed uniform), $\nabla_{\bot0} \equiv \partial/\partial \mathbf{x}_{\bot0}$ is a gradient operator with respect to the initial plasma coordinates, and the sum indicates that the total flux at any particular position on the detector can in general have contributions from protons with many different initial positions. A numerical illustration of the validity of equations \eqref{pathintfieldSec2} and \eqref{screenfluxSec2} can be found in Appendix \ref{KinThyPradNumSim}.

The \emph{plasma-image mapping} \eqref{divmappingSec2} can be related to the magnetic-energy spectrum \eqref{magengspecsec2} via the perpendicular-deflection field \eqref{pathintfieldSec2}. This is entirely equivalent to relating the path-integrated field to the magnetic-energy spectrum, because the former can be recovered from the perpendicular-deflection field directly by rearranging \eqref{pathintfieldSec2}:
\begin{equation}
\int_0^{l_z} \mathrm{d}s \; \mathbf{B}_\bot\!\left(\mathbf{x}\!\left(s\right)\right)  = - \frac{m_p c}{e} \hat{\mathbf{z}} \times  \mathbf{w}\!\left(\mathbf{x}_{\bot0}\right)\, . \label{linpathintfield}
\end{equation}
If the stochastic magnetic field is assumed to be statistically isotropic and homogeneous (except for some global variation in RMS field strength magnitude), with an additional assumption of zero mean-magnetic field ($\bar{\mathbf{B}} = 0$, $\mathbf{B} = \delta \mathbf{B}$), it can then be shown (see Appendix \ref{DeflfieldCorr}) that the deflection-field spectrum is related to the magnetic-energy spectrum by
\begin{equation}
E_{B}\!\left(k\right) = \frac{m_p^2 c^2} {4\pi^2 l_z e^2} k E_{W}\!\left(k\right) \, .\label{deffieldspec}
\end{equation} 
Here the deflection-field spectrum $E_{W}\!\left(k\right)$ is defined by
\begin{equation}
E_W\!\left(k_\bot\right) \equiv \int \mathrm{d}\theta \, k_\bot\left<\left|\hat{\mathbf{w}}\!\left(\mathbf{k}_\bot\right)\right|^2\right> \, , \label{deflfieldspecdef}
\end{equation}
where $\hat{\mathbf{w}}\!\left(\mathbf{k}_\bot\right)$ is the two-dimensional Fourier transform of the perpendicular-deflection field, $\mathbf{k}_\bot$ the perpendicular wavevector, and the integral is over the polar angle $\theta$. Spectral relation \eqref{deffieldspec} implies that if we can determine the perpendicular-deflection field from the image-flux distribution, then we can recover the magnetic-energy spectrum. 

The deflection-field spectral relation \eqref{deffieldspec} also allows for a simple calculation of the typical deflection angle $\delta \theta$ in terms of the magnetic field and initial proton speed. More specifically, integrating \eqref{deffieldspec} over all wavenumbers (see Appendix \ref{DeflfieldCorr}) gives, in the small-deflections limit,
\begin{equation}
\delta \theta \equiv \frac{w_{rms}}{V} = \frac{e B_{rms}}{m_p c V} \sqrt{l_z l_B} \, , \label{RMSdeflecangle}
\end{equation}
where $w_{rms} \equiv \left<\mathbf{w}^2\right>^{1/2}$ is the RMS of the perpendicular-deflection field.

As noted by previous authors~\cite{K12,GTLL16}, the perpendicular-deflection field has the property of being irrotational, provided the typical deflection angle $\delta \theta \lesssim l_B/l_z$ (this assumption is potentially more restrictive than the small-angle approximation $\delta \theta \ll 1$ -- as discussed in Section \ref{Assum}). Irrotationality of the perpendicular-deflection field follows from the solenoidality of the magnetic field:
\begin{equation}
\nabla_{\bot0} \times \mathbf{w}\!\left(\mathbf{x}_{\bot0}\right) = \hat{\mathbf{z}}\frac{e}{m_p c} \int_0^{l_z} \mathrm{d}z' \left( \frac{\partial \mathbf{x}_{\bot}\!\left(z'\right)}{\partial \mathbf{x}_{\bot0}} \cdot \nabla_\bot \right) \cdot \mathbf{B} \approx  \hat{\mathbf{z}}\frac{e}{m_p c} \int_0^{l_z} \mathrm{d}z'  \, \nabla \cdot \mathbf{B} = 0 \, ,
\end{equation}
where $\nabla \equiv \partial/\partial \mathbf{x}$ denotes the gradient operator with respect to the beam-proton position $\mathbf{x}\!\left(s\right)$. The approximation $ \partial \mathbf{x}_{\bot}\!\left(z'\right)/\partial \mathbf{x}_{\bot0} \approx \underline{\underline{\mathbf{I}}}$ holds if the proton trajectories do not cross inside the plasma (for a more detailed discussion of this result, see Appendix \ref{KinThyPradPerpDefl}). As a consequence, the perpendicular-deflection field can always be written as the gradient of the \textit{deflection-field potential}, defined by
\begin{equation}
\varphi\!\left(\mathbf{x}_{\bot0}\right) \equiv \int_C \mathrm{d} \mathbf{l} \cdot \mathbf{w}\!\left(\tilde{\mathbf{x}}_{\bot0}\right) \, , \label{deffieldpotdef}
\end{equation}
where $C$ is any path from the origin to the perpendicular coordinate $\mathbf{x}_{\bot0}$ and $\mathrm{d}\mathbf{l}$ is an infinitesimal line element along this path. Under the same assumption, the perpendicular-deflection field is given by the magnetic field integrated along the \emph{unperturbed trajectories}:
\begin{equation}
\mathbf{w}\!\left(\mathbf{x}_{\bot0}\right) \approx \frac{e}{m_p c} \hat{\mathbf{z}} \times \int_0^{l_z} \mathrm{d}z' \; \mathbf{B}\!\left(\mathbf{x}_{\bot0}\left(1+\frac{z'}{r_i}\right),z'\right) \, . \label{deffieldundefl}
\end{equation}
As we discuss in Sections 2.4 and 2.5, the irrotationality property of the perpendicular-deflection field is essential for attempts for reconstructing the path-integrated field from the image-flux distribution. 

\subsection{The contrast parameter $\mu$} \label{Contrast}

Equation \eqref{screenfluxSec2} is the desired Kugland image-flux relation between the image-flux distribution and the path-integrated magnetic field, via plasma-image mapping \eqref{divmappingSec2}; we now explore the properties of the Kugland image-flux relation. As is clear from the appearance of the Jacobian determinant of the plasma-image mapping in \eqref{screenfluxSec2}, the size of Jacobian-matrix elements of the plasma-image mapping will be of significance for characterising image-flux features. These elements have physical meaning, describing the relative size of initial gradients in perpendicular velocities of protons compared to gradients resulting from deflections due to magnetic forces. Mathematically, their size is quantified by the \emph{constrast parameter} $\mu$, which we define for a proton imaging set-up applied to a stochastic magnetic field with correlation length $l_B$ by 
\begin{equation}
\mu \equiv  \frac{d_s}{\mathcal{M} l_B}  = \frac{r_s \delta \theta}{\tilde{l}_B} = \frac{\delta \theta}{\delta \alpha} \frac{r_s}{r_s+r_i} \frac{l_\bot}{l_B} \, , \label{contrastdef}
\end{equation}
where $d_s \equiv r_s \delta \theta$ is the typical perpendicular displacement of a proton from its undeflected position on the detector, and $\tilde{l}_B = \mathcal{M} l_B$ is the magnified correlation length
\begin{equation}
\mathcal{M} = \frac{r_i+r_s}{r_i} \, . \label{screenmagfactorsec2}
\end{equation}
If $r_s \gg r_i$, \eqref{contrastdef} reduces to the definition of $\mu$ given in previous literature~\cite{K12,GTLL16}: $\mu = r_i \delta \theta/l_B$.

The universal dependence of image-flux features on $\mu$ -- irrespective of the particular magnetic field structure -- enables a systematic approach to the heuristic interpretation of proton-flux images. Since $\mu$ is a function of field strength via the typical proton deflection angle $\delta \theta$, its identification for a particular image-flux distribution is a useful way to estimate magnetic field strengths attained in experiments. Substituting definitions \eqref{paraxialdef} and \eqref{RMSdeflecangle} for $\delta \alpha$ and $\delta \theta$ respectively into \eqref{contrastdef} gives
\begin{equation}
\mu = \frac{r_s r_i}{r_s+r_i} \frac{e B_{rms}}{m_p c V} \sqrt{\frac{l_z}{l_B}} \, . \label{contrastmagfield}
\end{equation}
This shows that $\mu \propto B_{rms}/V$, the same as the RMS proton deflection angle $\delta \theta_{rms}$. By substituting appropriate values of physical constants into \eqref{contrastmagfield} we can explicity write the field strength associated with a particular $\mu$ as
\begin{equation}
B_{rms}\!\left(\mathrm{kG}\right) \approx 250 \left(\frac{\mathcal{M}}{\mathcal{M}-1}\right) \mu \left[\frac{W\!\left(\mathrm{MeV}\right)}{3.3 \, \mathrm{MeV}}\right]^{1/2} \left[\frac{r_i\!\left(\mathrm{cm}\right)}{1 \, \mathrm{cm}}\right]^{-1} \left[\frac{l_B\!\left(\mathrm{cm}\right)}{l_z\!\left(\mathrm{cm}\right)}\right]^{1/2} \, \mathrm{kG} \, , \label{contrastest}
\end{equation}
However, comparing \eqref{RMSdeflecangle} and \eqref{contrastmagfield}, it is clear that $\delta \theta_{rms}$ and $\mu$ have different dependences on the correlation scale of the field: $\delta \theta_{rms} \propto l_B^{1/2}$, while  $\mu \propto l_B^{-1/2}$. Thus a field with smaller-scale structures will give larger values for $\mu$, despite the typical deflection angle being reduced (a numerical example of this is given in Appendix \ref{ConstrastScale}). We discuss how to estimate $\mu$ from a proton image in Section \ref{Contrastregimes}. 

Beyond qualitative estimates, whether the path-integrated field (and hence magnetic-energy spectrum for isotropic stochastic magnetic fields) can be directly extracted from experimental data -- and if so, how this extraction is carried out -- changes depending on $\mu$. This change is best elucidated for stochastic magnetic fields in terms of four contrast regimes: \textit{linear} ($\mu \ll 1$), \textit{nonlinear injective} ($\mu$ below some critical value $\mu_c \sim 1$), \textit{caustic}  ($\mu \geq \mu_c$) and \textit{diffusive} ($\mu \gtrsim r_s /\!\left(r_s + r_i\right) \delta \alpha $, where $\delta \alpha$ is again the paraxial parameter). Here, $\mu_c$ is defined to be the smallest value of $\mu$ associated with the imaging of a particular stochastic magnetic field such that the plasma-image mapping \eqref{divmappingSec2} is not injective (one-to-one). These four contrast regimes are discussed in the next section. 

\section{Contrast regimes} \label{Contrastregimes}

We begin by providing a general characterisation of the linear (Section \ref{LinRgme}), nonlinear injective (Section \ref{NonLinInjRgme}), caustic (Section \ref{CauRgme}) and diffusive (Section \ref{HighConRgme}) regimes respectively. 

To help with this illustration we consider proton images associated with a particular stochastic magnetic-field configuration. These images are simulated numerically using the technique presented in Appendix \ref{NumSimFluxImageGenFull}. Multi-scale fields possessing a power-law spectrum are of interest in many physical situations~\cite{SCTM04,G15}
so, using methods described in Appendix \ref{NumSimMagFieldGen}, we generate an artificial Gaussian stochastic field with a magnetic-energy spectrum $E_B\!\left(k\right)$ of the form
\begin{equation}
E_B\!\left(k\right) = \frac{B_{rms}^2}{8 \pi} \left(p-1\right) \frac{k^{-p}}{k_l^{-p+1}-k_u^{-p+1}} \, , \quad k \in \left[k_l,k_u\right] \, , \label{powerlawspecdefSec2}
\end{equation}
where $k_l$ and $k_u$ represent the lower and upper wavenumber cutoffs and $p$ is the spectral index. Calculations of key quantities for this spectrum, such as correlation scale $l_B$, are given in Appendix \ref{ToySpecLinThyPowerLaw}. For this section, we set $p = -11/3$, corresponding to a Golitsyn spectrum~\cite{G60}. 
We make this choice for two reasons. Firstly, the magnetic-energy spectrum is thought to follow such a power law in a turbulent magnetised flow with a low Reynolds number~\cite{M61,SICMRY07,G15}. 
Secondly, a rapidly decaying power law of this form has the useful property that both the dominant magnetic and image-flux structures have similar spatial scales, avoiding certain complications important for more shallow power laws (discussed in Section \ref{TheoCom}). 

\begin{figure}[htbp]
\centering
    \begin{subfigure}{.49\textwidth}
        \centering
        \includegraphics[width=\linewidth]{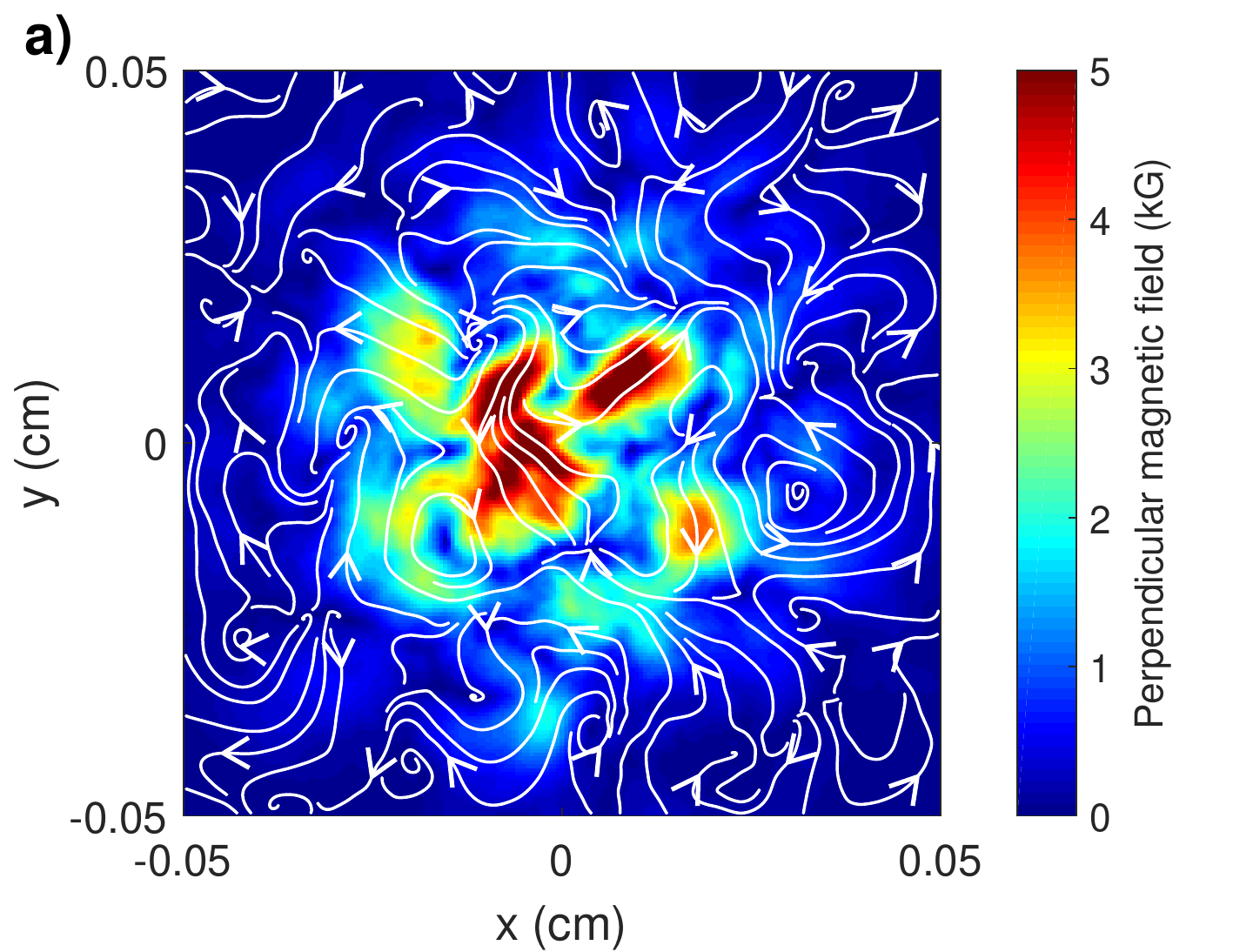}
    \end{subfigure} %
    \begin{subfigure}{.48\textwidth}
        \centering
        \includegraphics[width=\linewidth]{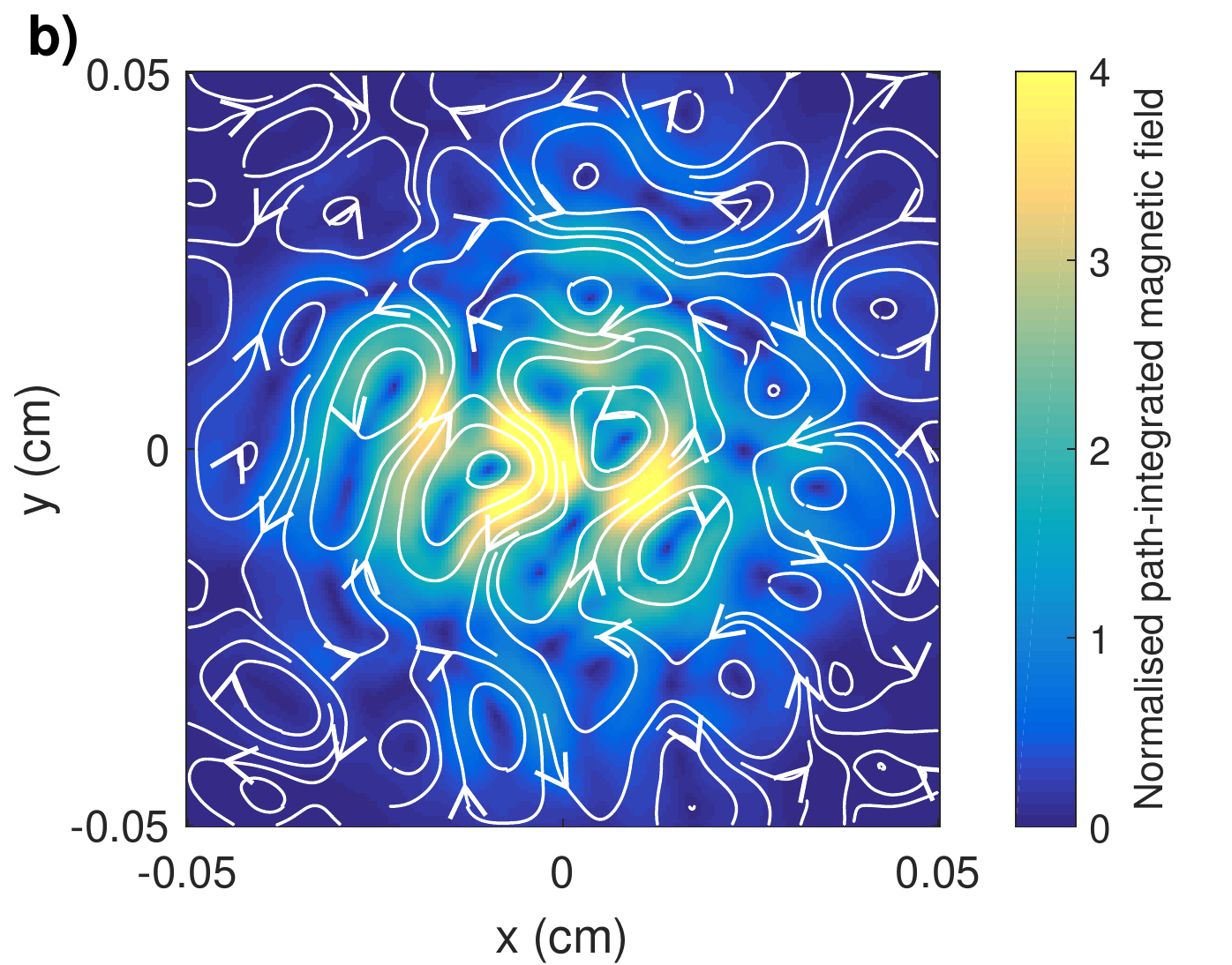}
    \end{subfigure} %
\caption{\textit{Sample of Gaussian compact stochastic magnetic field configuration for use in illustration of contrast regimes (Section \ref{Contrastregimes}).} The field sample is taken to have length scales $l_i = l_z = l_{\bot} = 0.1 \, \mathrm{cm}$, and is specified on a $201^3$ array (grid spacing $\delta x = l_i/201$). The sample is generated using the methods described in Appendix \ref{NumSimMagFieldGen}. The magnetic-energy spectrum is set to be a Golitsyn power law of the form \eqref{powerlawspecdefSec2}, 
 with index $p =-11/3$, spectral cutoffs $k_l = 6 \pi/l_i$, $k_u = 120 \pi/l_i$,  $B_{rms} = 1 \, \mathrm{kG}$, and $l_B = 80 \, \mu \mathrm{m}$. A Gaussian envelope of the form \eqref{magfieldwind} is applied, with $\sigma = 3$, giving $B_{rms,0} \approx 3 \, \mathrm{kG}$. In all plots of two-dimensional vector fields, colour variations denote the value of the labelled quantity, while the white lines with directional arrows are streamlines of the plotted vector field. \textbf{a)} Slice of perpendicular magnetic field taken along central perpendicular plane in configuration. \textbf{b)} Path-integrated perpendicular magnetic field experienced by 3.3 MeV protons originating from a point source located at a distance $r_i = 1 \, \mathrm{cm}$ from the magnetic field configuration. This field was calculated numerically using test protons (see Appendix \ref{NumSimFluxImageGen} for a description of this technique).} \label{GolitysncompactfieldSec2}
\end{figure}

\begin{figure}[htbp]
\centering
    \begin{subfigure}{.36\textwidth}
        \centering
        \includegraphics[width=0.95\linewidth]{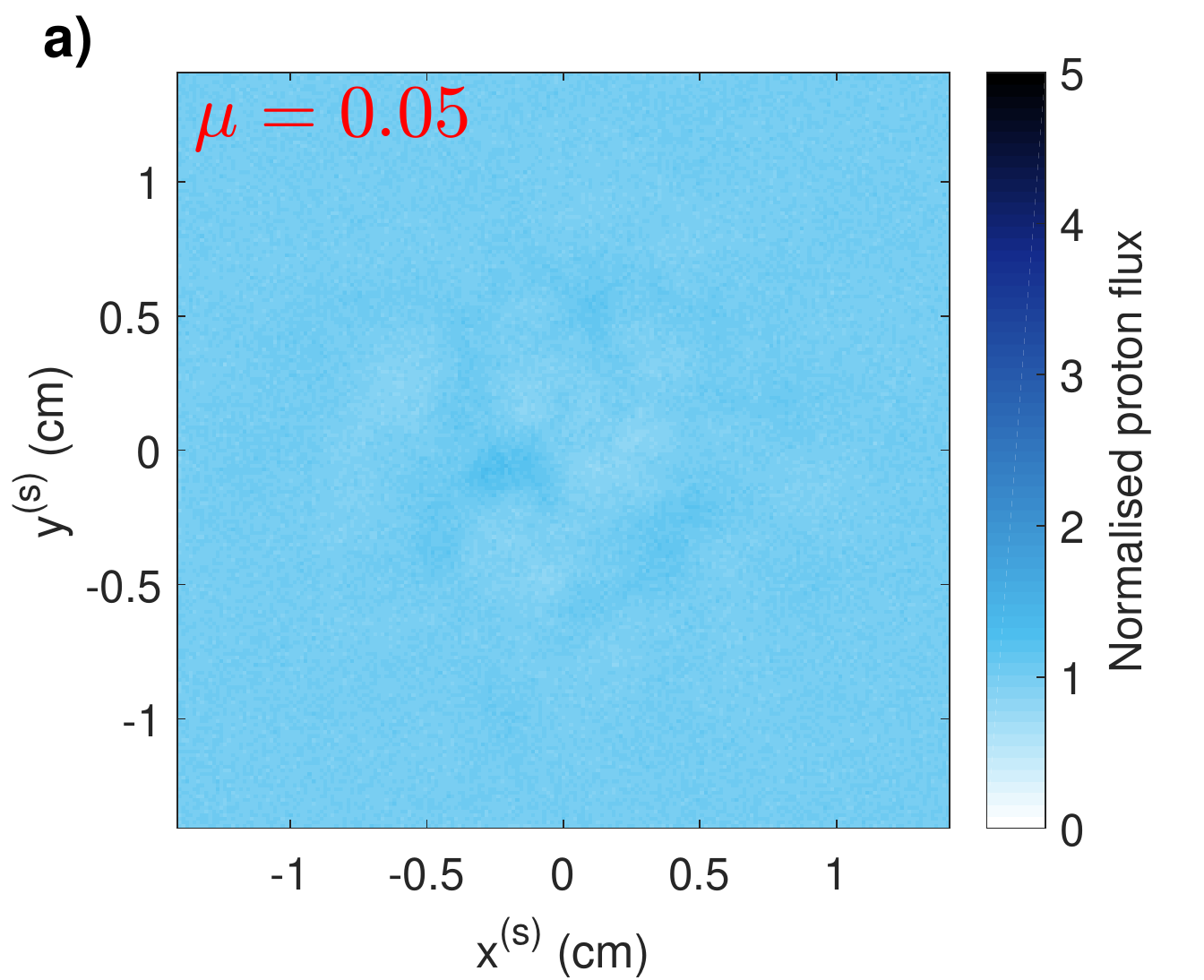}
    \end{subfigure} %
    \begin{subfigure}{.315\textwidth}
        \centering
        \includegraphics[width=0.95\linewidth]{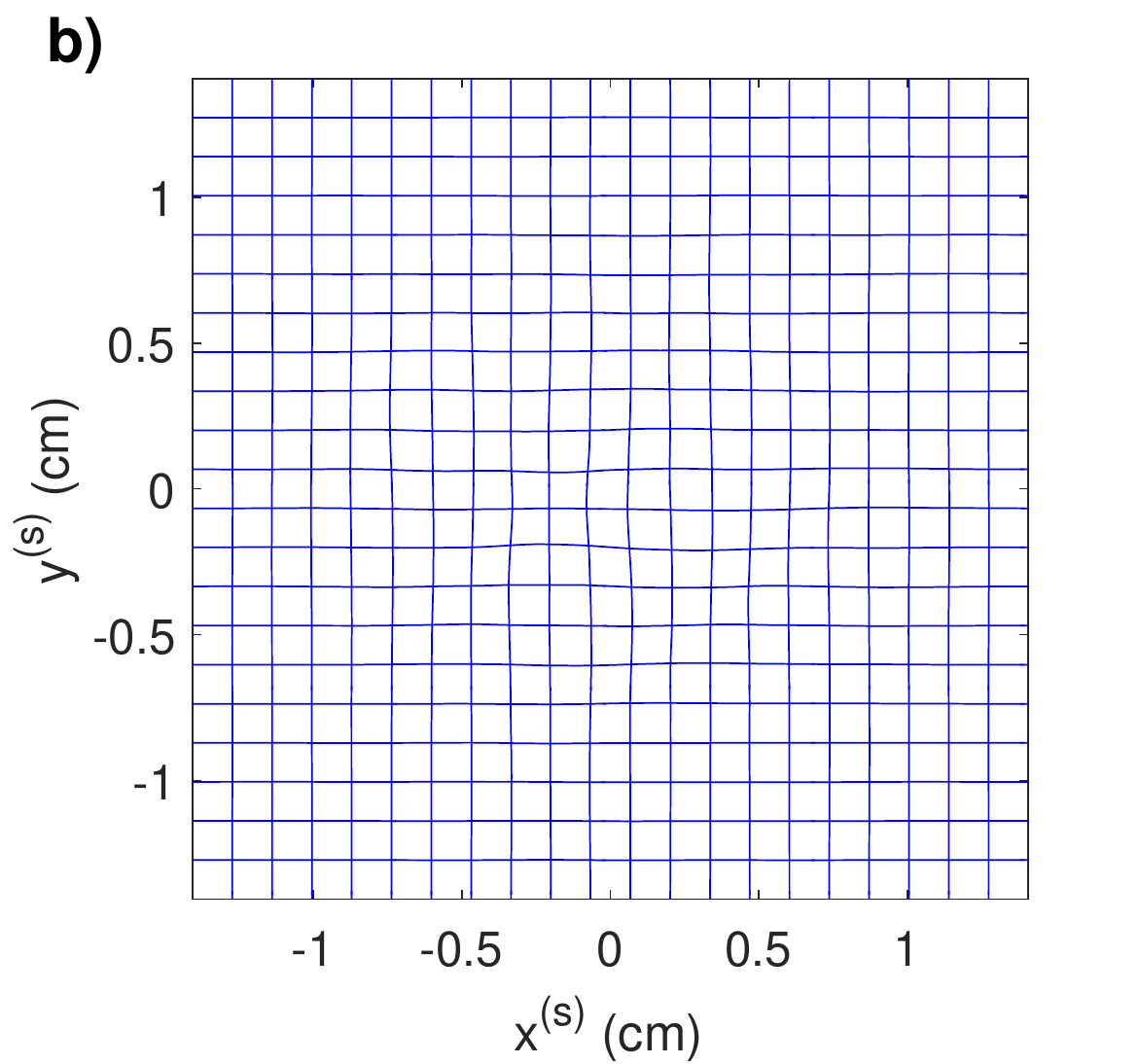}
    \end{subfigure} %
    \begin{subfigure}{.36\textwidth}
        \centering
        \includegraphics[width=0.95\linewidth]{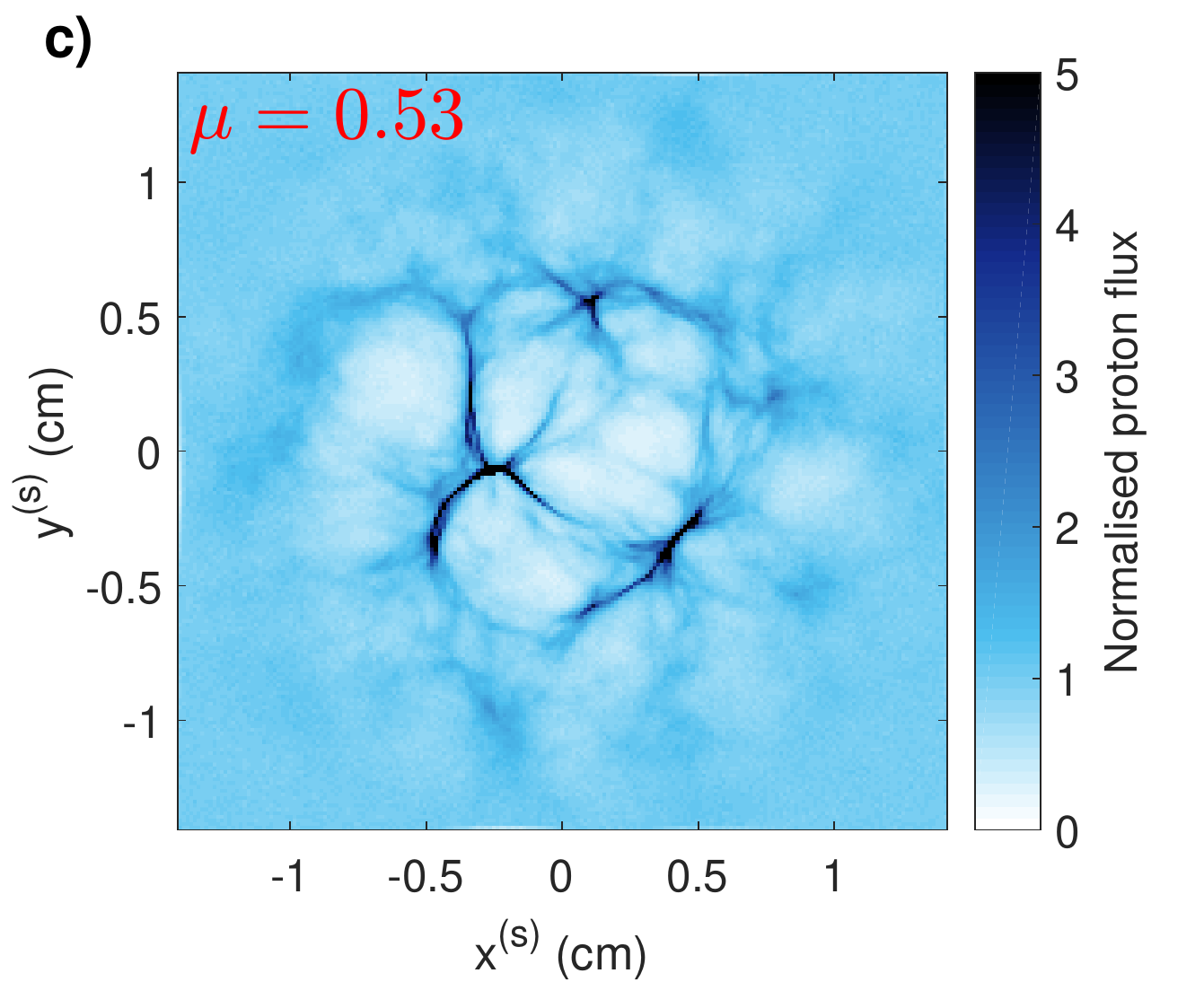}
    \end{subfigure} %
    \begin{subfigure}{.315\textwidth}
        \centering
        \includegraphics[width=0.95\linewidth]{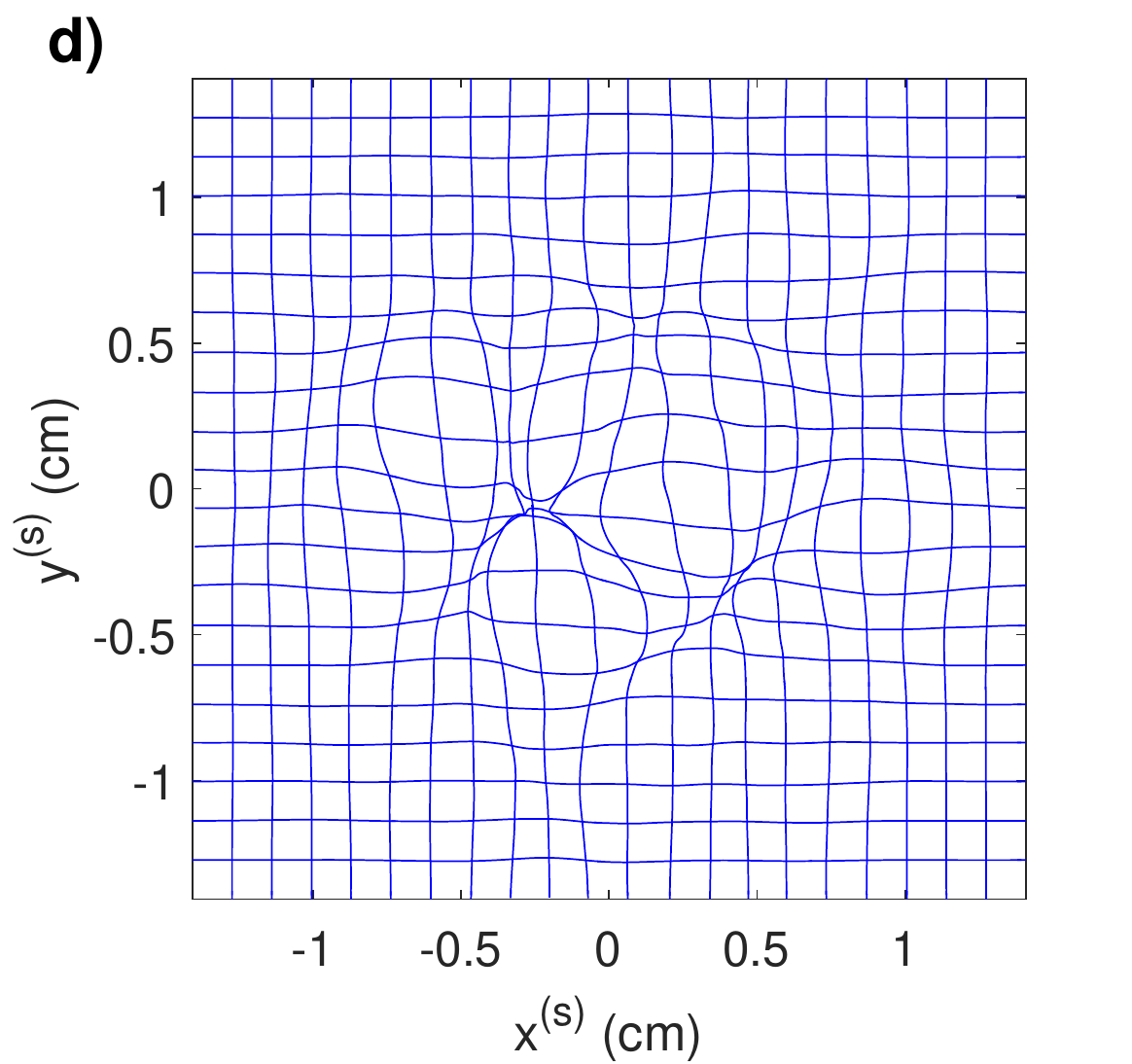}
    \end{subfigure} %
    \begin{subfigure}{.36\textwidth}
        \centering
        \includegraphics[width=0.95\linewidth]{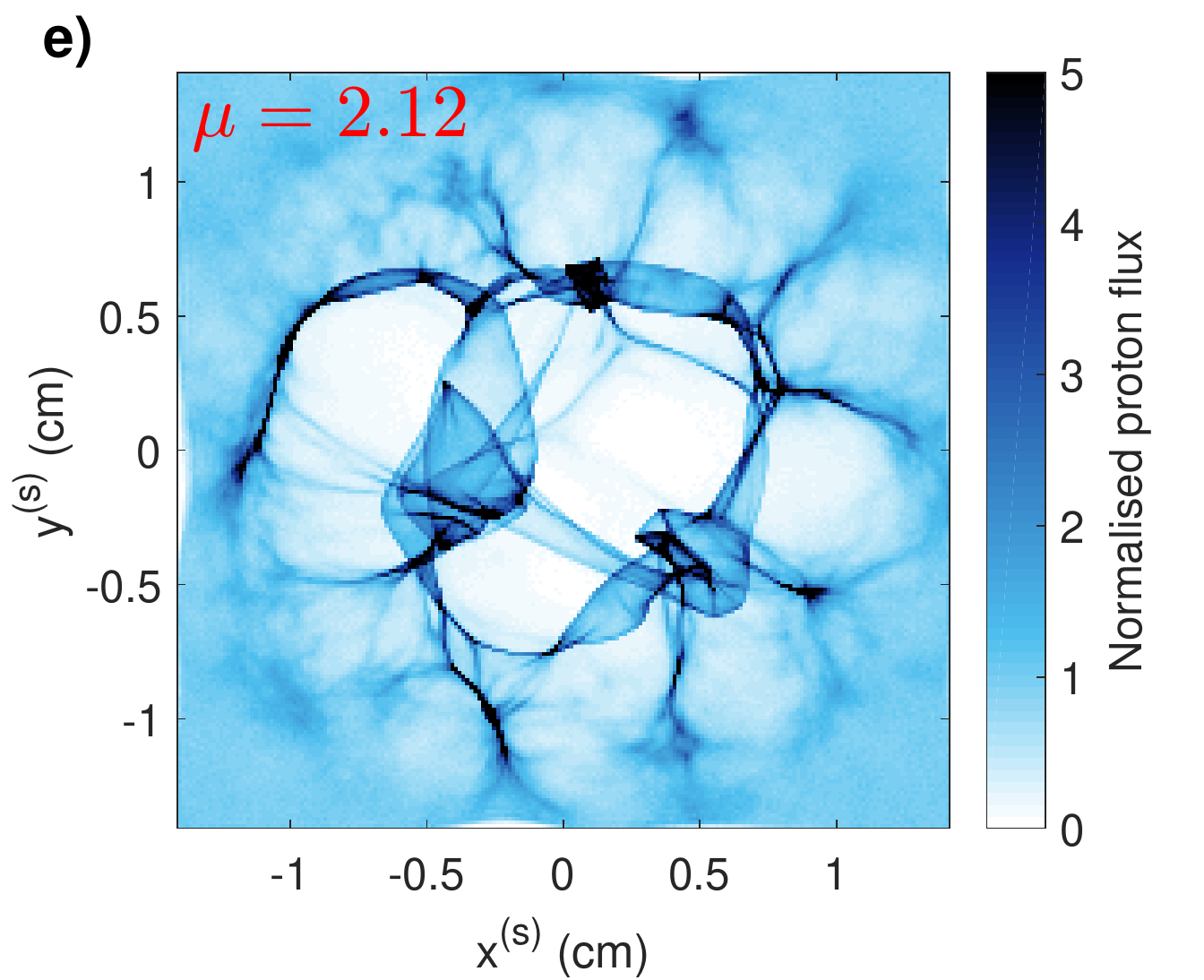}
    \end{subfigure} %
    \begin{subfigure}{.315\textwidth}
        \centering
        \includegraphics[width=0.95\linewidth]{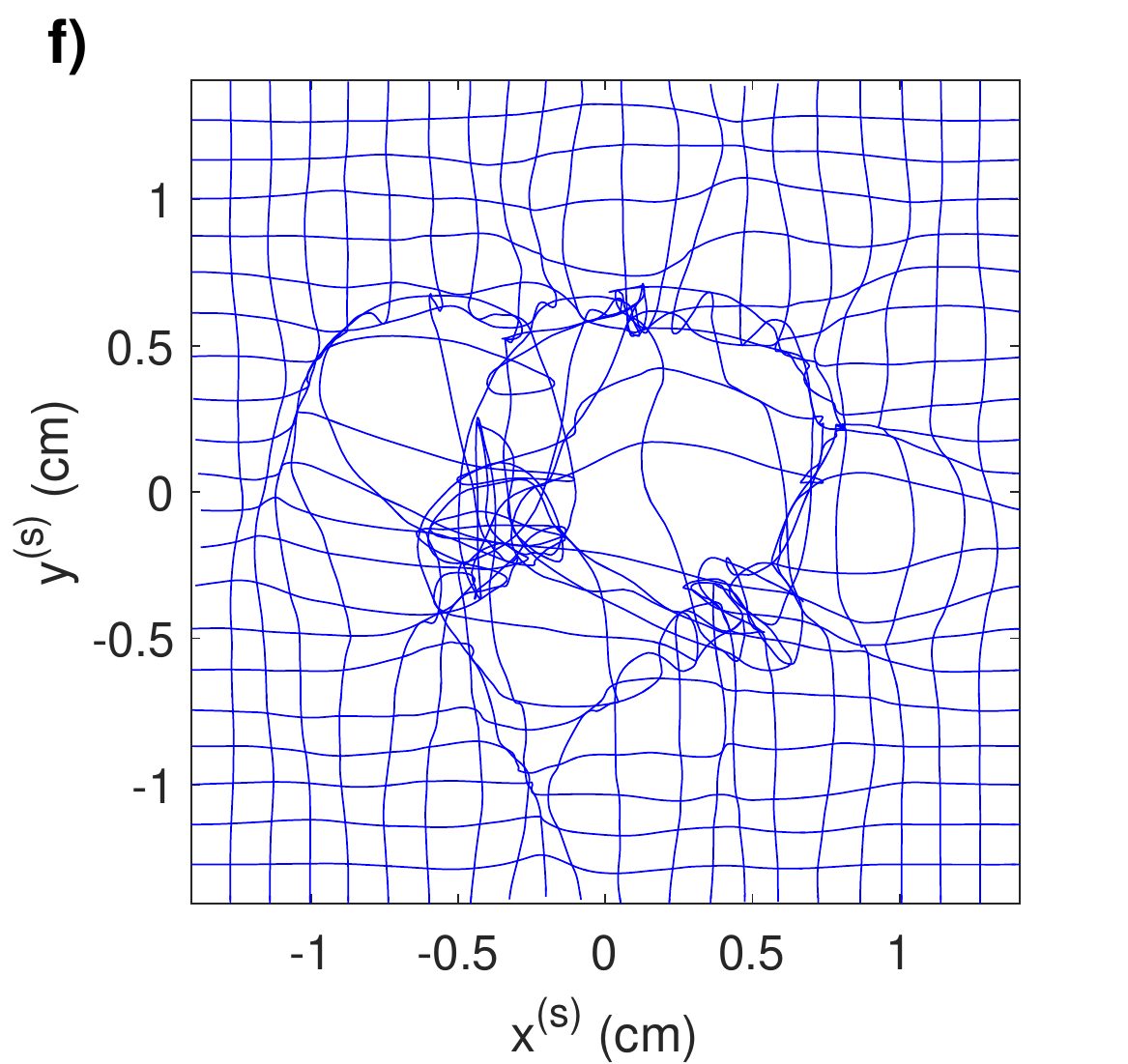}
    \end{subfigure} %
        \begin{subfigure}{.36\textwidth}
        \centering
        \includegraphics[width=0.95\linewidth]{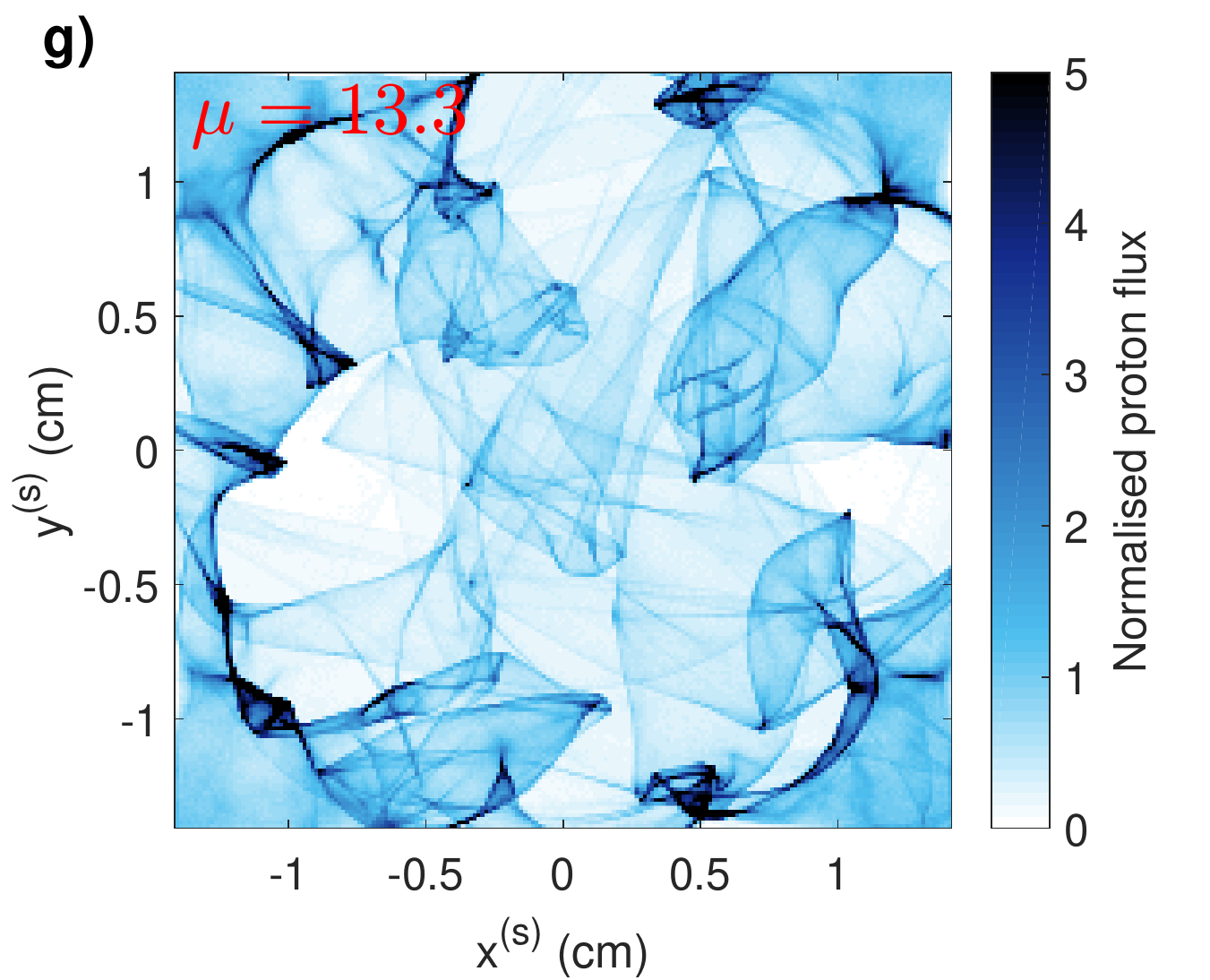}
    \end{subfigure} %
    \begin{subfigure}{.315\textwidth}
        \centering
        \includegraphics[width=0.95\linewidth]{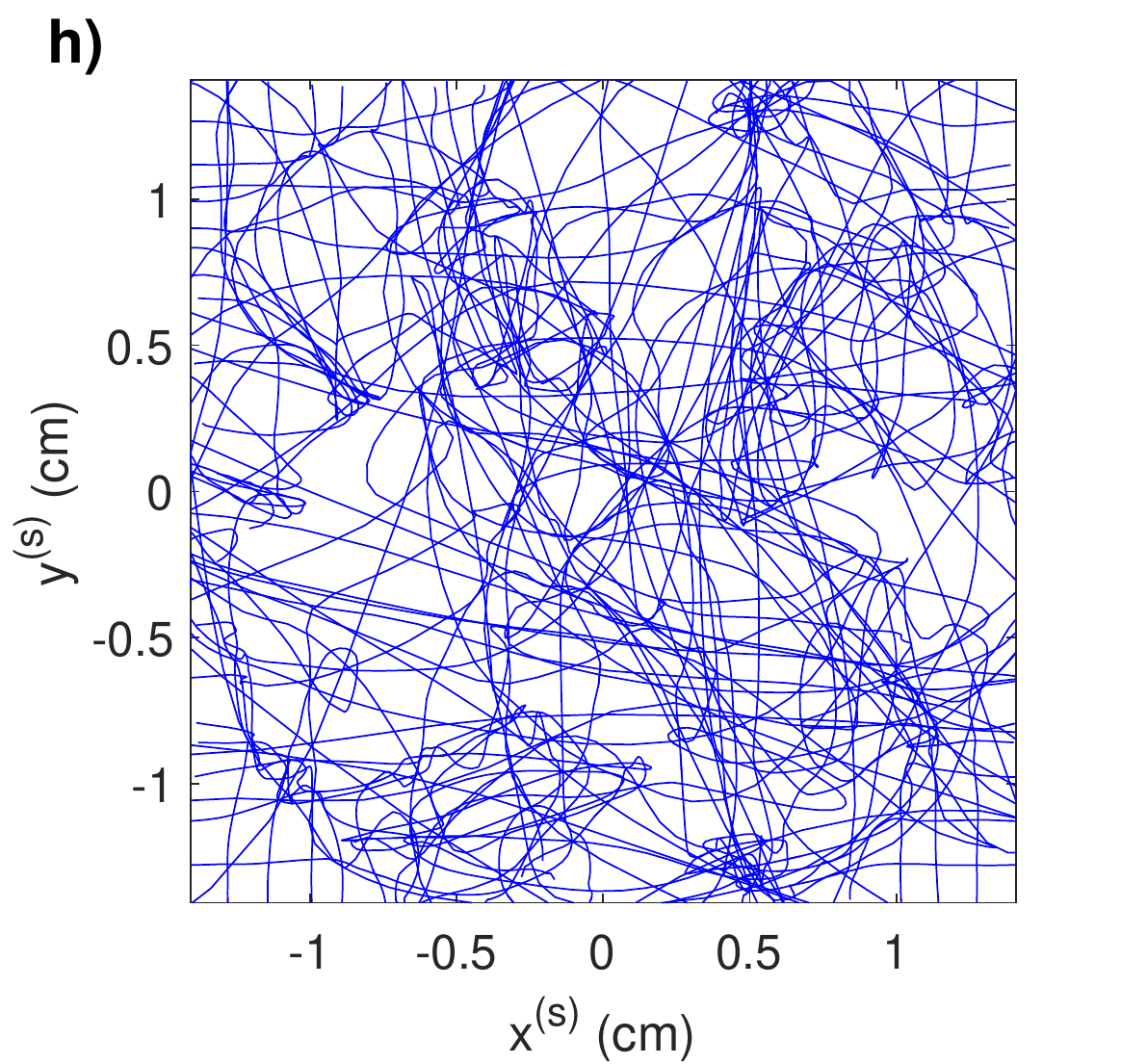}
    \end{subfigure} %
\caption{\textit{Characterisation of proton-flux images of stochastic magnetic fields by contrast regime.} 3.3-MeV-proton flux images generated on an artificial detector at a distance $r_s = 30 \, \mathrm{cm}$ from magnetic field configuration described in Figure \ref{GolitysncompactfieldSec2}, for a range of magnetic field strengths [and hence values of $\mu$ \eqref{contrastdef}]. The general arrangement of the imaging is the same as shown in Figure \ref{PRsetup}. The proton point-source was located at a distance $r_i = 1 \, \mathrm{cm}$ on the opposing side of the configuration to the detector, and 30,000,000 protons were used per image. The procedure used to generate these images is described in Appendix \ref{NumSimFluxImageGenFull}. \textbf{a)} Normalised proton-flux image in the linear regime, with $\mu \ll 1$ ($B_{rms,0} \approx 3 \, \mathrm{kG}$). \textbf{b)} Effective image-coordinate grid arising from magnetic perturbations to initial Cartesian grid in the linear regime. \textbf{c)} Normalised proton-flux image in the nonlinear injective regime, with $\mu < \mu_{c} \sim 1$ ($B_{rms,0} \approx 30 \, \mathrm{kG}$). \textbf{d)} Image-coordinate grid in the nonlinear injective regime. \textbf{e)} Normalised proton-flux image in the caustic regime, with $\mu \geq \mu_{c}$ ($B_{rms,0} \approx 120 \, \mathrm{kG}$). \textbf{f)} Image-coordinate grid in the caustic regime. \textbf{g)} Normalised proton-flux image in the diffusive regime, with $\mu \gtrsim 1/\delta \alpha$ ($B_{rms,0} = 750 \, \mathrm{MG})$. \textbf{h)} Image-coordinate system in a diffusive regime.} \label{Golitsynfluxrange}
\end{figure}

To imitate configurations realistic to actual experiments, we also introduce a overall Gaussian envelope
\begin{equation}
\tilde{\mathbf{B}} = \mathbf{B} \, \exp{\left[-4\sigma \left(\mathbf{x}-l_z \hat{\mathbf{z}}/2\right)^2\!/l_z^2\right]} \, , \label{magfieldwind}
\end{equation}
where $\sigma$ is an adjustable constant. While this does introduce a range of local field strengths, and hence effective contrasts regimes, due to the slowly varying inhomogeneity of the magnetic field RMS relative to the field-structure size, the central part of the proton-flux image still manifests a single contrast regime. For the particular Gaussian shape used, the effective RMS magnetic field strength $B_{rms,0}$ experienced by protons with trajectories close to the perpendicular origin can be analytically related to the global RMS values (see Appendix \ref{NumSimMagFieldGen}):
\begin{equation}
B_{rms,0} = B_{rms} \left[\frac{\pi}{8 \sigma} \mathrm{erf}\!\left(\sqrt{2 \sigma}\right)\right]^{-1/2} \, . \label{Brms0mod}
\end{equation}
Plots of such synthetic fields generated are shown in Figure \ref{GolitysncompactfieldSec2}.
The contrast parameter $\mu$, defined by \eqref{contrastdef}, is linear in the field strength, so the same field configuration can be used to explore all possible regimes. For each regime, we calculate the normalised image flux, shown in Figure \ref{Golitsynfluxrange}, with fixed scales for the sake of comparison. We also show the perturbed image-coordinate grid associated with the plasma-image mapping \eqref{divmappingSec2}. Throughout Sections \ref{LinRgme}, \ref{NonLinInjRgme}, \ref{CauRgme} and \ref{HighConRgme}, we will refer back to Figure \ref{Golitsynfluxrange} as a visual aid for typical features of proton images in each contrast regime.   

\subsection{Linear regime of proton imaging - $\mu \ll 1$} \label{LinRgme}

If $\mu$ is small, relating the magnetic field and the image flux becomes a much simpler problem. This follows from the result (derived in Appendix \ref{LinThyDev} - see also~\cite{GTLL16}) that if $\mu \ll 1$, the plasma-image mapping \eqref{divmappingSec2} becomes, to leading order in $\mu$,
\begin{equation}
\mathbf{x}_{\bot}^{\left(s\right)} = \frac{r_s+r_i}{r_i}  \mathbf{x}_{\bot0} \left[1 + \mathcal{O}\!\left(\mu\right)\right] \, , \label{linmapSec2}
\end{equation}
while the Kugland image-flux relation \eqref{screenfluxSec2} can be rewritten as
\begin{equation}
\frac{\delta \Psi\!\left(\mathbf{x}_\bot^{\left(s\right)}\right)}{\Psi_0^{\left(s\right)}} =  \frac{r_s r_i}{r_s+r_i} \frac{4 \pi e}{m_p c^2 V} \int_0^{l_z} j_z\!\left(\mathbf{x}_{\bot0}\left(1+\frac{z'}{r_i}\right),z'\right) \mathrm{d}z' \, , \label{linrelfluxgenSUM}
\end{equation}
where $\mathbf{j} = c \nabla \times \mathbf{B}/4 \pi$ is the MHD current, and $\delta \Psi\!\left(\mathbf{x}_\bot^{\left(s\right)}\right) \equiv \Psi\!\left(\mathbf{x}_\bot^{\left(s\right)}\right)-\Psi_0^{\left(s\right)}$ is the image-flux deviation from the initial mean image flux, which in turn is related to the initial flux $\Psi_0$ by the image-magnification factor $\mathcal{M}$, viz.,  $\Psi_0^{\left(s\right)} = \Psi_0/\mathcal{M}^2$.  It follows from \eqref{linrelfluxgenSUM} that proton-flux images in the linear regime have a simple physical interpretation: they display the \emph{undeflected path integrated} $z$-\emph{component $j_z$ of the MHD current}~\cite{GTLL16}. The linear regime is therefore so called, because the magnitude of image-flux deviations is linear in the magnetic field. 

Linear-regime image-flux relation \eqref{linrelfluxgenSUM} has another consequence: estimating the magnitude of the image-flux deviations compared to the mean image flux, we see that $\delta \Psi \sim \mu \Psi_0^{\left(s\right)} \ll \Psi_0^{\left(s\right)}$. The linear regime is therefore characterised by small relative image-flux deviations, providing a useful observational tool for recognising proton-flux images of stochastic magnetic fields in the linear regime. An example of this phenomenon is shown in Figure \ref{Golitsynfluxrange}a, a flux image of a Golitsyn field with parameters tuned to have small $\mu$: relative image-flux deviations indeed appear faint. The smallness of coordinate-grid perturbations relative to a Cartesian grid implied by \eqref{linmapSec2} is demonstrated in Figure \ref{Golitsynfluxrange}b for a Golitsyn field with $\mu \ll 1$.

More quantitatively, the RMS of relative image-flux variations is small in the linear regime. Linear-regime image-flux relation \eqref{linrelfluxgenSUM} enables the following relation between $\mu$ and the relative image-flux RMS to be derived analytically under the assumption of homogeneous and isotropic magnetic field statistics (see Appendix \ref{RelFluxRMSLinThy}):
\begin{equation}
\left(\frac{\delta \Psi}{\Psi_0}\right)_{rms} \equiv \left<\left(\frac{\delta \Psi}{\Psi_0^{\left(s\right)}}\right)^2\right>^{1/2} = \sqrt{\frac{\pi}{2}} \frac{r_i r_s}{r_s+r_i} \frac{e B_{rms}} {m c V} \sqrt{\frac{l_z}{l_{\Psi}}} =  \sqrt{\frac{\pi}{2} \frac{l_B}{l_\Psi}} \mu \label{linfluxscalsect3} \, ,
\end{equation}
where $l_\Psi$ is the relative image-flux correlation length (defined in Appendix \ref{RelFluxRMSLinThy}). This can be rearranged to give a formula for $\mu$ in terms of the RMS of relative image-flux deviations:
\begin{equation}
\mu = \mu_0 \left(\frac{\delta \Psi}{\Psi_0}\right)_{rms}  \, , \label{contrastlinrgme}
\end{equation}
where $\mu_0 \equiv \sqrt{2 l_\Psi/l_B \pi}$ depends on the particular stochastic field configuration, but is can be shown analytically that for any isotropic, homogeneous stochastic magnetic field, $\mu_0 \leq 2/\pi$ (Appendix \ref{RelFluxRMSLinThy}). The estimate \eqref{contrastlinrgme} of $\mu$ combined with RMS magnetic field strength estimate \eqref{contrastest} leads to a simple formula for the RMS field strength in terms of the RMS of the relative image-flux:
\begin{equation}
B_{rms}\!\left(\mathrm{kG}\right) \approx 40 \left[\frac{W\!\left(\mathrm{MeV}\right)}{3.3 \, \mathrm{MeV}}\right]^{1/2} \left[\frac{r_i\!\left(\mathrm{cm}\right)}{1 \, \mathrm{cm}}\right]^{-1} \left[\frac{l_z\!\left(\mathrm{cm}\right)}{1 \, \mathrm{mm}}\right]^{-1/2} \left[\frac{l_B\!\left(\mathrm{cm}\right)}{0.08\!\left(\mathrm{mm}\right)}\right]^{1/2} \left(\frac{\delta \Psi}{\Psi_0}\right)_{rms}  \, \mathrm{kG} \, .
\end{equation}
We note, however, that this expression only applies for small constrasts $\mu$; once $\mu$ approaches unity, the RMS of the relative image-flux begins to increase nonlinearly with $\mu$ (see Appendix \ref{RelFluxRMSLinThy}). 

In the linear regime, the path-integrated field can always be reconstructed uniquely from the image flux. This follows from the result \eqref{deffieldpotdef} that the perpendicular-deflection field can be written as the gradient of the deflection-field potential $\varphi$ provided $\delta \theta \lesssim l_B/l_z$:
\begin{equation}
\mathbf{w}\!\left(\mathbf{x}_{\bot0}\right) \approx \nabla_{\bot0} \varphi\!\left(\mathbf{x}_{\bot0}\right) \, . \label{lindeflfield}
\end{equation}
The assumption $\delta \theta \lesssim l_B/l_z$ is indeed valid in the linear regime, because $\delta \theta \, l_z/l_B \sim \mu \, \delta \alpha \left(r_s+r_i\right)/r_s \ll 1$. It can then be shown (Appendix \ref{LinThyDev}) that the image-flux deviation $\delta \Psi$ and $\varphi$ are related by a Poisson equation
\begin{equation}
\nabla_{\bot0}^2 \varphi\!\left(\mathbf{x}_{\bot0}\right) = - \Xi\!\left(\mathbf{x}_{\bot0}\right) \, , \label{linfluxref}
\end{equation} 
where source function $\Xi\!\left(\mathbf{x}_\bot\right)$ is proportional to relative image-flux deviations:
\begin{equation}
\Xi\!\left(\mathbf{x}_{\bot0}\right) = \mathcal{M}\frac{V}{r_s} \frac{\delta \Psi\!\left(\mathbf{x}_\bot^{\left(s\right)}\right)}{\Psi_0^{\left(s\right)}} \, .
\end{equation}
If suitable boundary conditions are applied, for example
\begin{equation}
\hat{\mathbf{n}} \cdot \nabla_{\bot0} \varphi\!\left(\mathbf{x}_{\bot0}\right) =  0 \, ,\label{linfluxBCs}
\end{equation}
\eqref{linfluxref} is a well-posed equation with a unique solution for $\nabla_{\bot0} \varphi$. For example, Kugland \textit{et. al.}~\cite{K12} observe that Poisson equation \eqref{linfluxref} combined with vanishing boundary conditions at infinity can be inverted analytically to give
\begin{equation}
\varphi\!\left(\mathbf{x}_{\bot0}\right) = \frac{1}{2 \pi} \int \mathrm{d}^2\tilde{\mathbf{x}}_\bot\log{\left(\frac{l_\bot}{\left|\mathbf{x}_\bot-\tilde{\mathbf{x}}_{\bot0}\right|}\right)} \Xi\!\left(\tilde{\mathbf{x}}_\bot\right) \, .
\end{equation}
The use of $l_\bot$ in this expression is arbitrary, since the integral of  $\Xi\!\left(\mathbf{x}_\bot\right)$ over the image vanishes by conservation of particles. The perpendicular-deflection field is then given by \eqref{lindeflfield}; for the case of infinite boundary conditions, we find
\begin{equation}
\mathbf{w}\!\left(\mathbf{x}_{\bot0}\right) = - \frac{1}{2\pi} \int \mathrm{d}^2\tilde{\mathbf{x}}_\bot \frac{\mathbf{x}_{\bot0}-\tilde{\mathbf{x}}_\bot}{\left|\mathbf{x}_{\bot0}-\tilde{\mathbf{x}}_\bot\right|^2} \Xi\!\left(\tilde{\mathbf{x}}_\bot\right) \, .
\end{equation}
The path-integrated magnetic field follows from \eqref{linpathintfield}. For finite regions, Poisson equation \eqref{linfluxref} can in principle be inverted numerically. However, Graziani \textit{et.~al.} report that such an approach when applied to reconstructing path-integrated fields from proton images quickly becomes unsuccessful for non-asymptotically small $\mu$~\cite{GTLL16}. For this reason, we suggest using the field reconstruction algorithm described in Section \ref{NonLinInjRgme} to reconstruct the path-integrated field instead. 

If the perpendicular-deflection field (and hence the path-integrated field) has been reconstructed, the magnetic-energy spectrum can be predicted using spectral relation \eqref{deffieldspec}. However, the simple form of the relation \eqref{linfluxscalsect3} between image flux and magnetic field in the linear regime allows for the application of statistical methods directly to proton-flux images to obtain properties of the fields creating that image. In particular, for homogeneous and isotropic magnetic-field statistics satisfying $l_B \ll l_z$ (Appendix \ref{RelFluxRMSLinThy}, and~\cite{GTLL16}), the 1D magnetic-energy spectrum \eqref{magengspecsec2} is related to the 2D spectrum of image-flux deviations $\hat{\eta}\!\left(k\right)$ by
\begin{equation}
E_B\!\left(k\right) = \frac{1}{2\pi} \frac{m_p^2 c^2 V^2}{e^2 r_s^2 l_z} \hat{\eta}\!\left( \frac{r_i}{r_s+r_i} k\right) \, . \label{linfluxspec3}
\end{equation}
Here, $\hat{\eta}\!\left(k\right)$ is defined by
\begin{equation}
\hat{\eta}\!\left(k_\bot\right) \equiv \frac{1}{2 \pi} \int \mathrm{d}\theta \, \frac{\mathcal{M}^4}{\Psi_0^2} \left<\left|\hat{\delta \Psi}\!\left(\mathbf{k}_\bot\right)\right|^2\right> \, , \label{fluxspecdef}
\end{equation}
where $\hat{\delta \Psi}\!\left(\mathbf{k}_\bot\right)$ is the Fourier transform of the relative image-flux deviation. 

To summarise, if a stochastic magnetic field with isotropic and locally homogeneous statistics is imaged in the linear regime, the path-integrated field and magnetic-energy spectrum can be reconstructed. However, we caution that unless $\mu$ is very small, distortions to results obtained using linear analysis manifest themselves, both in reconstructing perpendicular-deflection fields, and in the magnetic-energy spectrum using linear-regime flux spectral relation \eqref{linfluxspec3} (see Section \ref{NonLinInjRgme} for an example).

A detailed discussion of proton imaging in the linear regime -- including an alternative field-reconstruction algorithm to cope with weakly nonlinear effects -- is presented in~\cite{GTLL16}.

\subsection{Nonlinear injective regime:  $\mu < \mu_{c} \sim 1$} \label{NonLinInjRgme}

In the nonlinear injective regime, $\mu$ is sufficiently large that beam-focusing effects associated with the nonlinear term resulting from magnetic deflections in plasma-image mapping \eqref{divmappingSec2} have a non-trivial effect on the image-flux distribution. However, $\mu$ is not so great as to lead to the proton beam intersecting itself, and hence loss of injectivity of the plasma-image mapping, which first occurs at some critical $\mu = \mu_{c}$ (shown in Appendix \ref{CauExist}). The importance of nonlinearity for moderate $\mu$ and preservation of injectivity for $\mu < \mu_c$ are illustrated by the plasma-image coordinate mapping for the test Golitsyn field shown in Figure \ref{Golitsynfluxrange}d: the image-coordinate grid is visually distorted, but coordinate curves do not cross each other. The nonlinear injective regime can be distinguished from the linear regime simply by the presence of image-flux structures whose deviation from the mean image flux is similar in magnitude to the mean (see Figure \ref{Golitsynfluxrange}c). 

When injective, the plasma-image mapping tends to preserve the morphology of the proton-flux image obtained for the same magnetic field, but with small $\mu$ -- so image-flux structures can still be qualitatively interpreted in terms of path-integrated MHD current structure. However, structures with positive relative image-flux tend to be narrow due to beam focusing, and those with negative relative image-flux enlarged. This phenomenon is evident in the nonlinear injective proton-flux image of the Golitsyn field, Figure \ref{Golitsynfluxrange}c. 

Nonlinear effects mean that a different approach for reconstructing path-integrated fields must be adopted to that expounded for the linear regime. One such approach can be found by noting that since the plasma-image mapping \eqref{divmappingSec2} is injective, the sum in Kugland image-flux relation \eqref{screenfluxSec2} disappears, leaving 
\begin{equation}
\Psi\!\left(\mathbf{x}_{\bot}^{\left(s\right)}\!\left(\mathbf{x}_{\bot0}\right)\right) = \frac{\Psi_{0}}{\det{\nabla_{\bot0}\!\left[\mathbf{x}_{\bot}^{\left(s\right)}\!\left(\mathbf{x}_{\bot0}\right)\right]}} \, . \label{screenfluxnonlinnotpot}
\end{equation}
Our goal is to solve \eqref{screenfluxnonlinnotpot} for $\mathbf{x}_{\bot}^{\left(s\right)}\!\left(\mathbf{x}_{\bot0}\right)$ given an image-flux distribution $\Psi\!\left(\mathbf{x}_{\bot}^{\left(s\right)}\!\left(\mathbf{x}_{\bot0}\right)\right)$. We proceed by noting that the plasma-image mapping \eqref{divmappingSec2} can be rewritten in terms of a potential field $\phi\!\left(\mathbf{x}_{\bot0}\right)$:
\begin{equation}
\mathbf{x}_{\bot}^{\left(s\right)} \approx \nabla_{\bot0} \phi\!\left(\mathbf{x}_{\bot0}\right) \approx \nabla_{\bot0} \left[
\frac{r_s+r_i}{2 r_i} \mathbf{x}_{\bot0}^2 + \frac{r_s}{V} \, \varphi\!\left(\mathbf{x}_{\bot0}\right) \right] \, .\label{divmappingnonlin121}
\end{equation}
The existence of $\phi\!\left(\mathbf{x}_{\bot0}\right)$ follows from that of the deflection-field potential, which is defined in Section \ref{LinRgme}, equation \eqref{deffieldpotdef}. Equation \eqref{screenfluxnonlinnotpot} can then be restated as an equation for $\phi\!\left(\mathbf{x}_{\bot0}\right)$: 
\begin{equation}
\Psi\!\left(\nabla_{\bot0} \phi\!\left(\mathbf{x}_{\bot0}\right) \right) = \frac{\Psi_{0}}{\det{\nabla_{\bot0} \nabla_{\bot0} \phi\!\left(\mathbf{x}_{\bot0}\right) }} \, . \label{screenfluxnonlin121}
\end{equation}
Equation \eqref{screenfluxnonlin121} is an example of an Monge-Amp\`ere equation, which appear in numerous mathematical and physical contexts~\cite{GM96}. Despite its nonlinearity, it can be shown~\cite{B91} that there is a unique (up to a constant) solution for $\phi$ with Neumann boundary conditions
\begin{equation}
\hat{\mathbf{n}} \cdot \nabla_{\bot0} \phi\!\left(\mathbf{x}_{\bot0}\right) = \hat{\mathbf{n}} \cdot \mathbf{x}_{\bot0} \, .\label{nonlinfluxBCs}
\end{equation}
One approach for establishing this result comes from the observation that the solution of the Monge-Amp\`ere equation also solves the $L_2$ \emph{Monge-Kantorovich problem}~\cite{V08}. This correspondance is explained in Appendix \ref{MongeKantorovich}. This means that with an appropriate \textit{field-reconstruction algorithm}, $\nabla_{\bot0} \phi$ can be reconstructed from a given proton-flux image, which can then be used to calculate the perpendicular-deflection field
\begin{equation}
\mathbf{w}\!\left(\mathbf{x}_{\bot0}\right) = \frac{V}{r_s} \nabla_{\bot0} \left(\phi-\frac{r_s+r_i}{2 r_i} \mathbf{x}_{\bot0}^2\right) \, . \label{deflfieldfrommap}
\end{equation}
The path-integrated magnetic field can then be calculated using \eqref{linpathintfield}. There exist a number of possible algorithms for solving the Monge-Amp\`ere equation \eqref{screenfluxnonlin121}~\cite{DG06}, and some have recently been applied to the problem of recovering path-integrated magnetic fields from proton-flux images~\cite{K16}. In Appendix \ref{NumSimNonLinRecon}, we describe one such field-reconstruction algorithm, which is both simple to implement and computationally efficient~\cite{S11}. As explained in Section \ref{PlasMap}, the reconstructed path-integrated magnetic field can be combined with deflection-field spectral relation \eqref{deffieldspec} to deduce the magnetic-energy spectrum.

To summarise, like the linear regime the path-integrated magnetic field and magnetic-energy spectrum are always recoverable from an individual proton-flux image in the nonlinear injective regime. This perhaps counter-intuitive result essentially holds because of the irrotationality of the perpendicular-deflection field, discussed in Section \ref{PlasMap}. Furthermore, the techniques used to achieve this are more widely applicable than those derivable from linear theory. However, care must be taken when applying the field-reconstruction algorithm to arbitrary proton-flux images, because the results can be misleading if the plasma-image mapping is not injective. This is described in the next section.

\subsection{Caustic regime: $\mu \geq \mu_{c}$} \label{CauRgme}

For $\mu$ greater than $\mu_c$, gradients in the perpendicular-deflection field are sufficiently large that the plasma-image mapping \eqref{divmappingSec2} becomes multi-valued in places (see Figure \ref{Golitsynfluxrange}f) -- a phenomenon sometimes referred to as \textit{mesh-twisting}~\cite{K12}. 
Physically, in this regime some of the paths of imaging protons cross before reaching the detector, and hence there exist regions of image flux whose constituent protons originate from spatially-disconnected initial positions. That multi-valuedness of the plasma-image mapping only occurs at $\mu \geq \mu_c$ follows from the observation that such crossing requires that the determinant of the plasma-image mapping change sign. For this to happen, the magnitude of gradients of the perpendicular-deflection field must be comparable to gradients in the undeflected mapping -- which by the definition of $\mu$ \eqref{contrastdef} is precisely the criterion of $\mu$ sufficiently large. This alone does not guarantee the existence of $\mu_c$; however, the conditions required for the absence of mesh-twisting at all values of $\mu$ are typically incompatible with stochasticity (see Appendix \ref{CauExist}). The particular value of $\mu_{c}$ depends on the particular stochastic field, but typically is order unity (although heuristic arguments can be given implying that it likely decreases logarithmically with the field scale $l_B$; see Appendix \ref{CauExist}). 

The multi-valuedness of the plasma-image mapping is closely associated with caustics, defined as curves on which the determinant of the plasma-image mapping \eqref{divmappingSec2} vanishes~\cite{K12}: 
\begin{equation}
\det{\nabla_{\bot0}\!\left[\mathbf{x}_{\bot}^{\left(s\right)}\!\left(\mathbf{x}_{\bot0}\right)\right]} = 0 \, .
\end{equation}
This is (trivially) because a sign reversal of the determinant for a continuous mapping cannot occur without the value of that determinant passing through zero. If the determinant does indeed vanish, the denominator in Kugland image-flux relation \eqref{screenfluxSec2} also does, yielding a formally infinite local image-flux. In practice the local image-flux value is limited by finite resolution of proton images~\cite{K12} -- 
however, the morphology of proton-flux images is still dominated by caustics if there are present. This is evident in the caustic-regime proton-flux image of the test Golitsyn field, shown in Figure \ref{Golitsynfluxrange}e.

The relationship between caustics and magnetic fields creating them has been studied in great depth elsewhere~\cite{K12}; for our purposes, we simply note that the width, strength and distance between caustics do not necessarily reflect equivalent properties of the magnetic field, implying that great care must be taken when qualitatively assessing magnetic structures from proton-flux images containing caustics.

Similar care must be taken when attempting quantitative analysis. In the caustic regime the reconstruction of the path-integrated field from a proton-flux image is no longer a well-posed problem: for a given proton image, there exists a number of path-integrated fields (and hence plasma-image mappings) which give the same flux distribution via the Kugland image-flux relation \eqref{screenfluxSec2}. Intuitively this seems reasonable, because without the injectivity constraint, the Kugland image-flux relation insufficiently determines the path-integrated field. An explicit example of many different path-integrated fields corresponding to a simple image-flux distribution is given in Appendix \ref{IllPosNonLinRecon}. In addition to this simple analytical example, the impossibility of solving the Kugland image-flux relation \eqref{screenfluxSec2} for non-injective plasma-image mappings can be demonstrated numerically (and is done so in Section \ref{NumReconTest}). Attempts to reconstruct the magnetic-energy spectrum in the caustic regime are also prone to failure. In short, we emphasize that the caustic regime is much less ameanable to direct analysis than either the linear, or the nonlinear injective regime. 

This being the case, identifying the presence of caustics in a proton image is essential for a sensible interpretation of that image. In principle, caustics should typically be identifiable in proton-flux images from narrow curves of high proton flux. In practice, proton-flux images have a finite spatial resolution which can disguise these, making distinguishing between the nonlinear injective regime and caustic regime a non-trivial problem. The issue, and possible ways around it, is explored in Section \ref{SmearFinSource}.  

For a systematic review of caustic theory in the context of proton imaging for electromagnetic structures, see Kugland \emph{et. al.}~\cite{K12}.

\subsection{Diffusive regime: $ \mu \gtrsim r_s/\!\left(r_i + r_s\right)\delta \alpha \gg 1$} \label{HighConRgme}

For large $\mu$, the characteristic nature of the plasma-image mapping changes yet again. The position of protons in any given local region on the detector is dominated by the perpendicular-deflection field, rather than the projected position in the absence of magnetic fields. More specifically, if $\mu \gtrsim r_s/\!\left(r_i + r_s\right)\delta \alpha$, then proton trajectories have already crossed as they leave the plasma. This follows because the perpendicular displacement inside the plasma due to magnetic deflections $l_z \, \delta \theta$ is comparable to the typical size of magnetic structures $l_B$: $l_z \, \delta \theta \sim \mu \, \delta \alpha \left(r_s+r_i\right)\!/r_s l_B \gtrsim l_B$. As a consequence, the perpendicular-deflection field can no longer be written as the gradient of the deflection-field potential, and the magnetic field integrated along the actual proton trajectories is not equivalent to the magnetic field integrated along undeflected trajectories (in other words, equations \eqref{deffieldpotdef} and \eqref{deffieldundefl} do not apply in the diffusive regime). 

The diffusive regime is incompatible with the small-deflection assumption for regular fields with $l_B \sim l_z$, since then the deflection angle $\delta \theta \gtrsim 1$. Furthermore, successful imaging in such a parameter space is practically unfeasible, because it would require a prohibitively large detector. However, the diffusive regime can be reached without violating the small-angle approximation for sufficiently small-scale fields due to the opposite scalings of $\mu$ and the typical deflection angle $\delta \theta$ with correlation length $l_B$: $\mu \propto l_B^{-1/2}$, while $\delta \theta \propto l_B^{1/2}$.
 
In the diffusive regime, image-flux variations are typically much less pronouced in magnitude than seen in the caustic regime (Figure \ref{Golitsynfluxrange}g, central part of image). Qualitatively, this is due to chaotic tangling of proton trajectories from spatially uncorrelated regions of the imaged field (Figure \ref{Golitsynfluxrange}h). In a technical sense, a proton-flux image in the diffusive regime is still filled with caustics, the outlines of which are still visible in the central part of Figure \ref{Golitsynfluxrange}g; however, since even pairs of protons with close initial positions often end up at disparate spatial locations, the proportion of image flux concentrated into an individual caustic drops as $\mu$ increases. 

An alternative interpretation of the proton-flux image in the diffusive regime can be given in terms of diffusive-type models. As first shown by Dolginov and Toptygin~\cite{DT67}, the diffusion of a fast proton beam through compact, isotropic stochastic magnetic fields with correlation length $l_B \ll l_z, l_\bot$ can be described by perpendicular velocity diffusion coefficient
\begin{equation}
D_{w} = \frac{V e^2 B_{rms}^2 l_B}{4 m_p^2 c^2} \, . \label{difftensSection2}
\end{equation}
The result is valid in the small-deflections limit $\delta \theta \ll 1$, which in turn necessitates that the proton Larmor radius $\rho_p$ associated with the typical RMS magnetic field strength $B_{rms}$ satisfy $\rho_p \gg l_B$:
\begin{equation}
\frac{\rho_p}{l_B} = \frac{m c V}{e B l_B} \sim \delta \theta^{-1} \sqrt{\frac{l_i}{l_B}} \gg 1 \, . \label{Larmorrad}
\end{equation}
For the reader's convenience, Appendix \ref{StocMagDiffTensDiffCoeff} presents an alternative derivation of diffusion coefficient \eqref{difftensSection2}.

\begin{figure}[htbp]
\centering
    \begin{subfigure}{.48\textwidth}
        \centering
        \includegraphics[width=\linewidth]{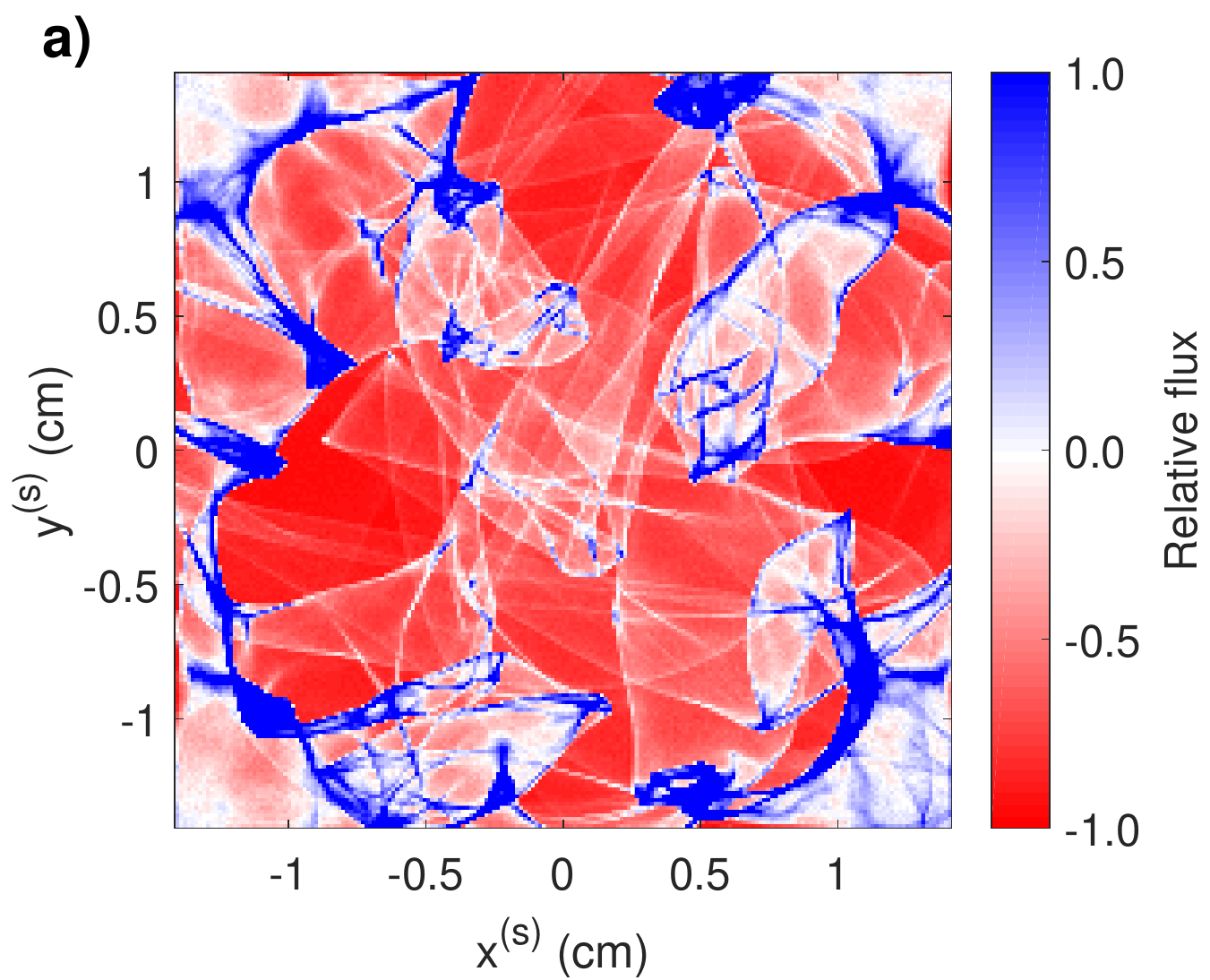}
    \end{subfigure} %
    \begin{subfigure}{.48\textwidth}
        \centering
        \includegraphics[width=\linewidth]{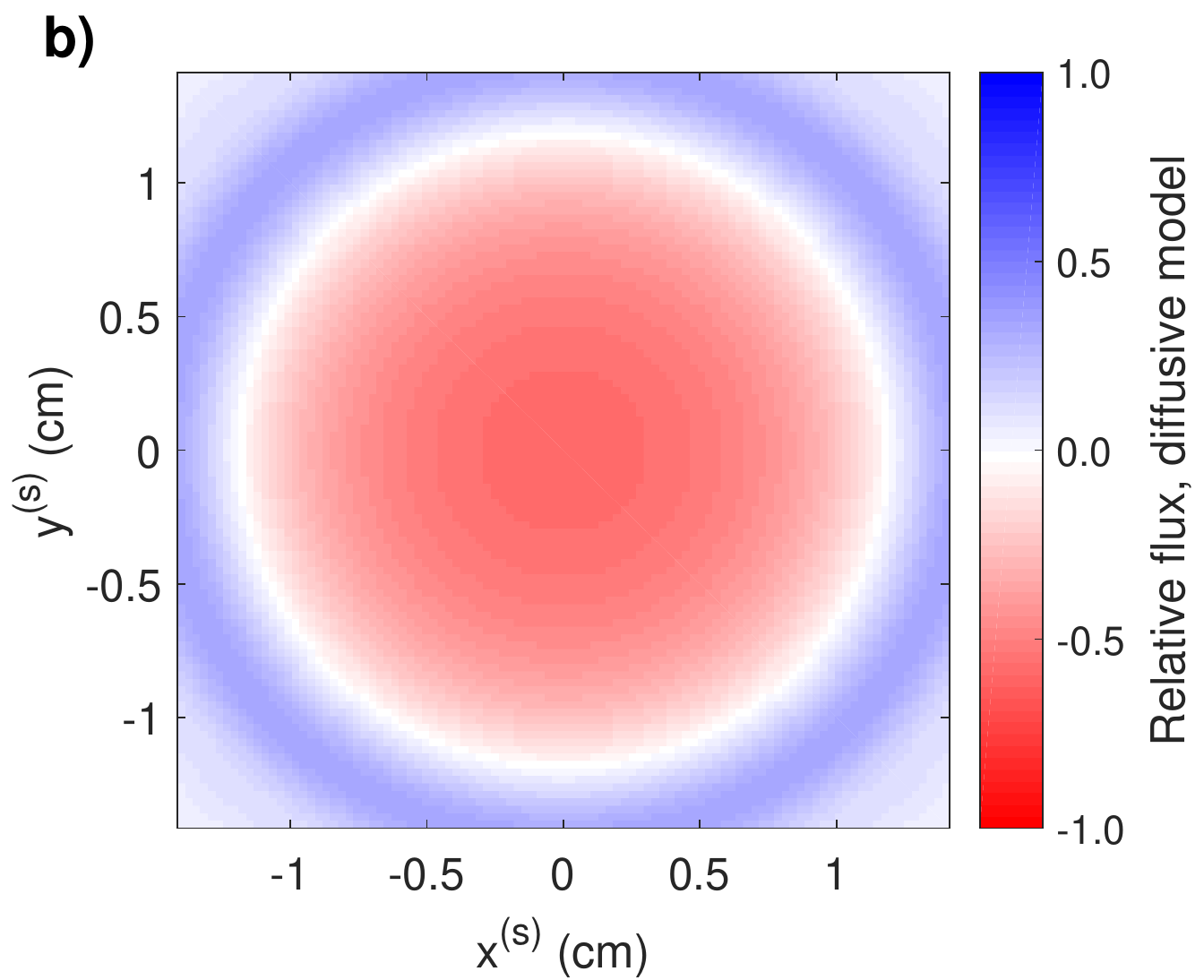}
    \end{subfigure} %
\caption{\textit{Comparison of relative flux image predicted by diffusive model compared with actual result in diffusive regime.} \textbf{a)} 3.3 MeV proton image-flux distribution associated with magnetic field described in Figure 2, with $B_{rms,0} = 750 \, \mathrm{kG}$. Here, relative flux refers to $\delta \Psi\!\left(\mathbf{x}_\bot^{\left(s\right)}\right)/\Psi_0^{\left(s\right)}$, where $\delta \Psi\!\left(\mathbf{x}_\bot^{\left(s\right)}\right) \equiv \Psi\!\left(\mathbf{x}_\bot^{\left(s\right)}\right)-\Psi_0^{\left(s\right)}$ is the flux deviation, and $\Psi_0^{\left(s\right)}$ the mean image flux. The same imaging parameters as those described in Figure 4 are used (see Figure 4g for normalised proton-flux image). 
\textbf{b)} Predicted relative proton-flux image assuming uniform initial flux, and diffusive model of imaging beam evolution \eqref{stocmagsmrdscreenflux}, with $\delta  = \delta\!\left(\mathbf{x}_\bot\right) = \delta_0 \exp{\left[-4 \sigma \mathbf{x}_\bot^2/l_i^2\right]}$. Here, $\delta_0$ is given by \eqref{RWvelexactsec2}, with $B_{rms} = 750 \, \mathrm{kG}$, and $l_i = l_z = l_\bot = 0.1 \, \mathrm{cm}$.
} \label{highcontrastdiffmodel}
\end{figure}

The uncertainty in perpendicular velocity $\Delta w$ acquired due to diffusion leads to a `smearing effect': under a diffusive model with coefficent $D_{w}$ given by \eqref{difftensSection2}, the image-flux distribution is related to the (here non-uniform) initial distribution by convolution
\begin{equation}
\Psi\!\left(\mathbf{x}_\bot^{\left(s\right)}\right) =  \frac{1}{\pi \delta^2} \int \mathrm{d}^2  \tilde{\mathbf{x}}_{\bot}^{\left(s\right)} \, \Psi_0^{\left(s\right)}\!\left(\tilde{\mathbf{x}}_\bot^{\left(s\right)}\right) \exp{\left[-\left(\frac{\mathbf{x}_\bot^{\left(s\right)}- \tilde{\mathbf{x}}_{\bot}^{\left(s\right)}}{\delta}\right)^2\right]} \, ,  \label{stocmagsmrdscreenflux}
\end{equation}
where
\begin{equation}
\delta = r_s \frac{\Delta w}{V} = \frac{e B_{rms} r_s}{m_p c V}\sqrt{l_z l_B} \, . \label{RWvelexactsec2}
\end{equation}
This result is derived in Appendix \ref{StocMagDiffTensScreenFlux}. We see that a diffusive model of proton-flux images predicts that an initially uniform flux distribution will not display relative image-flux deviations if the stochastic magnetic field being imaged has a uniform envelope. For a magnetic field with a spatially varying envelope, such as the Golitsyn field shown in Figure \ref{GolitysncompactfieldSec2}, \eqref{stocmagsmrdscreenflux} instead indicates that different sections of the initial flux distribution will be subject to different diffusion rates. For the Gaussian envelope used to create the stochastic field shown in Figure \ref{GolitysncompactfieldSec2}, this results in the central region of the proton image having a reduced image flux when compared to the mean initial image flux; the central region is surrounded by a ring of greater-than-average image flux (illustrated in Figure \ref{highcontrastdiffmodel}b).
Comparing the diffusive prediction of the relative image-flux distribution to the actual relative image-flux distribution shown Figure \ref{highcontrastdiffmodel}a, we see that the diffusive model fails to capture many features of the true proton-flux image: in particular, caustic structures. However, for stochastic fields with smaller-scale structures than the Golitsyn field, the diffusive model is more accurate, and caustic structures are suppressed (two examples are provided in Appendix \ref{StocMagDiffTensNumSim}).

The observation of decreased intensity of image-flux deviations relative to the mean image flux is not sufficient to identify uniquely the diffusive regime -- the same observation also holds for the linear regime (as discussed in Section \ref{LinRgme}). Operationally, the diffusive regime can usually be distinguished by noting that unless fields have an extremely small spatial correlation scale $l_B$, deflections of protons will be large enough to result in net loss of proton flux from an image. This loss of flux can be measured if the initial flux is known. Another, more conclusive test identifying the diffusive regime -- requiring practical modification to experimental platforms -- is to introduce a partial obstruction into the path of the imaging proton beam. This could include a sharp edge, a pinhole, or some type of grid. In the diffusive regime, the edges associated with any of these features will appear blurred by diffusive scattering of protons due to the stochastic fields. 

Quantitative analysis of proton images in the diffusive regime is much more restricted in scope that in other regimes. Whilst the plasma-image mapping \eqref{divmappingSec2} is still formally valid, in the diffusive regime, the perpendicular-deflection field is an extremely complicated object, and is not recoverable from an individual proton-flux image (for the same reason as described for the caustic regime for Section \ref{CauRgme}). That being said, there is some statistical information extractable from the proton-flux image: for sufficiently small-scale fields, the diffusion coefficient $D_{w}$ \eqref{difftensSection2} can be measured: this is most reliably done by using one of the partial-obstruction methods mentioned above, and calculating the extent of blurring effects. This gives a estimate of $B_{rms}^2 l_B$.

There is an extensive literature discussing diffusion due to stochastic magnetic fields in the context of cosmic rays~\cite{DT67,J72,HS67,P65,BE87}; however, the authors of this paper are not aware of any discussion directly concerned with the diffusive scattering of a proton imaging beam. The key positive result of this section, then, is to have derived a relation \eqref{stocmagsmrdscreenflux} between the image-flux distribution and the diffusive scattering of the proton beam by a stochastic magnetic field.

\subsection{Numerical demonstration of field-reconstruction algorithm} \label{NumReconTest}

\begin{figure}[htbp]
\centering
    \begin{subfigure}{.365\textwidth}
        \centering
        \includegraphics[width=0.95\linewidth]{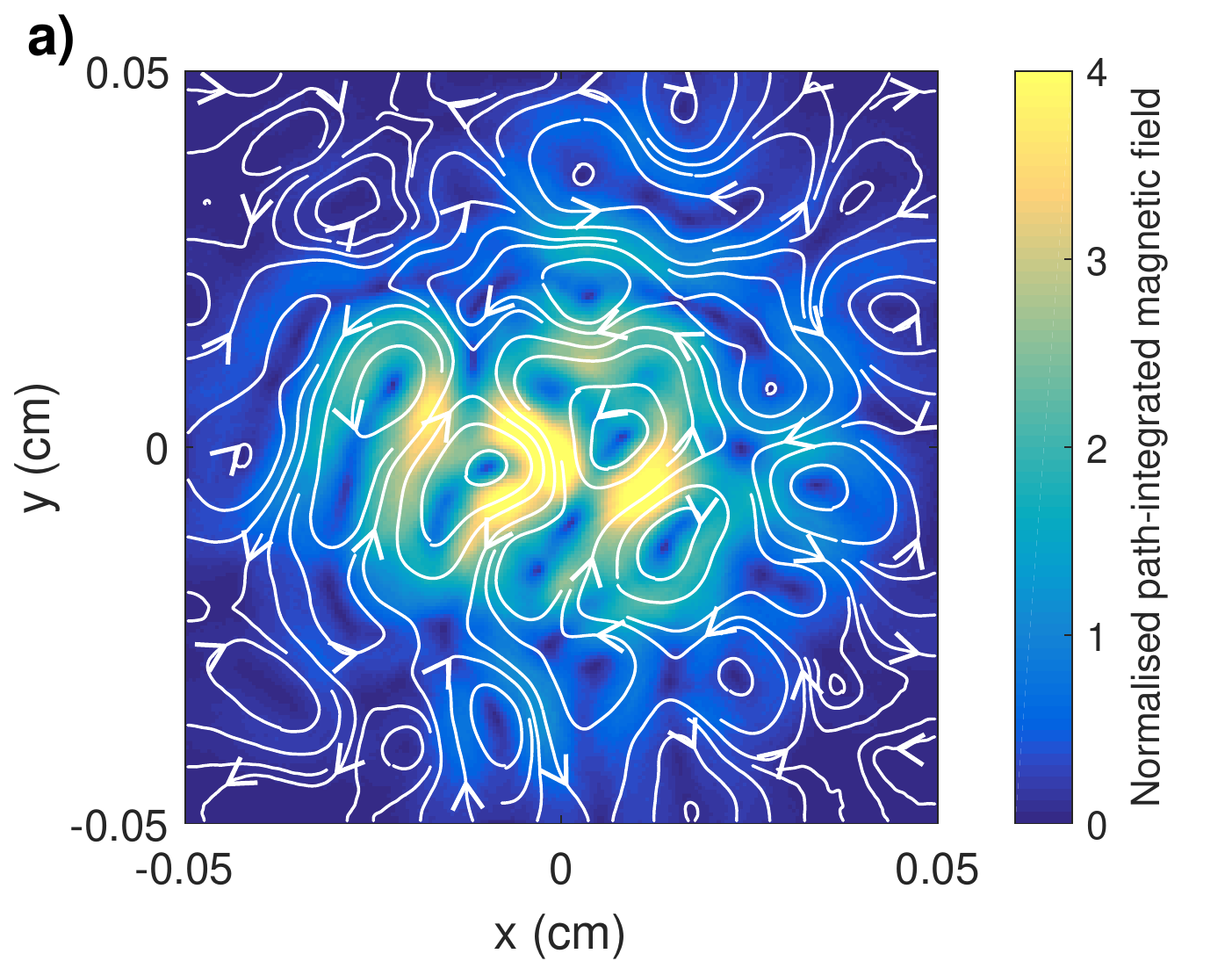}
    \end{subfigure} %
    \begin{subfigure}{.325\textwidth}
        \centering
        \includegraphics[width=0.95\linewidth]{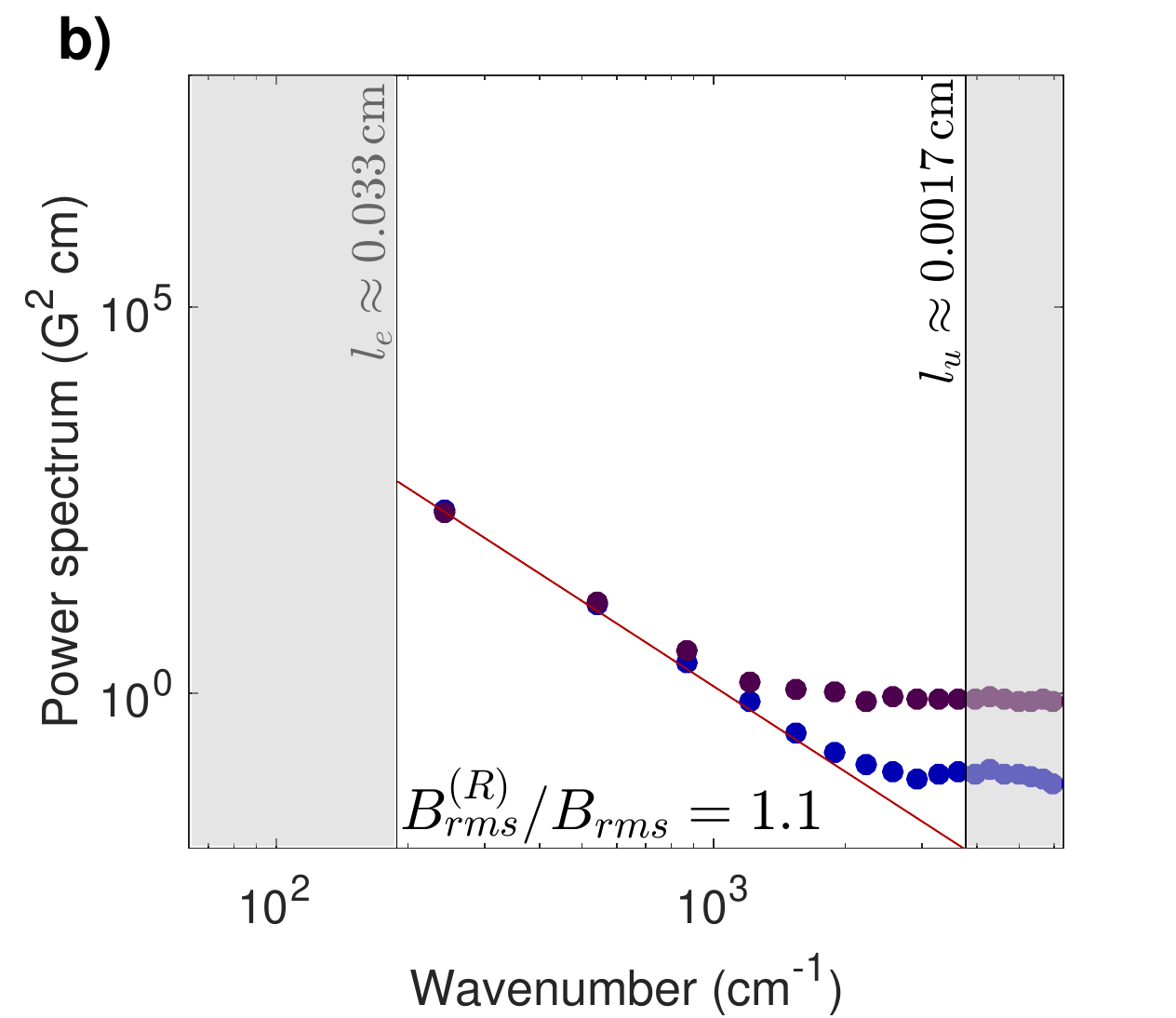}
    \end{subfigure} %
    \begin{subfigure}{.365\textwidth}
        \centering
        \includegraphics[width=0.95\linewidth]{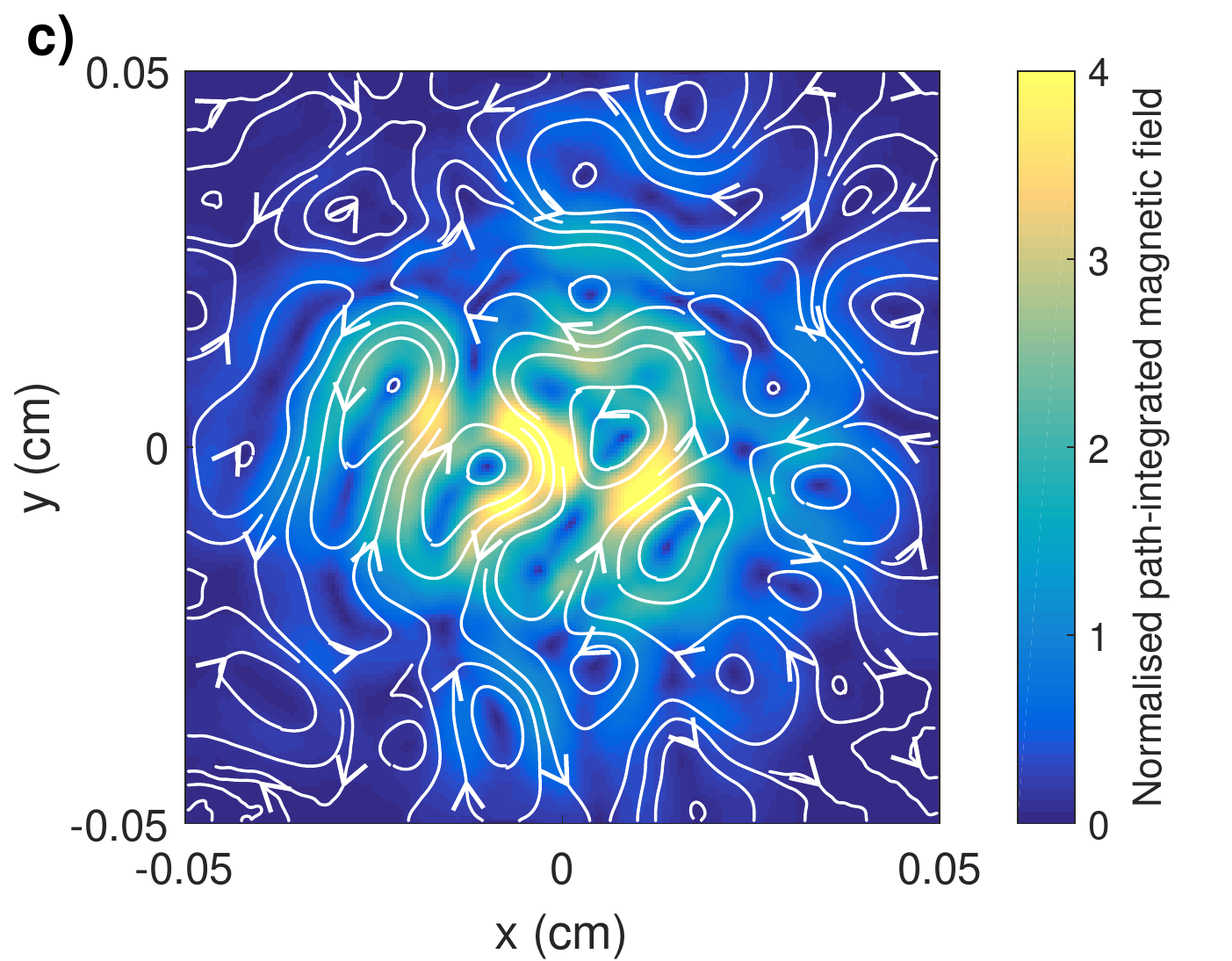}
    \end{subfigure} %
    \begin{subfigure}{.325\textwidth}
        \centering
        \includegraphics[width=0.95\linewidth]{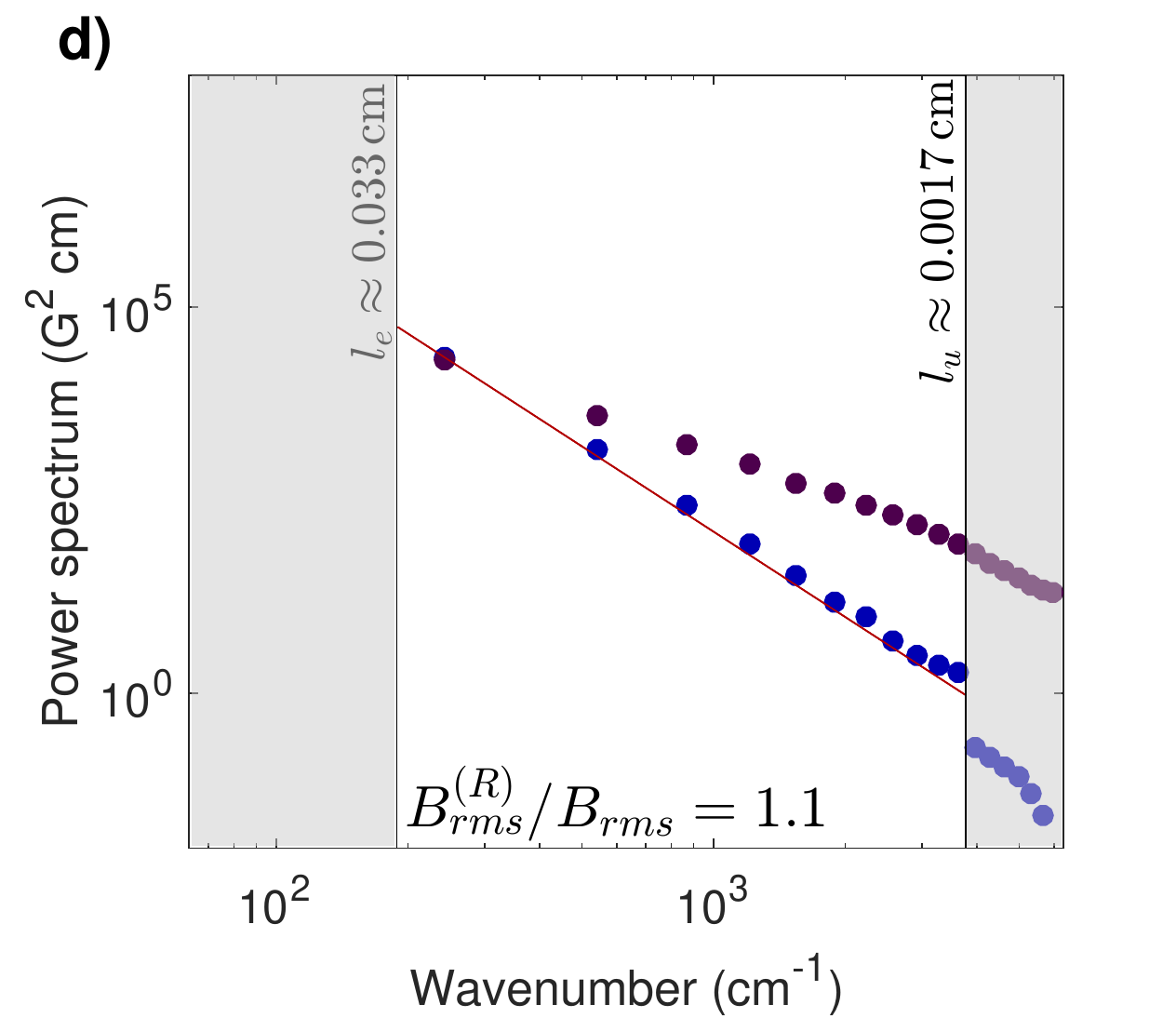}
    \end{subfigure} %
    \begin{subfigure}{.365\textwidth}
        \centering
        \includegraphics[width=0.95\linewidth]{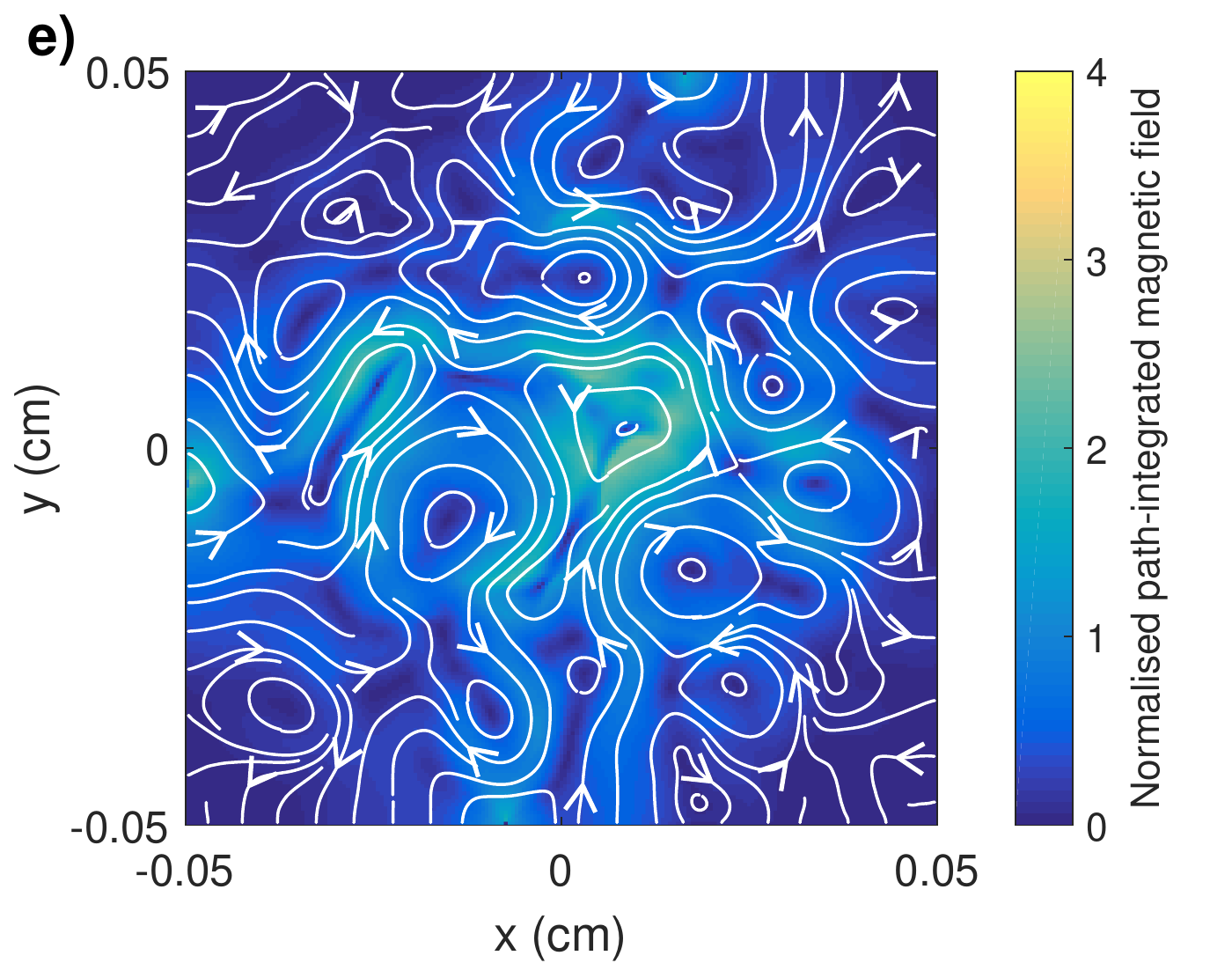}
    \end{subfigure} %
    \begin{subfigure}{.325\textwidth}
        \centering
        \includegraphics[width=0.95\linewidth]{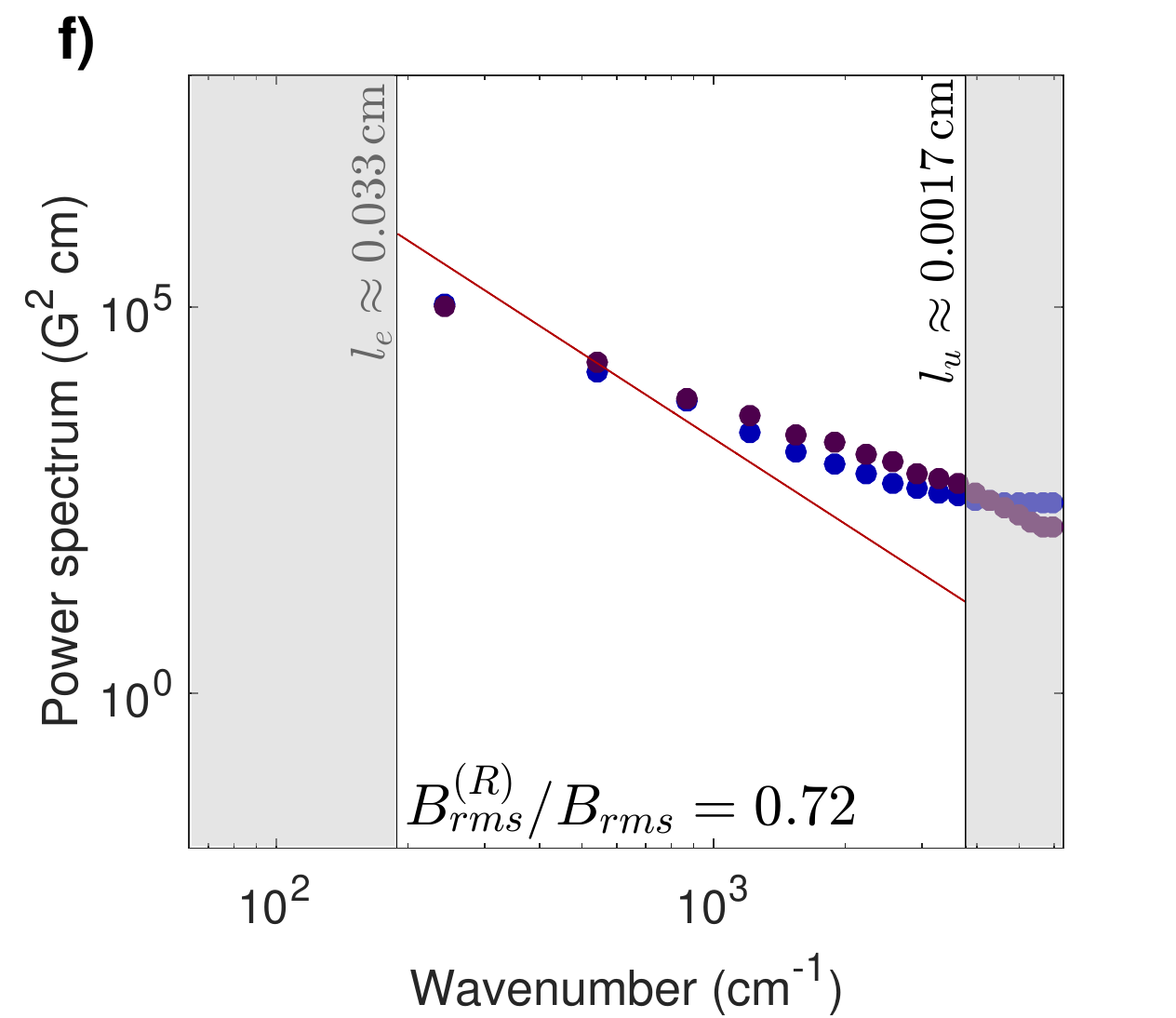}
    \end{subfigure} %
        \begin{subfigure}{.365\textwidth}
        \centering
        \includegraphics[width=0.95\linewidth]{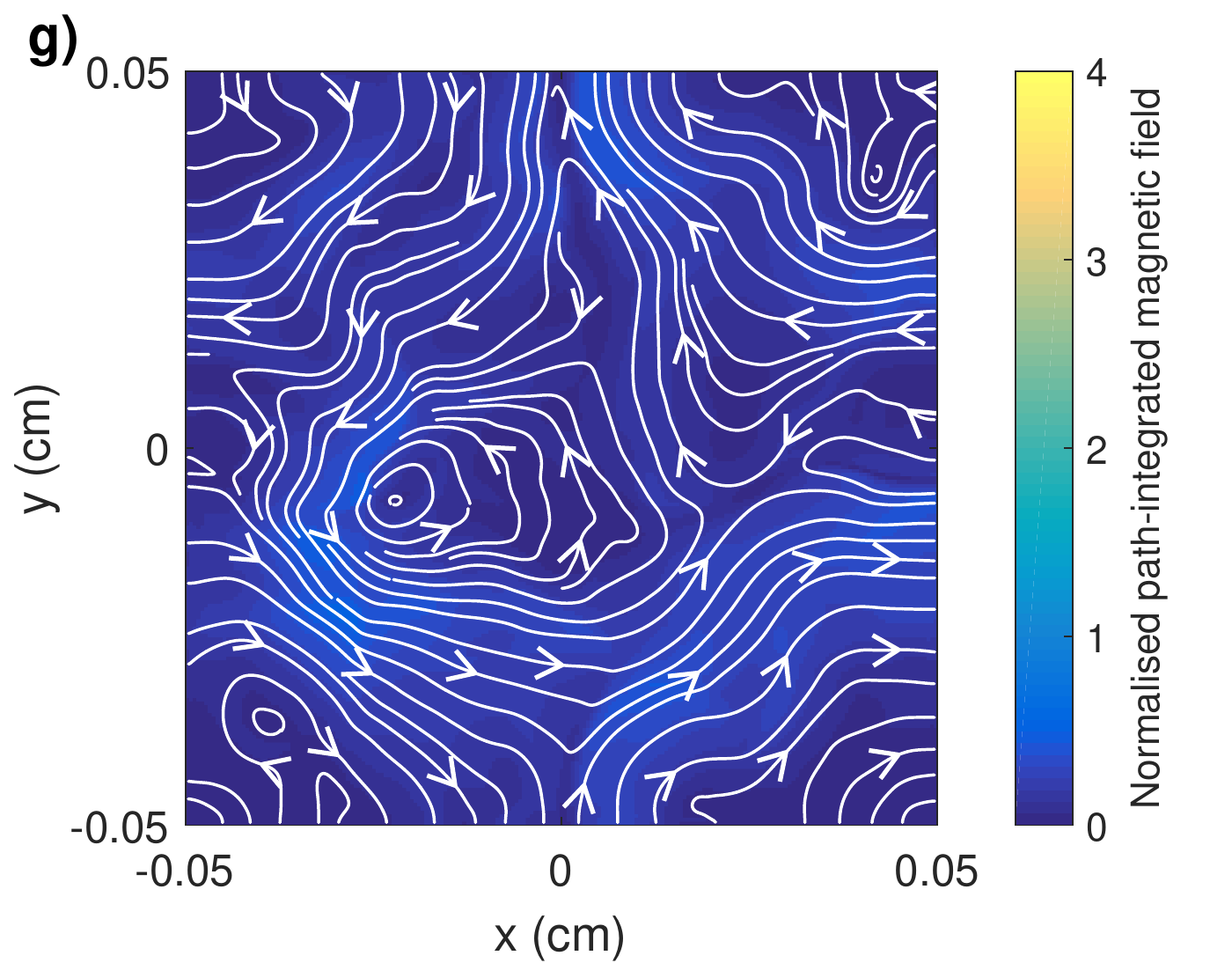}
    \end{subfigure} %
    \begin{subfigure}{.325\textwidth}
        \centering
        \includegraphics[width=0.95\linewidth]{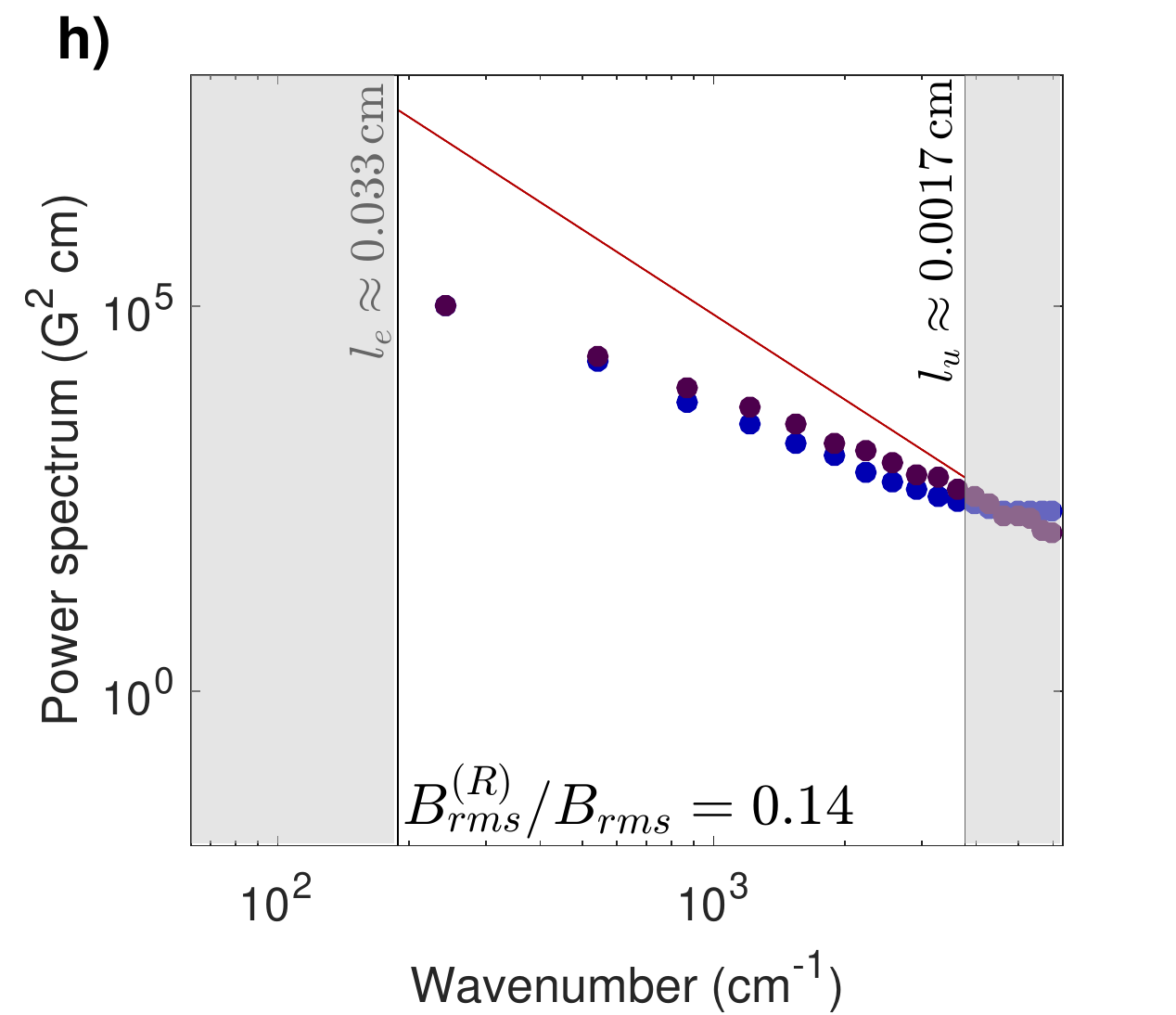}
    \end{subfigure} %
\caption{\textit{Efficacy of techniques for reconstructing magnetic field statistics directly from proton-flux images across contrast regimes}. For each proton-flux image shown in Figure \ref{Golitsynfluxrange}, a reconstructed path-integrated magnetic field was produced by applying the field-reconstruction algorithm (see Appendix \ref{NumSimNonLinRecon}) appropriate for inverting the Monge-Amp\`ere equation \eqref{screenfluxnonlin121} with boundary conditions \eqref{nonlinfluxBCs}, before recovering the perpendicular-deflection field using \eqref{deflfieldfrommap}, and finally the path-integrated field by \eqref{linpathintfield}. The true field is shown in Figure \ref{GolitysncompactfieldSec2}b. The magnetic-energy spectrum (true result shown in red) was predicted in two ways: using spectral relation \eqref{deffieldspec} applied to the reconstructed perpendicular-deflection field (blue), and linear-regime spectral relation \eqref{linfluxspec3} applied directly to the proton-flux images (purple). 
\textbf{a)} Path-integrated magnetic field reconstructed from linear-regime proton-flux image  ($\mu \ll 1$), Figure \ref{Golitsynfluxrange}a. 
\textbf{b)} Predicted magnetic energy spectra, derived from Figures \ref{Golitsynfluxrange}a and \ref{Golitsynreconmethods}a. 
\textbf{c)} Same as \textbf{a)}, but for nonlinear-injective-regime proton-flux image ($\mu < \mu_c$), Figure \ref{Golitsynfluxrange}c.
\textbf{d)} Same as \textbf{b)}, but spectra derived from Figures \ref{Golitsynfluxrange}c and \ref{Golitsynreconmethods}c.
\textbf{e)} Same as \textbf{a)}, but for caustic-regime proton-flux image ($\mu \geq \mu_c$), Figure \ref{Golitsynfluxrange}e.
\textbf{f)} Same as \textbf{b)}, but for Figures \ref{Golitsynfluxrange}e and \ref{Golitsynreconmethods}e.
\textbf{g)} Same as \textbf{a)}, but for diffusive-regime proton-flux image ($\mu \gtrsim 1/\delta \alpha \gg 1$), Figure \ref{Golitsynfluxrange}f.
\textbf{h)} Same as \textbf{b)}, but for Figures \ref{Golitsynfluxrange}f and \ref{Golitsynreconmethods}f. 
} \label{Golitsynreconmethods}
\end{figure}

To conclude this characterisation of the four contrast regimes, we illustrate the efficacy (or lack thereof) of the proposed field-reconstruction algorithm for recovering the path-integrated field numerically in each contrast regime. More specifically, we apply the field-reconstruction algorithm outlined in Appendix \ref{NumSimNonLinRecon} to the proton images presented in Figure \ref{Golitsynfluxrange}. The reconstructed path-integrated fields normalised to the RMS of the actual magnetic field in each case are shown in Figure \ref{Golitsynreconmethods}; these are compared with the true path-integrated field, shown in Figure \ref{GolitysncompactfieldSec2}b. Figure \ref{Golitsynreconmethods} also presents a calculation of magnetic-energy spectrum determined using spectral relation \eqref{deffieldspec} (blue circles) applied to the reconstructed perpendicular-deflection field in each case. For comparison, the true spectra (red line) of the Golitsyn fields are shown, along with the results of linear-regime flux spectral relation \eqref{linfluxspec3} applied directly to each image-flux distribution (purple circles).

The possibility of successful reconstruction of the path-integrated field in the linear regime (from the proton image shown in Figure \ref{Golitsynfluxrange}a) is illustrated in Figure \ref{Golitsynreconmethods}a; we see strong agreement with Figure \ref{GolitysncompactfieldSec2}b in terms of both field morphology and strength in the central region. The recovery of the magnetic-energy spectrum from the same proton-flux image is illustrated in Figure \ref{Golitsynreconmethods}b: both the linear-regime flux spectral relation \eqref{linfluxspec3} and deflection-field spectral relation \eqref{deffieldspec} recover the correct power law at the energetically dominant wavenumbers. However, at higher wavenumbers, a flattening of both predicted spectra is observed. This effect is likely due to Poisson noise, and is discussed further in Section \ref{ExpCom}. 

The result of the field-reconstruction algorithm in the nonlinear injective regime -- that is, the algorithm is applied to the image-flux distribution shown in Figure \ref{Golitsynfluxrange}c --  is shown in Figure \ref{Golitsynreconmethods}b. Similarly to the results for the linear regime, the predicted path-integrated magnetic field is a close match to the true field in terms of both its field strength and direction. Figure \ref{Golitsynreconmethods}d shows the predicted magnetic-energy spectrum using deflection-field spectral relation \eqref{deffieldspec}; the true spectrum is recovered over a wide range of wavenumbers. We note that despite the Poisson noise being the same in proton-flux images Figure \ref{Golitsynfluxrange}a and \ref{Golitsynfluxrange}c, the spectrum predicted from the latter is less distorted at high wavenumbers. This is because the magnitude of deviations in the image flux due to magnetic fields are larger in the latter case, so the relative effect of Poisson noise is reduced.

In contrast,  the predicted magnetic-energy spectrum from linear-regime flux spectral relation \eqref{linfluxspec3} does not follow the expected $k^{-11/3}$ power law, instead moving towards a $k^{-2}$ law. This distorted power law is obtained because the strong, narrow image-flux structures that appear irrespective of the underlying magnetic-energy spectrum have a characteristic `sharp-edge' spectrum~\cite{A12}. We conclude that direct application to images involving order-unity relative image-flux variations of linear-regime flux spectral relation \eqref{linfluxspec3} can lead to misleading results. 

Figure \ref{Golitsynreconmethods}e shows the results of the field-reconstruction algorithm described in Section \ref{NonLinInjRgme} applied to the proton-flux image Figure \ref{Golitsynfluxrange}e (an image containing caustics). While the streamlines of the path-integrated perpendicular magnetic field retain a reasonable agreement, the field strength distribution does not: the typical magnitude is reduced. The mapping recovered by the field-reconstruction algorithm is instead the unique injective mapping satisfying Monge-Amp\`ere equation \eqref{screenfluxnonlin121} given the image-flux distribution shown in Figure \ref{Golitsynfluxrange}e.

Figure \ref{Golitsynreconmethods}f demonstrates that the magnetic-energy spectrum predicted by spectral relation \eqref{deffieldspec} applied to the reconstructed perpendicular-deflection field is distorted to a $k^{-2}$ spectrum, much as the magnetic-energy spectrum predicted by linear-regime flux spectral relation \eqref{linfluxspec3} is distorted as nonlinearity of the plasma-image mapping becomes important.

Finally, attempts to reconstruct the path-integrated field in the diffusive regime lead to extremely inaccurate results. This failure is illustrated by the path-integrated field Figure \ref{Golitsynreconmethods}g reconstructed from the image-flux distribution shown in Figure \ref{Golitsynreconmethods}f. Predicted field strengths are orders of magnitude lower than the true values, and the recovered field's morphology resembles that of a regular field rather than of the actual stochastic one. This inaccuracy is also replicated in the predicted magnetic-energy spectrum shown in Figure \ref{Golitsynfluxrange}h: the Golitysn spectrum is again distorted to a $k^{-2}$ power law due to the caustic ring-like structure seen in Figure \ref{Golitsynfluxrange}g, and the predicted spectrum at the energetically dominant wavenumbers is strongly suppressed.

A simple quantitative way to compare the quality of the predicted reconstructions for different contrast regimes is to calculate the predicted RMS magnetic field strength from the recovered magnetic-energy spectra using \eqref{Brmsspec}. The results are shown in Table \ref{Brmstable}. 
\begin{table}[htbp]
\centering
{\renewcommand{\arraystretch}{1.28}
\renewcommand{\tabcolsep}{0.2cm}
\begin{tabular}[c]{c|c|c|c}
Contrast regime & $\mu$ & $B_{rms}^{\left(R\right)}$ (kG) & $B_{rms}$ (kG)\\
\hline
Linear & 0.05 & 1.10 & 1.00\\
Nonlinear injective & 0.53 & 10.5 & 10.0\\
Caustic & 2.12 & 29.0 & 40.0\\
Diffusive & 13.1 & 36.1 & 250\\
\end{tabular}}
\caption{Comparison with true result of RMS magnetic field strength predicted using magnetic-energy spectrum derived from \eqref{deffieldspec} in different contrast regimes.} \label{Brmstable}
\end{table}
It is clear that the predicted values for $B_{rms}$ are close to the actual ones in the linear and nonlinear injective regimes, but $B_{rms}$ is under-predicted in the caustic and diffusive regimes. That the typical field strengths predicted by the field-reconstruction algorithm are reduced is a manifestation of the general analytical result that the reconstructed perpendicular-deflection field will always provide a lower bound on the RMS deflection-field strength~\cite{GM96}, which can in turn be used to provide a lower bound on the RMS magnetic field strength. Explicitly, it can be shown (see Appendix \ref{MongeKantorovichBound}) that
\begin{equation}
B_{rms}^2 \geq \frac{m_p^2 c^2}{e^2 l_z^2}  \left<\left(\nabla_{\bot0} \varphi\right)^2\right> \, , \label{causticmaglowerbound}
\end{equation}
where $\nabla_{\bot0} \varphi$ is the deflection-field potential recovered from the solution to Monge-Amp\`ere equation \eqref{screenfluxnonlin121}. Furthermore, in Appendix \ref{MongeKantorovichBound}, it is demonstrated for the Golitsyn test field outlined in Figure \ref{GolitysncompactfieldSec2} that this lower bound property becomes very weak as $\mu$ increases further.

In short, this numerical example validates the claim that for proton-flux images of stochastic magnetic fields created in the linear and nonlinear injective regimes, the true path-integrated field is extractable using an appropriate field-reconstruction algorithm. Furthermore, the predicted magnetic-energy spectrum agrees well with the true result. Neither of these results are true in the caustic and diffusive regimes. 

\section{Technicalities and complications} \label{TechnicalComp}

\subsection{Assumptions} \label{Assum}

As mentioned in Section \ref{PlasMap}, the applicability of plasma-image mapping \eqref{divmappingSec2} and Kugland image-flux relation \eqref{screenfluxSec2} to the proton imaging set-up depends on various assumptions: a mono-energetic, instantaneous, uniform beam from a point source, paraxialilty, point projection, and small deflections. Here, we state each of these precisely, and explore their validity. In addition to the effects stated below, in deriving the stated plasma-image mapping, a range of physical processes are neglected in line with previous work on analytic models of proton imaging~\cite{K12}. These are discussed in Appendix \ref{NegPhysPros}.

\subsubsection{Mono-energetic, instantaneous uniform proton beam from point source}

Proton beams are typically generated in practice using one of two methods. The first is production of protons via the target normal sheath acceleration process (TSNA) using a high-intensity laser~\cite{W01,B06b}. The second is the laser implosion of a capsule containing $\mathrm{D}_2$ and $\mathrm{D}^{3}\mathrm{He}$ gas that leads to creation of fusion protons~\cite{L06,S04,M12}. In this paper, we only discuss briefly those properties which justify our assumption that a proton beam used for imaging can be well modelled as mono-energetic, instantaneous, uniform, and as originating from a point source. The interested reader is referred to the referenced papers for a more comprehensive discussion of the generation of proton beams to be used for imaging.

\subsubsection*{Mono-energetic}

An imaging beam can be approximated as mono-energetic for two reasons. First, proton detectors are typically designed to image different proton energies separately. For example, calibrated radiographic film (RCF) in a stack configuration can achieve energy resolution $\gtrsim 0.2 \, \mathrm{MeV}$~\cite{N09}. Thus, irrespective of the full distribution of proton speeds, for any particular experimental proton-flux image, the protons creating it are essentially mono-energetic. Second, for the case of capsule implosions, the fusion-generated protons have a characteristic energy determined by the nuclear reaction creating them, a value which is retained to $3\%$ accuracy in the imaging beam. In contrast, protons generated using high-intensity laser sources typically have a thermal spectrum. 

If this assumption were violated, plasma-image mapping \eqref{divmappingSec2} would still be valid individually for a given proton with initial speed $V$ -- but Kugland image-flux relation \eqref{screenfluxSec2} would have to be modified to take into account that protons at different initial speeds have distinct plasma-image mappings. Doing so leads to a significantly more complicated image-flux relation. 

\subsubsection*{Instantaneous}

The instanteneity of the imaging process is the result of the temporal pulse length of proton sources $t_{pulse}$ and the transit time of protons across the plasma $t_{path}$ being small compared to the evolution time of the magnetic-field configuration of interest. For fusion-produced protons,  $t_{pulse} \sim 100 \, \mathrm{ps}$~\cite{S04}, while for TSNA-produced protons, $t_{pulse} \sim 10 \, \mathrm{ps}$~\cite{W01}. For both, the transit time across the plasma can be estimated as
\begin{equation}
t_{path} \sim 40 \left[\frac{l_z\!\left(\mathrm{cm}\right)}{0.1 \, \mathrm{cm}}\right] \left[\frac{W\!\left(\mathrm{MeV}\right)}{3.3 \, \mathrm{MeV}}\right] \, \mathrm{ps} \, .
\end{equation}
Thus, provided the plasma dynamics to be studied are on nano-second timescales, this approximation of instataneity is a good one. This is indeed the case for many relevant experiments involving stochastic magnetic fields~\cite{T16,T17}, though by no means all~\cite{Mo12}. For time-varying fields, the effect of inductive electric fields is likely significant -- which in turn leads to the invalidation of expression \eqref{pathintfieldSec2} for the perpendicular-deflection field. In addition, magnetic fields varying on timescales shorter than $t_{pulse}$ would lead to the front part of the proton beam seeing different magnetic fields to the back part of the beam; the resulting proton image would then be the superposition of the proton images for both fields. 

\subsubsection*{Point source}

The point source approximation is the natural consequence of the size of the proton source $a$ being much smaller than the distance $r_i$ from the source to the plasma. For fusion protons, $a \approx 40 - 50 \, \mu \mathrm{m}$~\cite{L06}, while for TSNA protons this is even smaller: $a \approx 10 \, \mu \mathrm{m}$~\cite{W01}. To prevent the diagnostic interfering with experiment, one typically chooses $r_i \geq 1 \, \mathrm{cm}$, so $r_i \gg a$. However, the source's finite size prevents imaging of magnetic structures on scales smaller than the size of the source; the consequences of this for the extraction of magnetic field statistics are discussed in Section \ref{SmearFinSource}.

\subsubsection*{Initially uniform proton flux}

Finally, the uniformity of initial proton flux is a consequence of approximately isotropic emission of protons from the source across the solid angle encapsulating the experiment. In practice, this assumption is not always satisfied. Experimental characterisation of proton-flux images produced by a capsule implosion in the absence of imaging fields show relative flux deviations of up to $50 \, \%$ across an image~\cite{M12}. As a result, using this approximation to identify properties of magnetic fields may be misleading; this problem is discussed further in Section \ref{InhomInitFlux}. However, such variations are typically limited to longer wavelengths, which allows for the application of high-pass filters to isolate flux features resulting from stochastic magnetic fields from those due to variations in the initial flux~\cite{M12}. The problem is more acute for TSNA protons, whose initial distribution can display perpendicular spatial structuring~\cite{N09}. 

\subsubsection{Paraxality}

If the distance from the proton source to the plasma is much greater than the dimensions of the plasma, $r_i \gg l_\bot, l_z$, we can approximate the section of the beam passing through the plasma as planar, despite the fact that proton beams generated by fusion reactions in a $\mathrm{D}_2$ capsule implosion generally take the form of a uniformly expanding spherical shell~\cite{L06}.
For a given proton, this paraxial approximation is effectively an expansion of the position and velocity of the proton in terms of half of the paraxial parameter
\begin{equation}
\frac{\delta \alpha}{2} = \frac{l_\bot}{2 r_i} \ll 1\, , \label{paraxialparamdef}
\end{equation}
the ratio of the size of the region being imaged to the radius of curvature of the beam. The factor of two appears, because the approximating plane is chosen to be exact at the centre of the beam. For typical experimental set-ups, $\delta \alpha \approx 0.05-0.2 \ll 1$~\cite{K12}.

Increasing the paraxial parameter $\delta \alpha$ does not significantly change the nature of the four contrast regimes described in Section \ref{Contrastregimes} qualitatively, although for sufficiently large values a slight decay in proton flux towards the edges of images may be detectable. For quantitative analysis, higher-order corrections in $\delta \alpha$ to the plasma-image mapping \eqref{divmappingSec2} can be introduced~\cite{K12}. 

\subsubsection{Point projection}

The point-projection assumption requires that the distance $r_s$ from the plasma to the detector be much greater than $l_z$. This means that  displacements of protons from their undeflected trajectories acquired inside the plasma due to magnetic forces are negligible compared to those  displacements of protons resulting from their free-streaming motion beyond the plasma with an altered deflection velocity. Mathematically, the plasma-image mapping \eqref{divmappingSec2} is effectively an expansion in
\begin{equation}
\delta \beta \equiv \frac{l_z}{r_s} \ll 1 \, , \label{pointprojectionparamdef}
\end{equation}
retaining only the leading order term in $\delta \beta$. In actual proton-imaging set-ups as typically implemented, $l_z \ll r_i \ll r_s$, so $\delta \beta \ll 1$.

\subsubsection{Small proton-deflection angles}

To derive the expression \eqref{pathintfieldSec2} for the perpendicular-deflection field, the size of typical proton-velocity deflections are assumed small compared to the initial proton velocity. Since \eqref{pathintfieldSec2} shows that deflection velocities are perpendicular to the initial direction, that is, that $\mathbf{w} \cdot \mathbf{v}_0 \approx \mathbf{w} \cdot \hat{\mathbf{z}} \approx 0$, the deflection angle of the proton due to magnetic forces must therefore be small:
\begin{equation}
\delta \theta \equiv \cos^{-1}{\left[\frac{\mathbf{v}_0 \cdot \left(\mathbf{v}_0 + \mathbf{w}\right)}{\left|\mathbf{v}_0\right|\left|\mathbf{v}_0+\mathbf{w}\right|}\right]} \approx \frac{\left|\mathbf{w}\right|}{V} \ll 1 \, .
\end{equation}
The magnitude of this parameter can be practically estimated for stochastic fields imaged by protons with energy $W$ using deflection angle RMS \eqref{RMSdeflecangle}: 
\begin{equation}
\delta \theta \sim \frac{e B}{m_p c V} \sqrt{l_z l_B}\approx \left[\frac{B\!\left(\mathrm{MG}\right)}{2.6 \, \mathrm{MG}}\right] \left[\frac{W\!\left(\mathrm{MeV}\right)}{3.3 \, \mathrm{MeV}}\right]^{-1/2} \left[\frac{l_z\!\left(\mathrm{cm}\right)l_B\!\left(\mathrm{cm}\right)}{0.01 \, \mathrm{cm}^2}\right]^{1/2} \, . \label{deflangleest}
\end{equation} 
The same estimate can also be used to determine whether $\delta \theta \lesssim l_B/l_z$, and hence whether the perpendicular-deflection field is irrotational. Since $l_B < l_z$, this condition is typically more stringent than $\delta \theta \ll 1$. 
 
It is clear that if deflections are not small, $\delta \theta \sim 1$ inevitably implies that the system is in the diffusive regime $\mu \sim r_i/l_B \gg 1$. This has the consequence that, despite the technical lack of validity of the derivation of plasma-image mapping \eqref{divmappingSec2} in this case, proton-flux images in the large-deflection regime are likely to have many qualitatively similiar features to diffusive proton-flux images  -- albeit combined with a significant loss of total flux from the detector. 

\subsection{Theoretical complications} \label{TheoCom}

We have claimed in the previous sections that the imaging of stochastic magnetic fields can be classified into four general regimes depending on $\mu$, and that in two of these regimes (linear and nonlinear injective), the path-integrated field can be extracted directly from proton-flux images; under the further assumption of isotropy, this is sufficient to determine the magnetic-energy spectrum of the stochastic field. Whilst these statements are true for many stochastic fields, there are others for which the efficiacy of the proposed analysis techniques must be re-considered: these are discussed in this section. Note that we distinguish between inherent constraints on the technique, which are unavoidable, and constraints due to experimental effects, which can in principle be overcome with advances in diagnostic implementation. The former are discussed in this section, the latter in Section \ref{ExpCom}.

\subsubsection{Spatially inhomogeneous and anisotropic fields}

\begin{figure}[htbp]
\centering
    \begin{subfigure}{.48\textwidth}
        \centering
        \includegraphics[width=\linewidth]{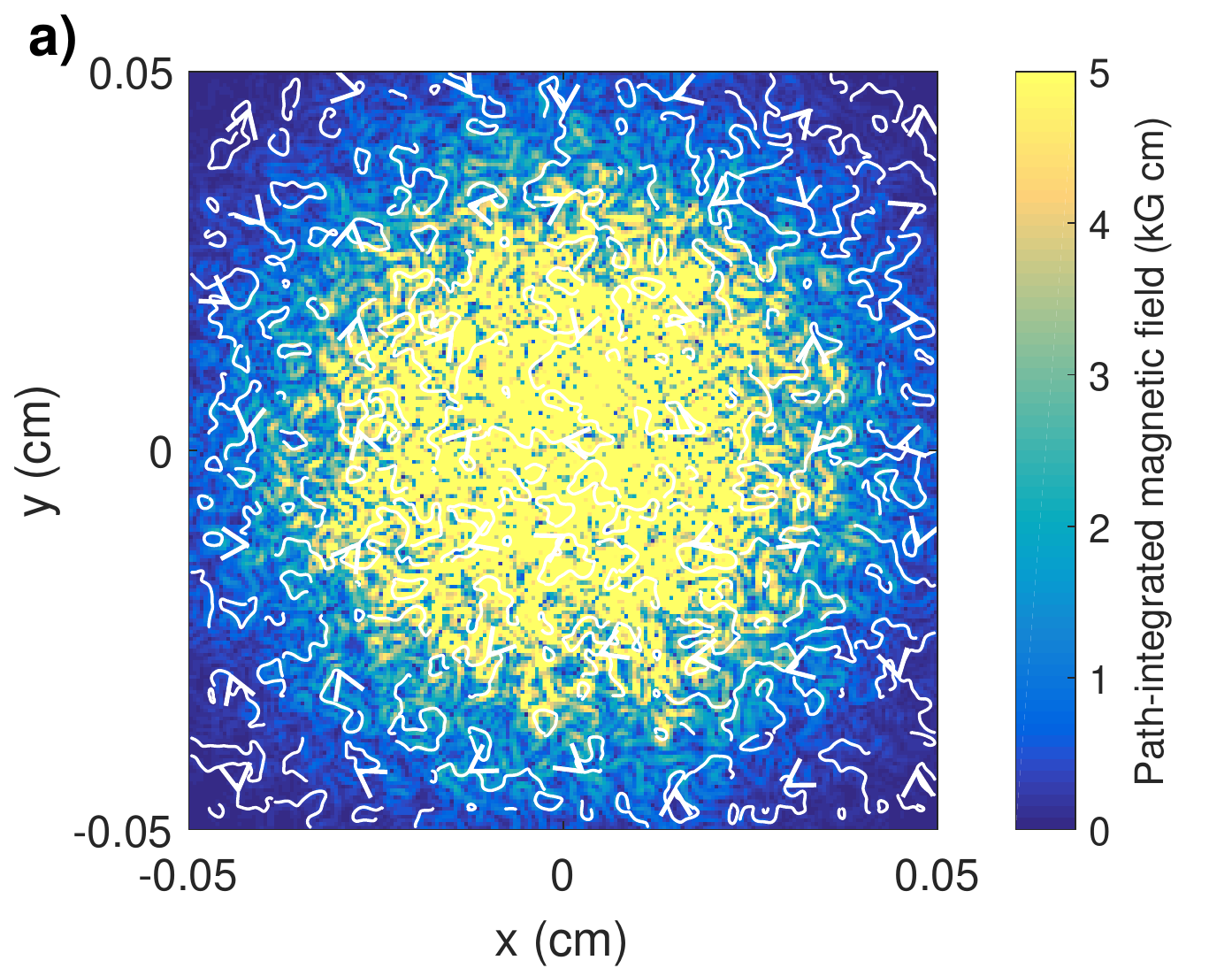}
    \end{subfigure} %
    \begin{subfigure}{.48\textwidth}
        \centering
        \includegraphics[width=\linewidth]{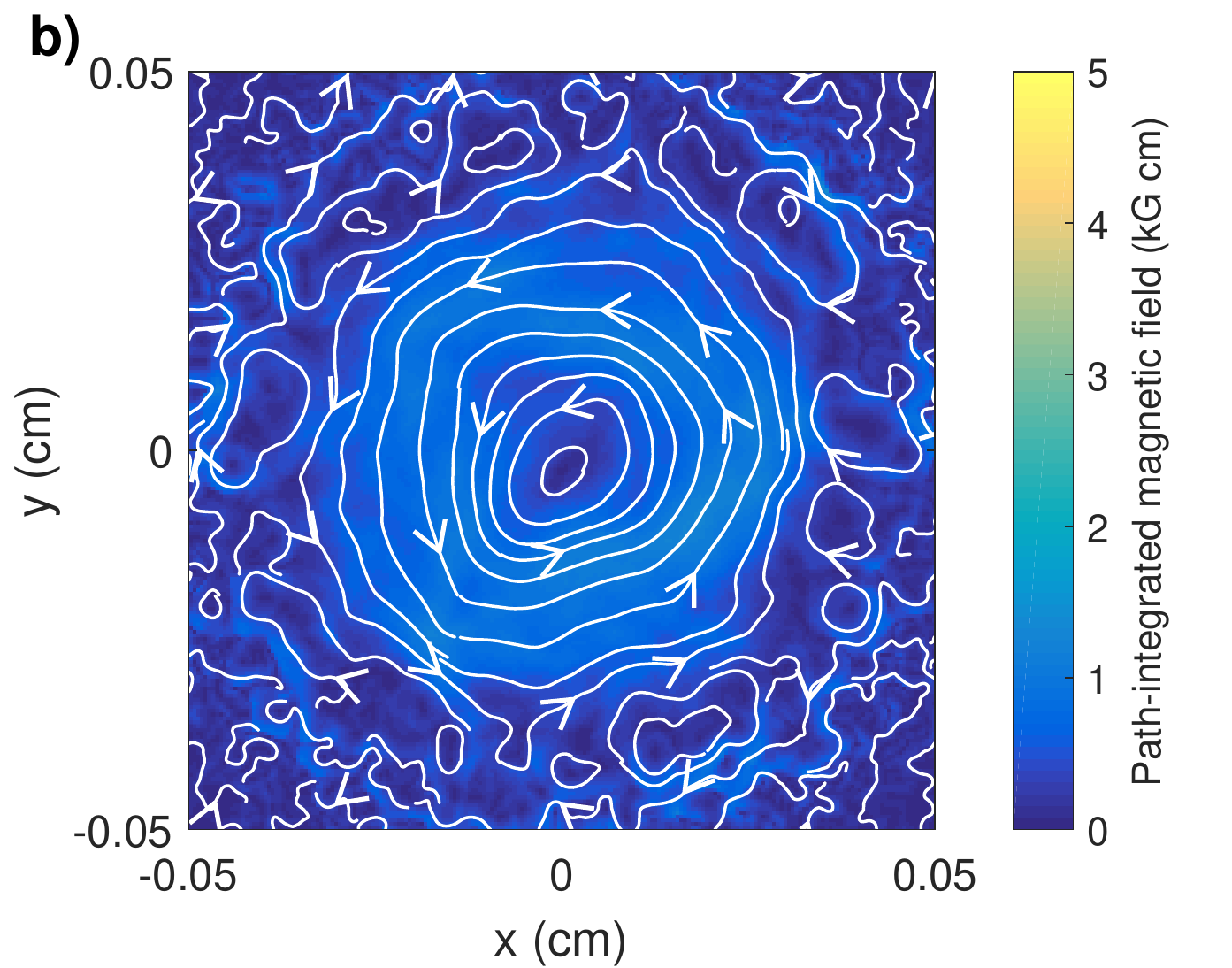}
    \end{subfigure} %
    \begin{subfigure}{.48\textwidth}
        \centering
        \includegraphics[width=\linewidth]{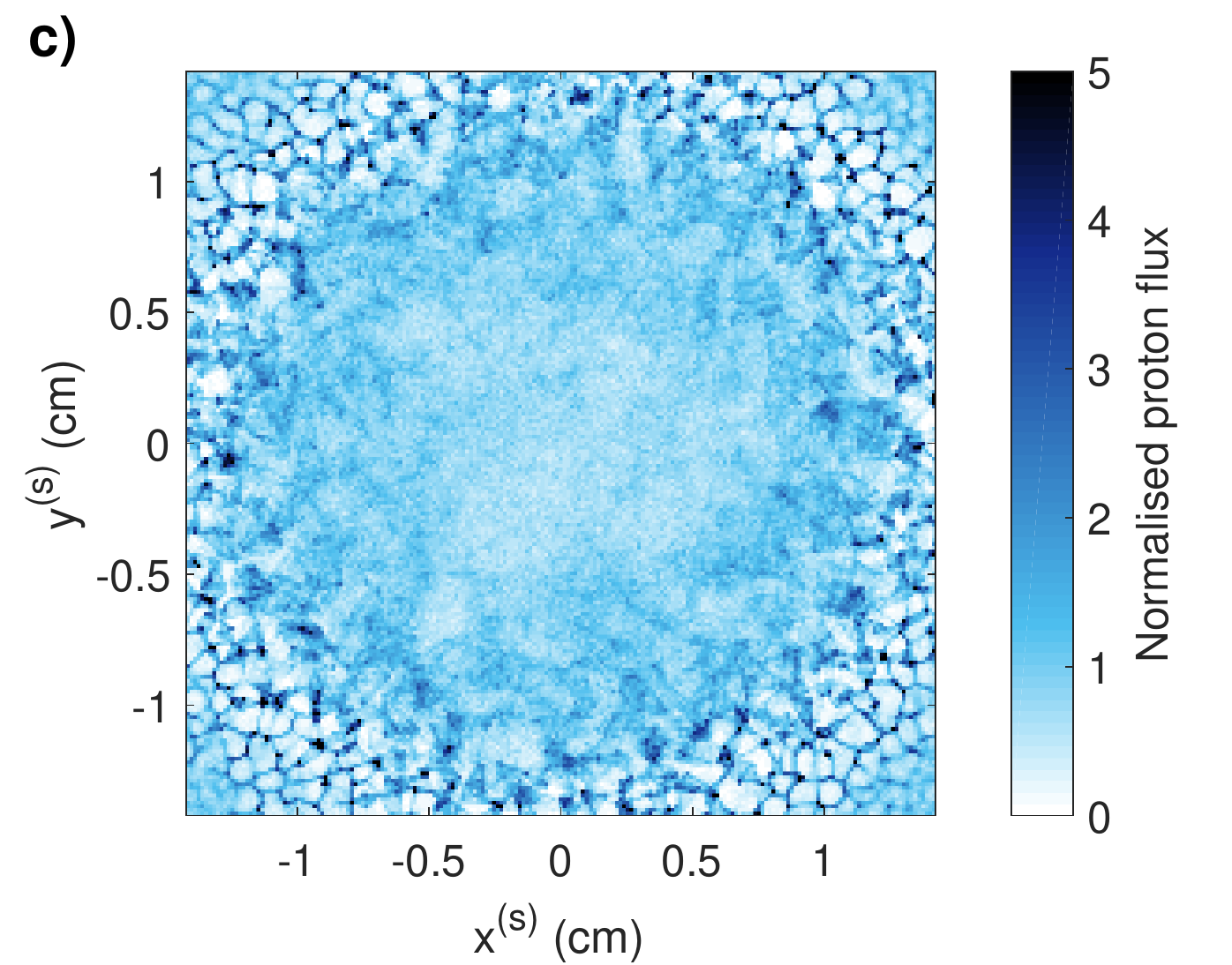}
    \end{subfigure} %
    \begin{subfigure}{.48\textwidth}
        \centering
        \includegraphics[width=\linewidth]{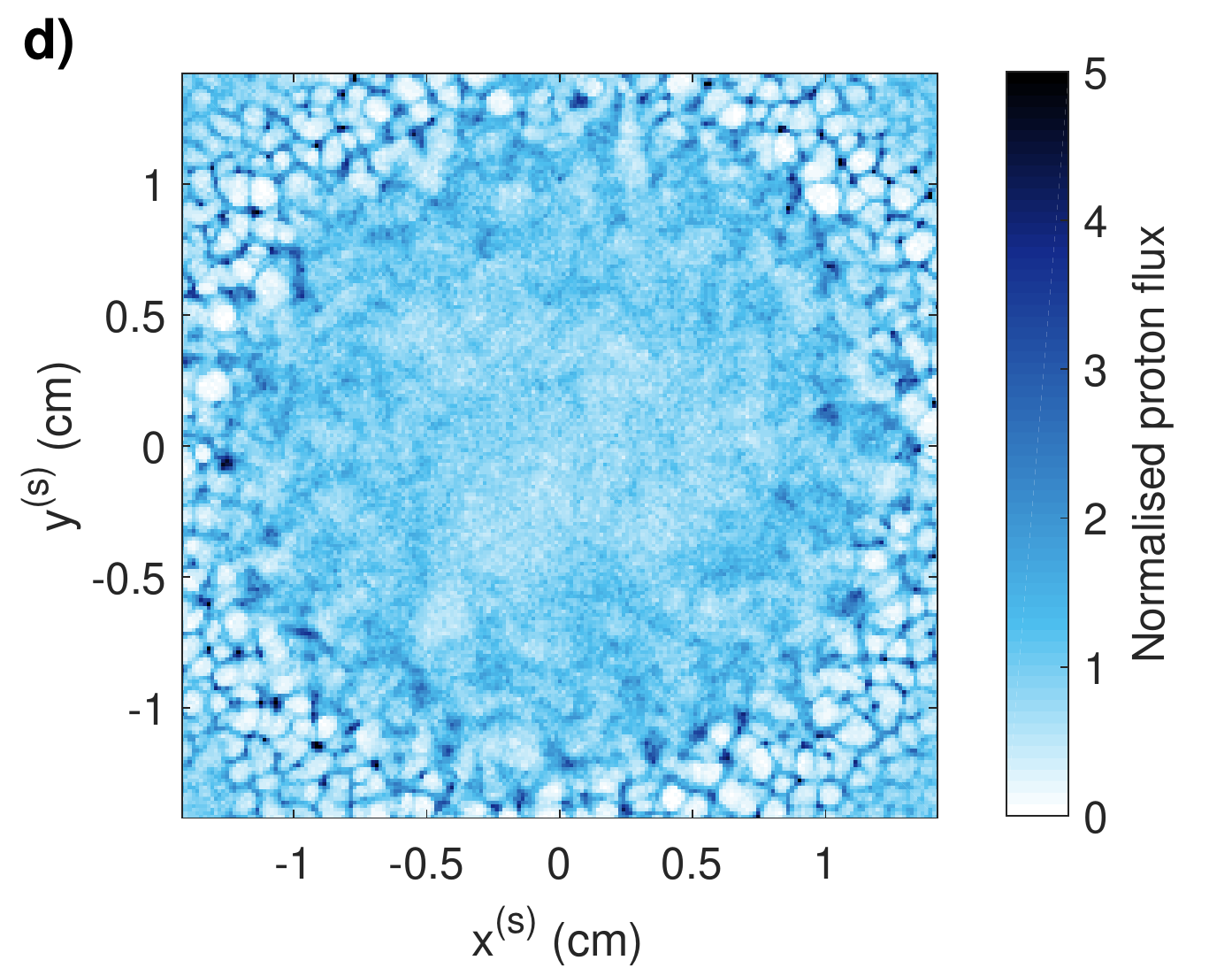}
    \end{subfigure} %
\caption{\textit{Illustration of ill-posedness of reconstruction for spatially inhomogeneous stochastic magnetic fields.} \textbf{a)} Path-integrated perpendicular magnetic field associated with small-scale magnetic cocoon configuration (see Appendix \ref{ToySpecLinThyCocoon}) with correlation scale $l_B = 0.8 \, \mu \mathrm{m}$. A Gaussian envelope of the form \eqref{magfieldwind} is applied to the magnetic field, with $\sigma = 3$ and $B_{rms,0} = 2.2 \, \mathrm{MG}$. \textbf{b)} Path-integrated magnetic field associated with perpendicular-deflection field reconstructed from proton-flux image c) generated from magnetic field described in a), assuming an injective mapping. Typical deflection velocities in the central region are reduced by a factor $\sim 25$. \textbf{c)} 3.3-MeV proton-flux image created by imaging field configuration a) with 3,000,000 15.0 MeV protons from point source located at $r_i = 1 \, \mathrm{cm}$ from the field configuration, and detector located on the opposing side, with $r_s = 30 \, \mathrm{cm}$. \textbf{d)} Predicted 3.3-MeV proton-flux image, assuming proton beam with same imaging parameters as c) experienced path-integrated magnetic field b) while traversing the plasma.} \label{varRMSsmlscleversusmeanfield}
\end{figure}

The classification into four contrast regimes that can be distinguished simply by particular characteristics of proton-flux images holds for locally homogeneous stochastic fields. However, this classification is not strictly valid for more general fields with inhomogeneous or anisotropic statistics. In particular, for an arbitrary proton-flux image, the contrast regime associated with that image cannot be uniquely identified. This is related to the non-uniqueness of the inversion problem associated with recovering path-integrated magnetic fields associated with a particular proton-flux image -- introduced in Section \ref{Contrast} and discussed in Sections \ref{CauRgme} and \ref{HighConRgme}. We further illustrate this phenomenon with an example consisting of two strikingly distinct path-integrated fields (incorporating different contrast regimes) giving near-identical image-flux distributions (Figure \ref{varRMSsmlscleversusmeanfield}).
Despite the mathematical impossibility of distinguishing between the path-integrated fields in Figures \ref{varRMSsmlscleversusmeanfield}a and \ref{varRMSsmlscleversusmeanfield}b without additional information, in practice there are various qualitative tests that could in principle be employed. For example, if there are multiple species of imaging protons, and the transit time and pulse length of all these species are much smaller than the evolution timescales of the system, the multiple-beam energy comparison technique described in Section \ref{CauRgme} to find caustics could also be used to distinguish between these path-integrated magnetic fields. More specifically, the path-integrated field could be reconstructed from a 3.3 MeV proton image (such as Figure \ref{varRMSsmlscleversusmeanfield}c), before predicting what a 15.0 MeV proton image would look like. This could be compared to the actual 15.0 MeV proton image to test the veracity of the reconstructed path-integrated field. 

Even in the situation when the path-integrated field is uniquely extractable from its associated proton-flux image (implying that the contrast regime of the imaging diagnostic must be linear, or nonlinear injective), for a spatially-inhomogeneous field, the magnetic-energy spectrum of the stochastic field is not necessarily extractable from the path-integrated field. More specifically, the derivations of equations for the magnetic-energy spectrum [deflection-field spectral relation \eqref{deffieldspec}, or linear-regime flux spectral relation \eqref{linfluxspec3}] given in Appendices \ref{DeflfieldCorr} and \ref{RelFluxRMSLinThy} rely on both the assumption that magnetic field statistics do not vary along the path of the proton beam, and that the statistics are the same perpendicular and parallel to the direction of proton motion. These assumptions can likely be relaxed to include the case when the field is statistically inhomogenenous, and anisotropic in the perpendicular direction~\cite{EV03}.

\subsubsection{Variable $\mu$ across scales}

It is clear from its definition \eqref{contrastdef} that $\mu$ is scale-dependent, increasing as the scale $l_B$ of magnetic structures decreases relative to the path length of the imaging protons. Magnetic structures imaged in a proton-imaging set-up are, therefore, in a potentially different contrast regime depending on their strength and size. For multi-scale stochastic fields, such as those with power law spectra of the form $E_B\!\left(k\right) \propto k^{-p}$, we can estimate $\mu$ at a particular scale $l_B$. The field strength at this scale for such a power law goes as
\begin{equation}
B \sim B_0 \left(\frac{l_B}{l_z}\right)^{\left(p-1\right)/2} \, ,\label{powerlawstrengthscal}
\end{equation}
and so $\mu$ is given by
\begin{equation}
\mu \sim \mu_0  \left(\frac{l_B}{l_z}\right)^{p/2-1} \, .
\end{equation}
We see that for $p > 2$, $\mu$ decreases with scale, whereas the opposite is true for $p < 2$. We discuss the consequences of both possibilities in turn.

The case of sufficiently steep power spectra ($p > 2$) -- of which the Golitsyn spectrum \eqref{powerlawspecdefSec2} with $p = -11/3$ is an example -- is generally much easier to investigate. The largest $\mu$ is at largest scales, which coincides with both the strongest magnetic and path-integrated structures. Whether the path-integrated field and magnetic-energy spectrum can be extracted successfully is therefore not altered by the presence of smaller-scale fields.

\begin{figure}[htbp]
\centering
    \begin{subfigure}{.48\textwidth}
        \centering
        \includegraphics[width=\linewidth]{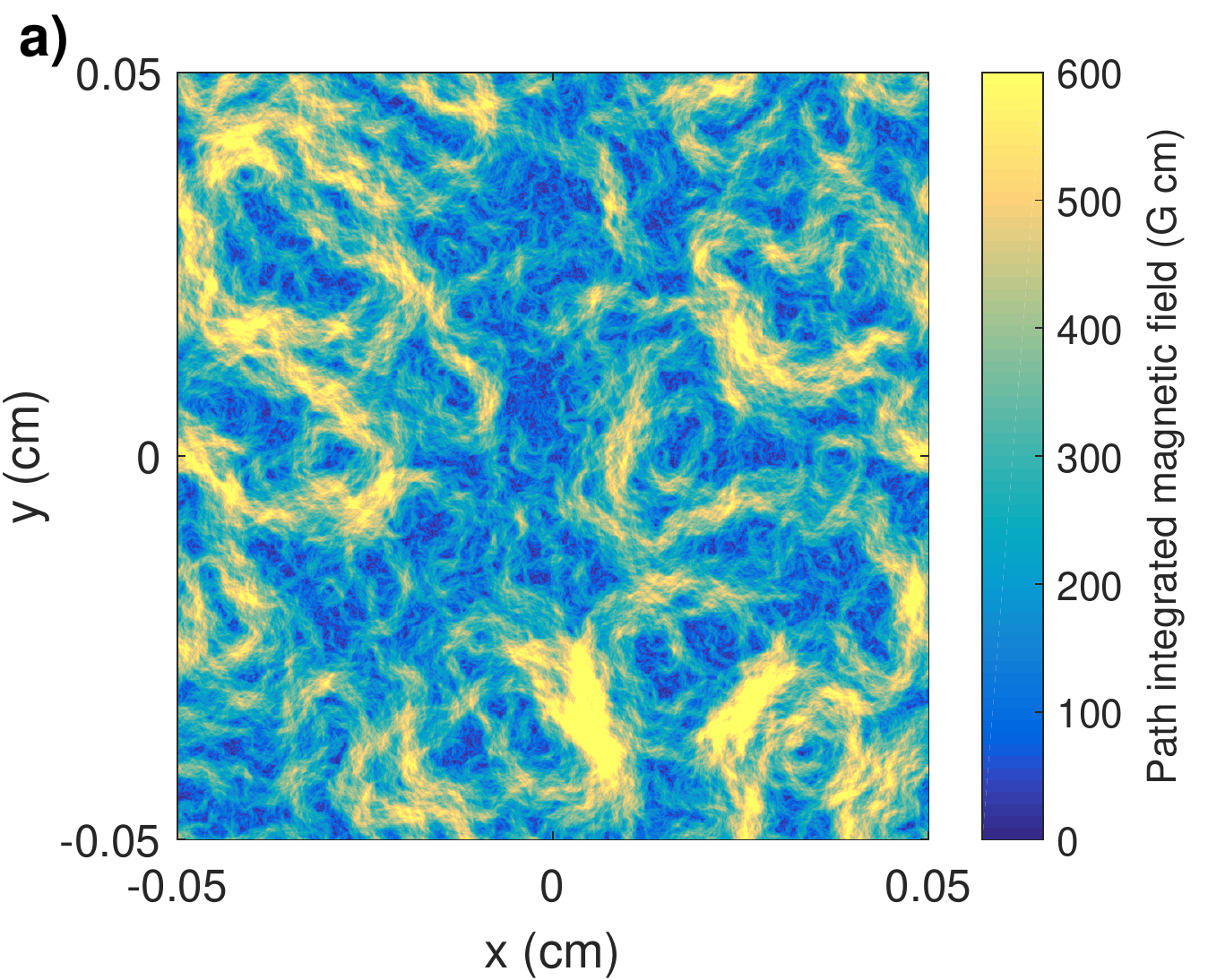}
    \end{subfigure} %
    \begin{subfigure}{.48\textwidth}
        \centering
        \includegraphics[width=\linewidth]{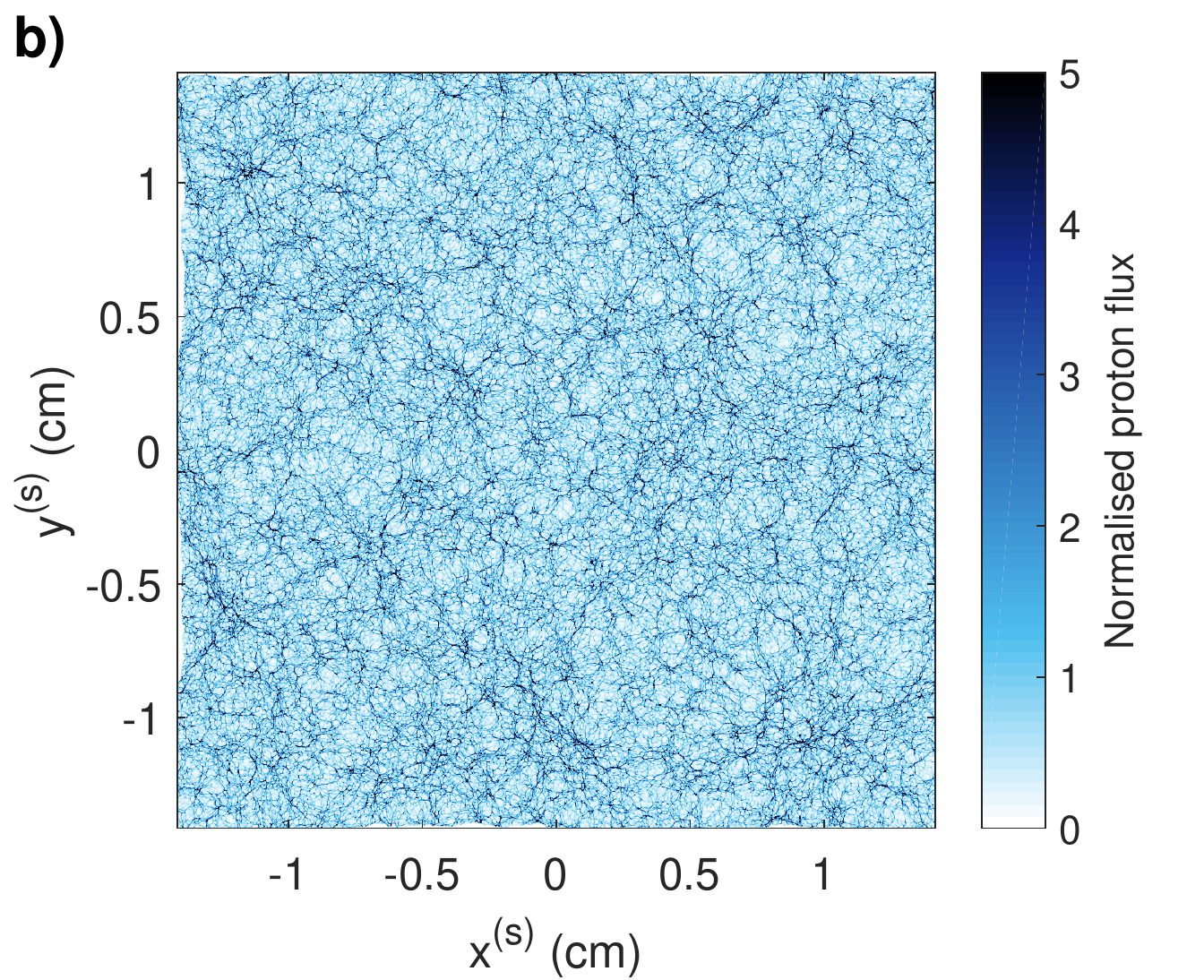}
    \end{subfigure} %
\caption{\textit{Illustration of variation in $\mu$ across scales, with consequent discrepancy in structure size between path-integrated magnetic field, and proton-flux image.} \textbf{a)} Predicted path-integrated field for periodic stochastic Gaussian magnetic field specified on $1001^2$ array of spatial extent $l_i = l_\bot = l_z = 0.1 \, \mathrm{cm}$, with power law spectrum of the form \eqref{kminusonespec}, spectral cutoffs $k_l = 4 \pi/l_i$, $k_u = 800 \pi/l_i$, and field RMS $B_{rms} = 10 \, \mathrm{kG}$. \textbf{b)} Proton-flux image generated with 3.3 MeV protons from a point source at distance $r_i = 1 \, \mathrm{cm}$ from the array, with imaging detector at distance $r_s = 30 \, \mathrm{cm}$ on the opposite side.} \label{variablecontrastscaleskminus1}
\end{figure}

On the contrary, for shallow spectra ($p < 2$), small-scale fields become extremely important for a proper understanding of proton-flux images, as well as their analysis. Firstly, the fact that $\mu$ increases with decreasing scale (in opposition to the path-integrated field) means that the dominant features in a proton image may be at much smaller scales than the dominant features in the path-integrated field. We illustrate this with an example. Figure \ref{variablecontrastscaleskminus1} shows a typical path-integrated magnetic field resulting from a $k^{-1}$ spectrum with a wavenumber range of a few decades. 
Creating a proton-flux image from this path-integrated field usin theg plasma-image mapping (as described in Appendix \ref{NumSimFluxImageGen}), we see that the strongest features are at the smallest scales, in contrast to the path-integrated magnetic field, where larger coherent structures are evident. This can be explained physically with recourse to an interpretation of proton-flux images as projections of path-integrated MHD current structures: for shallow enough spectra, the dominant MHD current structures are at the smallest scales. For actual proton-flux images, this fine structure would typically be masked by finite-resolution effects; that being said, for stochastic field configurations, Figure \ref{variablecontrastscaleskminus1} provides another example of the inadequacy of using image-flux structure as a proxy for magnetic field structure.

More quantitatively, the ability to reconstruct both path-integrated fields and magnetic energy spectra is limited by diffusive scattering at small scales. If $\mu\!\left(l_B\right) < \mu_c$ at all scales, for $\mu_c$ the critical value of $\mu$ at which the plasma-image mapping \eqref{divmappingSec2} loses injectivity, both can be recovered. However, this inequality being reversed at small scales can prevent successful reconstruction not only at those scales, but those at which injectivity is technically preserved. This is because large-scale structures are distorted by small-scale ones. 

\begin{figure}[htbp]
\centering
    \begin{subfigure}{.48\textwidth}
        \centering
        \includegraphics[width=0.95\linewidth]{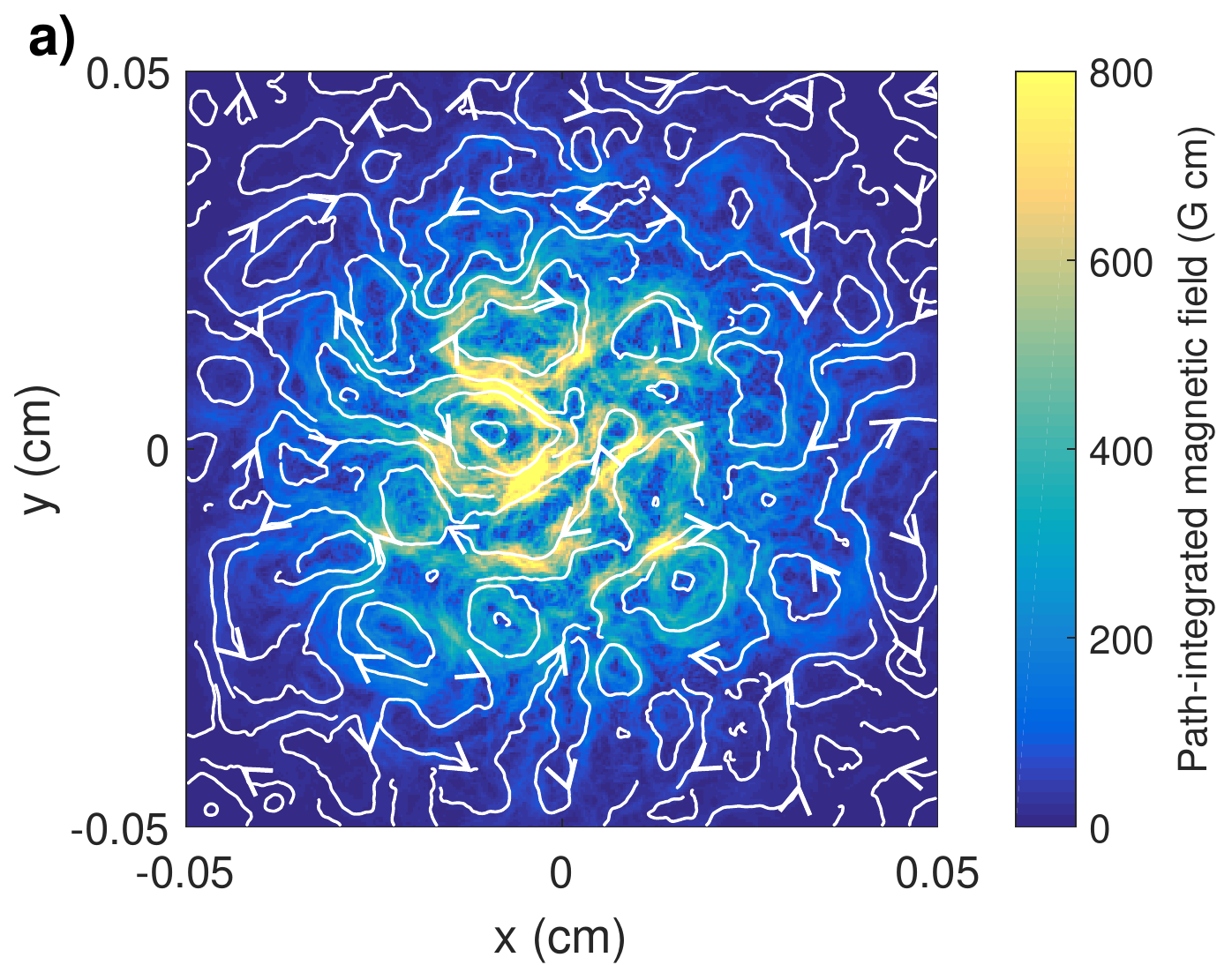}
    \end{subfigure} %
    \begin{subfigure}{.48\textwidth}
        \centering
        \includegraphics[width=0.95\linewidth]{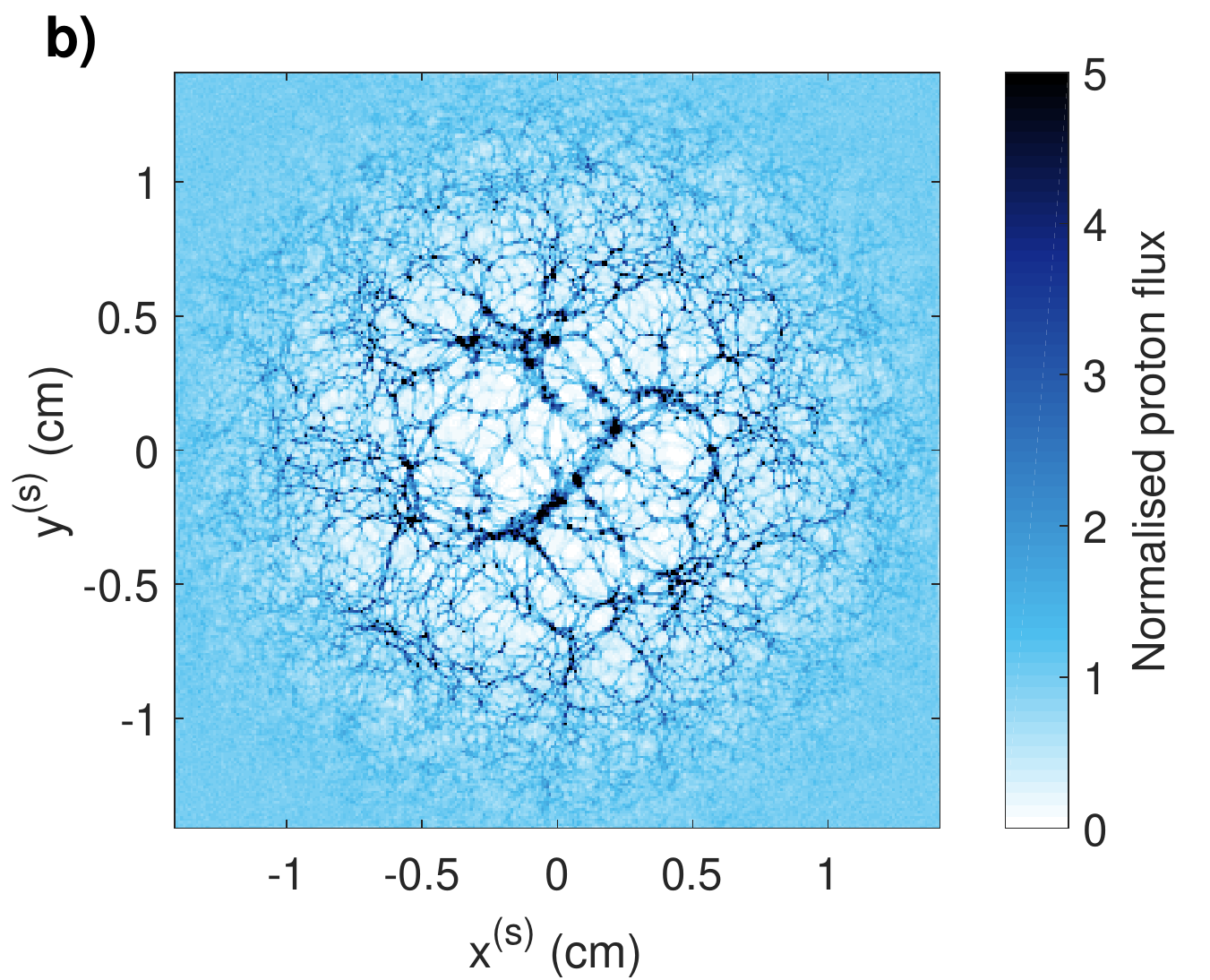}
    \end{subfigure} %
    \begin{subfigure}{.48\textwidth}
        \centering
        \includegraphics[width=0.95\linewidth]{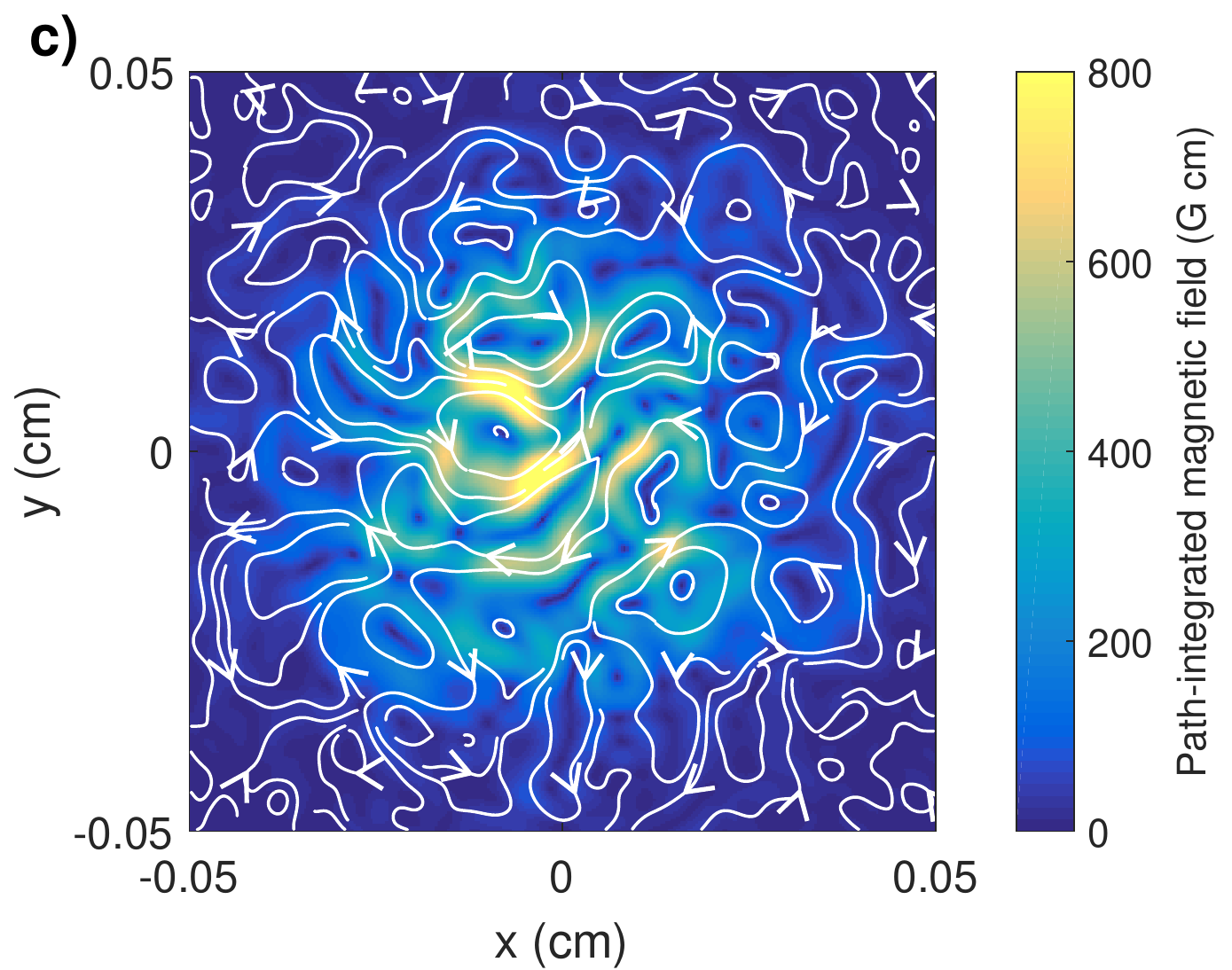}
    \end{subfigure} %
    \begin{subfigure}{.48\textwidth}
        \centering
        \includegraphics[width=0.95\linewidth]{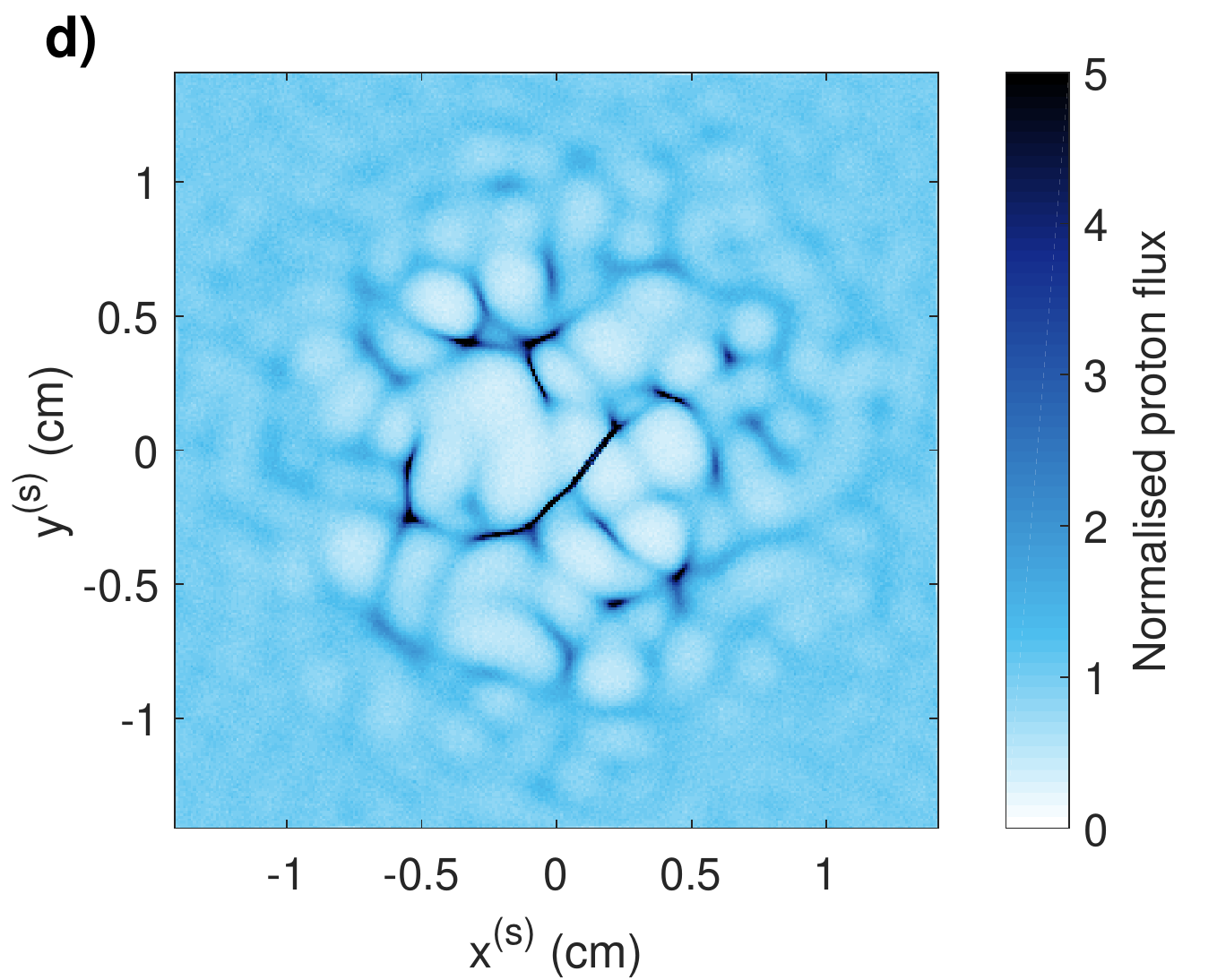}
    \end{subfigure} %
        \begin{subfigure}{.48\textwidth}
        \centering
        \includegraphics[width=0.95\linewidth]{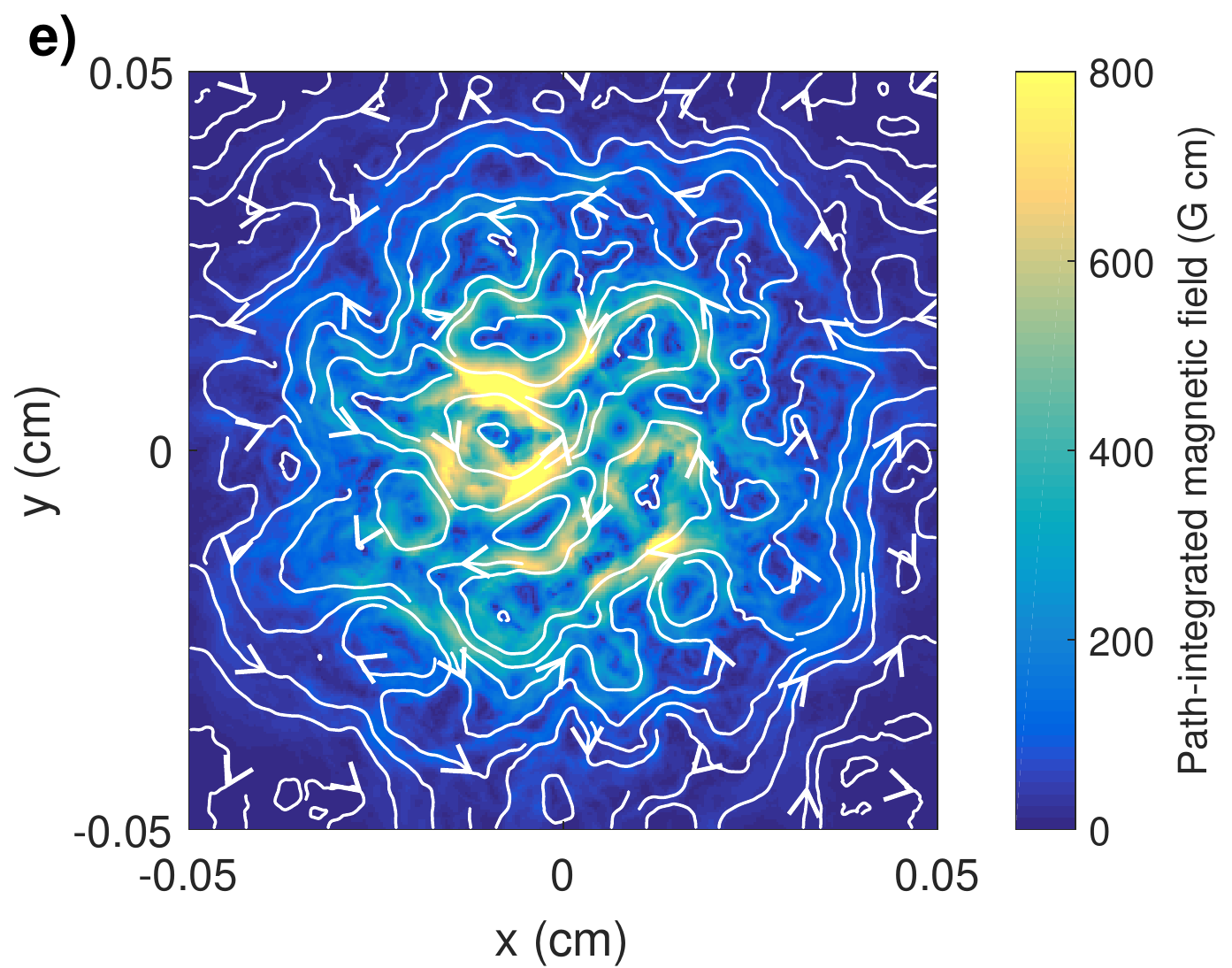}
    \end{subfigure} %
    \begin{subfigure}{.48\textwidth}
        \centering
        \includegraphics[width=0.95\linewidth]{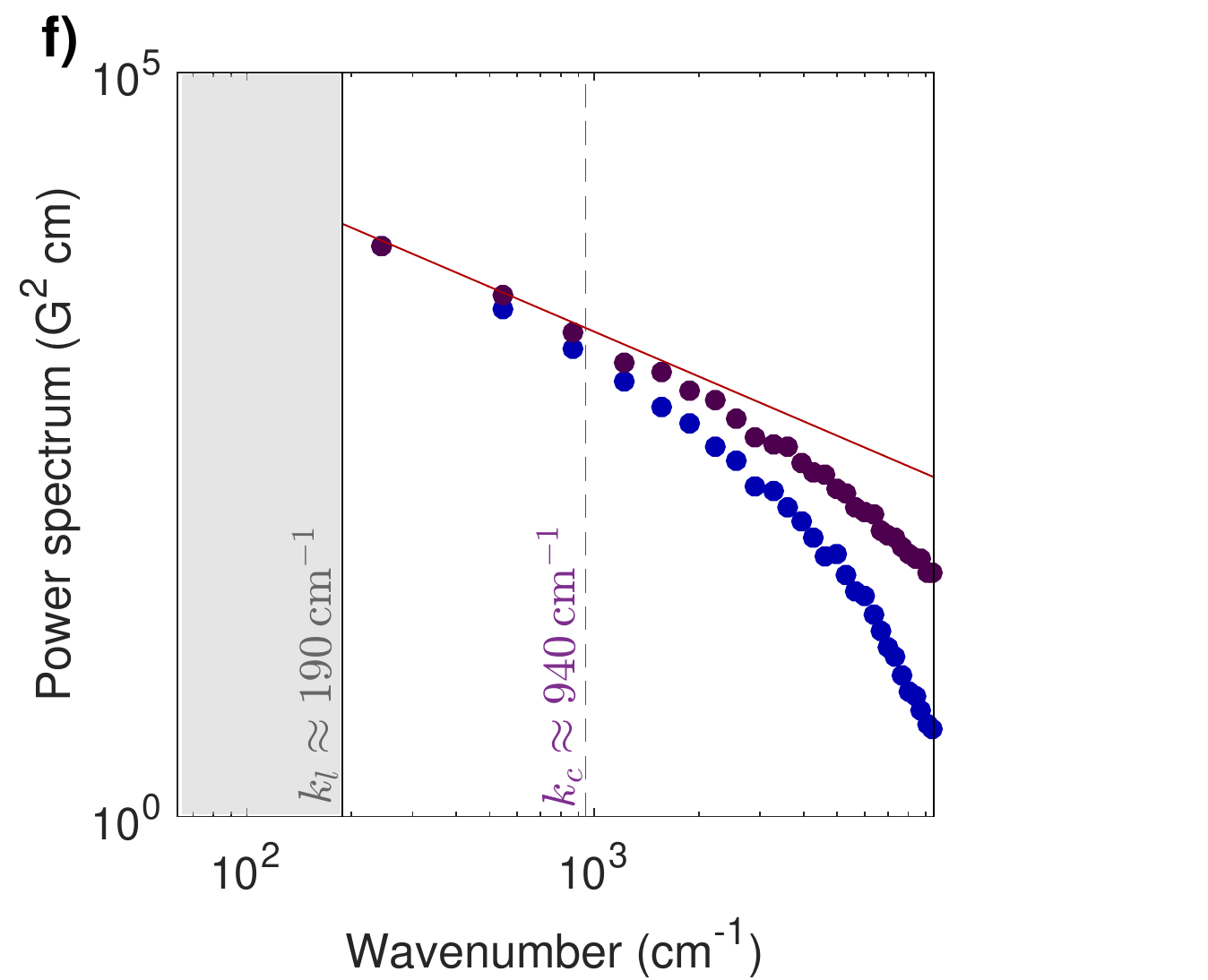}
    \end{subfigure} %
\caption{\textit{Analysing proton-flux images of stochastic magnetic fields with broad magnetic energy spectra.} A stochastic magnetic field sample, with magnetic-energy spectrum of the form \eqref{kminusonespec}, is generated in a region, with dimensions $l_i = l_z = l_\bot = 0.1 \, \mathrm{cm}$. Field values are specified on a $301^3$ array (grid spacing $\delta x = l_i/301$). The spectral cut-offs were set to be $k_l = 6 \pi/l_i$, $k_u = 2 \pi/\delta x$, and $B_{rms,0} = 40 \, \mathrm{kG}.$ A Gaussian envelope of the form \eqref{magfieldwind} is again applied to the magnetic field, with $\sigma = 3$. Imaging of the field is then implemented using the parameters given in Figure 4. \textbf{a)} Path-integrated perpendicular magnetic field as experienced by imaging 3.3 MeV protons. \textbf{b)} 3.3-MeV proton-flux image. \textbf{c)} Path-integrated perpendicular magnetic field, with magnetic field wavemodes whose wavenumber exceeds $k_c = 30\pi/l_i$ removed. \textbf{d)} 3.3-MeV proton-flux image of filtered magnetic field. \textbf{e)} Predicted path-integrated magnetic field found by applying field reconstruction algorithm. \textbf{f)} Magnetic energy spectra: true result (red), plotted with prediction using deflection-field spectral relation \eqref{deffieldspec} (blue) and linear-regime spectral relation \eqref{linfluxspec3} (purple).
} \label{variablecontrastscaleskminus1spec}
\end{figure}

The phenomenon of smaller-scale fields preventing successful reconstruction of larger-scale ones is best illustrated by considering an example. Figure \ref{variablecontrastscaleskminus1spec}a shows a stochastic magnetic field with a magnetic-energy spectrum of the form
\begin{equation}
E_B\!\left(k\right) = \frac{B_{rms}^2}{8 \pi}  \frac{\log{k_u/k_l}}{k} \, , \quad k \in \left[k_l,k_u\right] \, , \label{kminusonespec}
\end{equation}
combined with a Gaussian envelope of the form \eqref{magfieldwind}.
The wavenumber-range is chosen to be large -- that is, $k_l \ll k_u$ -- and field strength such that the proton imaging set-up applied to the largest structures has $\mu < \mu_c$. The proton-flux image (Figure \ref{variablecontrastscaleskminus1spec}b) has strong image-flux structures on many scales, with larger image-flux structures somewhat dispersed by smaller, caustic ones. Figure \ref{variablecontrastscaleskminus1spec}c, shows the same field, but with high wavenumbers filtered out at a scale $k_c$. The resulting proton-flux image (Figure \ref{variablecontrastscaleskminus1spec}d) now falls into the nonlinear injective regime, indicating that $\mu$ at larger length-scales is indeed below the critical value. Figure \ref{variablecontrastscaleskminus1spec}e shows the path-integrated magnetic field reconstructed from Figure \ref{variablecontrastscaleskminus1spec}b using the field reconstruction algorithm described in Section \ref{NonLinInjRgme}. While the reconstruction still captures the global morphology of the field, its fine structure is altered. This difference is shown more clearly by calculating the magnetic-energy spectra, shown in Figure \ref{variablecontrastscaleskminus1spec}f. The magnetic-energy spectrum resulting from spectral relation \eqref{deffieldspec} applied the reconstructed field is suppressed at high wavenumbers. Indeed, the effect is sufficiently strong to lead to spectral distortion at wavenumbers which would otherwise be correctly reconstructed if wavenumbers $k > k_c$ were removed. In contrast to the reconstruction of steeper power spectra, the magnetic-energy spectrum calculated from linear theory using formula \eqref{linfluxspec3} applied to the proton image shown in Figure \ref{variablecontrastscaleskminus1spec}b is closer to the true result, although the associated spectral curve is still distorted. 

In short, for shallow magnetic energy spectra, the presence of small-scale fields leads to the suppression of small-scale image-flux structures, which in turn hides high wavenumber modes when attempting to reconstruct both path-integrated fields and magnetic energy spectra. 

\subsection{Experimental complications} \label{ExpCom}

In addition to the theoretical restrictions placed on the use of a proton imaging diagnostic to assess magnetic field statistics, additional constraints arise from limitations to the implementation of the diagnostic. These are discussed more generally elsewhere,
but here we outline three effects which are particularly important when attempting to extract accurately path-integrated magnetic fields and energy spectra directly from proton images: Poisson noise, initial inhomogeneities in the initial flux distribution, and finite image resolution effects imposed by a finite source size.

\subsubsection{Poisson noise}

Poisson noise is an unavoidable effect associated with proton-flux images involving a finite number of imaging protons, and leads to spectral distortion. The finite number of protons in the imaging beam mean that locally the initial flux distribution is noisy. Since the sample is very large, these deviations are well described by Poisson statistics. Thus, if the mean flux per pixel is $N_\Psi$, then the standard deviation of fluctuations is $\sqrt{N_\Psi}$, giving signal-to-noise ratio $1/\sqrt{N_\Psi}$.

Since such noise is uncorrelated, its contribution to the two-dimensional relative image-flux spectrum is uniform. More specifically, the requirement that the integrated power spectrum of Poisson noise has variance $1/N_\Psi$ leads to
\begin{equation}
\hat{\eta}^{\left(PN\right)} \approx \left(\frac{\Delta x}{2 \pi}\right)^2 \frac{1}{N_\Psi} \, , 
\end{equation}
which in turn gives predicted magnetic magnetic  energy spectrum
\begin{equation}
E_B^{\left(PN\right)}\!\left(k\right) \approx \frac{1}{\left(2\pi\right)^3} \frac{m_p^2 c^2 V^2}{e^2 r_s^2 l_z} \Delta x^2 \frac{1}{N_\Psi} \, . \label{specPN}
\end{equation}
Thus for a multi-scale spectrum, values of the true spectrum will be dominated by the Poisson contribution if too small. This presents a potential difficulty for analysing proton-flux images in the linear regime if either the mean image-flux is low, or the physical signal is weak. The problem is evident for the predicted spectra shown in Figure \ref{Golitsynreconmethods}b (purple markers): the shallowing of the spectrum as predicted by linear-regime flux spectral relation \eqref{linfluxspec3} is due to Poisson noise. 

However, in practice the Poisson noise does not usually prevent successful analysis of proton-flux images. The effect on Poisson noise on the spectrum can be anticipated using \eqref{specPN}, and subsequently removed~\cite{C12}. Furthermore, nonlinear field reconstruction algorithms tend to reduce spectral distortion (Figure \ref{Golitsynreconmethods}b, blue markers). Thirdly, the restrictions on pixel size required to manage Poisson noise are usually less stringent that those already in place due to other factors limiting resolution~\cite{S04}. That being said, if possible it is preferable to design proton-imaging set-ups in such a way that the magnitude of Poisson noise be significantly less than image-flux features due to magnetic fields.

\subsubsection{Inhomogeneity of initial flux distribution} \label{InhomInitFlux}

The methods used to reconstruct magnetic field statistics directly from proton-flux images described in Sections \ref{LinRgme} and \ref{NonLinInjRgme} rely on prior knowledge of the initial flux distribution. For experimental implementations of proton imaging, this initial distribution is usually assumed to be uniform. However, as discussed in in Section \ref{Assum}, at the present time it is difficult in practice to achieve a completely uniform initial flux distribution. In the absence of alternative methods for determining the initial flux, there is consequently significant uncertainty over this quantity.

\begin{figure}[htbp]
\centering
    \begin{subfigure}{.48\textwidth}
        \centering
        \includegraphics[width=\linewidth]{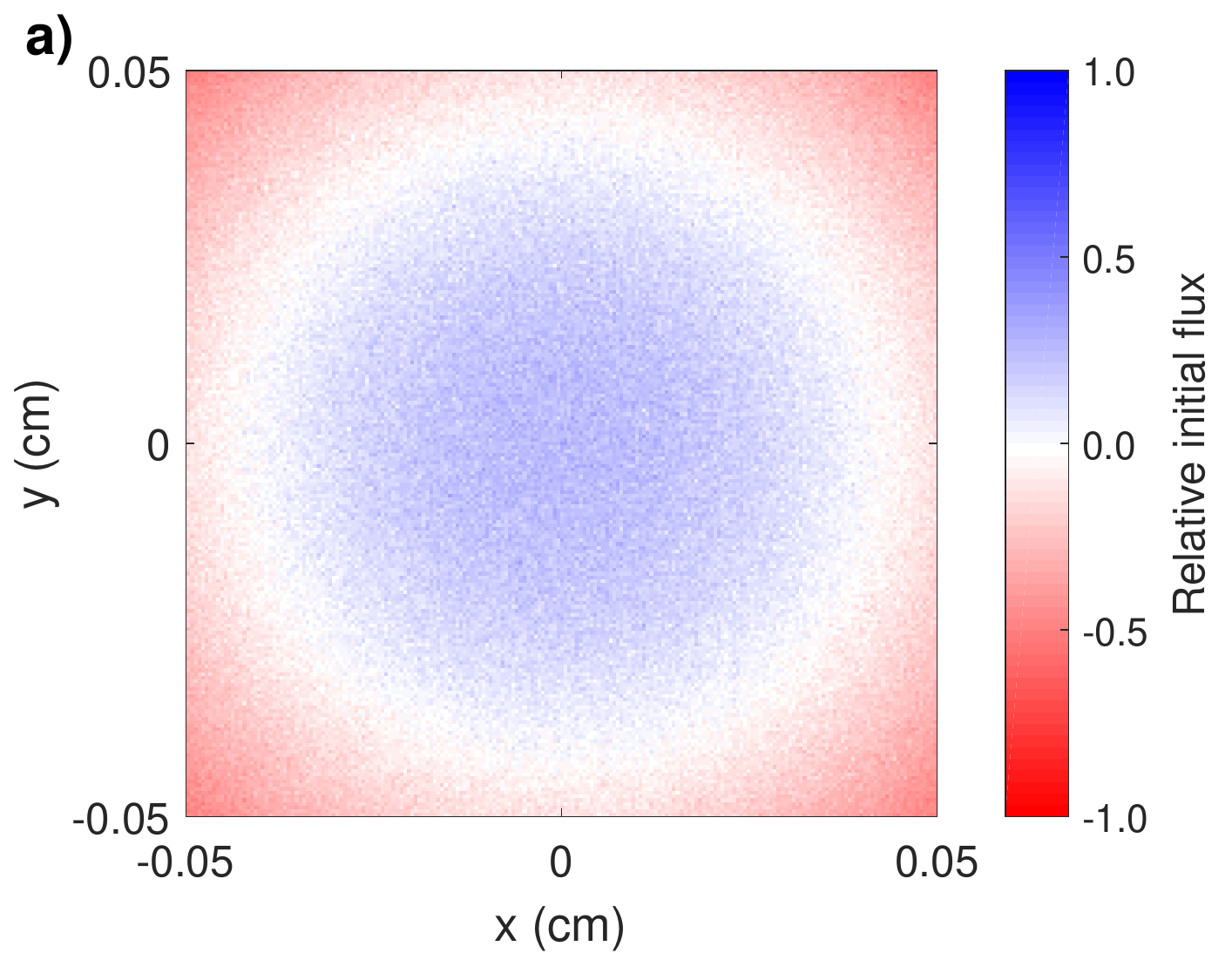}
    \end{subfigure} %
    \begin{subfigure}{.48\textwidth}
        \centering
        \includegraphics[width=\linewidth]{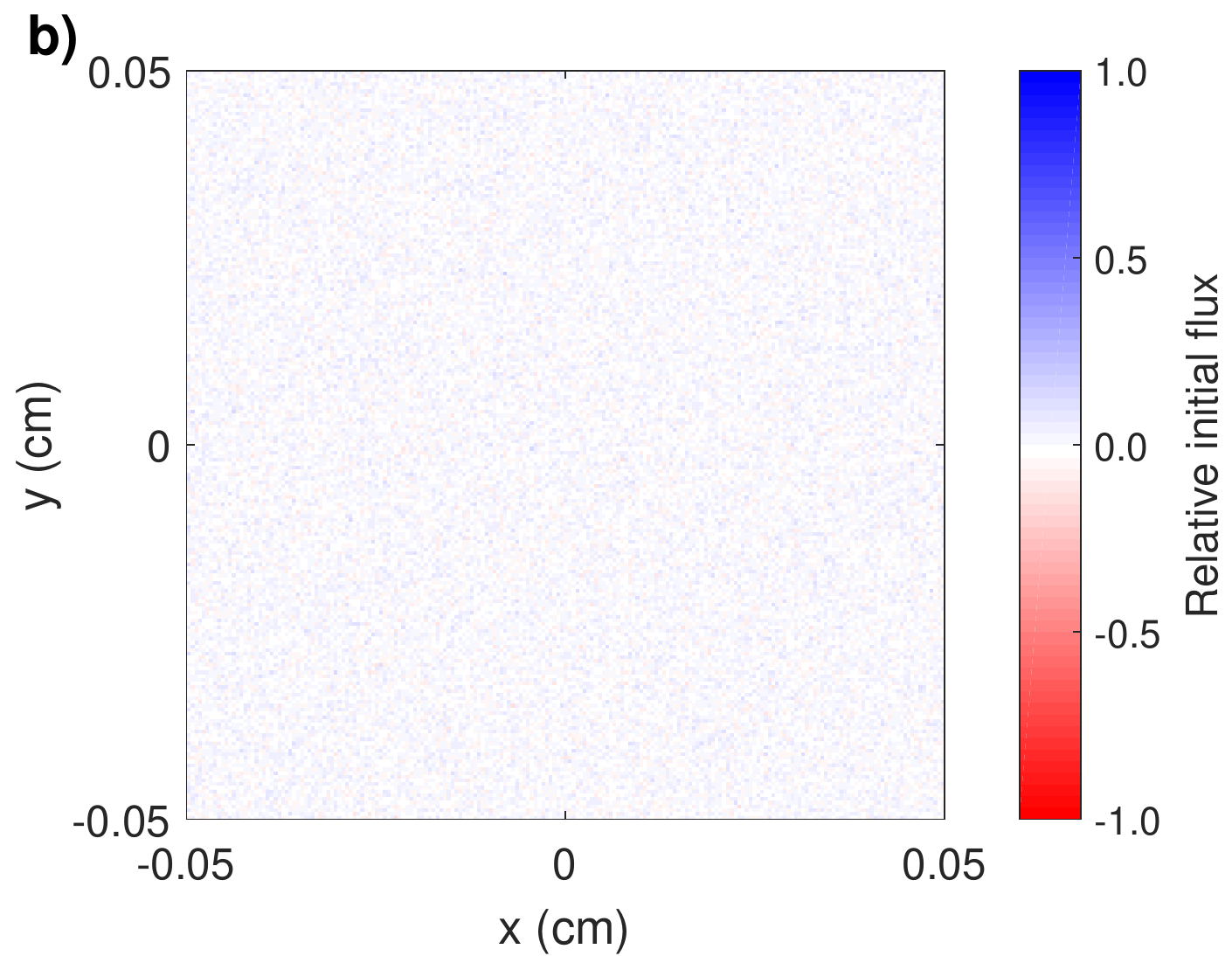}
    \end{subfigure} %
    \begin{subfigure}{.48\textwidth}
        \centering
        \includegraphics[width=\linewidth]{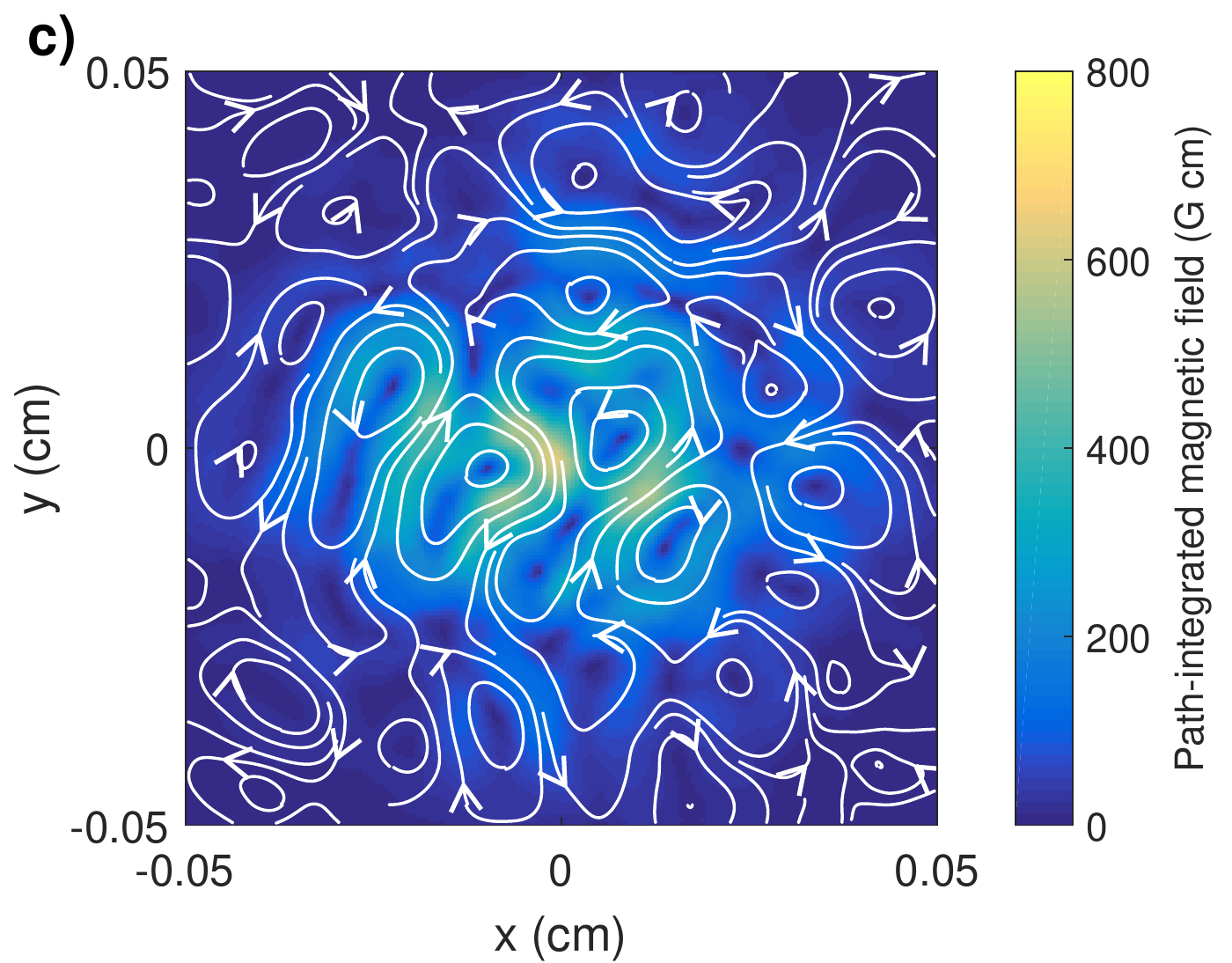}
    \end{subfigure} %
    \begin{subfigure}{.48\textwidth}
        \centering
        \includegraphics[width=\linewidth]{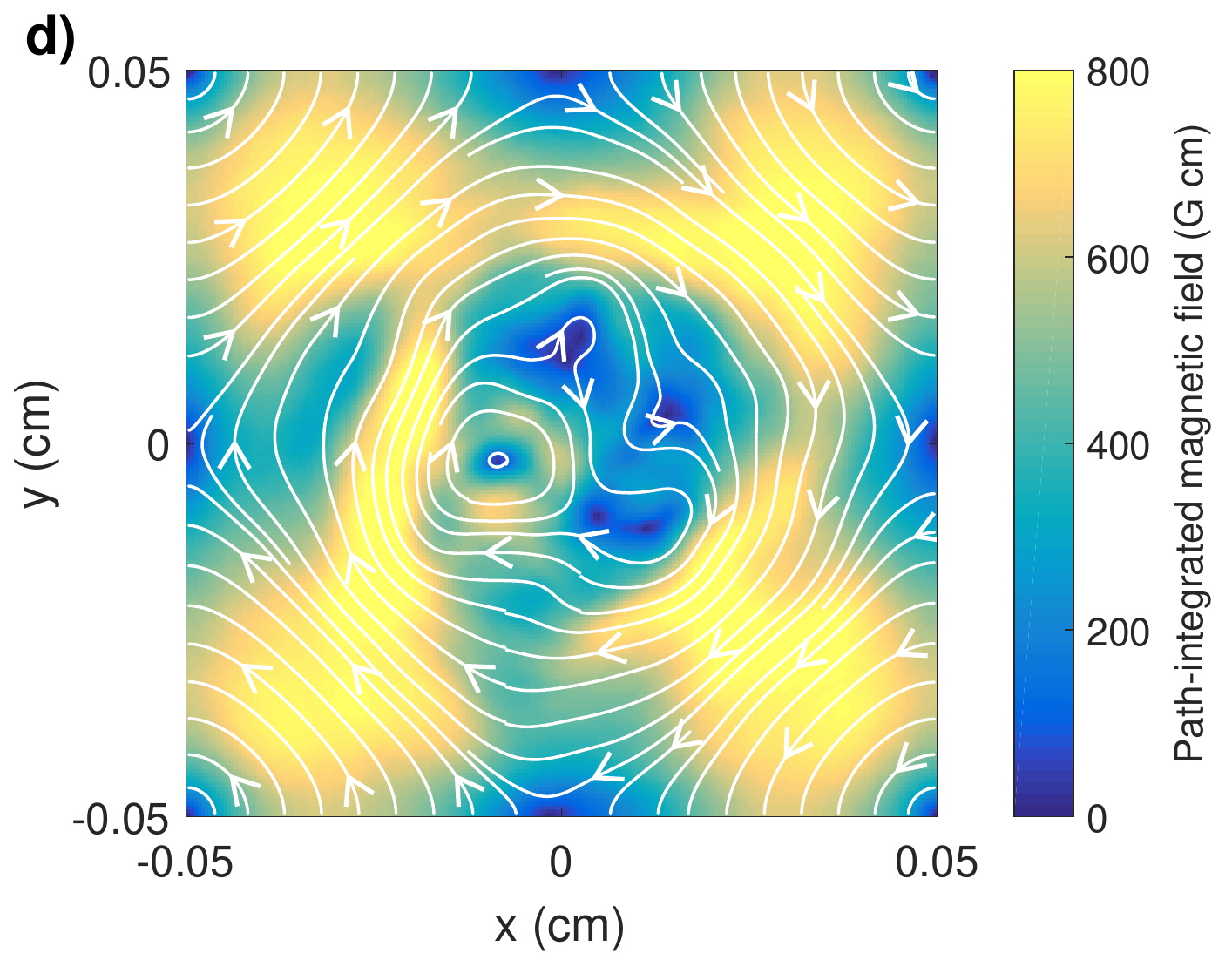}
    \end{subfigure} %
        \begin{subfigure}{.48\textwidth}
        \centering
        \includegraphics[width=\linewidth]{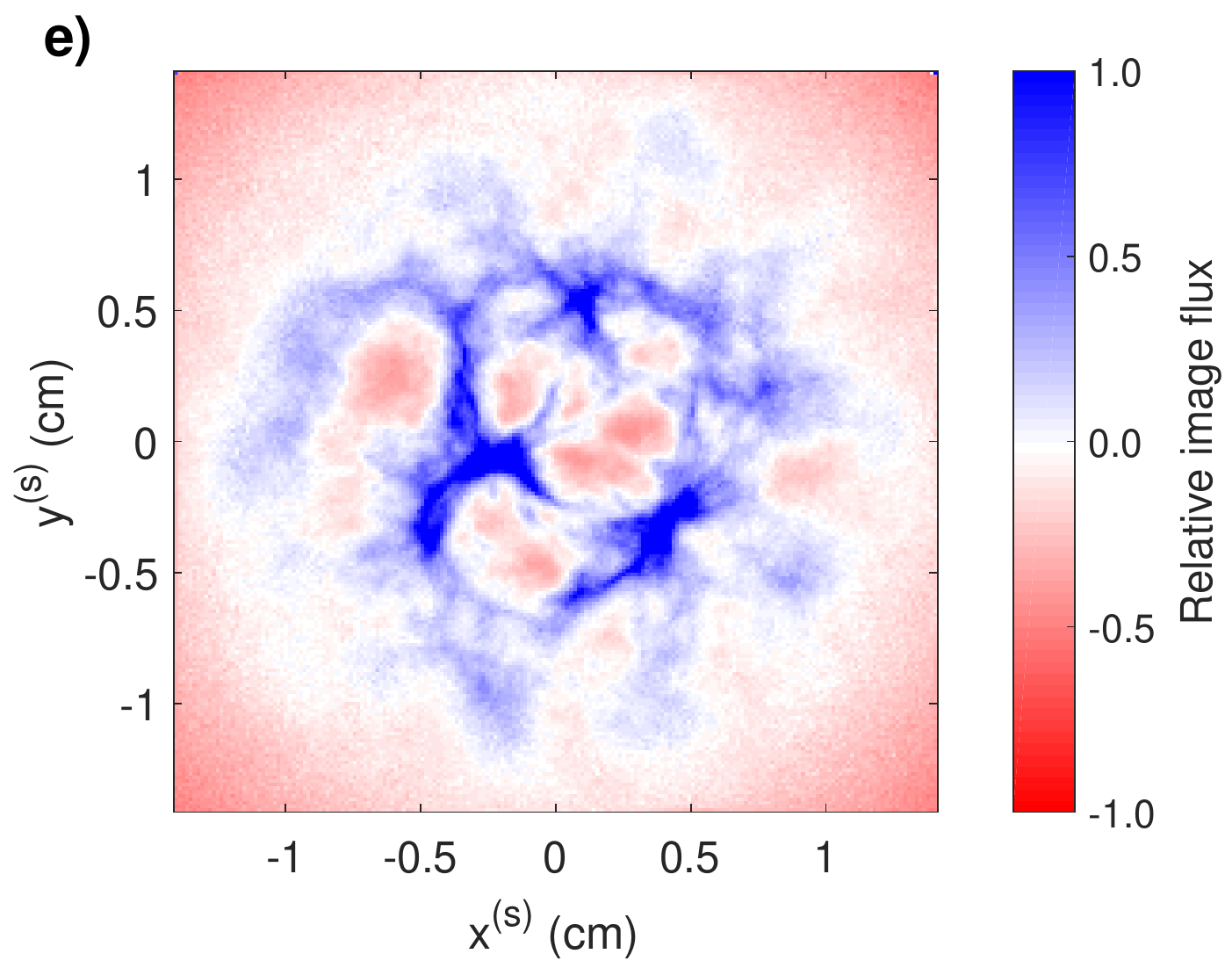}
    \end{subfigure} %
    \begin{subfigure}{.48\textwidth}
        \centering
        \includegraphics[width=\linewidth]{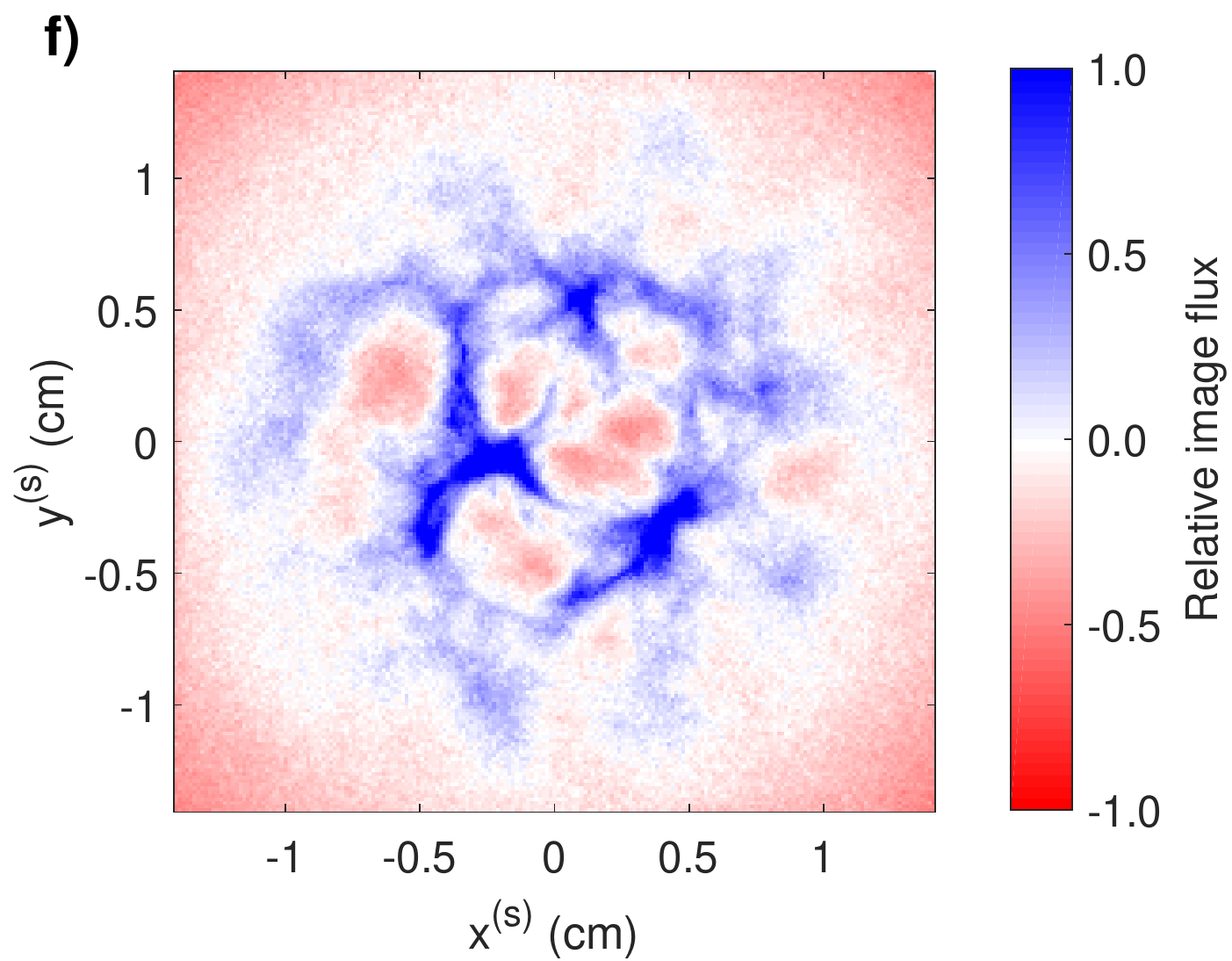}
    \end{subfigure} %
\caption{\textit{Effect of initial flux inhomogeneity on well-posedness of field reconstruction algorithms.} \textbf{a)} Non-uniform initial 3.3-MeV proton image-flux distribution, with $\Psi_0\!\left(\mathbf{x}_{\bot0}\right) = \Psi_0/4 \left(5-6 x_{\bot0}^2/l_i^2\right)$. Imaging parameters of the set-up are the same as those described in Figure 4. \textbf{b)} Uniform initial flux distribution. \textbf{c)} Path-integrated perpendicular magnetic field experienced by beam with initial flux distribution shown in a). The magnetic field is the same as that described in Figure 2, with $B_{rms,0} = 15 \, \mathrm{kG}$. \textbf{d)} Path-integrated perpendicular magnetic field experienced by initial flux distribution shown in b). The magnetic field is constructed by applying the field reconstruction algorithm to proton-flux image c). \textbf{e)} 3.3-MeV proton-flux image resulting from initial flux distribution a) experiencing path-integrated field c). \textbf{f)} 3.3-MeV proton-flux image resulting from initial flux distribution b) experiencing path-integrated field d). 
} \label{Variableinitialflux}
\end{figure}

Unfortunately, the ramifications of an unknown initial flux distribution for successful extraction of magnetic field statistics cannot be disregarded: in general, path-integrated fields are no longer uniquely determined. Conceptually this is because it becomes impossible to distinguish between variations in the image flux resulting from magnetic deflections, and those due to variations in the initial flux. We illustrate this phenomenon with a numerical example -- Figure \ref{Variableinitialflux}.
Taking two distinct path-integrated fields (Figures \ref{Variableinitialflux}c and \ref{Variableinitialflux}d), but also choosing two distinct initial flux distributions (Figures \ref{Variableinitialflux}a and \ref{Variableinitialflux}b), it can be seen that it is possible to produce a near identical image-flux distribution (Figures \ref{Variableinitialflux}e and \ref{Variableinitialflux}f). Naive assumption of a uniform initial flux distribution when applying a field reconstruction algorithm can therefore lead to prediction of non-physical field structures: a far from satisfactory outcome.

The distorting effect of initial flux inhomogeneities, however, can to a certain extent be countered. Variations in initial flux are typically much smaller than the mean flux over solid angles $\leq 1.1^{\circ}$~\cite{M12}.
As a result, short-scale image-flux deviations due to deflections, particularly strong deviations from nonlinear focusing, are distinguishable from potential initial flux structures. Difficulties then only arise when considering weaker, long-scale flux variation (strong long-scale variation is inevitably associated with the appearance of focused narrow image-flux features). Therefore, to isolate genuine path-integrated magnetic field structures, we propose applying a high-pass filter to the image-flux distribution to remove long-scale variation. Field reconstruction algorithms applied to the filtered image therefore give a reasonable estimate of fields below the scale of the filtering.

We do, however, note that long-scale path-integrated magnetic fields -- if present -- could in principle alter the positioning of short-scale image-flux structures, giving rise to the possibility that path-integrated fields reconstructed from filtered images are distorted compared to the true fields at that scale. The importance of this effect can be tested by comparing the results of the field reconstruction algorithm applied to the filtered image with the reconstructed path-integrated field deduced from the unfiltered image, but with filtering applied directly post-reconstruction. In any case, since long-scale variations in flux are typically not large, intuitively this effect should not be significant in practice.

\subsubsection{Smearing effects due to finite source size} \label{SmearFinSource}

As mentioned in Section \ref{Assum}, proton-imaging set-ups in practice have a small but finite source size, placing a lower limit on the size of magnetic structures which can be imaged. This statement can be made more precise by calculating the effect of the finite source of the perpendicular velocity distribution of the beam. It can then be shown using kinetic theory (Appendix \ref{ScreenDistFiniteSourceDev}) that when a magnetic field configuration is imaged with a proton beam generated by a finite source, the resulting image-flux distribution $\tilde{\Psi}\!\left(\mathbf{x}_\bot^{\left(s\right)}\right)$ is equal to the image-flux distribution generated by a point source (the `unsmeared flux' $\Psi$), but convolved with a point-spread function:
\begin{equation}
\tilde{\Psi}\!\left(\mathbf{x}_\bot^{\left(s\right)}\right) = \int \mathrm{d}^2  \tilde{\mathbf{x}}_{\bot}^{\left(s\right)} \, \Psi\!\left(\tilde{\mathbf{x}}_\bot^{\left(s\right)}\right) S\!\left(\mathbf{x}_{\bot}^{\left(s\right)}-\tilde{\mathbf{x}}_{\bot}^{\left(s\right)}\right) \, ,  \label{smearedscreenflux}
\end{equation}
where $S$ is given by
\begin{equation}
S\!\left(\mathbf{x}_{\bot}^{\left(s\right)}-\tilde{\mathbf{x}}_{\bot}^{\left(s\right)}\right) = \frac{V^2}{r_s^2} P\!\left(\frac{\tilde{\mathbf{x}}_{\bot}^{\left(s\right)}-\mathbf{x}_{\bot}^{\left(s\right)}}{r_s}V\right) \, .
\end{equation}
Here, $P =  P\!\left(\delta \mathbf{v}_{\bot0}\right)$ is the distribution function of perpendicular velocities associated with the source's finite size; it is a function of
\begin{equation}
\delta \mathbf{v}_{\bot0} = \mathbf{v}_{\bot0}-\frac{\mathbf{x}_{\bot0}}{r_i}V \, ,
\end{equation}
which quantifies the degree of deviation from the paraxial approximation for the initial beam-proton velocity. The form of $P$ depends on three-dimensional emission profile of the proton source (a calculation for $P$ is the case of a uniformly emitting sphere of radius $a$ is presented in Appendix \ref{ScreenDistFiniteSourceDev}). 

\begin{figure}[htbp]
\centering
    \begin{subfigure}{.36\textwidth}
        \centering
        \includegraphics[width=0.95\linewidth]{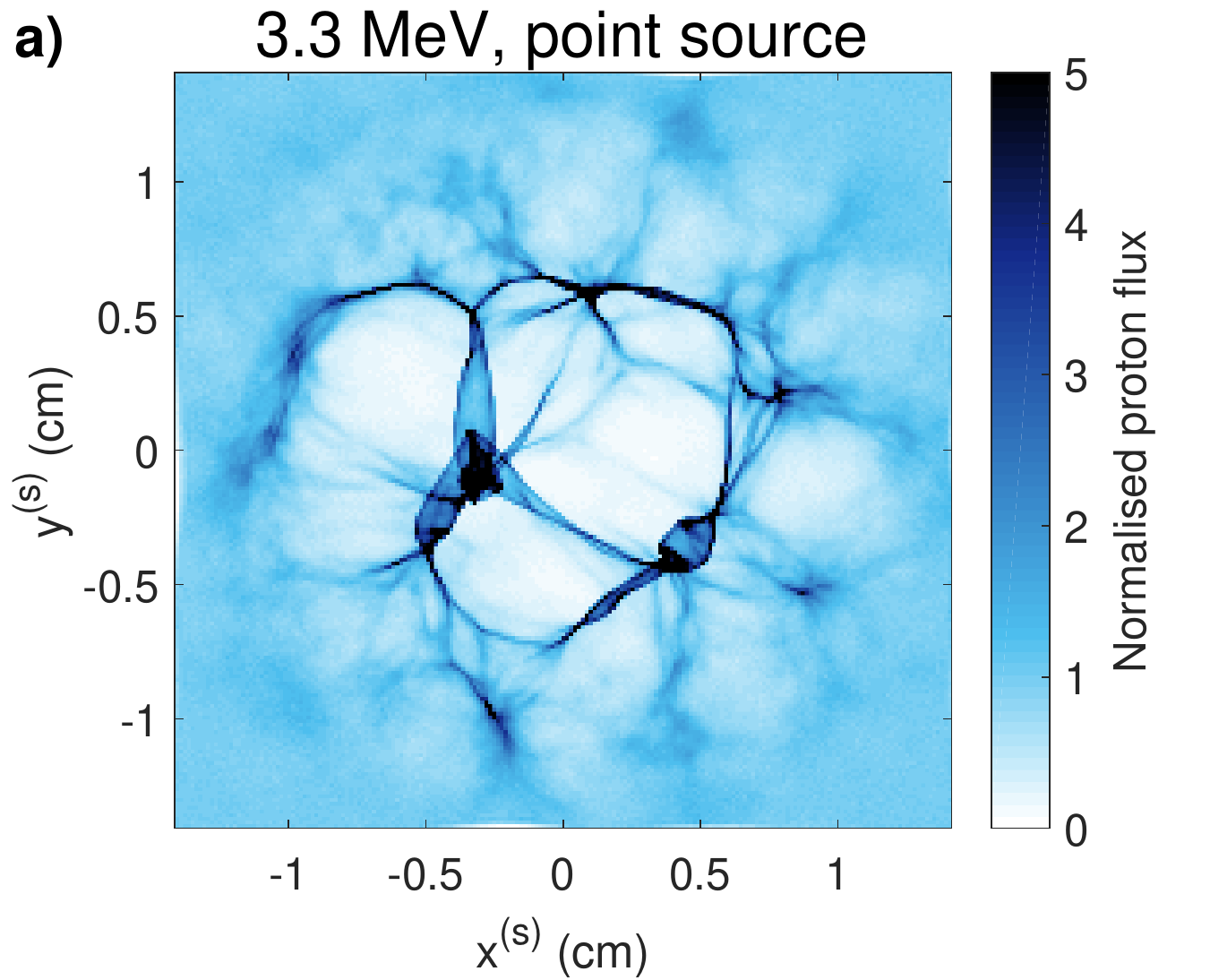}
    \end{subfigure} %
    \begin{subfigure}{.36\textwidth}
        \centering
        \includegraphics[width=0.95\linewidth]{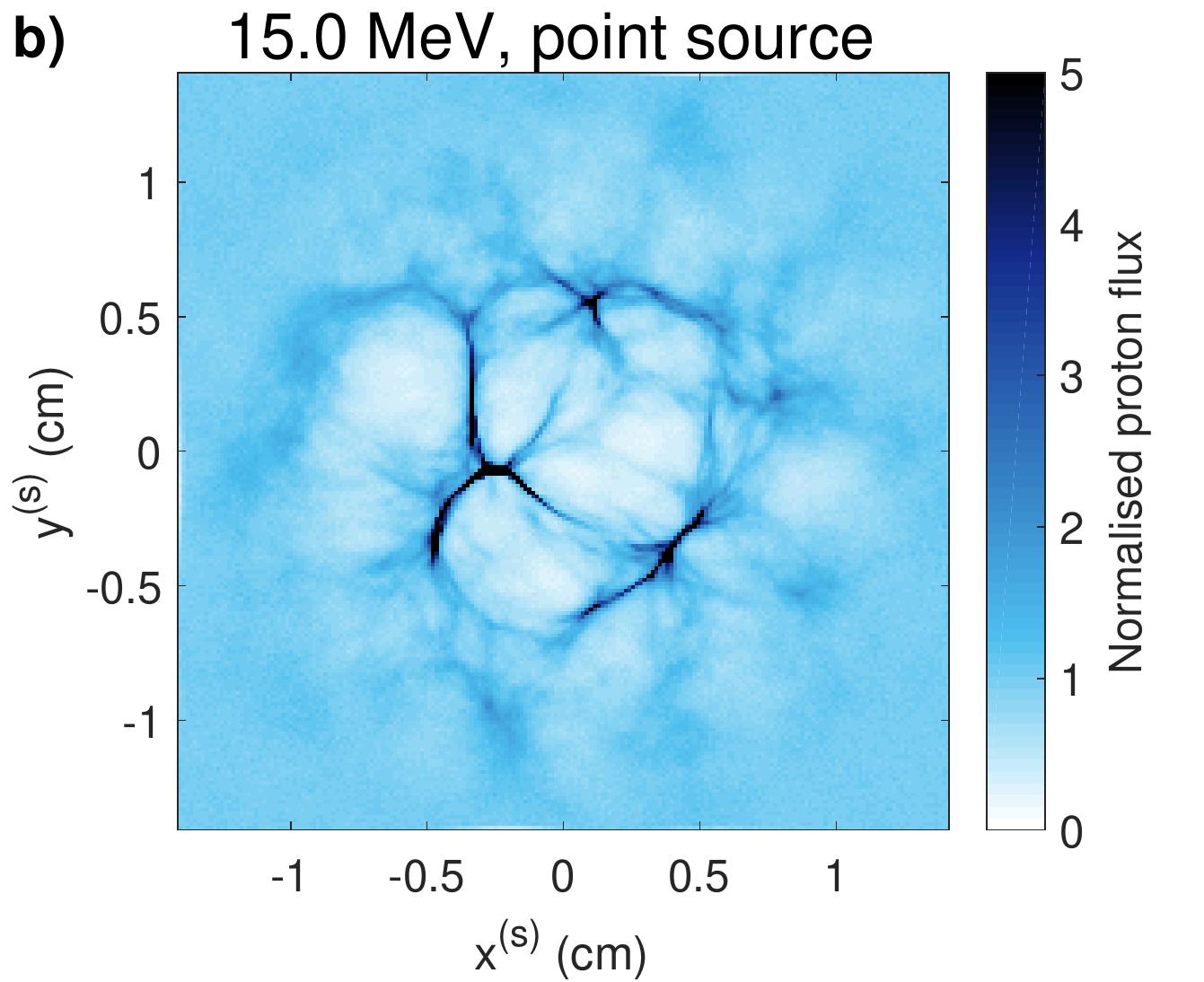}
    \end{subfigure} %
        \begin{subfigure}{.36\textwidth}
        \centering
        \includegraphics[width=0.95\linewidth]{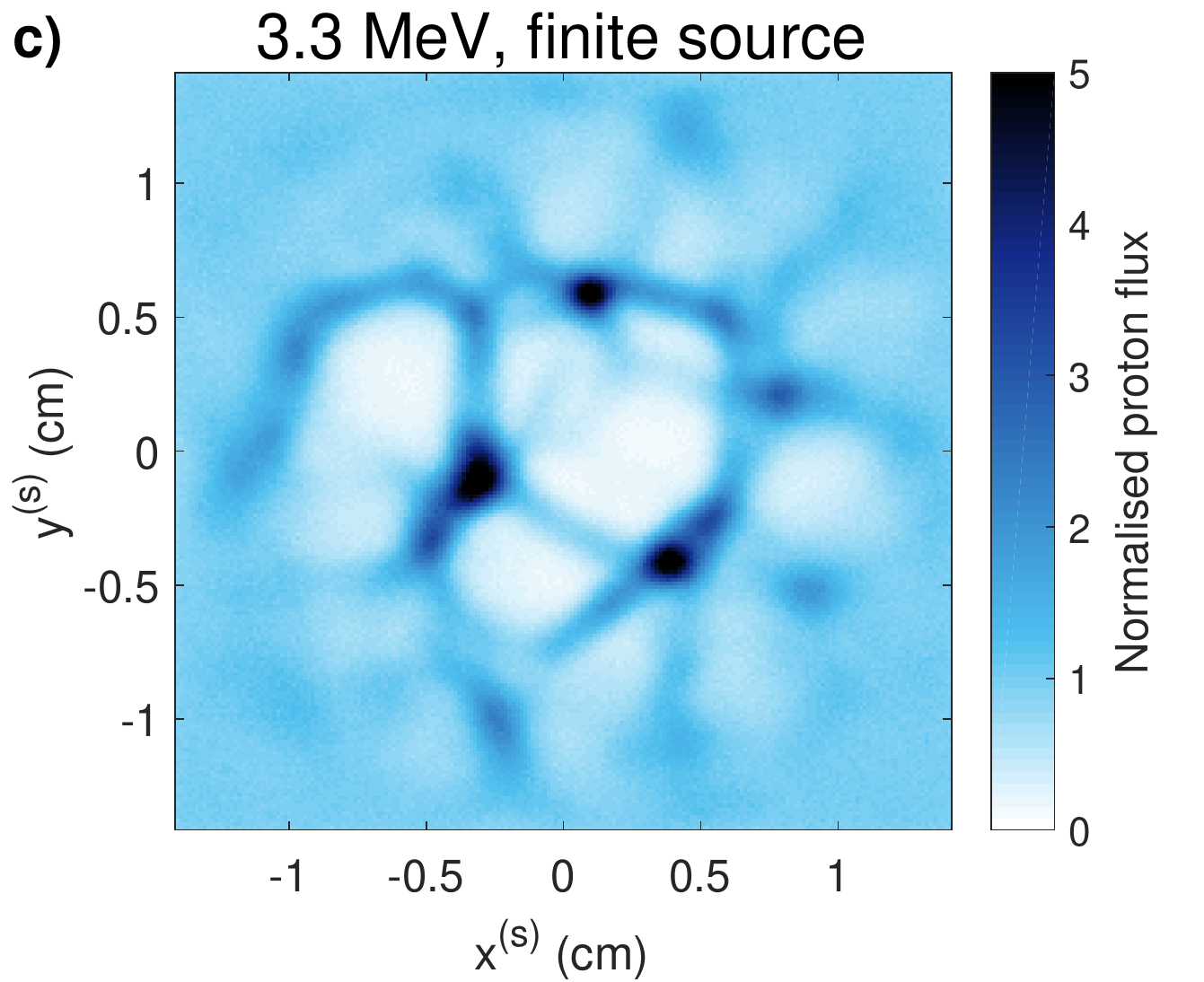}
    \end{subfigure} %
    \begin{subfigure}{.36\textwidth}
        \centering
        \includegraphics[width=0.95\linewidth]{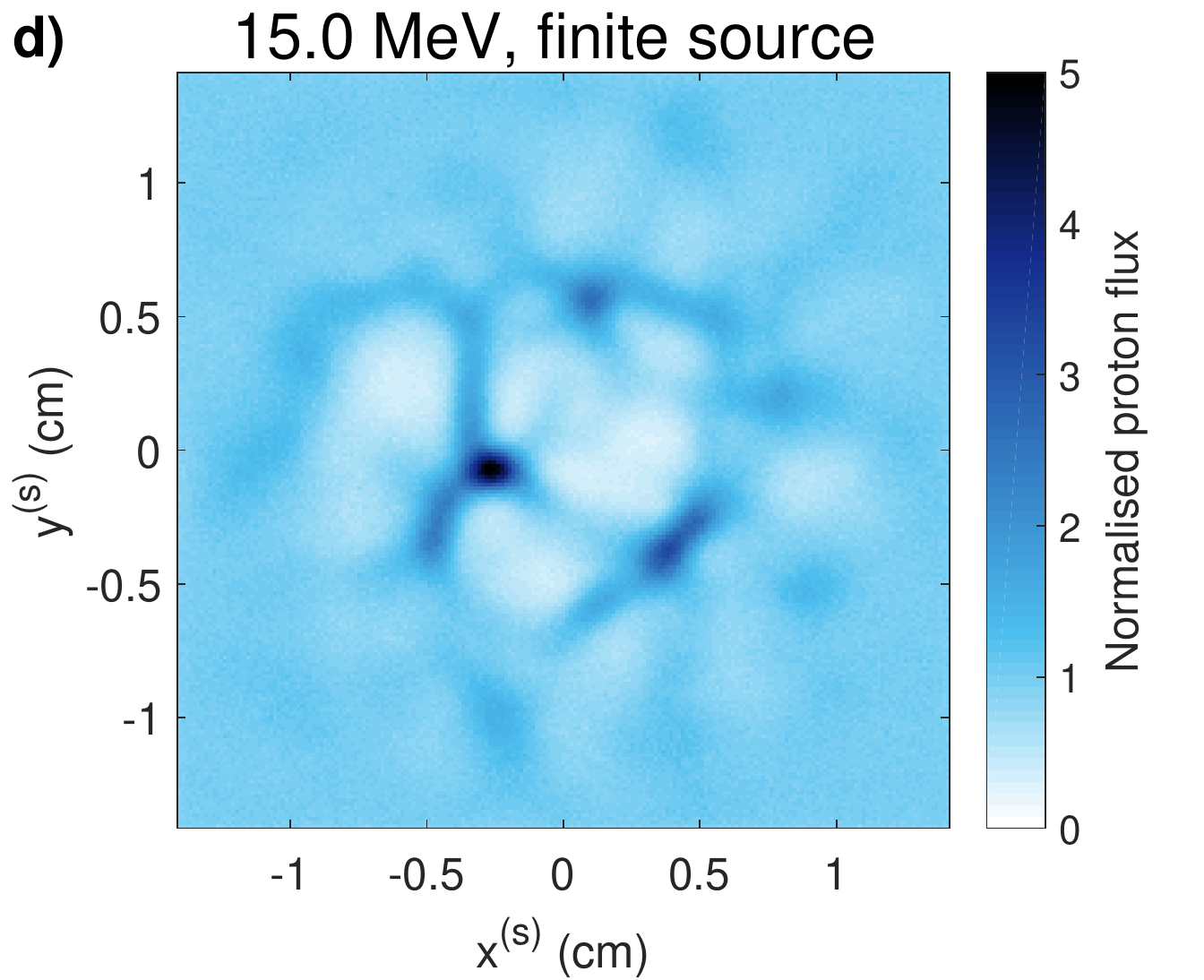}
    \end{subfigure} %
    \begin{subfigure}{.36\textwidth}
        \centering
        \includegraphics[width=0.95\linewidth]{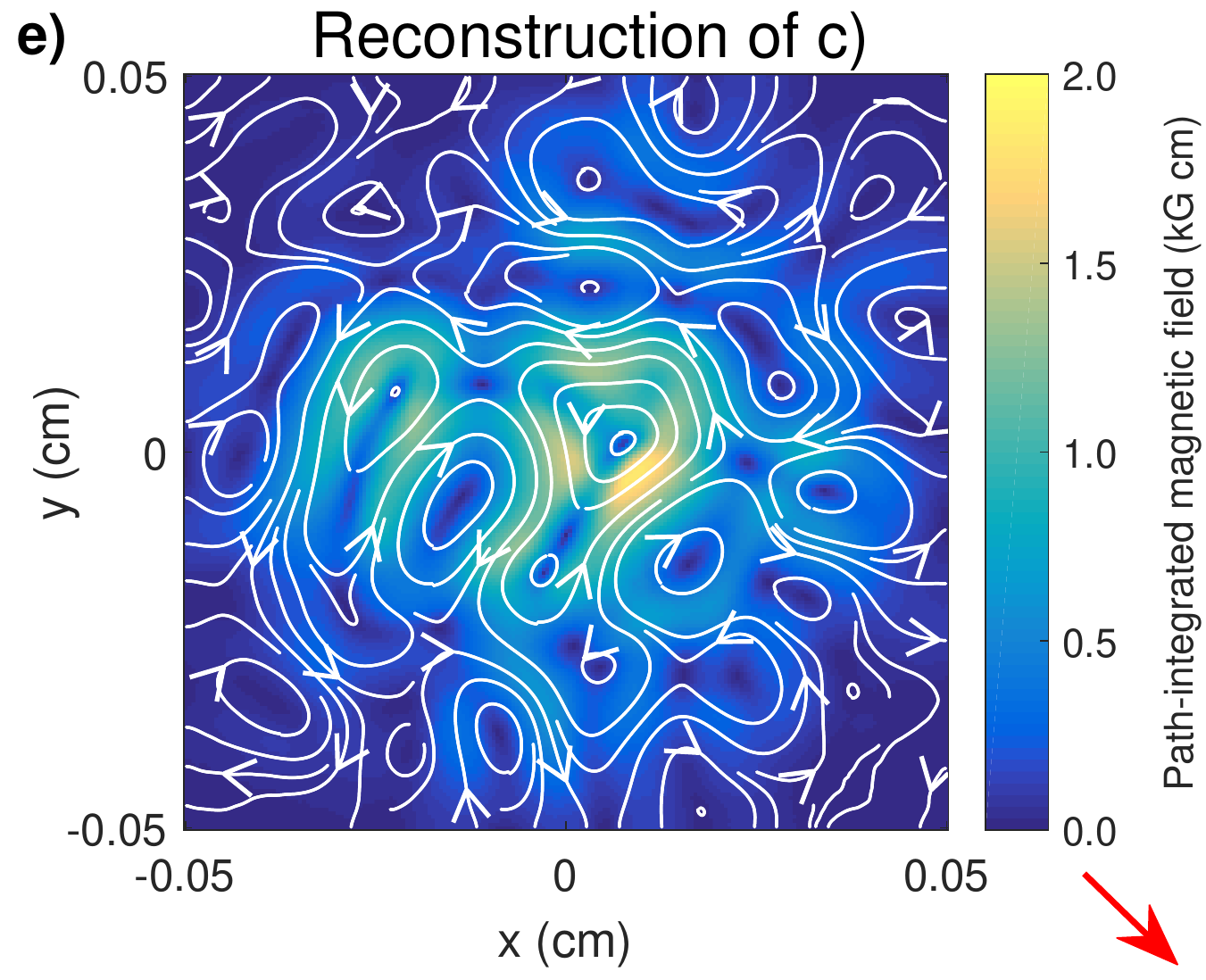}
    \end{subfigure} %
    \begin{subfigure}{.36\textwidth}
        \centering
        \includegraphics[width=0.95\linewidth]{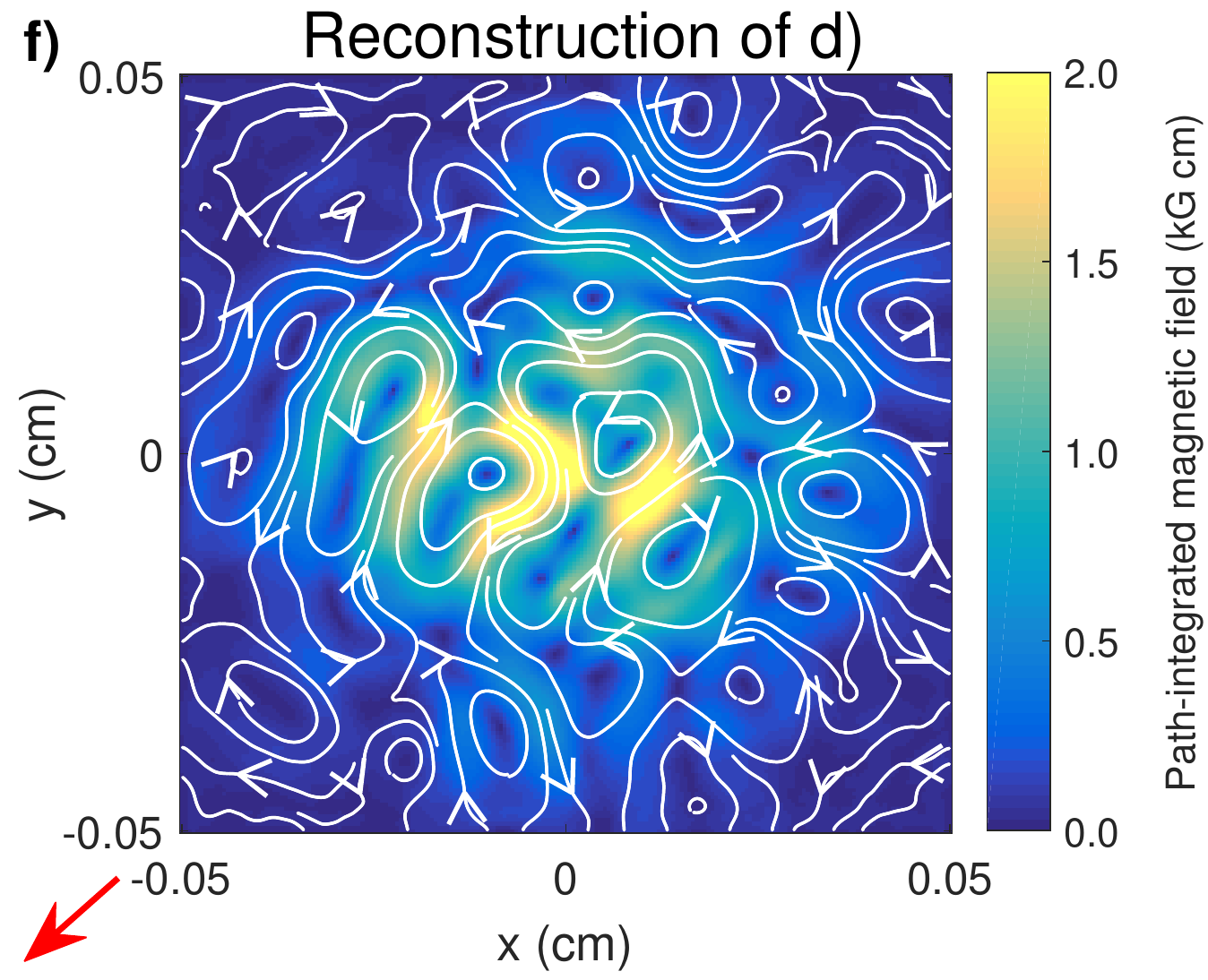}
    \end{subfigure} %
        \begin{subfigure}{.36\textwidth}
        \centering
        \includegraphics[width=0.95\linewidth]{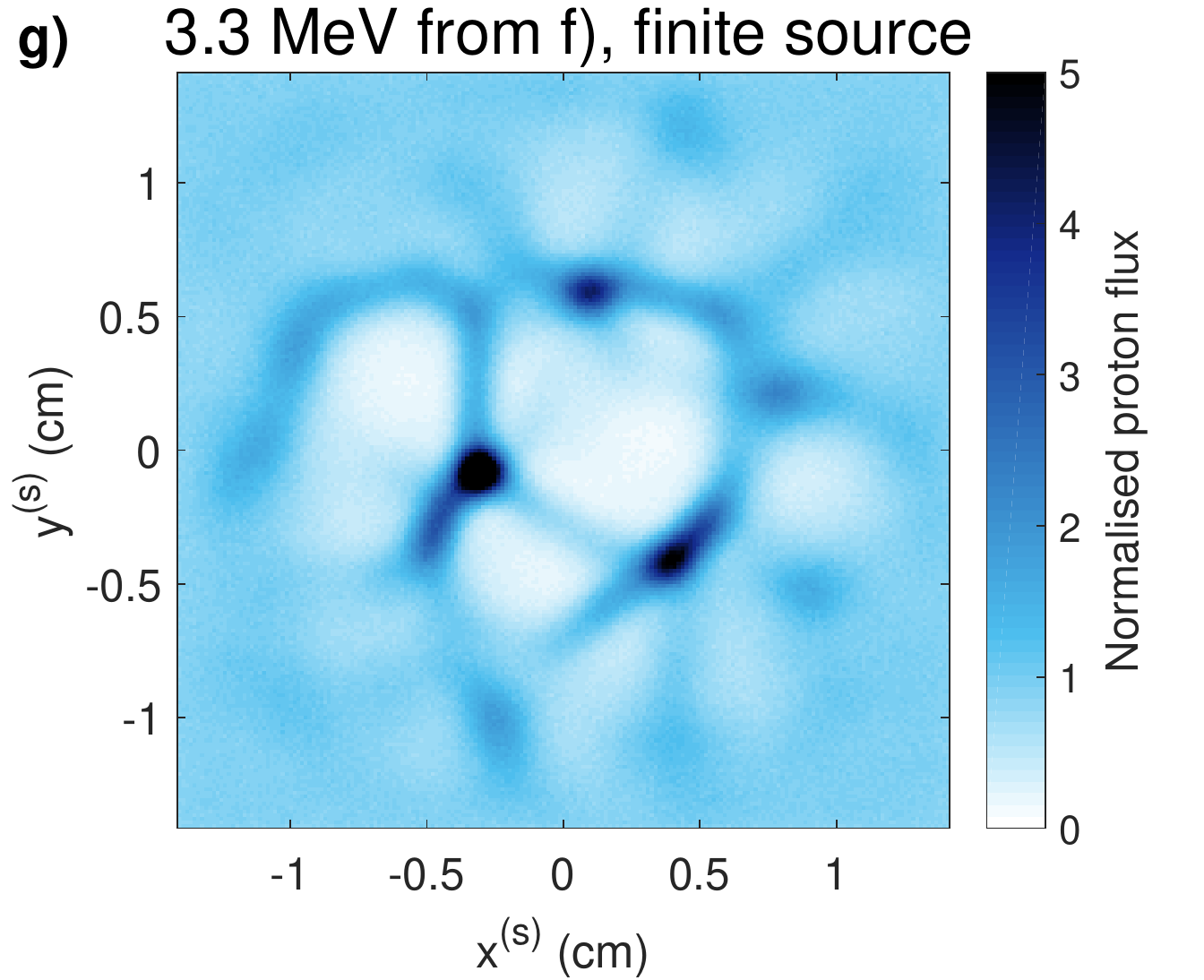}
    \end{subfigure} %
    \begin{subfigure}{.36\textwidth}
        \centering
        \includegraphics[width=0.95\linewidth]{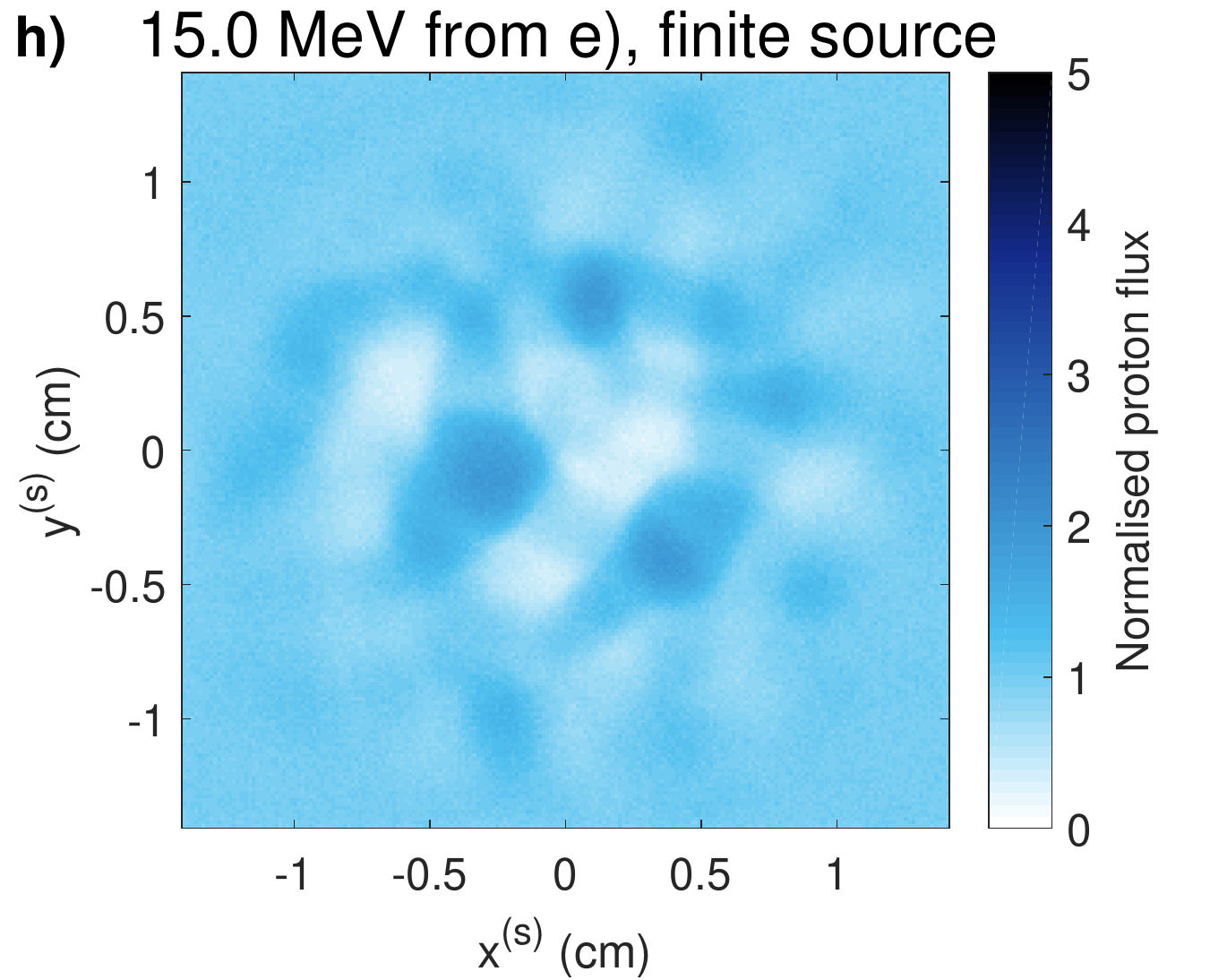}
    \end{subfigure} %
\caption{\textit{Distinguishing the nonlinear injective and caustic contrast regimes using nonlinear field-reconstruction algorithms and multiple beam energies}.  
\textbf{a)} 3.3 MeV Proton-flux image of magnetic field described in Figure 2, with $B_{rms,0} = 64 \, \mathrm{kG}$, and same imaging parameters as those described in Figure 3. 
\textbf{b)} Same as a), but with imaging beam composed of 15.0 MeV protons. 
\textbf{c)} Same as a), but now imaging implemented using a finite uniform spherical proton source with diameter $a = 50 \, \mu \mathrm{m}$.
\textbf{d)} Same as c), but with imaging beam composed of 15.0 MeV protons. 
\textbf{e)} Reconstructed deflection field obtained by applying the field-reconstruction algorithm (described in Section \ref{NonLinInjRgme}) to c).
\textbf{f)} Same as e), but algorithm applied to d).
\textbf{g)} Predicted 3.3 MeV proton-flux image, assuming imaging proton beam experienced reconstructed perpendicular-deflection field shown in f) as a result of traversing magnetic field (the implementation of this is discussed in Appendix \ref{NumSimFluxImageGen}). Under the caustic detection scheme outlined in the main text, this image is to be compared with c). 
\textbf{h)} Predicted 15.0 MeV proton-flux image, assuming imaging proton beam experienced reconstructed perpendicular-deflection field shown in e). This image is to be compared with d). 
} \label{Golitsynfluxident}
\end{figure}

The limitation of image resolution by a finite source size places a number of constraints on the possibility of reconstructing the magnetic-energy spectrum. Firstly, as mentioned in Section \ref{CauRgme}, smearing of fine image-flux structure can make the differentiation of the nonlinear injective and caustic regimes difficult. This is illustrated in Figure \ref{Golitsynfluxident}. 
Figures \ref{Golitsynfluxident}a and \ref{Golitsynfluxident}b show point-source proton-flux images of the same artificial test stochastic magnetic field introduced in Figure \ref{GolitysncompactfieldSec2} (with $B_{rms,0} \approx 60 \, \mathrm{kG}$), but created using two imaging proton species with different speeds: 3.3 MeV (Figure \ref{Golitsynfluxident}a) and 15.0 MeV (Figure \ref{Golitsynfluxident}b). Since $\mu$ is inversely proportional to the initial imaging proton speed, $\mu$ for the 15.0-MeV proton-flux image is lower -- with the consequence that its plasma-image mapping is injective, unlike the 3.3-MeV proton plasma-image mapping. The two point-source proton-flux images are qualitatively distinct -- pairs of caustics are evident in Figure \ref{Golitsynfluxident}a -- and hence can be identified as belonging to the caustic and nonlinear injective regimes respectively. Figures \ref{Golitsynfluxident}c and \ref{Golitsynfluxident}d show the same fields imaged with protons generated from a finite spherical source. This reduces the spatial resolution of the image -- and consequently, the sharp caustic structures disappear.  

To distinguish between the nonlinear injective and caustic regimes, we need more information than can be provided by a single proton image. Fortunately, typical experimental methods for generating imaging proton beams produce at least two beam energies simultaneously, which can be imaged independently~\cite{L06,W01}. If the transit and beam duration times of all proton beams are much shorter than the timescales on which dynamical evolution of the magnetic field occurs, then the proton-flux images from all beams will be of approximately the same magnetic field (see Section \ref{Assum} for a discussion of this). 

Comparing proton images of the same magnetic field at different imaging beam energies is a useful technique for identifying caustics. Most simply, if if proton images contain caustics, irrespective of the beam energy, then the width of strong, positive image-flux structures will typically increase with lowered beam-energy; in contrast, if all images fall into the nonlinear injective regime, then the narrowest structures will occur at the lowest energies. This principle is illustrated in Figures \ref{Golitsynfluxident}c and \ref{Golitsynfluxident}d, where strong, positive relative image-flux features are broader for the 3.3 MeV protons than the 15.0 MeV ones -- suggesting that the former contains smeared caustics. 

A more robust caustic detection scheme comes from application of the field-reconstruction algorithm outlined for $\mu < \mu_{c}$ in Section \ref{NonLinInjRgme} on the different beam-energy proton-flux images of the same magnetic structures. If a path-integrated field reconstructed from one proton species is close to the true path-integrated field, then all imaging protons species should have seen that reconstructed field: thus the proton-flux images generated using the plasma-image mapping combined with the predicted path-integrated field should be consistent with the actual proton-flux images (this procedure is discussed in Appendix \ref{NumSimFluxImageGen}). Figures \ref{Golitsynfluxident}g and \ref{Golitsynfluxident}h show the 3.3-MeV and 15.0-MeV proton-flux images predicted using the reconstructed perpendicular-deflection fields for 15.0 MeV (Figure \ref{Golitsynfluxident}f) and 3.3 MeV (Figure \ref{Golitsynfluxident}e) protons respectively, followed by artificial smearing with a point-spread function appropriate for a uniform finite spherical source (see Appendix \ref{FiniteSourcePSF}). Figure \ref{Golitsynfluxident}g is very similar to Figure \ref{Golitsynfluxident}c, verifying the validity of the reconstructed perpendicular-deflection field shown in Figure \ref{Golitsynfluxident}. However, Figure \ref{Golitsynfluxident}h disagrees substantially with Figure \ref{Golitsynfluxident}d: for example, the width of positive image-flux structures in \ref{Golitsynfluxident}h is much greater than the true widths in Figure \ref{Golitsynfluxident}d. We conclude that the field-reconstruction algorithm provides an effective method for detecting the presence of caustics.

\begin{figure}[htbp]
\centering
    \begin{subfigure}{.36\textwidth}
        \centering
        \includegraphics[width=0.95\linewidth]{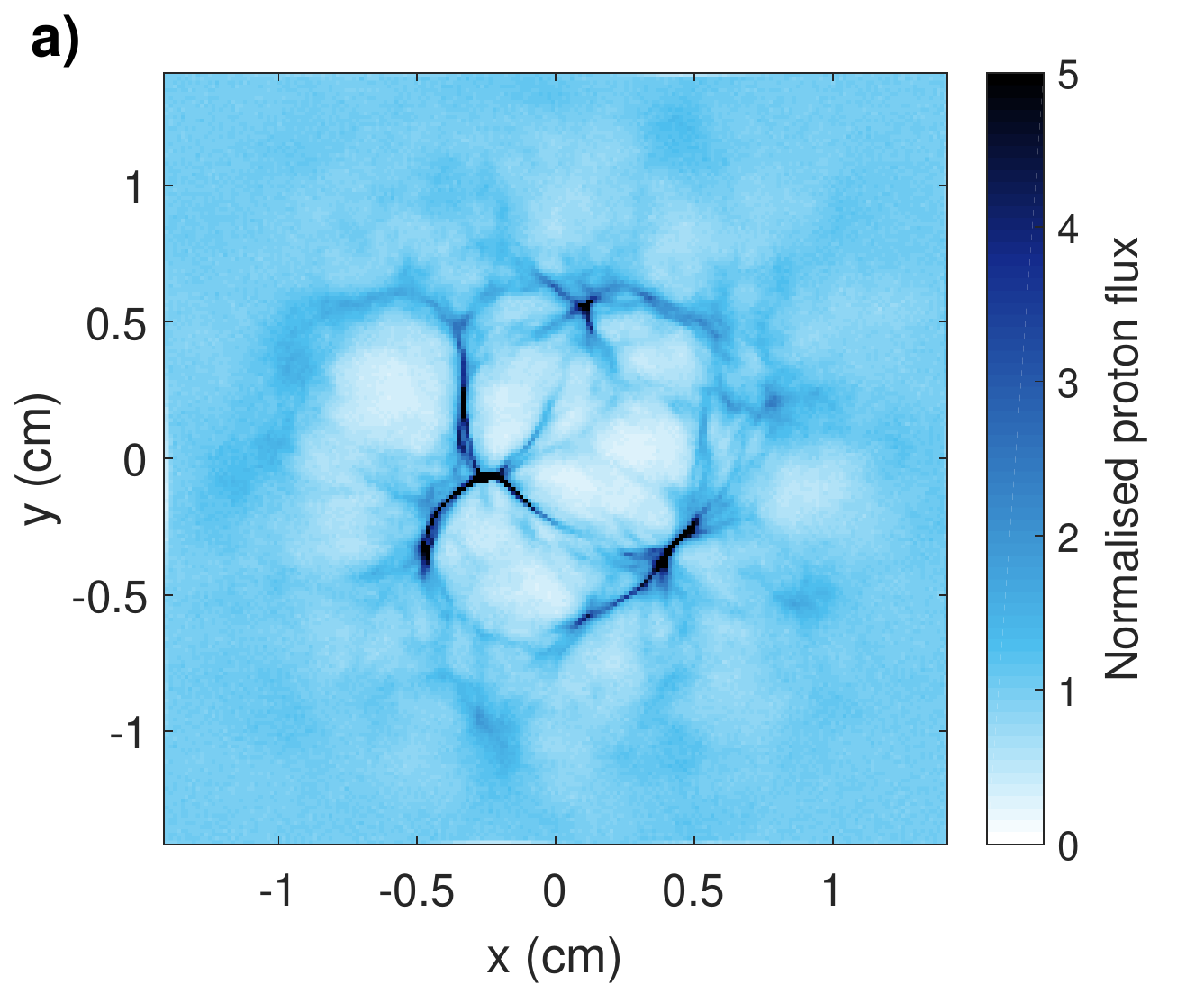}
    \end{subfigure} %
    \begin{subfigure}{.37\textwidth}
        \centering
        \includegraphics[width=0.95\linewidth]{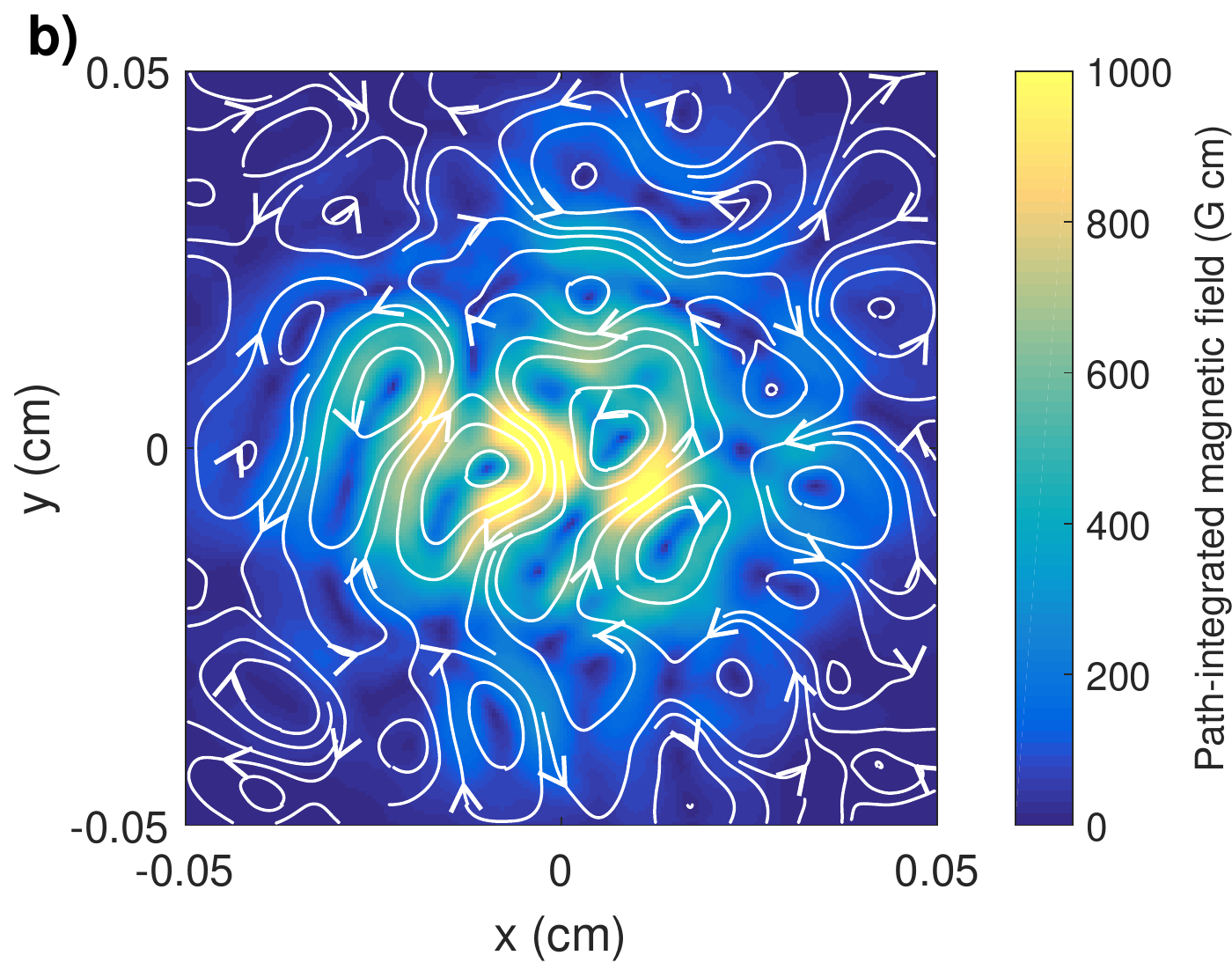}
    \end{subfigure} %
        \begin{subfigure}{.36\textwidth}
        \centering
        \includegraphics[width=0.95\linewidth]{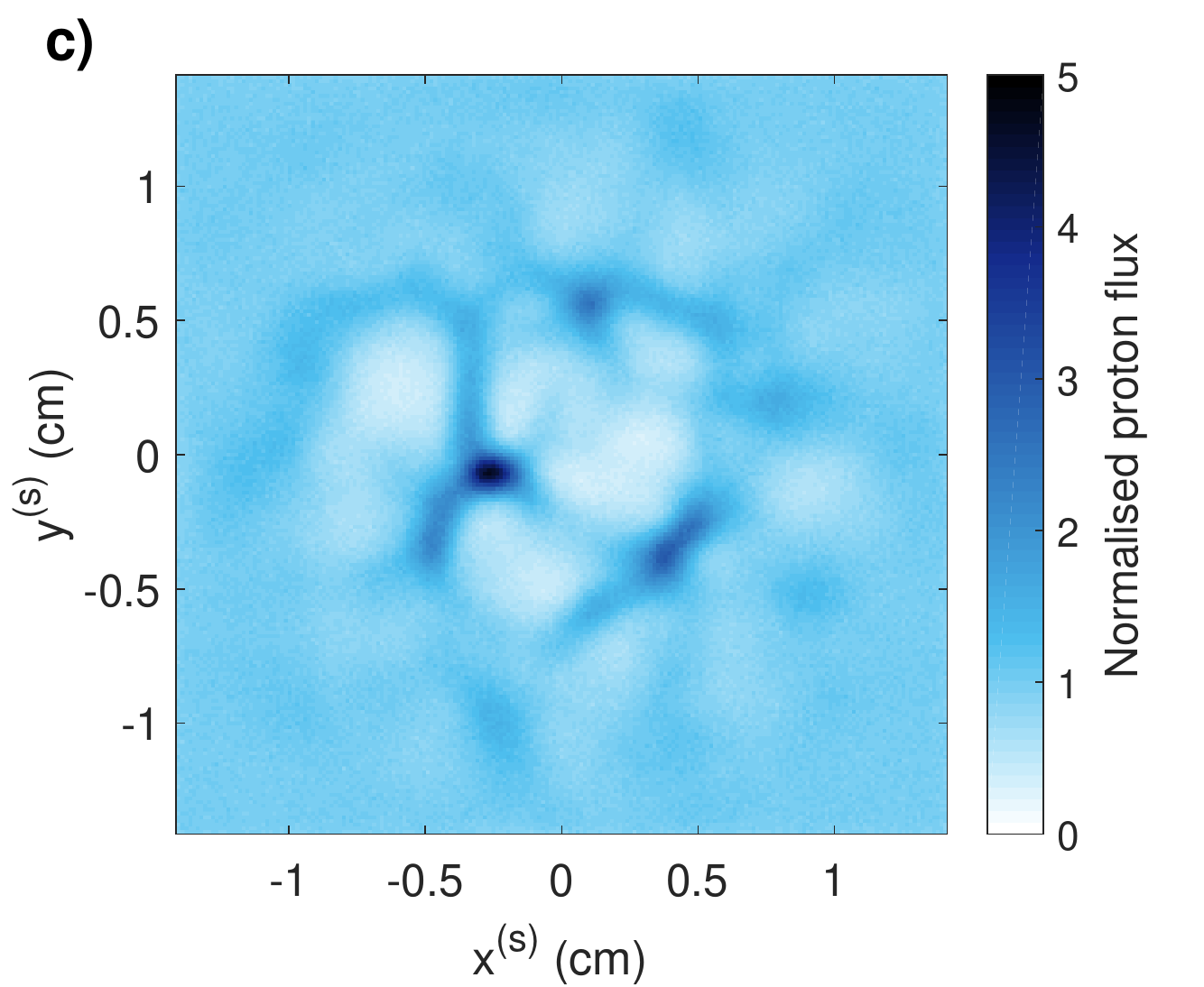}
    \end{subfigure} %
    \begin{subfigure}{.37\textwidth}
        \centering
        \includegraphics[width=0.95\linewidth]{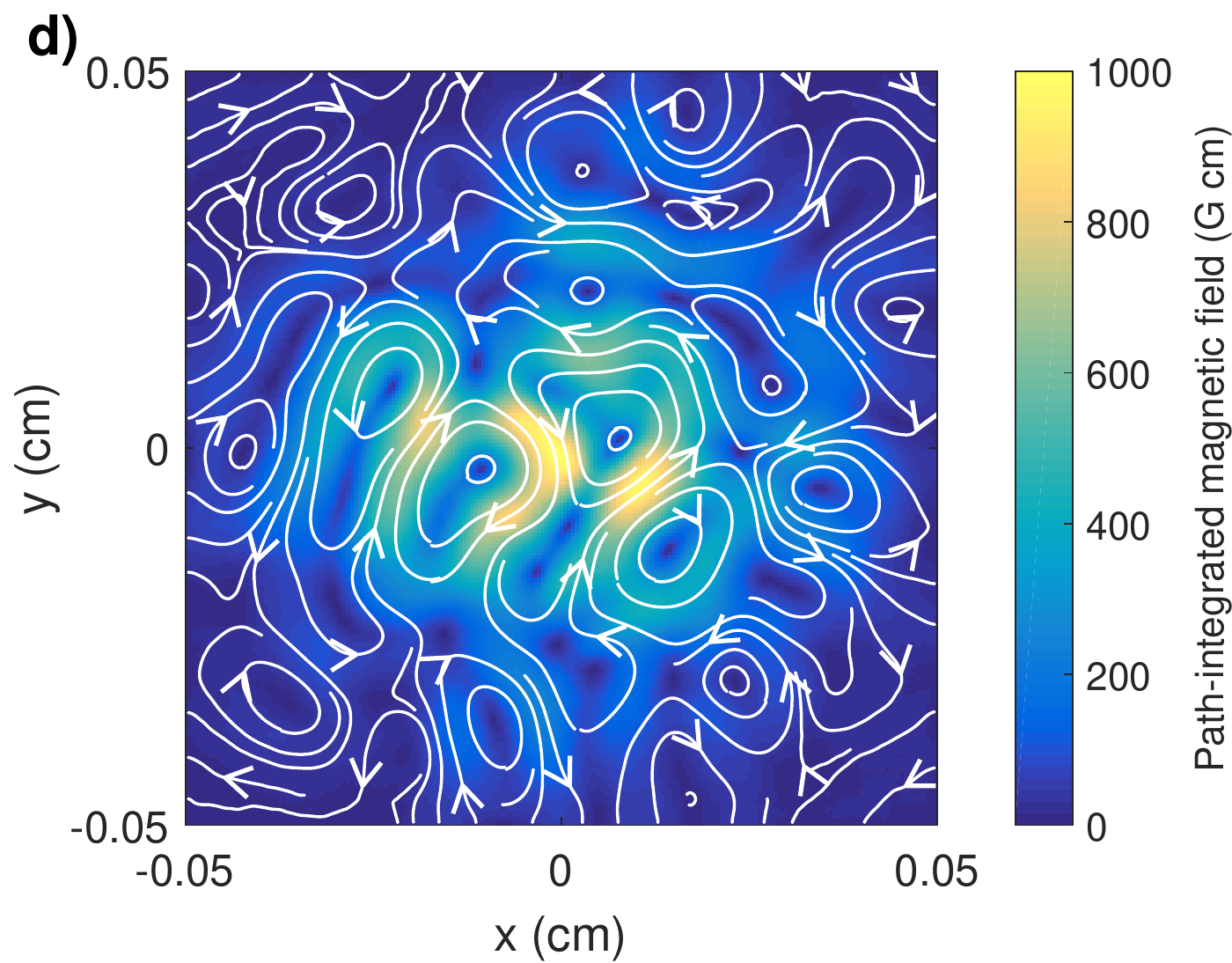}
    \end{subfigure} %
    \begin{subfigure}{.36\textwidth}
        \centering
        \includegraphics[width=0.95\linewidth]{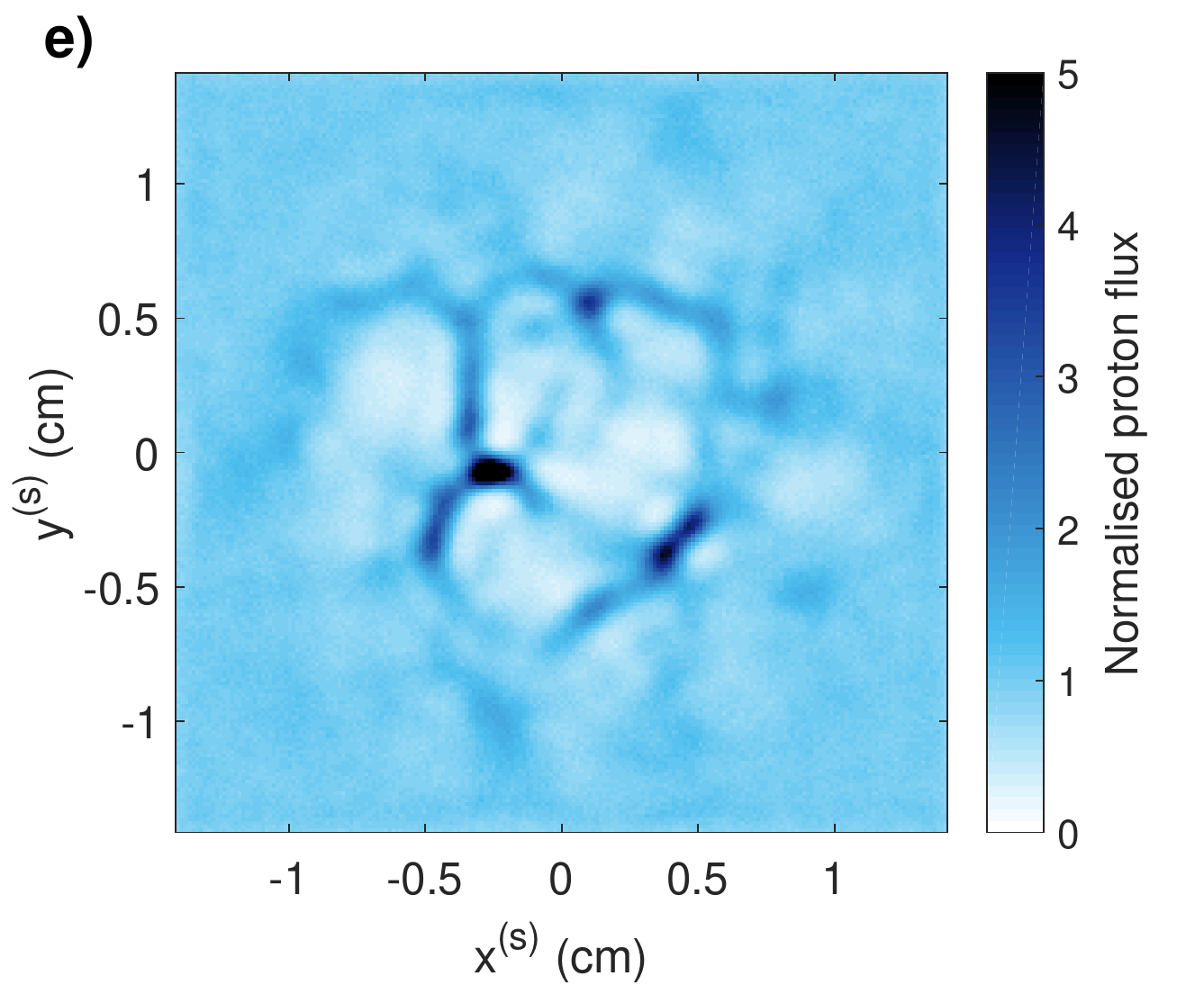}
    \end{subfigure} %
    \begin{subfigure}{.37\textwidth}
        \centering
        \includegraphics[width=0.95\linewidth]{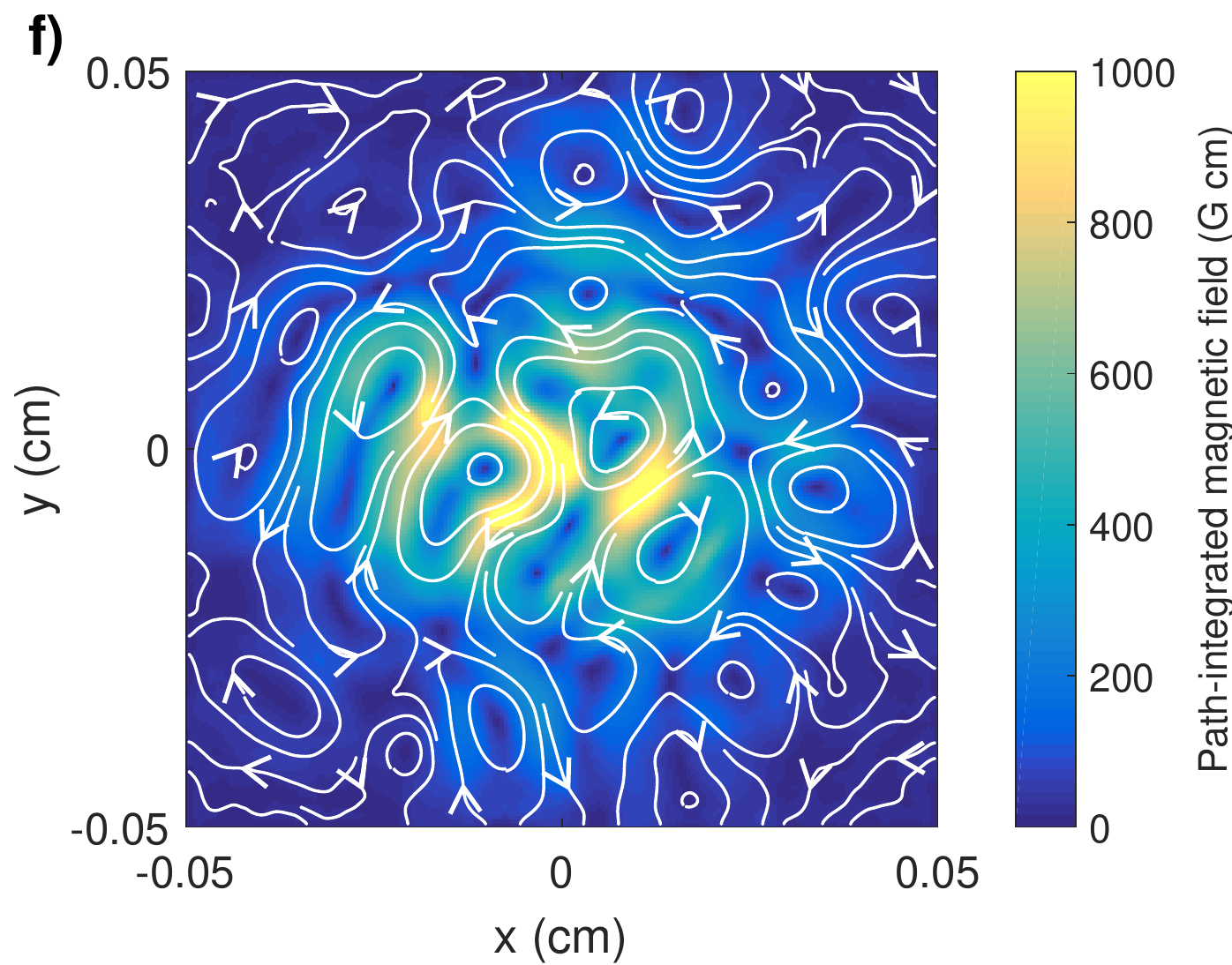}
    \end{subfigure} %
        \begin{subfigure}{.36\textwidth}
        \centering
        \includegraphics[width=0.95\linewidth]{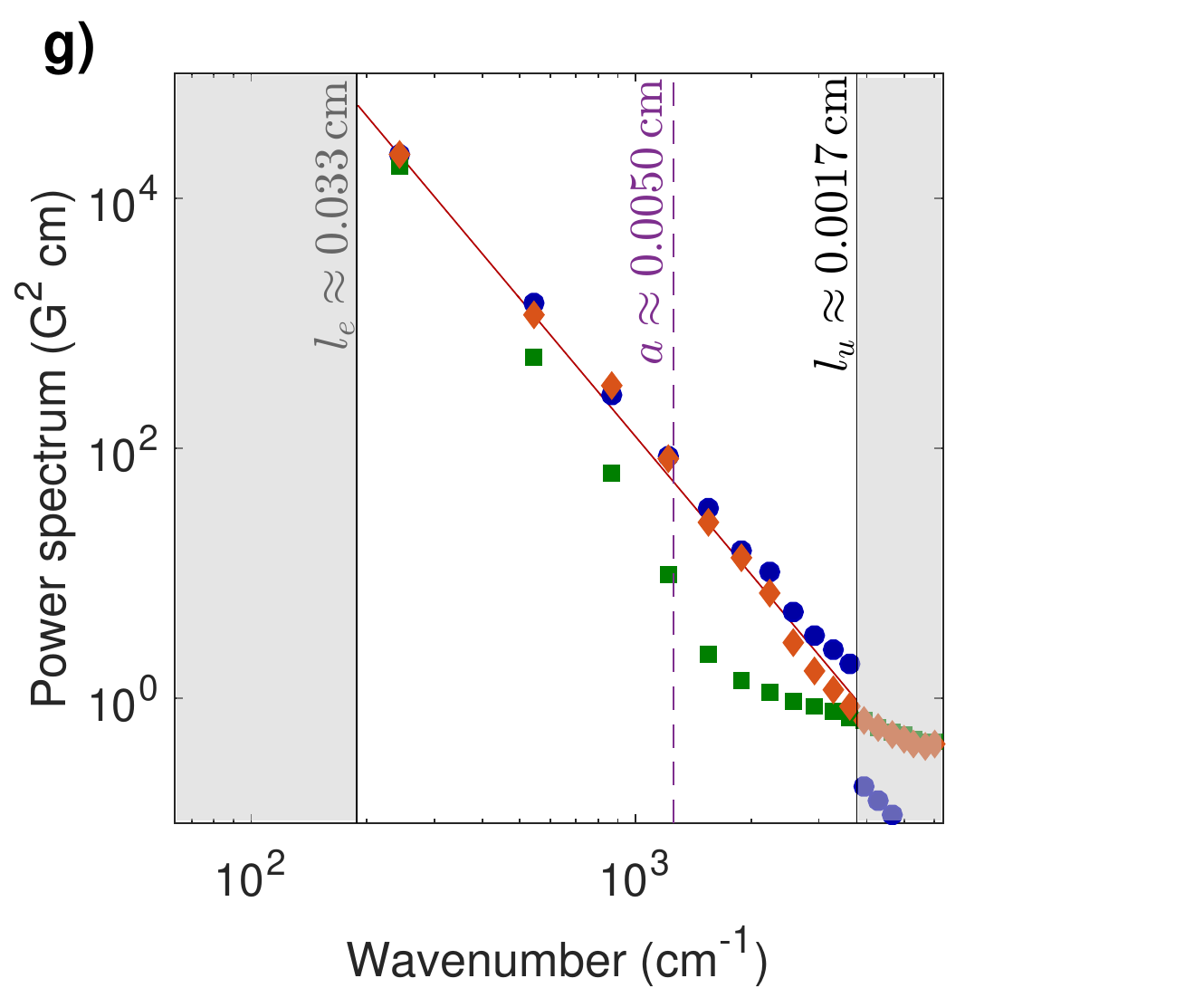}
    \end{subfigure} %
    \begin{subfigure}{.37\textwidth}
        \centering
        \includegraphics[width=0.95\linewidth]{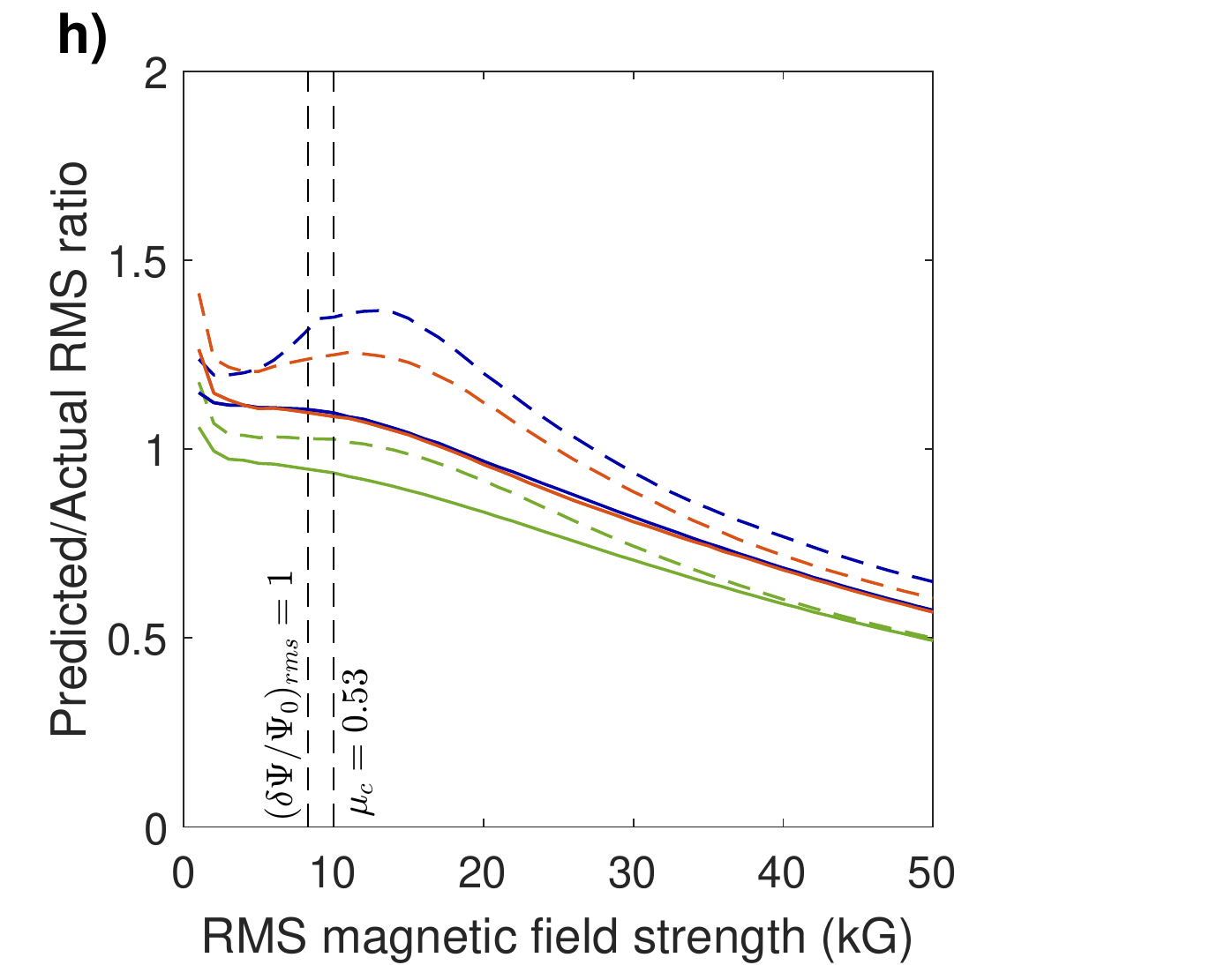}
    \end{subfigure} %
\caption{\textit{Effect of finite source size on successful extraction of magnetic field statistics.}  
\textbf{a)} `Unsmeared' 3.3-MeV proton-flux image created by imaging Golitsyn field, $B_{rms,0} = 30 \, \mathrm{kG}$ (reproduction of Figure 4c for convenience of reference). \textbf{b)} Predicted path-integrated magnetic field resulting from application of field reconstruction algorithm to Figure a). \textbf{c)} `Smeared' 3.3-MeV proton-flux image created by imaging same field as a), but with a finite proton source, with diameter $a = 50 \, \mu \mathrm{m}$. 
\textbf{d)} Same as b), but with field reconstruction algorithm applied to c). 
\textbf{e)} `De-smeared' 3.3-MeV proton image subsequent to application of Richardson-Lucy algorithm for ten iterations. 
\textbf{f)} Same as b), but with field reconstruction algorithm applied to e). 
\textbf{g)} Magnetic energy spectra: true result (red), plotted with predictions using deflection-field spectral relation \eqref{deffieldspec} applied to b) (blue), d) (green) and f) (orange).
\textbf{h)} Ratio of predicted to actual RMS magnetic field strength for a range of field strengths using both linear-regime flux spectral relation \eqref{linfluxspec3} (dashed) and reconstructed deflection-field spectral relation \eqref{deffieldspec} (solid), applied to unsmeared (blue), smeared (green) and de-smeared (orange).
} \label{Golitsynsmrfluxex}
\end{figure}

Smearing also has important consequences for accurately reconstructing both path-integrated fields and magnetic energy spectra. More specifically, smearing reduces the strength of path-integrated fields reconstructed directly from proton-flux images by reducing the sharpness of nonlinearly focused image-flux features. It also introduces a upper wavenumber cut-off on predicted magnetic energy spectra. Both these effects are illustrated in Figure \ref{Golitsynsmrfluxex} with a numerical example.
Figure \ref{Golitsynsmrfluxex}a shows a proton-flux image generated by imaging the Golitsyn field described in Figure \ref{GolitysncompactfieldSec2}, along with the (near-accurate) reconstructed field (Figure \ref{Golitsynsmrfluxex}b). Figure \ref{Golitsynsmrfluxex}c shows the equivalent image-flux distribution generated by a finite source. The reconstructed morphology, while similar, is reduced in strength. The effect on spectral reconstruction in this example is more pronouced (Figure \ref{Golitsynsmrfluxex}g): wavenumbers close to and beyond the smearing scale are suppressed in the spectrum derived from Figure \ref{Golitsynsmrfluxex}d. 

The impact of smearing on successful extraction of field statistics from proton-flux images can be somewhat reduced by use of a deconvolution algorithm. In particular, for a known point-spread function $S$, there exist algorithms (such as the Richardson-Lucy algorithm) for recovering the maximum likelihood solution for the unsmeared image-flux distribution given the smeared image-flux distribution~\cite{R72,L74}.
The procedure is illustrated in Figure \ref{Golitsynsmrfluxex}e, with associated reconstructed path-integrated field in Figure \ref{Golitsynsmrfluxex}f. It is clear that the maximum likelihood solution for the de-smeared image-flux distribution recovered by the Richardson-Lucy algorithm does not have identical morphology to the actual unsmeared image-flux distribution. However, the morphology and strength of the reconstructed path-integrated field is much more similar; furthermore, the magnetic-energy spectrum recovered from the path-integrated field (see \ref{Golitsynsmrfluxex}g, orange markers) is a much closer match to the true result ( \ref{Golitsynsmrfluxex}g, blue markers). 

More quantitatively, Figure \ref{Golitsynsmrfluxex}h gives the ratio of the predicted RMS magnetic field strength to the actual value for a range of field strengths, as calculated by integrating predictions of the magnetic-energy spectrum. The results from using both linear-regime flux spectral relation \eqref{linfluxspec3} and reconstructed deflection-field spectral relation \eqref{deffieldspec} applied to unsmeared (blue), smeared (green), and de-smeared (orange) proton-flux images are shown. It is clear that for both approaches, the smeared proton-flux image leads to reductions in predictions of field strength of 10-20 \% -- but that these can be removed quite effectively using a deconvolution algorithm. We also note that beyond the appearance of caustics, all methods increasingly under-estimate the predicted RMS field strength; this is related to the lower-bound property of the reconstructed perpendicular-deflection fields discussed in Section \ref{CauRgme}.

For experimental analysis with a somewhat limited resolution range, we conclude that deconvolution algorithms are helpful for deducing magnetic-energy spectra over a reasonable range of wavenumbers. 

\section{Conclusions}

In this paper, we have shown that the magnetic-energy spectrum of a wide range of stochastic magnetic fields can be determined using a proton imaging diagnostic, but only provided the parameters of the imaging are appropriately tuned. More specifically, we have used an analytic theory of proton imaging established under a set of simplifying assumptions typically valid when carrying out proton imaging to show that proton-flux images of stochastic magnetic fields can be classified into four regimes -- linear, nonlinear injective, caustic and diffusive  -- dependent on $\mu$ (which quantities the relative size of magnetic structures and proton displacements due to magnetic deflections). In the linear and nonlinear injective regimes, the magnetic energy is extractable from the image-flux distribution provided the stochastic fluctuations are homogeneous and isotropic: and we have described a procedure for achieving this. In the caustic and diffusive regimes, we have shown the the problem is not well posed -- at least, not without additional assumptions about the structure of the stochastic magnetic field -- in that stochastic magnetic fields with different magnetic energy spectra give proton image-flux distributions with identical statistics. We have investigated several complications to this description placed by particular types of stochastic fields, in particular showing that shallow power law spectras are often unreproducable using the proposed (or any other) proton imaging techniques. In addition, three limitations on the utility of the diagnostic due to experiemental constraints have been described, as well as techniques for minimising their impact.

This is by no means an exhaustive study. Magnetic stochasticity is often not isotropic, particularly in the presence of guiding fields, and much of the analysis techniques presented could be extended to take this into account. The techniques currently proposed for analysing caustic and diffusive regime images are significantly less comprehensive than other regimes; however, work carried out in the context of optics suggests that approaches such as statistical topology could be used to directly analyse caustic features more methodically, particularly with additional prescriptions for the sort of stochastic field being investigated~\cite{BU89}. 

However, we believe that the theory outlined provides a helpful conceptual framework for approaching analysis of stochastic magnetic fields using proton imaging, as well as enabling more information to be extracted from experimentally-obtained proton-flux images than previously possible. Furthermore, it might be hoped that insights of this paper might be useful for optimising the design of future experiments investigating magnetic stochasticity in plasma such that imaging regimes of maximal efficacy can be achieved. All these aspects will hopefully allow for proton imaging to be increasingly comprehensive as a diagnostic tool for assessing magnetic fields in laser-plasma experiments. 

\section*{Acknowledgements}

The research leading to these results has received funding from the Engineering and Physical Sciences Research Council (grant numbers EP/M022331/1 and EP/N014472/1) of the United Kingdom and AWE plc.

\newpage

\appendix

\section{Glossary of notation and mathematical conventions}

As an aid to reading, in this appendix we provide a glossary of all notation used throughout this paper in Table \ref{notationtable1}. The units system is Gaussian CGS.

\begin{table}[tbp]
\centering
{\renewcommand{\arraystretch}{1.28}
\renewcommand{\tabcolsep}{0.2cm}
\begin{tabular}[c]{c|c}
Notation & Quantity \\
\hline
$e$ & Elementary charge\\
$m_p$ & Beam-proton mass\\
$c$ & Speed of light\\
$\mathbf{B}$ & Total magnetic field \eqref{magfielddef}\\
$\delta \mathbf{B}$ & Fluctuating, stochastic magnetic field \eqref{magfielddef}\\
$\bar{\mathbf{B}}$ & Mean, regular magnetic field \eqref{magfielddef}\\
$\mathbf{j}$ & Current\\
$l_B$ & Magnetic field correlation length \eqref{corrfundefAppen}\\
$E_B\!\left(k\right)$ & Magnetic-energy spectrum \eqref{magengspecsec2}\\
$B_{rms}$ & RMS stochastic magnetic field strength \\
$B_{rms,0}$ & Effective RMS stochastic magnetic field strength \eqref{Brms0mod}\\
$r_i$ & Distance from beam source to magnetic field configuration/plasma\\
$l_z$ & Parallel size of magnetic field configuration/plasma\\
$l_\bot$ & Perpendicular size of magnetic field configuration/plasma\\
$r_s$ & Distance from magnetic field configuration/plasma to detector\\
$\mathcal{M}$ & image-magnification factor \eqref{screenmagfactorsec2}\\
$a$ & Finite proton-source radius\\
$\delta \theta$ & Typical proton deflection angle \eqref{deflangleest}\\
$\delta \alpha$ & Paraxial parameter \eqref{paraxialparamdef}\\
$\delta \beta$ & Point-projection parameter \eqref{pointprojectionparamdef}\\
$\mu$ & Contrast parameter \eqref{contrastdef}\\
$\left(\mathbf{x}_{\bot0}\right) \; \mathbf{x}_0$ & (Perpendicular) initial beam-proton position coordinate\\
$\left(\mathbf{x}_{\bot}\right) \; \mathbf{x}$ & (Perpendicular) beam-proton position coordinate\\
$(\mathbf{x}_{\bot}^{\left(s\right)}) \; \; \mathbf{x}^{\left(s\right)}$ & (Perpendicular) beam-proton image position coordinate\\
$\left(\mathbf{r}_{\bot}\right) \; \mathbf{r}$ & (Perpendicular) distance vector\\
$\left(\mathbf{v}_{\bot0}\right) \; \mathbf{v}_0$ & (Perpendicular) initial beam-proton velocity coordinate\\
$\mathbf{v}$ & Beam-proton velocity coordinate\\
$(\mathbf{v}_{\bot}^{\left(s\right)}) \; \; \mathbf{v}^{\left(s\right)}$ & (Perpendicular) beam-proton image velocity coordinate\\
$V$ & Beam-proton speed\\
$W$ & Beam-proton energy\\
$\mathbf{w}\!\left(\mathbf{x}_{\bot0}\right)$ & Perpendicular-deflection field \eqref{pathintfieldSec2}\\
$w_{rms}$ & RMS deflection-field strength \\
$E_W\!\left(k_\bot\right)$ & Deflection-field spectrum \eqref{deflfieldspecdef}\\
$\varphi\!\left(\mathbf{x}_{\bot0}\right)$ & Deflection-field potential \eqref{deffieldpotdef}\\
$\Psi_{0}\!\left(\mathbf{x}_{\bot0}\right)$ & Initial beam flux distribution\\
$\Psi\!\left(\mathbf{x}_{\bot}^{\left(s\right)}\right)$ & Image-flux distribution function \eqref{screenfluxSec2}\\
$\delta \Psi\!\left(\mathbf{x}_{\bot}^{\left(s\right)}\right)$ & Image-flux deviation from mean proton flux\\
$l_\Psi$ & Relative image-flux correlation length \eqref{screenfluxcorrlength}\\
$\Psi_0^{\left(s\right)}\!\left(\mathbf{x}_{\bot}^{\left(s\right)}\right)$ & Unperturbed image-flux distribution\\
$D_{w}$ & (Isotropic) perpendicular stochastic magnetic diffusion coefficient \eqref{difftensSection2}\\
$t_{pulse}$ & Temporal pulse-length of proton source \\
$t_{path}$ & Transit time of imaging protons across plasma \\
\end{tabular}}
\caption{Glossary of notation} \label{notationtable1}
\end{table}

We define the Fourier tranform $\hat{f}$ of a function $f$ and the inverse transform in $n$-dimensions according to the following convention:
\begin{IEEEeqnarray}{rCl}
\hat{f}\!\left(\mathbf{k}\right) & = & \frac{1}{\left(2 \pi \right)^{n}} \int \mathrm{d}^n \mathbf{k} \, \exp{\left[-i \mathbf{k} \cdot \mathbf{x}\right]} \, f\!\left(\mathbf{x}\right) \, ,\\
f\!\left(\mathbf{x}\right) & = & \int \mathrm{d}^n \mathbf{k} \, \exp{\left[i \mathbf{k} \cdot \mathbf{x}\right]} \hat{f}\!\left(\mathbf{k}\right) \, .
\end{IEEEeqnarray}
For isotropic functions $f = f\!\left(r\right)$ and $\hat{f} = \hat{f}\!\left(k\right)$ in three dimensions, this gives
\begin{equation}
f\!\left(r\right) = 4 \pi \int_0^{\infty} \mathrm{d}k \, k^{2} \hat{f}\!\left(k\right) \frac{\sin{kr}}{kr} \, ,
\end{equation}
which in turns implies that the value of $f$ at the origin under this convention is simply the integral of the 1D power spectrum of $f$:
\begin{equation}
f\!\left(0\right) = 4 \pi \int_0^{\infty} \mathrm{d}k \, k^{2} \hat{f}\!\left(k\right) \equiv \int_0^{\infty} \mathrm{d}k \, E_f\!\left(k\right) \, .
\end{equation}
The integral of $f$ over all radii (which for normalised correlation functions gives the correlation scale) is
\begin{equation}
\int_0^{\infty} \mathrm{d}r \, f\!\left(r\right) = 2 \pi^2 \int_0^{\infty} \mathrm{d}k \, k \hat{f}\!\left(k\right) \, .
\end{equation}
For two-dimensional isotropic functions,
\begin{equation}
f\!\left(r\right) = 2 \pi \int_0^{\infty} \mathrm{d}k \, k \hat{f}\!\left(k\right) J_0\!\left(kr\right) \, ,
\end{equation} 
where $J_0$ is the zeroth-order Bessel function of the first kind. Similarly to the three-dimensional case, we find
\begin{equation}
f\!\left(0\right) = 2 \pi \int_0^{\infty} \mathrm{d}k \, k \hat{f}\!\left(k\right) \equiv \int_0^{\infty} \mathrm{d}k \, E_f\!\left(k\right) \, ,
\end{equation}
and
\begin{equation}
\int_0^{\infty} \mathrm{d}r \, f\!\left(r\right) = 2 \pi \int_0^{\infty} \mathrm{d}k \, k \hat{f}\!\left(k\right) \, .
\end{equation}

\section{Negligible physical processes when deriving plasma-image mapping \eqref{divmappingSec2}} \label{NegPhysPros}

In deriving plasma-image mapping \eqref{divmappingSec2}, perpendicular-deflection field \eqref{pathintfieldSec2}, and Kugland image-flux relation \eqref{screenfluxSec2}, it was assumed that the evolution of the proton beam as it passes through the plasma is dominated by forces arising due to magnetic fields inherent in the plasma; in this appendix, we explain the justification for this simplification. In general, the interaction of a beam of protons with a plasma involves a broad range of physics. Under the general framework provided by kinetic theory, solving the problem without simplifications would require Maxwell-Boltzmann equations for particles of each of the constituent species of the plasma and the beam, combined with Maxwell's equations. Thankfully, due to the large velocity of the beam protons, and their low density, the governing physics can be simplified considerably.

Firstly, self-interaction of the beam can be taken to be negligible, on account of the beam's initial number density $n_{beam}$ taking values $n_{beam} \sim 10^{10} - 10^{12} \, \mathrm{cm}^{-3}$. This can be demonstrated by considering space-charge effects in the beam: it can be shown that the electric potential energy in an un-neutralised proton cloud (assuming the above values for the beam density) drops to a small fraction of the beam's kinetic energy well before reaching the plasma~\cite{K12}. Furthermore, once the protons reach the plasma, beam space charge is screened by plasma electrons.

Typically, a fast particle beam interacting with a plasma would be subject to collisionless kinetic effects, such as the beam-plasma instability. For initial beam velocity $V$, the maximum linear growth rate of the beam-plasma instability scales as
\begin{equation}
\gamma \sim \omega_{pi} \frac{n_{beam}}{n_i} \left(\frac{V}{\Delta V}\right)^2 \, ,
\end{equation}
where $\omega_{pi}$ is the plasma frequency, $n_i$ the density of ions in the plasma, and $\Delta V$ the initial velocity spread inherent in the proton beam~\cite{CL14}. 
For the beam-plasma instability to have a significant effect on the proton-flux image, we require that the transit time $T_{beam}$ through the plasma be much shorter than the typical time scale on which the beam-plasma instability grows: viz., 
\begin{equation}
\gamma \sim \frac{1}{T_{beam}} \approx \frac{V}{l_z} \, ,
\end{equation}
which can be rearranged to give an estimate for the typical uncertainty in deflection angle $\Delta \theta_{kin}$ resulting from collisionless kinetic effects:
\begin{equation}
\Delta \theta_{kin} \leq \frac{\Delta V}{V} \sim \left(\frac{n_{beam}}{n_i}\right)^{1/4} \left(\frac{Z}{A}\omega_{p,beam} T_{beam}\right)^{1/2} \, ,
\end{equation}
where $\omega_{p,beam}$ is the plasma frequency of beam protons in the beam, $Z$ the charge of the plasma ions, and $A$ their atomic mass. 
Substituting typical parameters in (for example) a carbon plasma, with $n_{i} \sim 10^{20} \, \mathrm{cm}^{-3}$, $n_{beam} \sim 10^{10} \, \mathrm{cm}^{-3}$ and $l_z \sim 0.1 \, \mathrm{cm}$, we have for 3.3 MeV protons that $\omega_{p,beam} \sim 1.3 \times 10^{8} \, \mathrm{s}^{-1}$, and $T_{beam} \sim 4 \times 10^{-11} \, \mathrm{s}$, which gives
\begin{equation}
\Delta \theta_{kin} \sim 2 \times 10^{-4} \, ,
\end{equation} 
a smaller effect that typical experimental uncertainties (and other asymptotic parameters). 

Collisional effects are also negligible, on account of $T_{beam}$ typically being larger than collisional relaxation times -- in a plasma, for fast ions with energy $W$ the stopping power is to a good approximation given by the Bohr formula
\begin{equation}
S\!\left(W\right) = - \frac{\mathrm{d} W}{\mathrm{d} z} = \frac{4 \pi e^{4} n_e}{m_p V^2} \log{\left(\frac{V}{\omega_p b_{min}}\right)} \, ,
\end{equation}
where $b_{min}$ is the impact parameter, and $z$ the distance travelled through the plasma~\cite{Z99}. For protons with energy $W \geq 3.3 \, \mathrm{MeV}$ passing through a carbon plasma, $n_e \sim 3.5 \times 10^{20} \, \mathrm{cm}^{-3}$, this gives
\begin{equation}
S\!\left(W\right) = 6 \times 10^4 \, \mathrm{eV}/\mathrm{cm} \, ,
\end{equation}
which implies that the energy $\Delta W$ lost by a typical proton during the transit of the plasma is
\begin{equation}
\Delta W \sim - l_z S\!\left(W\right) \, .
\end{equation}
For a $l_z \sim 1 \, \mathrm{mm}$ plasma, this gives
\begin{equation}
\frac{\Delta W}{W} \leq 2 \times 10^{-3} \, ,
\end{equation}
The resulting deflection angles are even smaller: for a Maxwellian plasma, it can be shown that for a suprathermal beam, the typical path length $\lambda_{\bot}$ over which the proton beam distribution spreads perpendicularly to the direction of travel of the beam is given by
\begin{equation}
\lambda_{\bot} = \frac{m_p^2}{16 \pi e^4} \frac{V^4}{n_e \log{\Lambda}} \, , 
\end{equation}
where $\log{\Lambda}$ is the Coulomb logarithm~\cite{CL14}. This then gives an estimate for the typical uncertainty in deflection angle $\Delta \theta_{coll}$ resulting from collisional relaxation:
\begin{equation}
\Delta \theta_{coll} \sim \frac{l_z}{\lambda_{\bot}} \sim \frac{16 \pi e^4}{m_p^2} \frac{l_z n_e \log{\Lambda}}{V^4} \, ,
\end{equation}
which has a strong beam-velocity dependence. Estimating $\Delta \theta_{coll}$ for the same carbon plasma (and $\log{\Lambda} \approx 10$) described previously, we find
\begin{equation}
\Delta \theta_{coll} \sim 1 \times 10^{-5} \, .
\end{equation}

Next, we can rule out the effect of electric fields in plasmas well described by MHD-type fluid models by noting that for such plasmas, 
\begin{equation}
E \sim \frac{UB}{c} \, ,
\end{equation}
where $U$ is the characteristic flow velocity associated with the plasma. Then the ratio of maximal electric to magnetic forces acting on $15 \, \mathrm{MeV}$ protons (velocity $V \sim 5.3 \times 10^{9} \, \mathrm{cm}/\mathrm{s}$) scales as
\begin{equation}
\frac{e E}{\frac{e}{c} \left|\mathbf{V} \times \mathbf{B} \right|} \sim \frac{U}{V} \, .
\end{equation}
Thus, provided $U \ll V$, the effect of electric fields on protons beams in such plasmas is negligible relative to magnetic fields.

Furthermore, whether the plasma can be described successfully by a fluid model or not, the parameter $U/V$ usually quantifies the extent to which the plasma flow (and hence collective electromagnetic fields) changes as the protons transit the plasma - and so if $U/V$ is small, the field sampled is effectively static. 

The validity of ignoring the above effects can also be tested experimentally, by examining the mean proton energy images typically extracted along with the proton-flux image. 
If the interaction region displays uniformity on a mean proton energy image, then it follows that electric fields, collisions or kinetic instabilties cannot be a significant factor in determining the beam dynamics: unlike magnetic forces, all these effects do not conserve proton energy.

\section{Further statistical characterisation of stochastic magnetic fields} \label{FurtherStatCharMagFields}

This appendix extends the discussion of the statistical characterisation of stochastic magnetic fields provided in Section \ref{StatCharStocMagField} by introducing the \emph{magnetic autocorrelation function} (defined in Appendix \ref{FurtherStatCharMagFieldsMagAuto}). For clarity of exposition, we do not directly refer to the magnetic autocorrelation function in the main text, because by the Wiener-Khinchin theorem it is equivalent to the magnetic-energy spectrum \eqref{magengspecsec2} for isotropic, homogeneous stochastic fields~\cite{EV03}: and the magnetic-energy spectrum is generally of greater interest physically, for reasons discussed in Section \ref{StatCharStocMagField}. However, as we state in Section  \ref{StatCharStocMagField}, we introduce the magnetic autocorrelation function in this appendix for two reasons. First, it provides the most natural definition of the correlation length $l_B$ of a stochastic magnetic field (Appendix \ref{FurtherStatCharMagFieldsCorrDef}). We show in Appendix \ref{FurtherStatCharMagFieldsCorrSpec} how the result \eqref{corrlengthspec} in the main text is derived from this definition. Second, using the magnetic autocorrelation function to derive deflection-field spectral relation \eqref{deffieldspec} (Appendix \ref{DeflfieldCorr}) and linear-regime flux spectral relation \eqref{linfluxspec3} (Appendix \ref{RelFluxRMSLinThy}) enables a simple quantification of asymptotic approximations made in doing so. In particular, in Appendix \ref{FurtherStatCharMagFieldsCorrTail} we state a result concerning the magnitude of the integrated tail of the magnetic autocorrelation function, a term whose neglect is necessary for the derivations of \eqref{deffieldspec} and \eqref{linfluxspec3}. 

\subsection{The magnetic autocorrelation function} \label{FurtherStatCharMagFieldsMagAuto}
 
The \emph{magnetic autocorrelation tensor} $M_{nn'}\!\left(\mathbf{x}, \tilde{\mathbf{x}}\right)$ of a general, zero-mean stochastic magnetic field ($\bar{\mathbf{B}} = 0$, $\mathbf{B} = \delta \mathbf{B}$) is defined by
\begin{equation}
M_{nn'}\!\left(\mathbf{x}, \tilde{\mathbf{x}}\right) \equiv \left< B_{n}\!\left(\mathbf{x}\right) B_{n'}\!\left(\tilde{\mathbf{x}}\right)\right> \, , \label{magcorrtensdef}
\end{equation}
for arbitary positions $\mathbf{x}$, $\tilde{\mathbf{x}}$ in the plasma. 
If the stochastic fields are isotropic and homogeneous, the autocorrelation function can be written as a function of distance coordinate $r = \left|\mathbf{x}-\tilde{\mathbf{x}}\right|$ only:
\begin{equation}
M_{nn'}\!\left(\mathbf{x}, \tilde{\mathbf{x}}\right) = M_{nn'}\!\left(\mathbf{x}-\tilde{\mathbf{x}}\right) = M_{nn'}\!\left(r\right) =  M_N\!\left(r\right) \delta_{nn'} + \left(M_L\!\left(r\right)-M_N\!\left(r\right)\right) \frac{r_n r_{n'}}{r^2} + M_H\!\left(r\right)\epsilon_{nn'l}r_l \, ,\label{isomagcoll}
\end{equation}
for longitudinal, normal and helical autocorrelation functions $M_L\!\left(r\right)$, $M_N\!\left(r\right)$ and $M_H\!\left(r\right)$~\cite{EV03}. The solenoidal condition on the magnetic field $\nabla \cdot \mathbf{B} = 0$ gives corresponding relation
\begin{equation}
\frac{\partial}{\partial r_i} M_{ij}\!\left(r\right) = 0 \, , \label{solenoidal}
\end{equation}
which can be used to relate $M_L$ to $M_N$~\cite{EV03}. The magnetic autocorrelation function $M\left(r\right)$ is then defined as trace of the magnetic autocorrelation tensor:
\begin{equation}
M\!\left(r\right) = M_{ii}\!\left(r\right) \, .
\end{equation}
Similarly to the magnetic-energy spectrum, the magnetic autocorrelation function provides information about typical magnetic field strengths. More specifically, the RMS magnetic field strength is given in terms of $M\!\left(r\right)$ by
\begin{equation}
B_{rms}^2 = M\!\left(0\right)\, .
\end{equation}
As discussed in Appendix \ref{FurtherStatCharMagFieldsCorrDef}, the autocorrelation function can also be used to estimate typical structure sizes via the correlation length. 

We now state the sense in which the magnetic-energy spectrum and magnetic autocorrelation function are equivalent: from the Wiener-Khinchin theorem~\cite{EV03}, it follows that the (assumed isotropic) magnetic-energy spectrum $E_{B}\!\left(k\right)$ defined by \eqref{magengspecsec2} and autocorrelation function $M\!\left(r\right)$ are related by
\begin{equation}
E_{B}\!\left(k\right) = \frac{k^2}{2\left(2\pi\right)^3} \int \mathrm{d}^3 \mathbf{r} \, \exp{\left(-i \mathbf{k} \cdot \mathbf{r} \right)} \, M\!\left(r\right) =  \frac{1}{\left(2\pi\right)^2} \int_0^{\infty} \mathrm{d}r \, kr \sin{kr} \, M\!\left(r\right) \, , \label{magengspeccorr}
\end{equation}
where we have assumed Fourier-transform normalisation conventions as specified in Appendix A. The Fourier-inversion relations specified in the same appendix can be used to deduce the magnetic autocorrelation function from the magnetic-energy spectrum: 
\begin{equation}
M\!\left(r\right) = 8 \pi \int_0^{\infty} \mathrm{d}k \, E_B\!\left(k\right) \frac{\sin{kr}}{kr} \label{magcorrfuneng}
\end{equation}
Thus if the magnetic-energy spectrum is known, the magnetic autocorrelation function can be determined, and visa versa. 

\subsection{Formal definition of correlation length} \label{FurtherStatCharMagFieldsCorrDef}

The correlation length of a stochastic magnetic field $l_B$ is defined in terms of the magnetic autocorrelation function $M\!\left(r\right)$ introduced in the previous sub-appendix:
\begin{equation}
l_B \equiv \frac{1}{B_{rms}^2}\int_0^{\infty} \mathrm{d} r \, M\!\left(r\right) \, . \label{corrfundefAppen}
\end{equation}
Thus, as the name suggests, the correlation length of a stochastic magnetic field is simply a measure of the typical distance over which that field decorrelates with itself.  

\subsection{Derivation of expression \eqref{corrlengthspec} for the correlation length in terms of the magnetic-energy spectrum} \label{FurtherStatCharMagFieldsCorrSpec}

The inversion relation \eqref{magcorrfuneng} enables a simple derivation of the expression \eqref{corrlengthspec} for the correlation length, that is
\begin{equation}
l_B = \frac{\pi}{2} \frac{\int_0^{\infty} \mathrm{d}k \, E_B\!\left(k\right)/k}{\int_0^{\infty} \mathrm{d}k \, E_B\!\left(k\right)} \, , \label{corrlengthspecAppend}
\end{equation} 
in terms of the magnetic-energy spectrum. Integrating \eqref{magcorrfuneng} over the interval $r \in \left[0, \infty \right)$ gives
\begin{equation}
\int_0^{\infty} \mathrm{d} r \, M\!\left(r\right) = 8 \pi \int_0^{\infty} \mathrm{d} r \, \int_0^{\infty} \mathrm{d}k \, E_B\!\left(k\right) \frac{\sin{kr}}{kr} \, . 
\end{equation}
Applying correlation-length definition \eqref{corrfundefAppen} and switching the order of integration leads to
\begin{equation}
B_{rms}^2 \, l_B = 8 \pi \int_0^{\infty} \mathrm{d}k \, E_B\!\left(k\right) \int_0^{\infty} \mathrm{d}r \frac{\sin{kr}}{kr} = \frac{\pi}{2} \int_0^{\infty} \mathrm{d}k \, \frac{E_B\!\left(k\right)}{k}  \, ,
\end{equation}
which on rearrangement and use of equation \eqref{Brmsspec}, viz.
\begin{equation}
B_{rms}^2 = 8 \pi \int_0^{\infty} \mathrm{d}k \, E_B\!\left(k\right) \, ,
\end{equation}
for the magnetic field strength RMS gives the desired result \eqref{corrlengthspecAppend}. 

\subsection{Bound on integrated tail of magnetic autocorrelation function} \label{FurtherStatCharMagFieldsCorrTail}

In this subsection we provide an estimate for integrated tail of the magnetic autocorrelation function, which will be used in subsequent derivations of the deflection-field spectral relations \eqref{deffieldspec} and linear-regime flux spectral relation \eqref{linfluxspec3} given in Appendices \ref{DeflfieldCorr} and \ref{RelFluxRMSLinThy} respectively. 

It can be shown that in a finite magnetised volume, the autocorrelation function must satisfy~\cite{EV03}
\begin{equation}
\int_{0}^{\infty} \mathrm{d} r \,  r^2 \, M\!\left(r\right) = 0 \label{reg1} \, .
\end{equation}
which implies that $M\!\left(r\right) = o\!\left[\left(r/l_B\right)^{-3}\right]$ as $r/l_B \rightarrow \infty$. It follows from this that for lengths $l \gg l_B$,
\begin{equation}
\frac{1}{B_{rms}^2 l_B} \int_l^{\infty} \mathrm{d} r \, M\!\left(r\right) = \mathcal{O}\!\left(\left(\frac{l_B}{l}\right)^2\right) \, . \label{magcorrfuntailbound}
\end{equation}
Thus we conclude that for any magnetic autocorrelation function, the integral of its tail is algebraically small in $l_B/l$. For some correlation functions (such as exponential or Gaussian), the tail of the integrated autocorrelation function is smaller still; for example, if
\begin{equation}
M\!\left(r\right) = B_{rms}^2 \, \exp{\left[-r/l_B\right]} \,
\end{equation}
then
\begin{equation}
\int_l^{\infty} \mathrm{d} r \, M\!\left(r\right) = B_{rms}^2 \, l_B  \, \exp{\left[-l/l_B\right]} \, 
\end{equation}
which is exponentially small in $l_B/l$.

\section{Derivation of plasma-image mapping \eqref{divmappingSec2} and Kugland image-flux relation \eqref{screenfluxSec2} using kinetic theory of proton imaging} \label{KinThyPrad}

In the case of small typical particle deflections, the general relationship between the initial and image-flux distributions has previously been described by Kugland \emph{et. al.}~\cite{K12}, and is quoted in the main text in Section \ref{PlasMap} as equation \eqref{screenfluxSec2}. The approach used by Kugland \emph{et. al.} to derive this result depends on real-space particle conservation, combined with a simplified form for the plasma-image mapping \eqref{divmappingSec2} obtained using various asymptotic approximations. However, the evolution of such a proton beam can alternatively be described in terms of kinetic theory. This latter approach, working in phase space, will prove helpful for characterising non-trivial proton-beam velocity distributions. This will lead to an analytic method for including various \textit{smearing effects} not previously described analytically on simulated proton-flux images -- most relevantly to this paper, that of small-scale stochastic magnetic fields, but also finite proton-source extent. These smearing effects are so described, because they smear image-flux structures, effectively leading to decreased resolution of proton-flux images. Thus, in this appendix we develop a kinetic theory of proton imaging, which will re-derive plasma-image mapping \eqref{divmappingSec2}, and Kugland image-flux relation \eqref{screenfluxSec2} under the assumptions of a mono-energetic, instantaneous beam originating from a point source. We then derive a modified image-flux relation which includes smearing effects. 

\subsection{The image-flux distribution resulting from the interaction of a proton beam with a classical plasma} \label{KinThyPradGen}

Consider a system consisting of a proton beam, a plasma and a detector. The spatial arrangement of the system is chosen to imitate typical proton-imaging set-ups (see Figure \ref{PRsetup}). In particular, we choose a Cartesian coordinate system such that the normal direction to the detector is parallel to the $z$-direction. The plasma is presumed to be contained inside a cuboid region, which has centre $\left(0,0,l_z/2\right)$, and dimensions $l_\bot \times l_\bot \times l_z$; the detector plane is located at $z = l_z + r_s$. 

We describe the imaging proton beam with a beam distribution function $f\!\left(\mathbf{x},\mathbf{v},t\right)$. It is assumed that the beam distribution function has a known compact initial form $f\!\left(\mathbf{x},\mathbf{v},0\right) = f_0\!\left(\mathbf{x},\mathbf{v}\right)$ determined by the proton-beam generation method at an initial time $t = 0$, where the time origin is chosen to correspond to the first instance of the beam intersecting the cuboid region containing the plasma. The domain of the system is assumed infinite, with no boundaries. The equation governing the evolution of the beam distribution function as the beam passes through a plasma is a Fokker-Planck equation of the form
\begin{equation}
\frac{\partial f}{\partial t} + \mathbf{v} \cdot \nabla f + \frac{e}{m_p} \left(\mathbf{E}^{\left(T\right)}\!\left(\mathbf{x},t\right)+\frac{\mathbf{v} \times \mathbf{B}^{\left(T\right)}\!\left(\mathbf{x},t\right)}{c}\right) \cdot \frac{\partial f}{\partial \mathbf{v}} = C\!\left(f\right) \, . \label{Fokkerplanckinit}
\end{equation}
Here, $C\!\left(f\right)$ is a operator describing collisional interactions with the plasma, and $\mathbf{E}^{\left(T\right)}\!\left(\mathbf{x},t\right)$ and $\mathbf{B}^{\left(T\right)}\!\left(\mathbf{x},t\right)$ are time-dependent electromagnetic fields satisfying Maxwell's equations, with current and charge density terms determined self-consistently by the beam distribution function, as well as by the distribution functions of the plasma constituent ions and electrons. These in turn also satisfy Fokker-Planck equations. 

Our ultimate goal is then to determine the time-integrated flux distribution $\Psi$ of the proton beam through the detector plane subsequent to the interactions of that proton beam with a plasma. We refer to this distribution as the image-flux distribution. Explicitly, we have that flux of particles through a detector located at $z = r_s+l_z$ at perpendicular location $\mathbf{x}_\bot^{\left(s\right)}$ and time $t$ is given by
\begin{equation}
\psi\!\left(\mathbf{x}_\bot^{\left(s\right)},t\right) = \int \mathrm{d}^3 \mathbf{v} \, v_z \, f\!\left(\mathbf{x}_{\bot}^{\left(s\right)}\!,r_s+l_z,\mathbf{v},t\right) \, . \label{fluxtime}
\end{equation}
The image-flux distribution is then
\begin{equation}
\Psi\!\left(\mathbf{x}_\bot^{\left(s\right)}\right) \equiv \int_0^{\infty} \mathrm{d} t \, \psi\!\left(\mathbf{x}_\bot^{\left(s\right)},t\right) \, . \label{fluxtot}
\end{equation} 
Solving the problem as currently stated is seemingly impossible analytically for two reasons: first, the complicated set of physical processes governing the collective interaction of the beam with the plasma; second, the wide range of proton trajectories possible for arbitrary electromagnetic fields and proton-beam parameters. However, proton beams typically used for imaging in laser-plasma experiments have a high (though non-relativistic) energy, a low density, and a characteristic spatial configuration. As we show, this can be used to simplify both the governing equations and the initial conditions, before then finding a tractable solution for the image-flux distribution in terms of the electromagnetic fields already present in the plasma prior to the introduction of the imaging beam. 

\subsection{Simplifications to governing physics} \label{KinThyPradSimpPhys}

First, we simplify the equations governing the proton beam, and find a general solution for the beam distribution function $f\!\left(\mathbf{x},\mathbf{v},t\right)$ in terms of the initial condition $f\!\left(\mathbf{x},\mathbf{v},0\right) = f_0\!\left(\mathbf{x},\mathbf{v}\right)$. As justified in Appendix \ref{NegPhysPros} (also see~\cite{K12}), the following physical effects on the proton beam are often small: collisions, kinetic beam-instabilities, self-interactions of the proton beam, electric fields in the plasma, and time-varing magnetic fields. We can re-write \eqref{Fokkerplanckinit} explicitly in terms of fields associated with each of these effects:
\begin{eqnarray}
\frac{\partial f}{\partial t} + \mathbf{v} \cdot \nabla f + \frac{e}{m_p c} \mathbf{v} \times \mathbf{B}\!\left(\mathbf{x}\right) \cdot \frac{\partial f}{\partial \mathbf{v}} & = & C\!\left(f\right) - \frac{e}{m_p} \mathbf{E}^{\left(p\right)}\!\left(\mathbf{x},t\right) \cdot \frac{\partial f}{\partial \mathbf{v}}  -\frac{e}{m_p c} \mathbf{v} \times \left(\mathbf{B}^{\left(p\right)}\!\left(\mathbf{x},t\right)-\mathbf{B}\!\left(\mathbf{x}\right)\right) \cdot \frac{\partial f}{\partial \mathbf{v}} \nonumber \\
&& - \frac{e}{m_p} \left(\mathbf{E}^{\left(i\right)}\!\left(\mathbf{x},t\right)+\frac{\mathbf{v} \times \mathbf{B}^{\left(i\right)}\!\left(\mathbf{x},t\right)}{c}\right) \cdot \frac{\partial f}{\partial \mathbf{v}}\, , \label{FokkerplanckinitB}
\end{eqnarray}
where $\mathbf{B}\!\left(\mathbf{x}\right)$ is the static magnetic field at $t = 0$, $\mathbf{B}^{\left(p\right)}\!\left(\mathbf{x},t\right)$, $\mathbf{E}^{\left(p\right)}\!\left(\mathbf{x},t\right)$ the time-dependent magnetic and electric fields present in the plasma in the absence of the beam, and $\mathbf{B}^{\left(i\right)}\!\left(\mathbf{x},t\right)$, $\mathbf{E}^{\left(i\right)}\!\left(\mathbf{x},t\right)$ the electromagnetic fields resulting from both self-interactions of the beam, and interactions of the beam with the plasma. We note that the total electric and magnetic fields $\mathbf{E}^{\left(T\right)}\!\left(\mathbf{x},t\right)$, $\mathbf{B}^{\left(T\right)}\!\left(\mathbf{x},t\right)$ are related to the electric and magnetic fields introduced by
\begin{IEEEeqnarray}{rCl}
\mathbf{B}^{\left(T\right)}\!\left(\mathbf{x},t\right) & = &\mathbf{B}\!\left(\mathbf{x}\right) + \left(\mathbf{B}^{\left(p\right)}\!\left(\mathbf{x},t\right)-\mathbf{B}\!\left(\mathbf{x}\right)\right) + \mathbf{B}^{\left(i\right)}\!\left(\mathbf{x},t\right) \, , \\
\mathbf{E}^{\left(T\right)}\!\left(\mathbf{x},t\right) & = &\mathbf{E}^{\left(p\right)}\!\left(\mathbf{x},t\right) + \mathbf{E}^{\left(i\right)}\!\left(\mathbf{x},t\right) \, .
\end{IEEEeqnarray}
We then neglect all of the terms on the right-hand-side of \eqref{FokkerplanckinitB}, leaving a Vlasov equation with a static magnetic field:
\begin{equation}
\frac{\partial f}{\partial t} + \mathbf{v} \cdot \nabla f + \frac{e}{m_p c} \mathbf{v} \times \mathbf{B}\!\left(\mathbf{x}\right) \cdot \frac{\partial f}{\partial \mathbf{v}} = 0 \, . \label{Vlasov}
\end{equation}
This is a linear, first-order partial differential equation in phase space, which can be solved exactly using the method of characteristics. 
Provided the mapping associated with characteristic rays between initial and final points in phase space is one-to-one, this gives a solution for the beam distribution function:
\begin{equation}
f\!\left(\mathbf{x},\mathbf{v},t\right) = f_{0}\!\left(\mathbf{x}_0\!\left(\mathbf{x},\mathbf{v},t\right)\!,\mathbf{v}_0\!\left(\mathbf{x},\mathbf{v},t\right)\right) \left|\frac{\partial\!\left(\mathbf{x}_0,\mathbf{v}_0\right)}{\partial\!\left(\mathbf{x},\mathbf{v}\right)}\right| \, ,\label{characsol}
\end{equation}
where $\mathbf{x}_0$ and $\mathbf{v}_0$ are initial points in phase space which map to $\left(\mathbf{x},\mathbf{v}\right)$ at time $t$. In fact, the characteristic rays are simply particle trajectories in phase space: thus in a sense the problem of evolving the beam distribution function is equivalent to that of single-proton motion. 
The proton trajectories are given explicitly by
\begin{IEEEeqnarray}{rCl}
\mathbf{x}\!\left(\mathbf{x}_0, \mathbf{v}_0, t\right) & = & \mathbf{x}_0+\int_0^t \mathbf{v}\!\left(\mathbf{x}_0, \mathbf{v}_0,t'\right) \mathrm{d}t' \, , \label{protpositionfull}\\
\mathbf{v}\!\left(\mathbf{x}_0, \mathbf{v}_0, t\right) & = & \mathbf{v}_0+\int_0^t \frac{e}{m_p c}\mathbf{v}\!\left(\mathbf{x}_0, \mathbf{v}_0,t'\right) \times \mathbf{B}\!\left(\mathbf{x}\!\left(\mathbf{x}_0, \mathbf{v}_0,t'\right)\right) \mathrm{d}t' \, . \label{protvelfull}
\end{IEEEeqnarray}

\subsection{Initial conditions} \label{KinThyPradInitCon}

Next, we characterise the initial beam distribution function $f_0$. More specifically, we aim to show that the initial spatial form of the proton beam is that of a two-dimensional near-planar sheet. 

In general, the appropriate choice for the initial beam distribution function depends on the method used to generate the beam, as well as the geometry of the proton-imaging set-up. As discussed in Section \ref{Assum} of the main text, proton beams are produced either using a high-intensity laser, or by the implosion of a spherical fusion capsule.  
It has been experimentally observed that both methods tend to produce a mono-energetic proton beam whose structure is approximately a spherical shell. This follows intuitively from the fact that the source size $a$ is typically much smaller than the distance from the source to the plasma $r_i$. 
For a proton with initial speed $V$ and position $\mathbf{x}_0$, this assumption gives initial velocity
\begin{equation}
\mathbf{v}_0\!\left(\mathbf{x}_0\right) = V \frac{\mathbf{x}_0-\mathbf{x}_s}{\left|\mathbf{x}_0-\mathbf{x}_s \right|} + \mathbf{v}_{s}= V\left(\hat{\mathbf{z}} \, \cos{\Theta}+ \hat{\mathbf{x}}_{\bot0}\,  \sin{\Theta}\right) +\delta \mathbf{v}_{0}  \, ,\label{protbminitvelfull}
\end{equation}
where $\mathbf{x}_s$ is the centre of the proton source, $\Theta$ the polar angle in a spherical coordinate system, with fixed zenith direction $\hat{\mathbf{z}}$, and $\delta \mathbf{v}_{0}$ a velocity perturbation associated with the finite size of the capsule. Since $a \ll r_i$, the velocity perturbation due to finite source size is perpendicular to $\mathbf{v}_0$; mathematically,
\begin{equation}
\frac{\mathbf{v}_{0} \cdot \delta \mathbf{v}_{0}}{\left|\mathbf{v}_{0}\right|\left|\delta \mathbf{v}_{0}\right|} = \mathcal{O}\!\left(\frac{a^2}{r_i^2}\right) \, .
\end{equation}
  
Throughout this paper we will be concerned with the paraxial limit, where the size of the plasma $l_\bot$ in the direction perpendicular to $z$ is much greater than the distance from the source to the detector $r_i$. More formally, introduce small paraxial parameter
\begin{equation}
\frac{\delta \alpha}{2} \equiv \frac{l_\bot}{2 r_i} \ll 1 \, .
\end{equation}
where the factor of 2 is introduced on account of the proton beam imaging symmetrically with respect to the centre of the plasma~\cite{K12}. Then, for the part of the imaging beam passing through the magnetic configuration, $\Theta \lesssim \delta \alpha$. As a result, we can expand \eqref{protbminitvelfull} in $\Theta$, to give
\begin{equation}
\mathbf{v}_0\!\left(\mathbf{x}_0\right)  = V\left(\hat{\mathbf{z}} + \frac{\mathbf{x}_{\bot0}}{r_i}+\delta \mathbf{v}_{\bot0}\right) \left[1+\mathcal{O}\left(\delta \alpha^2,\frac{a^2}{r_i^2}\right)\right]  \, .\label{protbminitvelparax}
\end{equation}
This is independent of $z_0$ -- and we previously stated that the initial condition $f\!\left(\mathbf{x},\mathbf{v},0\right) = f_0\!\left(\mathbf{x},\mathbf{v}\right)$ is imposed at the first time at which the first proton reaches the cuboid containing the plasma. The nearest side of this cuboid to the proton source is $z = 0$; thus in the paraxial approximation we conclude that the initial distribution function of the proton beam has general form 
\begin{equation}
f_0\!\left(\mathbf{x}_{0},\mathbf{v}_{0}\right) = f_{\bot0}\!\left(\mathbf{x}_{\bot0},\mathbf{v}_{\bot0}\right) \delta\!\left(z_0\right) \delta\!\left(v_{z0}-V\right) \, . \label{gendistinit}
\end{equation}
This form of initial distribution function matches with the stated expectation that the initial distribution has the form of a two-dimensional, planar sheet, since all particles in the distribution have the same $z$-position and velocity. 

For subsequent calculations in this appendix, we will simplify the form of the initial distribution function of the beam further, to one of two forms. The first is that of an initially diverging beam from a source at distance $r_i$ from the plasma, with perpendicular spatial structuring in the initial flux distribution of the beam -- an example still including initial proton distributions produced by high-intensity lasers or from capsule implosions with negligible sources sizes (see Section \ref{Assum}). In this case, the initial distribution function can be written as 
\begin{equation}
f_0\!\left(\mathbf{x}_{0},\mathbf{v}_{0}\right) = \Psi_{0}\!\left(\mathbf{x}_{\bot0}\right) \delta\!\left(z_0\right) \delta\!\left(\delta \mathbf{v}_{\bot0}\right)\delta\!\left(v_{z0}-V\right) \, ,\label{bmdistinit}
\end{equation}
where the perpendicular velocity perturbation $\delta \mathbf{v}_{\bot0}$ in the paraxial limit due to the finite source size is explicitly given by 
\begin{equation}
\delta \mathbf{v}_{\bot0} = \mathbf{v}_{\bot0}-\frac{\mathbf{x}_{\bot0}}{r_i}V \, . 
\end{equation} 
The second is the diverging beam associated with a uniform capsule implosion from a sphere of finite radius $a \ll r_i$. The precise form of the distribution function depends on relative proportion of emission from different parts of the capsule; however, for a sphere emitting isotropically, we can write 
\begin{equation}
f_0\!\left(\mathbf{x}_{0},\mathbf{v}_{0}\right) = \Psi_{0} \, \delta\!\left(z_0\right) \delta\!\left(v_{z0}-V\right) P\!\left(\delta \mathbf{v}_{\bot0}\right) \, ,
\label{spheredistinit}
\end{equation}
where $P$ is the (presumed homogeneous) distribution of perpendicular velocities associated with the finite capsule size. 
An example of such a distribution -- for a sphere emitting uniformly from all points -- is given by
\begin{equation}
P\!\left(\delta \mathbf{v}_{\bot0}\right) = \frac{3}{2 \pi } \left(\frac{r_i}{aV}\right)^2 \sqrt{1-\left(\frac{r_i}{aV}\right)^2 \left|\delta \mathbf{v}_{\bot0}\right|^2} H\!\left(\frac{aV}{r_i}-\left|\delta \mathbf{v}_{\bot0}\right|\right) \, .
\label{sphereunidistinit}
\end{equation}
The derivation of this distribution can be found in Appendix \ref{FiniteSourcePSF}. In practice, for capsules it has been experimentally observed that the spread function is typically Gaussian, with full-half-width maximum $a \approx 40 - 50 \, \mu\mathrm{m}$~\cite{S04,L06}. As mentioned in the introduction to this appendix, the ability to deal with such distributions quantitatively is an advantage of a kinetic approach to proton imaging over real space methods. 

\subsection{Solution for single-proton motion, and deviation of plasma-image mapping \eqref{divmappingSec2}} \label{KinThyPradSingProt}

The solution \eqref{characsol} of Vlasov equation \eqref{Vlasov} shows that determining the motion of a single particle gives the beam distribution function. Therefore, we consider a proton with initial velocity $\mathbf{v}_0$ and position $\mathbf{x}_0$, and aim to characterise its motion both inside the plasma, and subsequent to leaving it. This will enable us to derive the final position on the detector $\mathbf{x}_\bot^{\left(s\right)}$ of a proton with initial perpendicular position $\mathbf{x}_{\bot0}$: in other words, the plasma-image mapping \eqref{divmappingSec2}.

For any given magnetic field, the position $\mathbf{x}\!\left(t\right)$ and velocity $\mathbf{v}\!\left(t\right)$ can in principle be found using  \eqref{protpositionfull} and \eqref{protvelfull} respectively; however, for a general magnetic field the phase-space trajectory of a proton can be extremely complicated, which in turns makes the interpretation of the resulting image-flux distribution a challenge. To arrive at simplified expressions for the position and velocity of individual particles, we invoke the large speed of the protons, and assume that typical proton deflection angles induced by magnetic fields in the plasma are small. This is equivalent to assuming the proton deflection velocity, defined for a proton passing through by
\begin{equation}
\mathbf{w}\!\left(\mathbf{x}_0, \mathbf{v}_0, t\right) = \mathbf{v}\!\left(\mathbf{x}_0, \mathbf{v}_0, t\right) - \mathbf{v}_0  \, ,
\end{equation}
satisfies ordering $\left|\mathbf{w}\right| \ll V$ with respect to the initial proton speed. The equivalence can be seen by noting that the typical deflection angle $\delta \theta$ is
\begin{equation}
\delta \theta \equiv \cos^{-1}{\frac{\mathbf{v}_0 \cdot \left(\mathbf{v}_0 + \mathbf{w}\right)}{\left|\mathbf{v}_0\right|\left|\mathbf{v}_0+\mathbf{w}\right|}} \approx \left|\mathbf{w}\right|/V \ll 1 \, ,
\end{equation}
as required. As discussed in Section \ref{Assum}, the validity of this approximation can be checked, and is often reasonable. 

We then look for a perturbation solution for the position $\mathbf{x}\!\left(\mathbf{x}_0, \mathbf{v}_0, t\right)$ and velocity $\mathbf{v}\!\left(\mathbf{x}_0, \mathbf{v}_0, t\right)$ of a proton passing passing through the plasma of the form
\begin{IEEEeqnarray}{rCl}
\mathbf{x}\!\left(\mathbf{x}_0, \mathbf{v}_0, t\right) & = & \sum_{n = 0}^{\infty} \mathbf{x}^{\left(n\right)}\!\left(\mathbf{x}_0, \mathbf{v}_0, t\right) \, \delta \theta^n \, , \label{protpossum}\\ 
\mathbf{v}\!\left(\mathbf{x}_0, \mathbf{v}_0, t\right) & = & \sum_{n = 0}^{\infty} \mathbf{v}^{\left(n\right)}\!\left(\mathbf{x}_0, \mathbf{v}_0, t\right) \, \delta \theta^n \, , \label{protvelsum}
\end{IEEEeqnarray}
where $\mathbf{x}^{\left(n\right)}\!\left(\mathbf{x}_0, \mathbf{v}_0, t\right)$ and $\mathbf{v}^{\left(n\right)}\!\left(\mathbf{x}_0, \mathbf{v}_0, t\right)$ are vector functions to be determined. Substituting \eqref{protvelsum} into \eqref{protvelfull}, we obtain zeroth-order term in $\delta \theta$
\begin{equation}
\mathbf{v}^{\left(0\right)}\!\left(\mathbf{x}_0, \mathbf{v}_0, t\right) = \mathbf{v}_0 \, , \label{zerothordvel}
\end{equation}
and the first-order term 
\begin{equation}
\mathbf{v}^{\left(1\right)}\!\left(\mathbf{x}_0, \mathbf{v}_0, t\right) \delta \theta = \frac{e}{m_p c}\mathbf{v}_0 \times \int_0^t \mathbf{B}\!\left(\mathbf{x}\!\left(\mathbf{x}_0, \mathbf{v}_0,t'\right)\right) \mathrm{d}t' \, . \label{firstordvel}
\end{equation}
Now expanding $\mathbf{v}_0$ in paraxial parameter $\delta \alpha$ and source-size parameter $a/r_i$ using \eqref{protbminitvelparax}, we deduce that the proton deflection velocity at time $t$ can be approximated as
\begin{equation}
\mathbf{w}\!\left(\mathbf{x}_0, \mathbf{v}_0, t\right) = \frac{e}{m_p c}\hat{\mathbf{z}} \times \int_0^t \mathbf{B}\!\left(\mathbf{x}\!\left(\mathbf{x}_0, \mathbf{v}_0,t'\right)\right) \mathrm{d}t' \left[1+\mathcal{O}\!\left(\delta \alpha, \delta \theta, \frac{a}{r_i}\right)\right]\, .
\end{equation}
Note that this perturbation is perpendicular to the direction of motion (as expected given the form of the Lorentz force). Substituting \eqref{zerothordvel} and \eqref{firstordvel} into \eqref{protpositionfull} gives leading order expression for the position
\begin{equation}
\mathbf{x}^{\left(0\right)}\!\left(\mathbf{x}_0, \mathbf{v}_0, t\right) = \mathbf{x}_0 + \mathbf{v}_0 \, t \, . \label{firstordpos}
\end{equation}
Once again expanding in both the paraxial parameter $\delta \alpha$ and $a/r_i$ implies that the position of the particle as it moves through the plasma satisfies
\begin{IEEEeqnarray}{rCl}
\mathbf{x}_\bot\!\left(\mathbf{x}_0, \mathbf{v}_0, t\right) & = & \mathbf{x}_{\bot0}\left[1 +\frac{Vt}{r_i}+\mathcal{O}\!\left(\delta \theta, \delta \alpha^2, \frac{a}{r_i}\right)\right] \, , \label{perpcoord} \\
z\!\left(\mathbf{x}_0, \mathbf{v}_0, t\right) & = & z_0 + Vt\left[1 +\mathcal{O}\!\left(\delta \theta^{2},\delta \alpha^2,  \frac{a^2}{r_i^2}\right) \right] \, ,\label{zcoord}
\end{IEEEeqnarray}
Since $z_0 = 0$ for all protons, we see from \eqref{zcoord} that to quadratic order in all asymptotic quantities, the particle leaves the plasma region at $t = l_z/V$. The total deflection velocity acquired across the plasma by a particle entering the interaction region at position $\mathbf{x}_{0}$ and initial velocity $\mathbf{v}_{0}$ -- denoted $\mathbf{w}\!\left(\mathbf{x}_{0},\mathbf{v}_0\right)$ -- is then given by
\begin{equation}
\mathbf{w}\!\left(\mathbf{x}_{0},\mathbf{v}_0\right)  \equiv \mathbf{w}\!\left(\mathbf{x}_0, \mathbf{v}_0,\frac{l_z}{V}\right)  = \frac{e}{m_p c} \,\hat{\mathbf{z}} \times \int_0^{l_z} \mathbf{B}\!\left(\mathbf{x}_{\bot}\!\left(\mathbf{x}_{0},\mathbf{v}_0,z'/V\right),z'\right) \mathrm{d}z' \; \left[1+\mathcal{O}\!\left(\delta \alpha, \delta \theta, \frac{a}{r_i}\right)\right] \, . \label{deffieldgen}
\end{equation} 
We have replaced time integration with integration along the path length, which to quadratic order is the $z$ coordinate (by \eqref{zcoord}). The final velocity of the proton on leaving the plasma is given by
\begin{equation}
\mathbf{v}\!\left(\mathbf{x}_{0},\mathbf{v}_0\right) = \mathbf{v}_0 + \mathbf{w}\!\left(\mathbf{x}_{0},\mathbf{v}_0\right) \, , \label{finalplasmavel}
\end{equation}
while the final position is
\begin{equation}
\mathbf{x}\!\left(\mathbf{x}_{0},\mathbf{v}_0\right) = l_z \hat{\mathbf{z}} + \mathbf{x}_{\bot0} \left[1+\mathcal{O}\!\left(\delta \alpha, \delta \theta, \frac{a}{r_i}\right)\right] \, . \label{finalplasmaposition}
\end{equation}

Beyond the plasma, protons undergo free streaming, until they come into contact with the detector (located in the plane $z = r_s+l_z$). To distinguish from the position $\mathbf{x}\!\left(\mathbf{x}_{0},\mathbf{v}_0\right)$ and velocity $\mathbf{v}\!\left(\mathbf{x}_{0},\mathbf{v}_0\right)$ of a proton on leaving the plasma, we denote the position and velocity of a proton during this free-streaming phase -- in other words, for $t \geq l_z/V$ -- by $\mathbf{x}^{\left(s\right)}\!\left(\mathbf{x}_{0},\mathbf{v}_0,t\right) $ and $\mathbf{v}^{\left(s\right)}\!\left(\mathbf{x}_{0},\mathbf{v}_0,t\right) $ respectively. $\mathbf{x}^{\left(s\right)}\!\left(\mathbf{x}_{0},\mathbf{v}_0,t\right) $ and $\mathbf{v}^{\left(s\right)}\!\left(\mathbf{x}_{0},\mathbf{v}_0,t\right) $ are then related to $\mathbf{x}\!\left(\mathbf{x}_{0},\mathbf{v}_0\right)$ and $\mathbf{v}\!\left(\mathbf{x}_{0},\mathbf{v}_0\right)$ by
\begin{IEEEeqnarray}{rCl}
\mathbf{v}^{\left(s\right)}\!\left(\mathbf{x}_{0},\mathbf{v}_0,t\right)   & = & \mathbf{v}\!\left(\mathbf{x}_{0},\mathbf{v}_0\right) \, , \label{genmap1a} \\
\mathbf{x}^{\left(s\right)}\!\left(\mathbf{x}_{0},\mathbf{v}_0,t\right) & = &\mathbf{x}\!\left(\mathbf{x}_{0},\mathbf{v}_0\right) + \mathbf{v}\!\left(\mathbf{x}_{0},\mathbf{v}_0\right) \left(t-\frac{l_z}{V}\right)   \, .\label{genmap1b}
\end{IEEEeqnarray}
When combined with \eqref{deffieldgen}, \eqref{finalplasmavel} and \eqref{finalplasmaposition}, these phase-space coordinates become 
\begin{IEEEeqnarray}{rCl}
\mathbf{v}^{\left(s\right)}\!\left(\mathbf{x}_{0},\mathbf{v}_0,t\right)   & = & \mathbf{v}_0 + \mathbf{w}\!\left(\mathbf{x}_{0},\mathbf{v}_0\right) \, , \label{genmap1A} \\
\mathbf{x}^{\left(s\right)}\!\left(\mathbf{x}_{0},\mathbf{v}_0,t\right) & = & l_z \hat{\mathbf{z}} + \mathbf{x}_{\bot0} \left[1+\mathcal{O}\!\left(\delta \alpha, \delta \theta, \frac{a}{r_i}\right)\right] + \left[\mathbf{v}_0 + \mathbf{w}\!\left(\mathbf{x}_{0},\mathbf{v}_0\right)\right] \left(t-\frac{l_z}{V}\right)   \, .\label{genmap1B}
\end{IEEEeqnarray}
Equations \eqref{genmap1A} and \eqref{genmap1B} provide the desired expressions for a single-proton trajectory in phase space as a function of time and initial phase-space coordinates $\left(\mathbf{x}_{0},\mathbf{v}_0\right)$ .

To derive plasma-image mapping \eqref{divmappingSec2} from \eqref{genmap1a} and \eqref{genmap1b}, we first find the \emph{intersection time} $t_s$ at which a given proton reaches the detector. This can be done geometrically by considering the $z$-component of \eqref{genmap1b}
\begin{equation}
z^{\left(s\right)} =  l_z + V \left(t-\frac{l_z}{V}\right) \left[1 +\mathcal{O}\!\left(\delta \theta^{2},\delta \alpha^2,  \frac{a^2}{r_i^2}\right) \right] \, ,
\end{equation}
and substituting $z^{\left(s\right)} = l_z+r_s$ for $t = t_s$. Re-arranging the result gives
\begin{equation}
t_s = \frac{l_z + r_s}{V}\left[1 +\mathcal{O}\!\left(\delta \theta^{2},\delta \alpha^2,  \frac{a^2}{r_i^2}\right) \right] \, . \label{streamtime}
\end{equation}
Thus to quadratic asymptotic order in $\delta \theta$, $\delta \alpha$ and $a/r_i$, all particles arrive at the detector simultaneously.

Next, we substitute \eqref{deffieldgen}, \eqref{finalplasmavel}, and \eqref{streamtime} into the perpendicular component of \eqref{genmap1b}
\begin{equation}
\mathbf{x}_{\bot}^{\left(s\right)}\!\left(\mathbf{x}_{0},\mathbf{v}_0,\frac{l_z+r_s}{V}\right)  = \mathbf{x}\!\left(\mathbf{x}_{0},\mathbf{v}_0\right) +\left[\mathbf{v}_{\bot0} +\mathbf{w}\!\left(\mathbf{x}_{0},\mathbf{v}_0\right)\right] \frac{r_s}{V} \left[1+\mathcal{O}\!\left(\delta \theta,\delta \alpha, \frac{a}{r_i}\right)\right] \, , \label{plasmapdevA}
\end{equation}
before noting that 
\begin{equation}
\frac{\left|\mathbf{x}_\bot\!\left(\mathbf{x}_{0},\mathbf{v}_0\right)-\mathbf{x}_{\bot0}\right|}{r_s \left|\mathbf{w}\!\left(\mathbf{x}_{0},\mathbf{v}_0\right)\right|/V} \sim \frac{l_z \delta \theta}{r_s \delta \theta} = \delta \beta \ll 1 \, ,
\end{equation}
where $\delta \beta = l_z/r_s$ is the point-projection parameter (defined in Section \ref{PlasMap}, discussed in Section \ref{Assum}). Thus, displacements acquired across the plasma due to magnetic forces are negligible compared to displacements due to free-streaming at a perturbed velocity. \eqref{plasmapdevA} becomes
\begin{equation} 
\mathbf{x}_{\bot}^{\left(s\right)}\!\left(\mathbf{x}_{0},\mathbf{v}_0,\frac{l_z+r_s}{V}\right) = \left(\mathbf{x}_{\bot0} + \left[\mathbf{v}_{\bot0} +\mathbf{w}\!\left(\mathbf{x}_{0},\mathbf{v}_0\right)\right] \frac{r_s}{V}\right)\left[1+\mathcal{O}\!\left(\delta \theta,\delta \alpha, \delta \beta, \frac{a}{r_i}\right)\right]  \, . \label{posmap1c}
\end{equation}
Finally, we observe that under the paraxial approximation, the initial perpendicular velocity of a proton (neglecting the finite source size) is given by \eqref{protbminitvelparax}, that is
\begin{equation}
\mathbf{v}_{\bot0} = \mathbf{v}_{\bot0}\!\left(\mathbf{x}_{\bot0}\right)  = \frac{V \mathbf{x}_{\bot0}}{r_i} \left[1+\mathcal{O}\left(\delta \alpha^2,\frac{a}{l_\bot}\right)\right]  \, .\label{protbminitvelparaxB}
\end{equation}
Then defining the perpendicular-deflection field $\mathbf{w}\!\left(\mathbf{x}_{\bot0}\right)$ by
\begin{equation}
\mathbf{w}\!\left(\mathbf{x}_{\bot0}\right) \equiv \mathbf{w}\!\left(\mathbf{x}_{\bot0},\mathbf{v}_0\!\left(\mathbf{x}_{\bot0}\right)\right)
\end{equation}
allows for the final position of a proton on the detector $\mathbf{x}_{\bot}^{\left(s\right)}\!\left(\mathbf{x}_{\bot0}\right)$ to be written as a function of that proton's initial position:
\begin{IEEEeqnarray}{rCl}
\mathbf{x}_{\bot}^{\left(s\right)}\!\left(\mathbf{x}_{\bot0}\right) & \equiv & \mathbf{x}_{\bot}^{\left(s\right)}\!\left(\mathbf{x}_{\bot0} + \left(l_z+r_s\right)\hat{\mathbf{z}},\mathbf{v}_0\!\left(\mathbf{x}_{\bot0}\right),\frac{l_z+r_s}{V}\right) \nonumber \\
& = &\left(
\frac{r_s+r_i}{r_i} \mathbf{x}_{\bot0} + \frac{r_s}{V} \, \mathbf{w}\!\left(\mathbf{x}_{\bot0}\right) \right) \left[1+\mathcal{O}\!\left(\delta \theta,\delta \alpha, \delta \beta, \frac{a}{r_i}\right)\right]  \, , \label{posmap1c}
\end{IEEEeqnarray}
the plasma-image mapping \eqref{divmappingSec2}.

\subsection{Relating initial and final beam distribution functions} \label{KinThyPradBeamDist}

The phase spacing mapping now specified between the initial and final phase space coordinates by \eqref{genmap1a} and \eqref{genmap1b} can be used in conjunction with \eqref{characsol} to determine the particle distribution function. Re-writing \eqref{characsol} in terms of phase-space image coordinates  $\left(\mathbf{x}^{\left(s\right)},\mathbf{v}^{\left(s\right)}\right)$ leads to 
\begin{equation}
f\!\left(\mathbf{x}^{\left(s\right)},\mathbf{v}^{\left(s\right)},t\right) = f_{0}\!\left[\mathbf{x}_0\!\left(\mathbf{x}^{\left(s\right)},\mathbf{v}^{\left(s\right)},t\right),\mathbf{v}_0\!\left(\mathbf{x}^{\left(s\right)},\mathbf{v}^{\left(s\right)},t\right)\right] \left|\frac{\partial\!\left(\mathbf{x}^{\left(s\right)},\mathbf{v}^{\left(s\right)}\right)}{\partial\!\left(\mathbf{x}_0,\mathbf{v}_0\right)}\right|^{-1} \, .\label{characsolB}
\end{equation}
First evaluating the phase space determinant determinant, we have
\begin{equation}
\frac{\partial\!\left(\mathbf{x}^{\left(s\right)}\!,\mathbf{v}^{\left(s\right)}\right)}{\partial\!\left(\mathbf{x}_0,\mathbf{v}_0\right)} = 
\renewcommand\arraystretch{1.5}
\begin{vmatrix} 
\; \frac{\partial \mathbf{x}^{\left(s\right)}}{\partial \mathbf{x}_0} & \frac{\partial \mathbf{x}^{\left(s\right)}}{\partial \mathbf{v}_0}  \; \\
\; \frac{\partial \mathbf{v}^{\left(s\right)}}{\partial \mathbf{x}_0} & \frac{\partial \mathbf{v}^{\left(s\right)}}{\partial \mathbf{v}_0} \;
\end{vmatrix}
=
\begin{vmatrix} 
\; \underline{\underline{\mathbf{I}}}+\frac{\partial \mathbf{w}}{\partial \mathbf{x}_0} \left(t-\frac{l_z}{V}\right) \; & \; \left(t-\frac{l_z}{V}\right) \underline{\underline{\mathbf{I}}} \; \\
\frac{\partial \mathbf{w}}{\partial \mathbf{x}_0} & \underline{\underline{\mathbf{I}}} 
\end{vmatrix}
= 1+\mathcal{O}\!\left(\delta \theta,\delta \alpha, \frac{a}{r_i}\right) \, ,
\end{equation}
where for the last identity we subtract scalar multiples of the bottom rows from the top ones (a procedure which leaves the value of the determinant unchanged). The beam distribution function is then
\begin{equation}
f\!\left(\mathbf{x}^{\left(s\right)},\mathbf{v}^{\left(s\right)},t\right) = f_{0}\!\left[\mathbf{x}_0\!\left(\mathbf{x}^{\left(s\right)},\mathbf{v}^{\left(s\right)},t\right)\!,\mathbf{v}_0\!\left(\mathbf{x}^{\left(s\right)},\mathbf{v}^{\left(s\right)},t\right)\right]  \, .\label{characsol2}
\end{equation}

We are now able to write done an expression for the time-dependent flux, and image-flux distribution, in terms of the initial beam distribution function. Recalling general expression \eqref{fluxtime} for the time-dependent flux of protons through the detector at perpendicular location $\mathbf{x}_\bot^{\left(s\right)}$ and time $t$, we substitute \eqref{characsol2} to give
\begin{equation}
\psi\!\left(\mathbf{x}_\bot^{\left(s\right)},t\right) = \int \mathrm{d}^3 \mathbf{v}^{\left(s\right)} \, v_z \,  f_{0}\!\left[\mathbf{x}_0\!\left(\mathbf{x}_{\bot}^{\left(s\right)}+\left(r_s+l_z\right) \hat{\mathbf{z}},\mathbf{v}^{\left(s\right)},t\right)\!,\mathbf{v}_0\!\left(\mathbf{x}_{\bot}^{\left(s\right)}+\left(r_s+l_z\right) \hat{\mathbf{z}},\mathbf{v}^{\left(s\right)},t\right)\right] \, . \label{fluxtimeB}
\end{equation}
The image-flux distribution follows as before from \eqref{fluxtot}:
\begin{equation}
\Psi\!\left(\mathbf{x}_\bot^{\left(s\right)}\right) = \int_0^{\infty} \mathrm{d} t \, \psi\!\left(\mathbf{x}_\bot^{\left(s\right)},t\right) \, . \label{fluxtotB}
\end{equation} 
However, the result \eqref{streamtime} for the intersection time of an arbitrary proton with the detector -- in particular, its identical value for all beam protons in the paraxial limit -- suggests that the time-dependent flux will be related to the image-flux distribution by a delta function.

Time-dependent flux experession \eqref{fluxtimeB} demonstrates that the relation between the image-flux distribution and the magnetic fields is crucially dependent on the form of the initial distribution. In the next two sub-appendices, we study the two particular forms of initial distribution specified in Appendix \ref{KinThyPradInitCon}: a proton point-source \eqref{bmdistinit}, and a finite, uniformly emitting proton source \eqref{sphereunidistinit}. 

\subsection{Image-flux distribution for a point proton source - derivation of Kugland image-flux relation \eqref{screenfluxSec2}} \label{KinThyPradPointSource}

In the case of a proton beam emitted from a point source ($a = 0$), the initial distribution function is given by \eqref{bmdistinit}, that is, 
\begin{equation}
f_0\!\left(\mathbf{x}_{0},\mathbf{v}_{0}\right) = \Psi_{0}\!\left(\mathbf{x}_{\bot0}\right) \delta\!\left(z_0\right) \delta\!\left(\delta \mathbf{v}_{\bot0}\right)\delta\!\left(v_{z0}-V\right) \, .\label{bmdistinitB}
\end{equation}
The beam distribution function at a time $t \geq l_z/V$ is then given by recourse to \eqref{characsol2}, combined with :
\begin{IEEEeqnarray}{rCl}
f\!\left(\mathbf{x}^{\left(s\right)},\mathbf{v}^{\left(s\right)},t\right) & = & \Psi_{0}\!\left(\mathbf{x}_{\bot0}\right) \delta\!\left(z^{\left(s\right)}-Vt\right) \delta\!\left(\mathbf{v}_{\bot}^{\left(s\right)}-\mathbf{w}\!\left(\mathbf{x}_{\bot0},\mathbf{v}_{\bot0}+V\hat{\mathbf{z}}\right) -\frac{\mathbf{x}_{\bot0}}{r_i}V\right)\delta\!\left(v_z^{\left(s\right)}-V\right) \nonumber \\ 
 & = & \Psi_{0}\!\left(\mathbf{x}_{\bot0}\right) \delta\!\left(z^{\left(s\right)}-Vt\right) \delta\!\left(\mathbf{v}_{\bot}^{\left(s\right)}-\mathbf{w}\!\left(\mathbf{x}_{\bot0}\right) -\frac{\mathbf{x}_{\bot0}}{r_i}V\right)\delta\!\left(v_z^{\left(s\right)}-V\right) \, ,\label{distfuncbeam}
\end{IEEEeqnarray}
where $\mathbf{x}_{\bot0} = \mathbf{x}_{\bot0}\!\left(\mathbf{x}_{\bot}^{\left(s\right)},\mathbf{v}_{\bot}^{\left(s\right)},t\right)$ implicitly depend on the relevant perpendicular phase-space coordinates. In the second line, we have noted that since $\mathbf{v}_{\bot0} = \mathbf{v}_{\bot0}\!\left(\mathbf{x}_{\bot0}\right)$ in the initial distribution function, we can regard $\mathbf{w}\!\left(\mathbf{x}_{\bot0},\mathbf{v}_{\bot0}+V\hat{\mathbf{z}}\right)$ as a function of $\mathbf{x}_{\bot0}$ alone, that is 
\begin{equation}
\mathbf{w}\!\left(\mathbf{x}_{\bot0},\mathbf{v}_{\bot0}+V\hat{\mathbf{z}}\right) = \mathbf{w}\!\left(\mathbf{x}_{\bot0},\mathbf{v}_{\bot0}\!\left(\mathbf{x}_{\bot0}\right)+V\hat{\mathbf{z}}\right) = \mathbf{w}\!\left(\mathbf{x}_{\bot0}\right) \, .
\end{equation}

Evaluating the time-dependent flux using \eqref{fluxtimeB}, the $v_z^{\left(s\right)}$ integral drops out to give
\begin{equation}
\psi\!\left(\mathbf{x}_{\bot}^{\left(s\right)},t_s\right) = V \Psi_{0}\!\left(\mathbf{x}_{\bot0}\right) \delta\!\left(r_s+l_z-Vt\right) \int \mathrm{d}^2  \mathbf{v}_{\bot}^{\left(s\right)} \delta\!\left(\mathbf{v}_{\bot}^{\left(s\right)}-\mathbf{w}\!\left(\mathbf{x}_{\bot0}\right)-\frac{\mathbf{x}_{\bot0}}{r_i}V\right) \, . \label{timedepfluxpointsource}
\end{equation}
Since the latter two terms in the delta function are implicit functions of $\mathbf{v}_{\bot}^{\left(s\right)}$, we make use of the general result
\begin{equation}
 \int \mathrm{d}\mathbf{y}\, \delta\!\left(\mathbf{g}\!\left(\mathbf{y}\right)\right) =  \sum_{\tilde{\mathbf{y}}}\frac{1}{\left|\det{\frac{\partial \mathbf{g}}{\partial \mathbf{y}}}\!\left(\tilde{\mathbf{y}}\right)\right|} \, , \label{deltaintegral}
\end{equation}
where $\tilde{\mathbf{y}}$ are the set of all values of $\mathbf{y}$ satisfying $\mathbf{g}\!\left(\mathbf{y}\right) = 0$. Setting the argument of the delta function in \eqref{timedepfluxpointsource} equal to zero gives 
\begin{equation}
\mathbf{v}_{\bot}^{\left(s\right)} =\mathbf{w}\!\left(\mathbf{x}_{\bot0}\right)-\frac{\mathbf{x}_{\bot0}}{r_i}V \, . \label{deltaansatz}
\end{equation}
where we have substituted for the first two terms of the argument using the perpendicular component of \eqref{genmap1A}. From the perpendicular component of \eqref{genmap1b} combined with \eqref{genmap1a} it follows that
\begin{equation}
\frac{\partial \mathbf{x}_{\bot0}}{\partial \mathbf{v}_{\bot}^{\left(s\right)}}\bigg|_{\mathbf{x}_\bot^{\left(s\right)}\!,\,t} = -\underline{\underline{\mathbf{I}}} \left(t-\frac{l_z}{V}\right) \, ,
\end{equation}
The derivative of the argument of the delta function in \eqref{deltaintegral} is then
\begin{equation}
\frac{\partial}{\partial \mathbf{v}_\bot^{\left(s\right)}}\bigg|_{\mathbf{x}_\bot^{\left(s\right)}\!,\,t}\!\left[\mathbf{v}_{\bot}^{\left(s\right)}-\mathbf{w}\!\left(\mathbf{x}_{\bot0}\right)-\frac{\mathbf{x}_{\bot0}}{r_i}V\right] = \underline{\underline{\mathbf{I}}}\left(1+\frac{t}{r_i}V-l_z\right)+  \left(t-\frac{l_z}{V}\right) \frac{\partial \mathbf{w}\!\left(\mathbf{x}_{\bot0}\right)}{\partial \mathbf{x}_{\bot0}} = \frac{\partial \mathbf{x}_{\bot}^{\left(s\right)}}{\partial \mathbf{x}_{\bot0}} \, ,
\end{equation}
where $\mathbf{x}_{\bot}^{\left(s\right)} = \mathbf{x}_{\bot}^{\left(s\right)}\!\left(\mathbf{x}_{\bot0},t\right)$ is now the purely spatial map defined by 
\begin{equation}
\mathbf{x}_{\bot}^{\left(s\right)}\!\left(\mathbf{x}_{\bot0},t\right) \equiv \mathbf{x}_{\bot0} + \left[\frac{\mathbf{x}_{\bot0}}{r_i}V +\mathbf{w}\!\left(\mathbf{x}_{\bot0}\right)\right]  \left(t-\frac{l_z}{V}\right) \, .
\end{equation}
By construction, this satisfies the vanishing delta-function ansatz requirement \eqref{deltaintegral}. The time-dependent flux is then
\begin{equation}
\psi\!\left(\mathbf{x}_\bot^{\left(s\right)},t_s\right) = V \delta\!\left(r_s+l_z-Vt\right)  \sum_{\mathbf{x}_\bot^{\left(s\right)} = \mathbf{x}_\bot^{\left(s\right)}\!\left(\mathbf{x}_{\bot0}\right)}\frac{\Psi_{0}\!\left(\mathbf{x}_{\bot0}\right)}{\left|\det{\frac{\partial \mathbf{x}_{\bot}^{\left(s\right)}}{\partial \mathbf{x}_{\bot0}}}\right|} \, .
\end{equation}
Finally, integrating over all time, we obtain the image-flux distribution
\begin{equation}
\Psi\!\left(\mathbf{x}_\bot^{\left(s\right)}\right) = \sum_{\mathbf{x}_\bot^{\left(s\right)} = \mathbf{x}_\bot^{\left(s\right)}\!\left(\mathbf{x}_{\bot0}\right)}\frac{\Psi_{0}\!\left(\mathbf{x}_{\bot0}\right)}{\left|\det{\frac{\partial \mathbf{x}_{\bot}^{\left(s\right)}}{\partial \mathbf{x}_{\bot0}}}\right|} \, , \label{screenfluxA}
\end{equation}
as well as the intersection time $t_s$ at which the proton beam reaches the detector: 
\begin{equation}
t_s = \frac{r_s+l_z}{V} \, ,
\end{equation}
in agreement with result \eqref{streamtime} as derived using geometric arguments. The mapping between perpendicular coordinates on the detector and in the plasma becomes
\begin{equation}
\mathbf{x}_{\bot}^{\left(s\right)} = \left(
\frac{r_s+r_i}{r_i} \mathbf{x}_{\bot0} + \frac{r_s}{V} \, \mathbf{w}\!\left(\mathbf{x}_{\bot0}\right) \right) \left[1 + \mathcal{O}\!\left(\delta \theta,\delta \alpha, \delta \beta \right)\right] \, ,\label{divmapping}
\end{equation}
recovering the plasma-image mapping \eqref{divmappingSec2} once again. 

The image-flux relation \eqref{screenfluxA} agrees with that stated in the main text, that is \eqref{screenfluxSec2}, as well as that given by Kugland \emph{et. al.}~\cite{K12}, in the case of an initially uniform distribution $\Psi_{0}\!\left(\mathbf{x}_{\bot0}\right) = \Psi_0$. The contribution of magnetic deflections becomes clearer on expanding the determinant in \eqref{screenfluxA}
\begin{IEEEeqnarray}{rCl}
\det{\frac{\partial \mathbf{x}_{\bot}^{\left(s\right)}}{\partial \mathbf{x}_{\bot0}}} & = & \det{\left(\frac{r_s+r_i}{r_i} \underline{\underline{\mathrm{I}}}+\frac{r_s}{V} \frac{\partial \mathbf{w}\!\left(\mathbf{x}_{\bot0}\right)}{\partial \mathbf{x}_{\bot0}}\right)} \nonumber \\
& = &  \left(\frac{r_s+r_i}{r_i}\right)^2 \left\{1+\frac{r_s r_i}{\left(r_s+r_i\right)V} \nabla_{\bot0} \cdot \mathbf{w}\!\left(\mathbf{x}_{\bot0}\right) + \left[\frac{r_s r_i}{\left(r_s+r_i\right)V}\right]^2 \det{\frac{\partial \mathbf{w}\!\left(\mathbf{x}_{\bot0}\right)}{\partial \mathbf{x}_{\bot0}}}\right\} \label{fluxdet}
\end{IEEEeqnarray}
This gives
\begin{equation}
\Psi\!\left(\mathbf{x}_\bot^{\left(s\right)}\right) = \sum_{\mathbf{x}_\bot^{\left(s\right)} = \mathbf{x}_\bot^{\left(s\right)}\!\left(\mathbf{x}_{\bot0}\right)}\frac{\Psi_{0}\!\left(\mathbf{x}_{\bot0}\right)}{\left| 1+\frac{r_s r_i}{\left(r_s+r_i\right)V} \nabla_{\bot0} \cdot \mathbf{w}\!\left(\mathbf{x}_{\bot0}\right) + \left(\frac{r_s r_i}{\left[r_s+r_i\right]V}\right)^2 \det{\frac{\partial \mathbf{w}\!\left(\mathbf{x}_{\bot0}\right)}{\partial \mathbf{x}_{\bot0}}}\right|} \, .\label{screenfluxB}
\end{equation}

\subsection{Image-flux distribution for a finite proton source} \label{ScreenDistFiniteSourceDev}

Having dealt with the point source case, we now consider distributions resulting from a finite source, with initial distribution given by \eqref{spheredistinitB} -- for convenience we reproduce it here:
\begin{equation}
f_0\!\left(\mathbf{x}_{0},\mathbf{v}_{0}\right) = \Psi_{0} \, \delta\!\left(z_0\right) \delta\!\left(v_{z0}-V\right) P\!\left(\delta \mathbf{v}_{\bot0}\right) \, ,
\label{spheredistinitB}
\end{equation}
The final beam distribution function is again given by \eqref{characsol2}:
\begin{IEEEeqnarray}{rCl}
f\!\left(\mathbf{x}_{\bot}^{\left(s\right)}\!,z^{\left(s\right)}\!,\mathbf{v}^{\left(s\right)}\!,t\right) & = & \Psi_{0} \, \delta\!\left(z^{\left(s\right)}-Vt\right) \delta\!\left(v_{z}^{\left(s\right)}-V\right) P\!\left(\mathbf{v}_{\bot}^{\left(s\right)}-\mathbf{w}\!\left(\mathbf{x}_{\bot0}\right)-\frac{\mathbf{x}_{\bot0}}{r_i}V\right) \, ,
\end{IEEEeqnarray}
where $\mathbf{x}_{\bot0} = \mathbf{x}_{\bot0}\!\left(\mathbf{x}_{\bot}^{\left(s\right)},\mathbf{v}_{\bot}^{\left(s\right)},t\right)$ as before.

Next, we evaluate the time-dependent flux using \eqref{fluxtimeB}; similarly to the point source case, the $v_z^{\left(s\right)}$ integral in the time-dependent flux can be evaluated separately to give
\begin{equation}
\psi\!\left(\mathbf{x}_{\bot}^{\left(s\right)},t\right) = V \Psi_{0} \, \delta\!\left(z^{\left(s\right)}-Vt\right) \int \mathrm{d}^2  \mathbf{v}_{\bot}^{\left(s\right)} P\!\left(\mathbf{v}_{\bot}^{\left(s\right)}-\mathbf{w}\!\left(\mathbf{x}_{\bot0}\right)-\frac{\mathbf{x}_{\bot0}}{r_i}V\right) \, .\label{smearedflux}
\end{equation} 
To compute this integral, define new integration variable
\begin{equation}
\tilde{\mathbf{x}}_{\bot}^{\left(s\right)}\!\left(\mathbf{v}_{\bot}^{\left(s\right)}\right) \equiv  \mathbf{x}_{\bot0}\!\left(\mathbf{v}_{\bot}^{\left(s\right)}\right) + \left\{\frac{\mathbf{x}_{\bot0}\!\left(\mathbf{v}_{\bot}^{\left(s\right)}\right)}{r_i}V +\mathbf{w}\!\left[\mathbf{x}_{\bot0}\!\left(\mathbf{v}_{\bot}^{\left(s\right)}\right)\right]\right\} \left(t-\frac{l_z}{V}\right)\, ,
\end{equation}
and substitute for $\mathbf{v}_{\bot}^{\left(s\right)}$ using the perpendicular component of \eqref{genmap1a} and \eqref{genmap1b}:
\begin{equation}
\mathbf{v}_{\bot}^{\left(s\right)} = \frac{\tilde{\mathbf{x}}_{\bot}^{\left(s\right)}-\mathbf{x}_{\bot0}}{t-l_z/V} \, .
\end{equation}
Then, \eqref{smearedflux} becomes 
\begin{equation}
\psi\!\left(\mathbf{x}_{\bot}^{\left(s\right)},t\right) = V \Psi_{0} \, \delta\!\left(z^{\left(s\right)}-Vt\right) \int \mathrm{d}^2  \tilde{\mathbf{x}}_{\bot}^{\left(s\right)} \left|\det{\frac{\partial \mathbf{v}_{\bot}^{\left(s\right)}}{\partial \tilde{\mathbf{x}}_{\bot}^{\left(s\right)}}}\right|P\!\left(\frac{\tilde{\mathbf{x}}_{\bot}^{\left(s\right)}-\mathbf{x}_{\bot}^{\left(s\right)}}{t-l_z/V}\right) \, . \label{smearedflux2}
\end{equation}
Evaluating the Jacobian matrix, we find
\begin{equation}
\frac{\partial \tilde{\mathbf{x}}_{\bot}^{\left(s\right)}}{\partial \mathbf{v}_{\bot}^{\left(s\right)}}\bigg|_{\mathbf{x}_\bot^{\left(s\right)}} =  \frac{\partial \mathbf{x}_{\bot0}}{\partial \mathbf{v}_{\bot}^{\left(s\right)}}\bigg|_{\mathbf{x}_\bot^{\left(s\right)}} \cdot \frac{\partial \tilde{\mathbf{x}}_{\bot}^{\left(s\right)}}{\partial \mathbf{x}_{\bot0}} = - \left(t-\frac{l_z}{V}\right) \frac{\partial \tilde{\mathbf{x}}_{\bot}^{\left(s\right)}}{\partial \mathbf{x}_{\bot0}} \, ,
\end{equation}
and so
\begin{equation}
\left|\det{\frac{\partial \tilde{\mathbf{x}}_{\bot}^{\left(s\right)}}{\partial \mathbf{v}_{\bot}^{\left(s\right)}}}\right| =  \left(t-\frac{l_z}{V}\right)^2 \left|\det{\frac{\partial \tilde{\mathbf{x}}_{\bot}^{\left(s\right)}}{\partial \mathbf{x}_{\bot0}}}\right| \, .
\end{equation}
The time-dependent flux is then
\begin{equation}
\psi\!\left(\mathbf{x}_{\bot}^{\left(s\right)},t\right) = V \Psi_{0} \, \delta\!\left(z^{\left(s\right)}-Vt\right) \left(t-\frac{l_z}{V}\right)^{-2} \int \mathrm{d}^2  \tilde{\mathbf{x}}_{\bot}^{\left(s\right)} \left|\det{\frac{\partial \mathbf{x}_{\bot0}}{\partial \tilde{\mathbf{x}}_{\bot}^{\left(s\right)}}}\right| P\!\left(\frac{\tilde{\mathbf{x}}_{\bot}^{\left(s\right)}-\mathbf{x}_{\bot}^{\left(s\right)}}{t-l_z/V}\right) \, , \label{smearedflux2}
\end{equation}
This leads to image-flux distribution for a finite proton source
\begin{equation}
\tilde{\Psi}\!\left(\mathbf{x}_\bot^{\left(s\right)}\right) = \frac{V^2}{r_s^2}\int \mathrm{d}^2  \tilde{\mathbf{x}}_{\bot}^{\left(s\right)} \frac{\Psi_{0}}{\left|\det{\frac{\partial \tilde{\mathbf{x}}_{\bot}^{\left(s\right)}}{\partial \mathbf{x}_{\bot0}}}\right|} P\!\left(\frac{\tilde{\mathbf{x}}_{\bot}^{\left(s\right)}-\mathbf{x}_{\bot}^{\left(s\right)}}{r_s}V\right) \, . \label{smearedscreenfluxa}
\end{equation}
where we have again noted that the detector is located at $z^{\left(s\right)} = r_s+ l_z$, and then neglected $\mathcal{O}\!\left(\delta \beta\right)$ terms. \eqref{smearedscreenfluxa} can be rewritten in terms of the flux from a point source as
\begin{equation}
\tilde{\Psi}\!\left(\mathbf{x}_\bot^{\left(s\right)}\right) = \int \mathrm{d}^2  \tilde{\mathbf{x}}_{\bot}^{\left(s\right)} \, \Psi\!\left(\tilde{\mathbf{x}}_\bot^{\left(s\right)}\right) S\!\left(\mathbf{x}_{\bot}^{\left(s\right)}-\tilde{\mathbf{x}}_{\bot}^{\left(s\right)}\right) \, ,  \label{smearedscreenfluxb}
\end{equation}
where $S$ is a point-spread function given by
\begin{equation}
S\!\left(\mathbf{x}_{\bot}^{\left(s\right)}-\tilde{\mathbf{x}}_{\bot}^{\left(s\right)}\right) = \frac{V^2}{r_s^2} P\!\left(\frac{\tilde{\mathbf{x}}_{\bot}^{\left(s\right)}-\mathbf{x}_{\bot}^{\left(s\right)}}{r_s}V\right) \, .
\end{equation}
Thus we see that the image-flux distribution resulting from a finite source is simply the convolution of the flux from a point source with a point-spread function, the form of which is closely related to the initial distribution function of perpendicular velocities at each point. For example, the uniformly emitting sphere (with $P$ given by \eqref{sphereunidistinit}) gives
\begin{equation}
S\!\left(\mathbf{x}_{\bot}^{\left(s\right)}-\tilde{\mathbf{x}}_{\bot}^{\left(s\right)}\right) = \frac{3}{2 \pi }\left(\frac{r_i}{a r_s}\right)^2 \sqrt{1-\left(\frac{r_i}{a}\right)^2 \left(\frac{\tilde{\mathbf{x}}_{\bot}^{\left(s\right)}-\mathbf{x}_{\bot}^{\left(s\right)}}{r_s}\right)^2} H\!\left(\frac{ar_s}{r_i}-\left|\mathbf{x}_{\bot}^{\left(s\right)}-\tilde{\mathbf{x}}_{\bot}^{\left(s\right)}\right|\right) \, .  \label{smearedscreenfluxufmPSF}
\end{equation}

\subsection{Properties of the perpendicular-deflection field for $\delta \theta \lesssim l_B/l_z$} \label{KinThyPradPerpDefl}

In Section \eqref{PlasMap}, it is stated that provided deflections are sufficiently small -- that is, $\delta \theta \lesssim l_B/l_z$ -- then the perpendicular-deflection field could be written as the gradient of the deflection-field potential $\varphi$, defined by
\begin{equation}
\varphi\!\left(\mathbf{x}_{\bot0}\right) \equiv \frac{r_s}{V} \int_C \mathrm{d} \mathbf{l} \cdot \mathbf{w}\!\left(\tilde{\mathbf{x}}_{\bot0}\right) \, , \label{deffieldpotdefAppend}
\end{equation}
where $C$ is any path from origin to the perpendicular coordinate $\mathbf{x}_{\bot0}$ and $\mathrm{d}\mathbf{l}$ is an infinitesimal line element associated with the path $C$. Under the same assumption, the perpendicular-deflection field became
\begin{equation}
\mathbf{w}\!\left(\mathbf{x}_{\bot0}\right) \approx \frac{e}{m_p c} \hat{\mathbf{z}} \times \int_0^{l_z} \mathrm{d}z' \; \mathbf{B}\!\left(\mathbf{x}_{\bot0}\left(1+\frac{z'}{r_i}\right),z'\right) \, . \label{deffieldundeflAppend}
\end{equation}
In this appendix, we use the formalism developed in sub-appendix \ref{KinThyPradSingProt} to derive \eqref{deffieldpotdefAppend} and \eqref{deffieldundeflAppend}, and specify more precisely the asymptotic errors associated with these results. 

To demonstate the existence of the deflection-field potential \eqref{deffieldpotdefAppend}, we show that the curl of the perpendicular-deflection field is asymptotically small. The perpendicular-deflection field \eqref{deffieldgen} derived in sub-appendix \ref{KinThyPradSingProt} using asymptotic expansions in the typical deflection angle $\delta \theta$, $\delta \alpha$ and $a/r_i$ can be written in terms of the perpendicular initial coordinate $\mathbf{x}_{\bot0}$ alone as
\begin{equation}
\mathbf{w}\!\left(\mathbf{x}_{\bot0}\right) = \frac{e}{m_p c} \,\hat{\mathbf{z}} \times \int_0^{l_z} \mathbf{B}\!\left(\mathbf{x}_{\bot}\!\left(\mathbf{x}_{\bot0},\mathbf{v}_0\!\left(\mathbf{x}_{\bot0}\right),z'/V\right),z'\right) \mathrm{d}z' \; \left[1+\mathcal{O}\!\left(\delta \alpha, \delta \theta, \frac{a}{r_i}\right)\right] \, . \label{deffieldgenAppend}
\end{equation}
Taking the two-dimensional curl with respect to the initial perpendicular coordinates gives
\begin{equation}
\nabla_{\bot0} \times \mathbf{w}\!\left(\mathbf{x}_{\bot0}\right) = \hat{\mathbf{z}}\frac{e}{m_p c} \int_0^{l_z} \mathrm{d}z' \left( \frac{\partial \mathbf{x}_{\bot}\!\left(z'\right)}{\partial \mathbf{x}_{\bot0}} \cdot \nabla_\bot \right) \cdot \mathbf{B} \, . \label{curldeflfieldAppend}
\end{equation}
where $\nabla_\bot = \partial/\partial \mathbf{x}_\bot$ is the gradient operator with respect to the perpendicular proton postion $\mathbf{x}_\bot$. Recalling from \eqref{perpcoord} that the perpendicular proton position has asymptotic expansion
\begin{equation}
\mathbf{x}_{\bot}\!\left(z'\right) \equiv \mathbf{x}_\bot\!\left(\mathbf{x}_0, \mathbf{v}_0\!\left(\mathbf{x}_{\bot0}\right)\!, z'/V\right) =  \mathbf{x}_{\bot0}\left[1 +\mathcal{O}\!\left(\delta \theta, \delta \alpha, \frac{a}{r_i}\right)\right] \, , \label{perpcoordAppend}
\end{equation}
we see that
\begin{equation}
\frac{\partial \mathbf{x}_{\bot}\!\left(z'\right)}{\partial \mathbf{x}_{\bot0}} = \underline{\underline{\mathbf{I}}} \left[1 +\mathcal{O}\!\left(\delta \theta \frac{l_z}{l_B}, \delta \alpha, \frac{a}{r_i}\right)\right] \, . \label{gradplascoord}
\end{equation}
Note the appearance of the asymptotic parameter $\delta \theta \, l_z/l_B$, which arises due to gradients of the magnetic field -- and hence the perpendicular particle position -- being on the scale $l_B$, which may well be different to the global transverse plasma scale $l_z$. Substituting \eqref{gradplascoord} into \eqref{curldeflfieldAppend} gives
\begin{equation}
\nabla_{\bot0} \times \mathbf{w}\!\left(\mathbf{x}_{\bot0}\right) = \hat{\mathbf{z}}\frac{e}{m_p c} \int_0^{l_z} \mathrm{d}z'  \, \nabla \cdot \mathbf{B} \approx 0 \, . \label{curldeflfieldAppend}
\end{equation}
Thus the perpendicular-deflection field is irrotational to this level of approximation, and thus a deflection-field potential of the form \eqref{deffieldpotdefAppend} must exist. 

In fact, the deflection-field potential can be simply related to the vector potential for the magnetic field $\mathbf{A}$. Using vector identity $\hat{\mathbf{z}} \times \mathbf{B} = \nabla \! \left(\hat{\mathbf{z}}  \cdot \mathbf{A}\right) - \left(\hat{\mathbf{z}} \cdot \nabla\right) \! \mathbf{A}$, we can deduce that
\begin{equation}
\mathbf{w}\!\left(\mathbf{x}_{\bot0}\right) = \nabla_{\bot0}\!\left(\frac{e}{m_p cV} \int_0^{l_z} \hat{\mathbf{z}} \cdot \mathbf{A}\!\left(\mathbf{x}\!\left(z'\right),z'\right) \mathrm{d}z'  \right)\; \left[1+\mathcal{O}\!\left(\delta \alpha, \delta \theta, \frac{a}{r_i}\right)\right] \, , \label{deffieldpotgenB}
\end{equation}
so
\begin{equation}
\varphi\!\left(\mathbf{x}_{\bot0}\right) = \frac{e}{m_p cV} \int_0^{l_z} \hat{\mathbf{z}} \cdot \mathbf{A}\!\left(\mathbf{x}\!\left(z'\right),z'\right) \mathrm{d}z' 
\end{equation}
gives an alternative expression for the deflection-field potential. 

We can illustrate the accuracy of the deflection-field potential by estimating the magnitude of the curl relative to typical perpendicular-deflection field gradients:
\begin{equation}
\frac{\nabla_{\bot0} \times \mathbf{w}\!\left(\mathbf{x}_{\bot0}\right)}{\mathbf{w}\!\left(\mathbf{x}_{\bot0}\right)/l_B} = \mathcal{O}\!\left(\delta \theta \frac{l_z}{l_B}, \delta \alpha, \frac{a}{r_i}\right) \, .
\end{equation}
Thus, we see that the deflection-field potential is accurate, provided the previously stated condition $\delta \theta \lesssim l_B/l_z$ is satisfied. The condition breaks down when the displacements acquired across the plasma become similar in magnitude to the size of magnetic structures themselves. This is equivalent to particle paths crossing inside the plasma.

To derive \eqref{deffieldundeflAppend}, we expand the argument of the magnetic field in \eqref{deffieldgenAppend} around the perpendicular proton position \eqref{perpcoordAppend}: 
\begin{equation}
\mathbf{B}\!\left(\mathbf{x}_{\bot}\!\left(z'\right),z\right) =  \mathbf{B}\!\left( \mathbf{x}_{\bot0} \left[1 +\frac{z'}{r_i}\right]\!,z\right) \left\{1 + \mathcal{O}\!\left[\delta \theta \frac{l_z}{l_B}, \, \delta \alpha, \frac{a}{r_i} \right] \right\} \, . \label{magfieldexpand1}
\end{equation}
to give
\begin{equation}
\mathbf{w}\!\left(\mathbf{x}_{\bot0}\right) =  \frac{e}{m_p c}\hat{\mathbf{z}} \times \int_0^{l_z} \mathbf{B}\!\left(\mathbf{x}_{\bot0}\left(1+\frac{z'}{r_i}\right)\!,z'\right) \mathrm{d}z' \;\left[1+\mathcal{O}\!\left(\frac{l_z}{l_B}\delta \theta,\delta \alpha, \frac{a}{r_i}\right)\right] \, , 
\end{equation}
as required.

\subsection{Numerical testing of perpendicular-deflection field and image-flux relations} \label{KinThyPradNumSim}

\begin{figure}[htbp]
\centering
\includegraphics[width=0.9\textwidth]{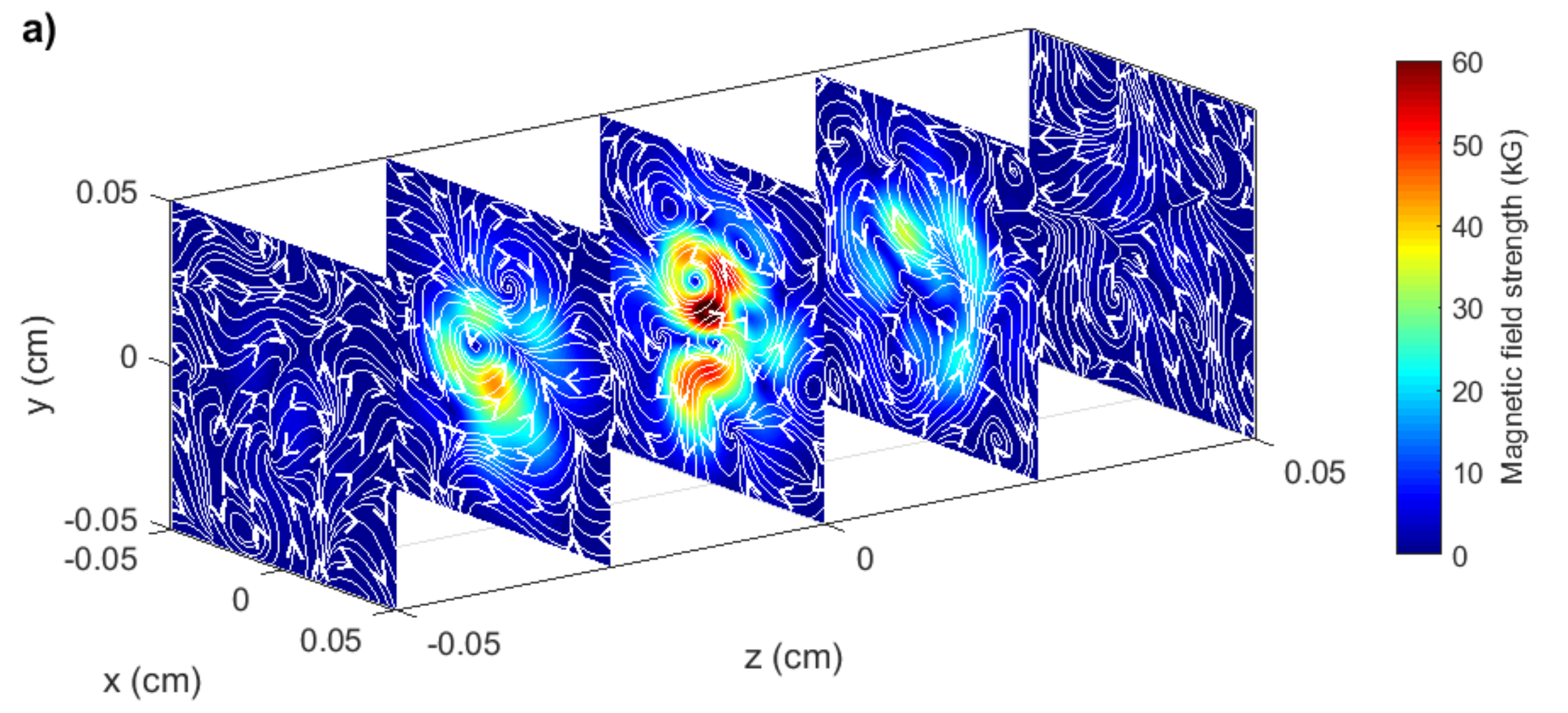}
        \begin{subfigure}{.48\textwidth}
        \centering
       \includegraphics[width=0.9\linewidth]{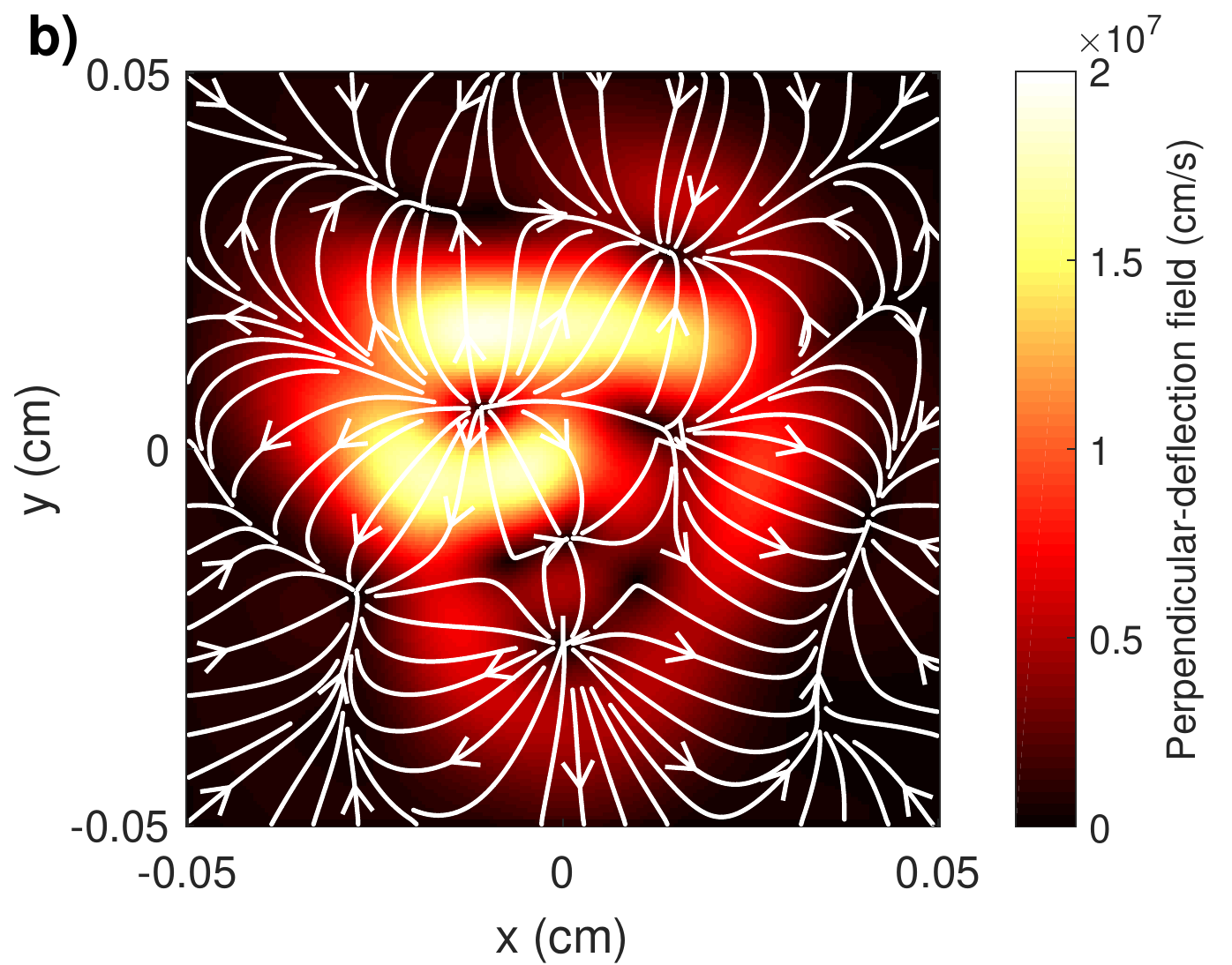}
    \end{subfigure} %
    \begin{subfigure}{.48\textwidth}
        \centering
        \includegraphics[width=0.9\linewidth]{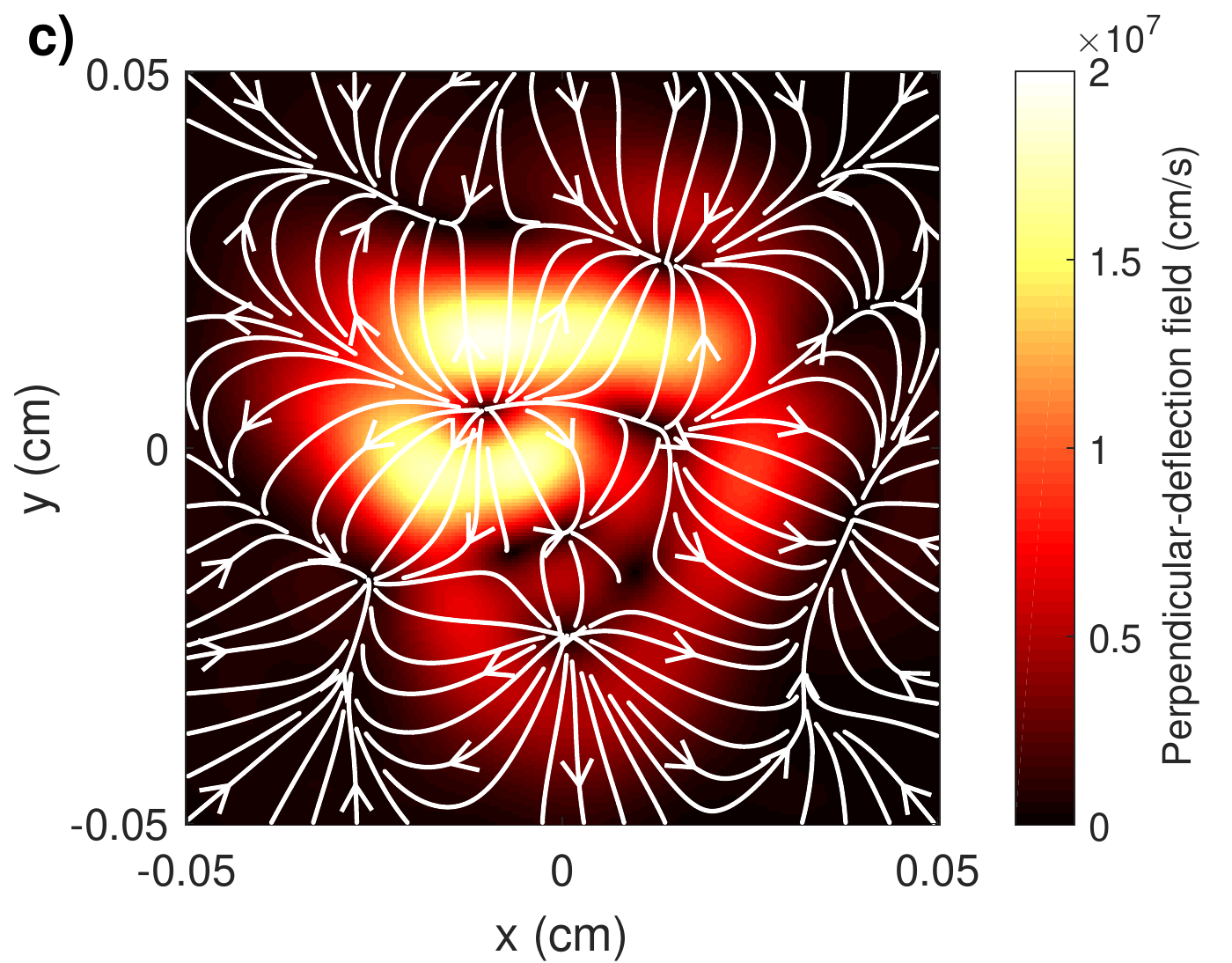}
    \end{subfigure} %
        \begin{subfigure}{.48\textwidth}
        \centering
        \includegraphics[width=0.9\linewidth]{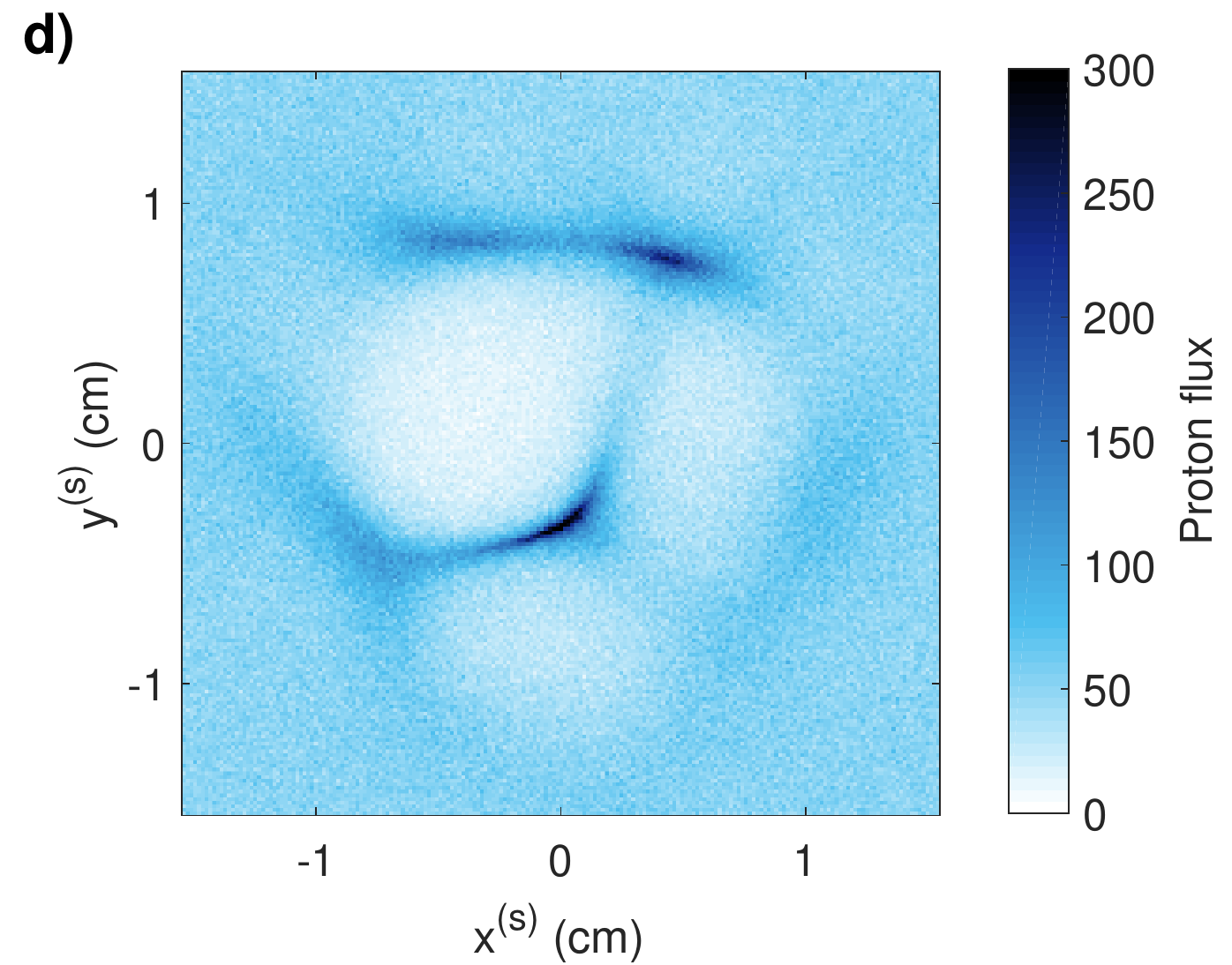}
    \end{subfigure} %
    \begin{subfigure}{.48\textwidth}
        \centering
        \includegraphics[width=0.9\linewidth]{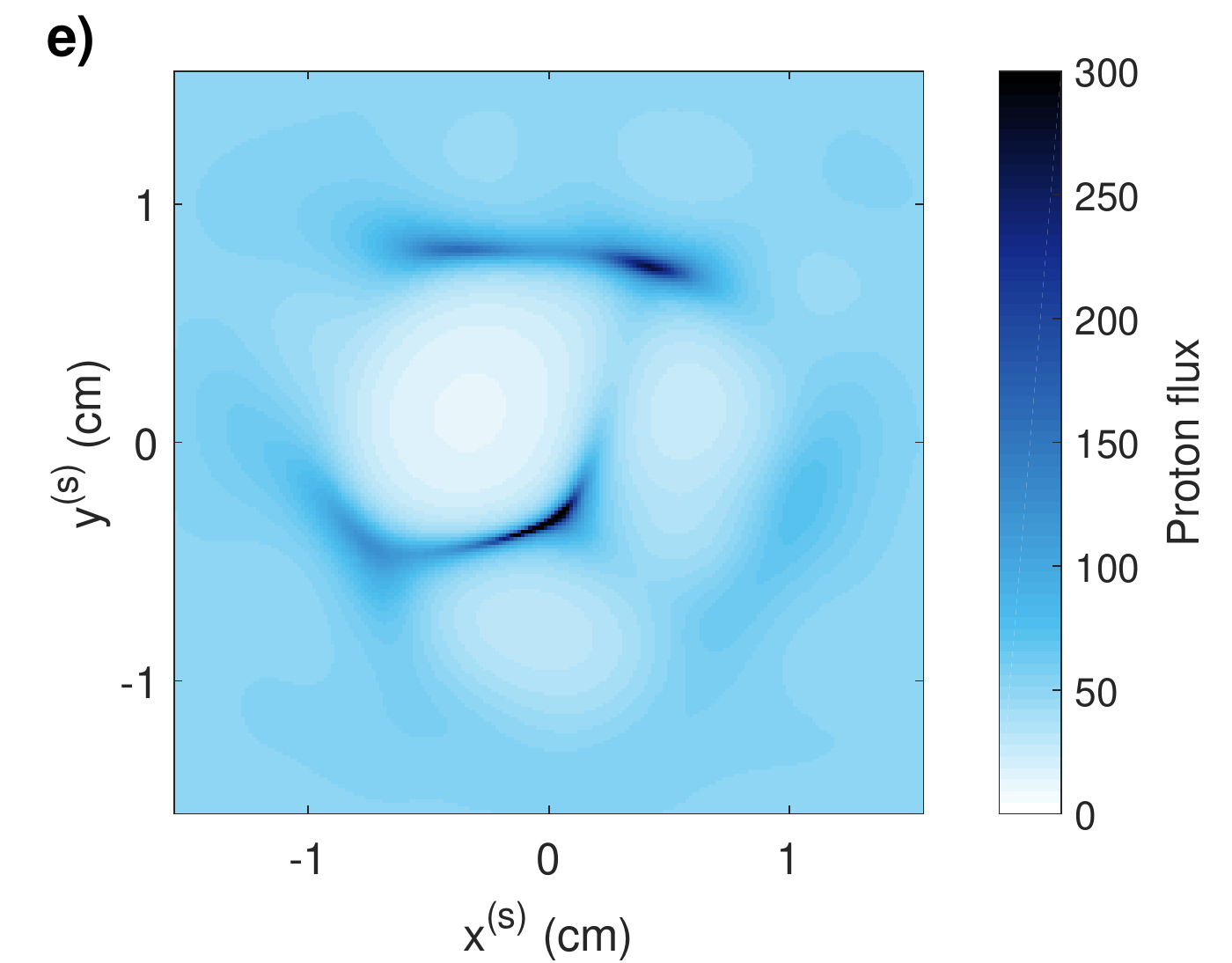}
     \end{subfigure} %
\caption{\textit{Comparison of analytic theory of proton imaging with results from a numerical ray-tracing code for a compact magnetic field configuration.}  The magnetic field -- generated using the technique described in Appendix \ref{NumSimMagFieldGen} is contained in a box, side length $l_i = l_z = l_{\bot} = 0.1 \, \mathrm{cm}$, and specified on a $201^3$ array (grid spacing $\delta x = l_i/201$); imaging is carried out using an artificial 3.3 MeV proton point-source at distance $r_i = 0.9 \, \mathrm{cm}$ from the near side of the interaction region. These artificial test protons are propagated through the field using a ray-tracing code, and then to a detector located at $r_s = 30 \, \mathrm{cm}$. \textbf{a)} Slice plot of magnetic field configuration perpendicular to direction of beam throughout the interaction region (stretched in $z$-direction). The image colour denotes perpendicular magnetic field strength, while the white arrows are the two-dimensional streamlines of the perpendicular magnetic field. \textbf{b)} Numerically measured perpendicular-deflection field using the test protons. Image colour represents the magnitude of the perpendicular-deflection field, while arrows are (two-dimensional) streamlines. \textbf{c)} Analytic prediction of perpendicular-deflection field \eqref{deffieldundefl}. \textbf{d)} Scneen-flux distribution created numerically using $2,000,000$ test protons. \textbf{e)} Prediction of image-flux distribution using analytic prediction \eqref{screenfluxSec2}, with same total proton flux.} \label{largescalecompactfieldA}
\end{figure}

As a complement to the analytic derivations of expressions for the image-flux distribution, this sub-appendix provides a numerical test of the key results: in particular, equation \eqref{deffieldundefl} for the perpendicular-deflection field when $\delta \theta \lesssim l_B/l_z$, and equation \eqref{screenfluxSec2} for the image-flux distribution. The parameters of the numerical test are described subsequently; results are shown in Figure \ref{largescalecompactfieldA}.

First, we generate a compact magnetic field configuration inside a cube-shaped region of side length $l_i$ (that is, $l_i = l_z = l_\bot$) using the technique described in Appendix \ref{NumSimMagFieldGen}. More specifically, the particular magnetic field used for this test is Gaussian, and has a magnetic-energy spectrum of an isotropic random field of magnetic cocoons of fixed size $l_e$ (described more thoroughly in Appendix \ref{ToySpecLinThyCocoon}~\cite{D04}):
\begin{equation}
E_B\!\left(k\right) = \frac{B_{rms}^2 l_e}{12\sqrt{2} \pi^{3/2}} l_e^4 k^4 \exp{\left(-l_e^2 k^2/2\right)} \, , \label{magcocoonspecAppendscaleGen}
\end{equation}
The typical cocoon size in this case is chosen to be $l_e = l_i/5$. Finally, we impose compactness of the magnetic field $\mathbf{B}$ using the Gaussian envelope 
\begin{equation}
\mathbf{B} \mapsto \mathbf{B} \, \exp{\left[-12 \mathbf{x}^2/l_i^2\right]} \, . \label{magfieldwindAppend}
\end{equation}
A slice plot of this magnetic field is presented in Figure \ref{largescalecompactfieldA} (with slices shown perpendicular to $\hat{\mathbf{z}}$ as defined using Figure \ref{PRsetup}). 

For the specified magnetic field configuration, a proton-imaging set-up is implemented via a ray-tracing code (more details given in Appendix \ref{NumSimFluxImageGenFull}), which enables numerical measurement of the perpendicular-deflection field, as well as the generation of artificial proton-flux images without using analytic theory. The numerically derived perpendicular-deflection field $\mathbf{w}^{\left(N\right)}\!\left(\mathbf{x}_{\bot0}\right)$ is determined as follows: record the initial and final velocities of each proton used for the imaging set-up, as well as their initial perpendicular position $\mathbf{x}_{\bot0}$ on intersection with the cuboid region containing the magnetic field, enabling exact numerical calculation of the deflection velocity experienced by any proton with the same initial perpendicular position. This is then be converted into a regular grid of initial positions on which the perpendicular-deflection field can be estimated using a scattered interpolation algorithm. The result is compared to the perpendicular-deflection field $\mathbf{w}\!\left(\mathbf{x}_{\bot0}\right)$ predicted analytically using \eqref{deffieldundefl}. The analytic prediction is calculated by interpolating the magnetic field at perpendicular locations specified by \eqref{perpcoord} before performing the $z$-integration in \eqref{deffieldundefl} over the grid spacing of the artificial magnetic field. 

The artificial proton-flux images are generated by propagating a large number of particles through the magnetic field using the ray-tracing code, before allowing them to move free beyond the field to the detector. Protons are then binned into pixels, resulting in a numerically determined image-flux distribution $\Psi^{\left(N\right)}\!\left(\mathbf{x}_{\bot0}\right)$. The resulting images can be compared with the analytic prediction $\Psi\!\left(\mathbf{x}_{\bot0}\right)$ from Kugland image-flux relation \eqref{screenfluxSec2}. The latter is calculated practically by first calculating the plasma-image mapping $\mathbf{x}_{\bot}^{\left(s\right)}\!\left(\mathbf{x}_{\bot0}\right)$ at each initial perpendicular position $\mathbf{x}_{\bot0}$ using \eqref{divmappingSec2}, before calculating the determinant of that mapping. The resulting values of the image-flux distribution are then interpolated on the pixel locations. 

For the analytic predictions of both the perpendicular-deflection field \eqref{deffieldundefl} and the image-flux distribution \eqref{screenfluxSec2}, the results of the numerical test shown in Figure \ref{largescalecompactfieldA} are positive. Figures \ref{largescalecompactfieldA}b and \ref{largescalecompactfieldA}c compare the numerically determined perpendicular-deflection field, and the analytical prediction \eqref{deffieldundefl}. The agreement is qualitatively (and quantitatively) strong, with an perpendicular-deflection field $L_2$ norm value $L_2\!\left[\mathbf{w}\right] = 0.05$. Here, $L_2\!\left[\mathbf{w}\right]$ is defined by
\begin{equation}
L_2\!\left[\mathbf{w}\right] = \sqrt{\frac{\sum_{\mathbf{x}_{\bot0}} \left|\mathbf{w}\!\left(\mathbf{x}_{\bot0}\right)-\mathbf{w}^{\left(N\right)}\!\left(\mathbf{x}_{\bot0}\right)\right|^2}{\sum_{\mathbf{x}_{\bot0}}  \left|\mathbf{w}^{\left(N\right)}\!\left(\mathbf{x}_{\bot0}\right)\right|^2}} \, .
\end{equation}  
Figures \ref{largescalecompactfieldA}d and Figure \ref{largescalecompactfieldA}e gives the associated proton-flux images generated by the ray-tracing code, and analytic prediction \eqref{screenfluxSec2}: agreement again seems close, with image-flux distribution $L_2$ norm value $L_2\!\left[\Psi\right] = 0.13$. This $L_2$ norm is defined by
\begin{equation}
L_2\!\left[\Psi\right] =  \sqrt{\frac{\sum_{\mathbf{x}_{\bot0}} \left|\delta \Psi\!\left(\mathbf{x}_{\bot0}\right)-\delta \Psi^{\left(N\right)}\!\left(\mathbf{x}_{\bot0}\right)\right|^2}{\sum_{\mathbf{x}_{\bot0}}  \left|\delta \Psi^{\left(N\right)}\!\left(\mathbf{x}_{\bot0}\right)\right|^2}} \, , \label{Lnormflux}
\end{equation}
for $\delta \Psi\!\left(\mathbf{x}_{\bot0}\right)$ and $\delta \Psi^{\left(N\right)}\!\left(\mathbf{x}_{\bot0}\right)$ the absolute image-flux deviations from the mean proton flux. We conclude that analytic theory provides a reasonable estimate of both the perpendicular-deflection field and the image-flux distribution. 

\begin{figure}[htbp]
\centering
    \begin{subfigure}{.48\textwidth}
        \centering
        \includegraphics[width=\linewidth]{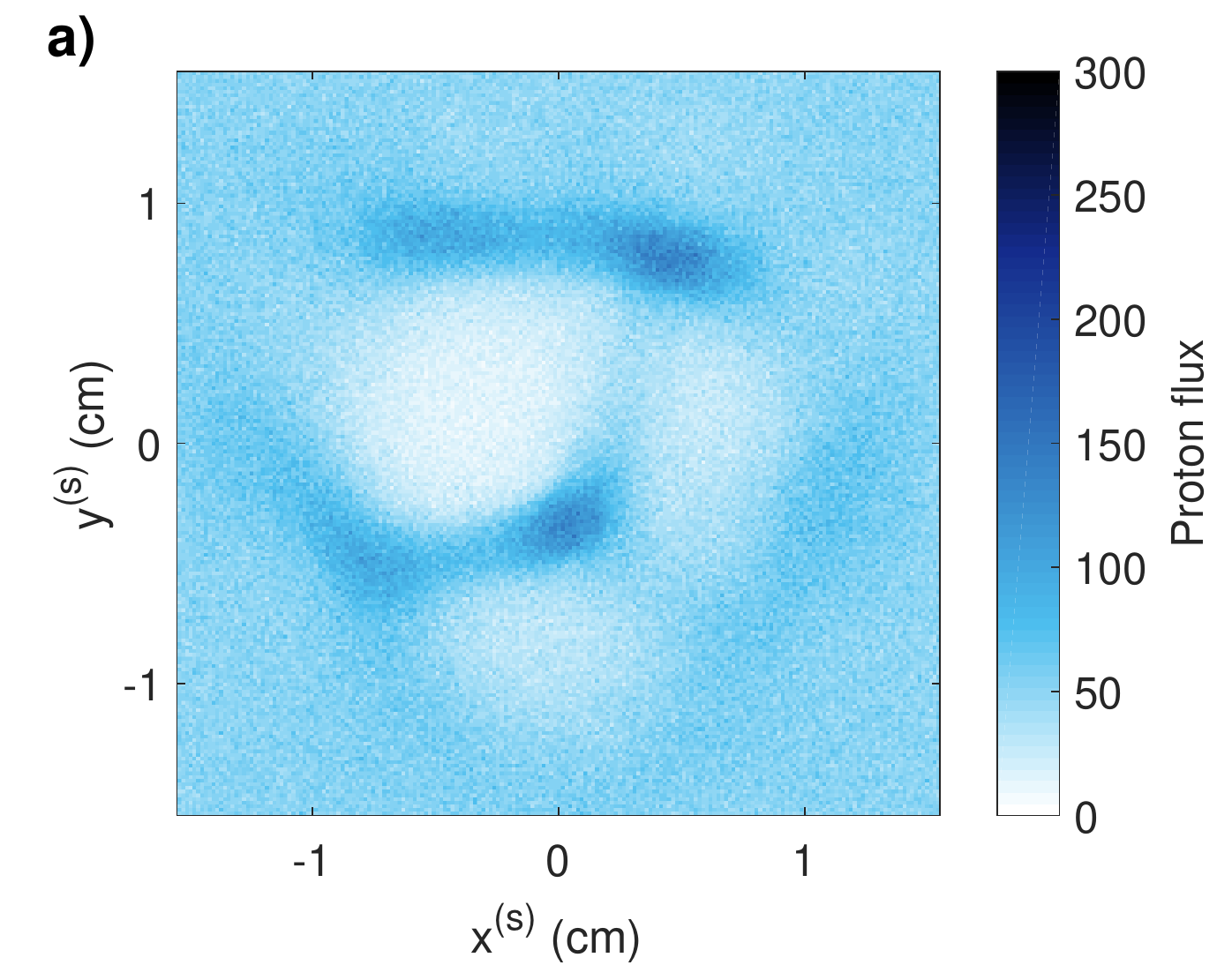}
    \end{subfigure} %
    \begin{subfigure}{.48\textwidth}
        \centering
        \includegraphics[width=\linewidth]{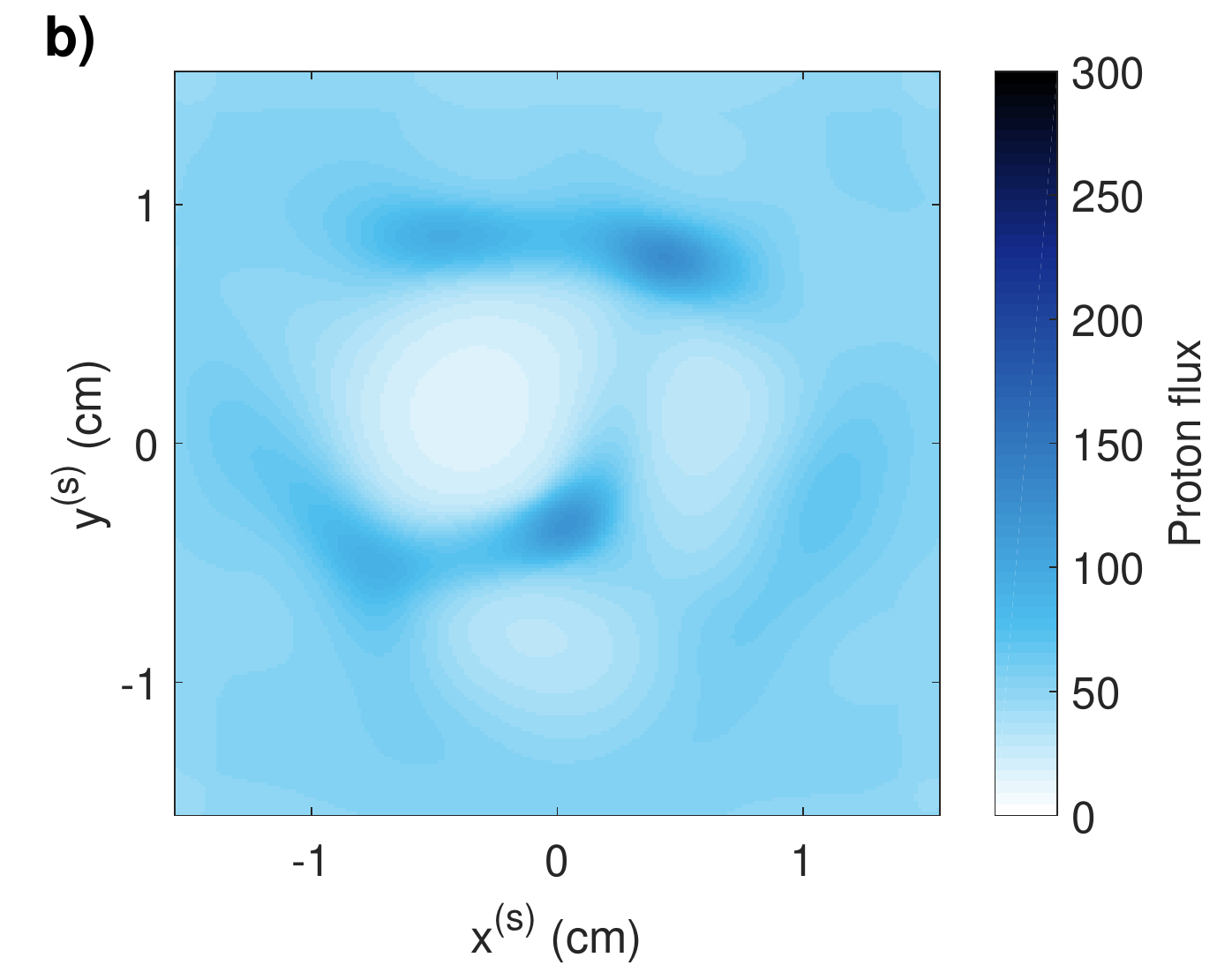}
    \end{subfigure} %
\caption{\textit{Comparison of analytic prediction \eqref{smearedscreenflux} with numerical result from a ray-tracing code of image-flux distribution  generated by imaging the field configuration described in Figure \ref{largescalecompactfieldA} using a finite proton source.} As with Figure \ref{largescalecompactfieldA}, the detector is located at $r_s = 30 \, \mathrm{cm}$, leading to a $31\times$ magnification. \textbf{a)} image-flux distribution generated using $2 \times 10^6$ test protons uniformly generated from a finite source, radius $a = 0.005 \, \mathrm{cm}$. \textbf{b)} Analytic prediction of image-flux distribution using \eqref{smearedscreenflux}, with point-spread function given by \eqref{smearedscreenfluxufmPSF}, and image-flux distribution from point proton-source shown in Figure \ref{largescalecompactfieldA}e.} \label{largescalecompactfieldC}
\end{figure}

We also present a numerical test of the analytic prediction \eqref{smearedscreenflux} of the image-flux distribution in the case of a finite source of protons. We carry out this test by imaging the same magnetic field configuration described in Figure \ref{largescalecompactfieldA}a with the same imaging parameters, except using test protons generated from a finite uniformly emitting point source, radius $a = 0.005 \, \mathrm{cm}$. In a similar manner to that described in the previous paragraph, we generate an artificial image-flux distribution. This is compared to analytic prediction \eqref{smearedscreenflux}, which in tunr is calculated by starting with the image-flux distribution shown in Figure \ref{largescalecompactfieldC}e, and then applying numerically the point-spread function \eqref{smearedscreenfluxufmPSF} as appropriate for a uniformly emitting sphere. The results are shown in Figure \ref{largescalecompactfieldC}a and Figure \ref{largescalecompactfieldC}b are are seemingly consistent with each other. The $L_2$ norm defined by \eqref{Lnormflux} gives $L_2\!\left[\Psi\right] = 0.11$, which is also reasonably small. 

\section{Velocity distribution of protons from a uniformly emitting sphere} \label{FiniteSourcePSF}

In Appendix \ref{KinThyPradInitCon}, we claim that the distribution of perpendicular velocities of a proton beam generated from a finite, uniformally and isotropically emitting source is given by equation \eqref{sphereunidistinit}, viz.,
\begin{equation}
P\!\left(\delta \mathbf{v}_{\bot0}\right) = \frac{3}{2 \pi } \left(\frac{r_i}{aV}\right)^2 \sqrt{1-\left(\frac{r_i}{aV}\right)^2 \left|\delta \mathbf{v}_{\bot0}\right|^2} H\!\left(\frac{aV}{r_i}-\left|\delta \mathbf{v}_{\bot0}\right|\right) \, 
\label{sphereunidistinitAppendix}
\end{equation}
in the asymptotic limit of small source size $a$ relative to the distance $r_i$ from the source to the plasma to be imaged by the source. As in the main text, $\delta \mathbf{v}_{\bot0}$ denotes the deviation of the perpendicular velocity from that arising from a point source in the paraxial limit
\begin{equation}
\delta \mathbf{v}_{\bot0} = \mathbf{v}_{\bot0}-\frac{\mathbf{x}_{\bot0}}{r}V \, ,
\end{equation}
and $H\!\left(x\right)$ is the Heaviside step function. In this appendix, we derive \eqref{sphereunidistinitAppendix}.

\begin{figure}[htbp]
\centering
\includegraphics[width=0.8\textwidth]{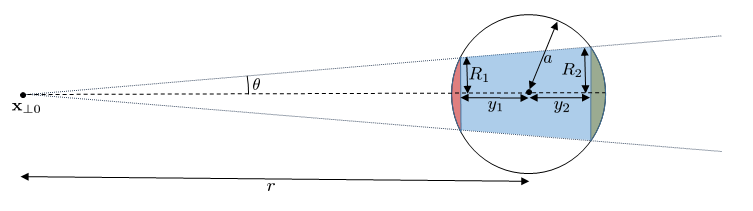}
\caption{Geometry of a finite source, with required lengths and angles identified for determining the volume resulting from the intersection between a cone, opening angle $2\theta$, and the source, which in turns gives the PDF of particle speeds at $\mathbf{x}_{\bot0}$. \label{finitesourceparam}}
\end{figure}

Finding the distribution of particles emitted from an arbitrarily shaped capsule with variable emission is in general a complex problem; however, in the special case of a sphere of radius $a$ emitting uniformly from all points, we are greatly assisted by symmetry. Consider some point $\mathbf{x}_0$ at a radius $r \gg a$ from the centre of sphere, and take an infinite cone with half angle $\theta < \arcsin{a/r}$, apex at $\mathbf{x}_0$, with axis passing through the centre of the sphere. Since the sphere is emitting uniformly, the volume enclosed inside both the sphere and the cone as a fraction of the total sphere volume gives the fraction of particles emitted with a deflection velocity at an angle less than or equal to $\theta$. 
To find this volume, introduce lengths as shown in Figure \ref{finitesourceparam}.

Then, $R_1$, $R_2$, $y_1$ and $y_2$ satisfy
\begin{IEEEeqnarray}{rCl}
R_1^2+y_1^2 & = & a^2 \, ,\\
R_2^2+y_2^2 & = & a^2 \, ,
\end{IEEEeqnarray}
and
\begin{IEEEeqnarray}{rCl}
R_1 & = & \left(r-y_1\right) \tan{\theta} \, , \\
R_2 & = & \left(r+y_2\right) \tan{\theta} \, .
\end{IEEEeqnarray}
Solving for $y_1$ and $y_2$, we find
\begin{IEEEeqnarray}{rCl}
y_1 & = & r \sin^2{\theta} + \cos{\theta} \sqrt{a^2-r^2\sin^2{\theta}} \, ,\\
y_2 & = & -r \sin^2{\theta} + \cos{\theta} \sqrt{a^2-r^2\sin^2{\theta}} \, .
\end{IEEEeqnarray}
Since $\sin{\theta} \leq a/r \ll 1$, we can expand these solutions in $\theta$ to find
\begin{equation}
y_1 = y_2 = y = \sqrt{a^2-r^2 \theta^2}\left[1+\mathcal{O}\!\left(\frac{a}{r}\right)\right] \, ,
\end{equation}
along with
\begin{equation}
R_1 = R_2 = R = r \theta \left[1+\mathcal{O}\!\left(\frac{a}{r}\right)\right] \, .
\end{equation}
The relevant volume is that of two spherical caps, with heights $a-y_1$, $a-y_2$ and base widths $R_1$, $R_2$ respectively, and a conical frustum of length $y_1+y_2$, base heights $R_1$ and $R_2$. These have volumes
\begin{IEEEeqnarray}{rCl}
\mathrm{Vol}_1 & = & \frac{\pi}{3} \left(a-y_1\right)^2 \left(2a+y_1\right) \, ,\\
\mathrm{Vol}_2 & = & \frac{\pi}{3} \left(a-y_2\right)^2 \left(2a+y_2\right)\, ,\\
\mathrm{Vol}_{frus} & = & \frac{\pi}{3} \left(y_1+y_2\right)\left(R_1^2+R_1 R_2 + R_2^2\right) \, ,
\end{IEEEeqnarray}
giving total volume
\begin{equation}
\mathrm{Vol}\!\left(\theta\right) =  \frac{\pi}{3} \left[\left(a-y_1\right)^2 \left(2a+y_1\right) + \left(a-y_2\right)^2 \left(2a+y_2\right)+\left(y_1+y_2\right)\left(R_1^2+R_1 R_2 + R_2^2\right)\right] \, .
\end{equation}
Evaluating this expression generally is somewhat messy; however, we have an asymptotic expression
\begin{equation}
\mathrm{Vol}\!\left(\theta\right) =  \frac{2\pi}{3} \left[\left(a- \sqrt{a^2-r^2 \theta^2}\right)^2 \left(2a+ \sqrt{a^2-r^2 \theta^2}\right) +3 r^2 \theta^2 \sqrt{a^2-r^2 \theta^2}\right] \left[1+\mathcal{O}\!\left(\frac{a}{r}\right)\right] \, .
\end{equation}
The 1D probability density function of proton approach angles $f_\theta\!\left(\theta\right)$ is then
\begin{equation}
f_\theta\!\left(\theta\right) = \frac{3}{4 \pi a^3} H\!\left(a-r\theta\right)\frac{\mathrm{d}}{\mathrm{d}\theta} \mathrm{Vol}\!\left(\theta\right) =  3 H\!\left(a-r\theta\right) \left(\frac{r}{a}\right)^2 \theta \sqrt{1-\theta^2\left(\frac{r}{a}\right)^2} \left[1+\mathcal{O}\!\left(\frac{a}{r}\right)\right] \, .
\end{equation}
Finally, since the perpendicular speed $\left|\mathbf{v}_{\bot0} \right| \ll V$, the angle $\theta$ of the source particle can be written
\begin{equation}
\theta = \frac{\left|\delta \mathbf{v}_{\bot0}\right|}{V} \, ,
\end{equation}
The distribution of perpendicular speeds $\tilde{P}\!\left(\left|\delta \mathbf{v}_{\bot0} \right|\right)$ becomes
\begin{equation}
\tilde{P}\!\left(\left|\delta \mathbf{v}_{\bot0} \right|\right) = \frac{1}{V} f_\theta\!\left(\frac{\left|\delta \mathbf{v}_{\bot0}\right|}{V}\right) = 3 H\!\left(aV-r\left|\delta \mathbf{v}_{\bot0}\right|\right) \left|\delta \mathbf{v}_{\bot0}\right| \left(\frac{r}{aV}\right)^2 \sqrt{1-\left(\frac{r}{aV}\right)^2  \left|\delta \mathbf{v}_{\bot0}\right|^2} \, , \label{perpradspeeddist}
\end{equation}
and so the perpendicular part of the velocity distribution function is then
\begin{IEEEeqnarray}{rCl}
P\!\left(\delta \mathbf{v}_{\bot0}\right) & = & \frac{1}{2 \pi \left|\delta \mathbf{v}_{\bot0}\right|} \tilde{P}\!\left(\left|\delta \mathbf{v}_{\bot0} \right|\right) \\
& = & \frac{3}{2 \pi} H\!\left(aV-r\left|\delta \mathbf{v}_{\bot0}\right|\right) \left(\frac{r}{aV}\right)^2 \sqrt{1-\left(\frac{r}{aV}\right)^2  \left|\delta \mathbf{v}_{\bot0}\right|^2} \, ,
\end{IEEEeqnarray}
in agreement with \eqref{sphereunidistinitAppendix}

\begin{figure}[htbp]
\centering
\includegraphics[width=0.6\textwidth]{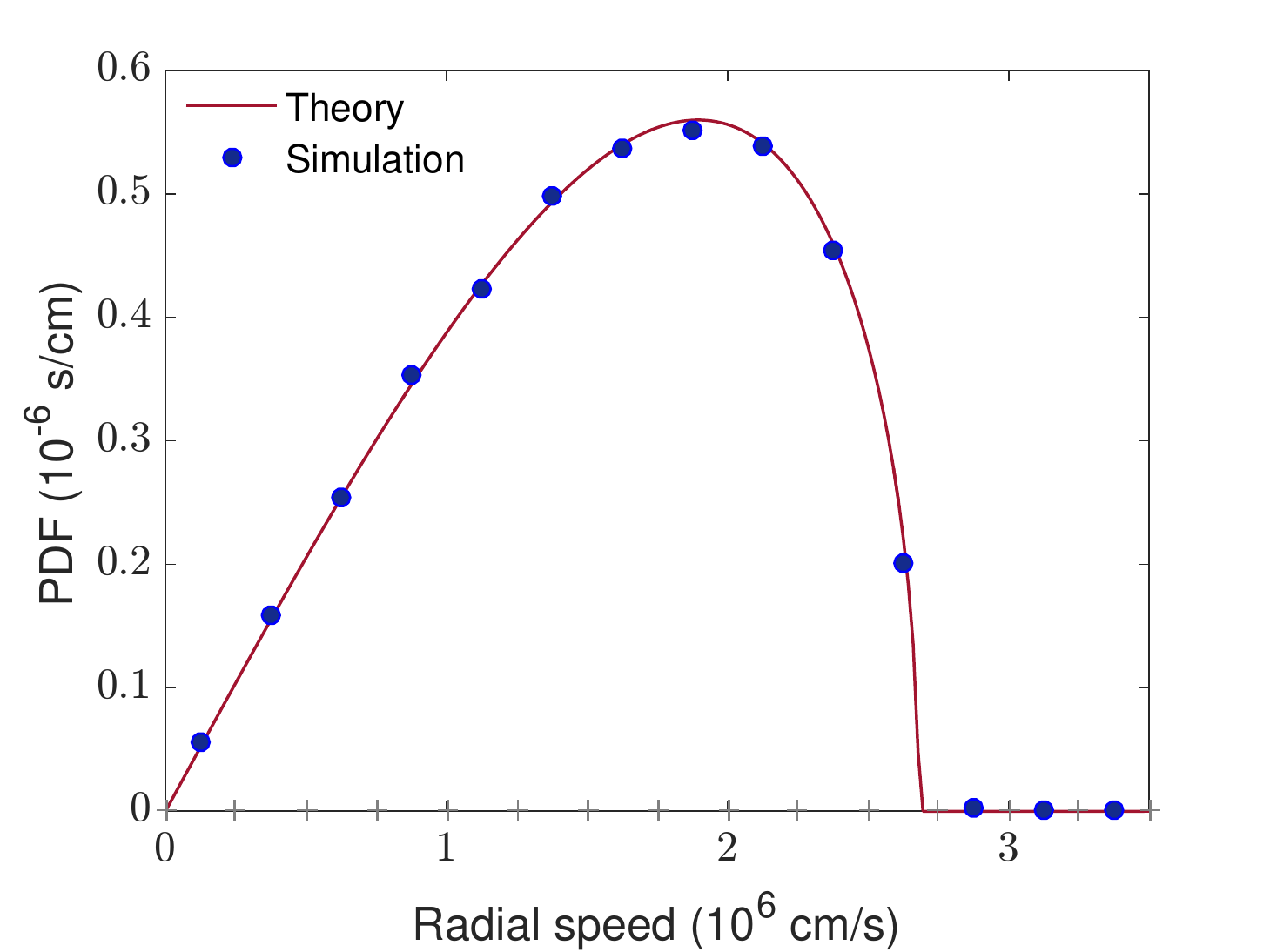}
\caption{PDF of perpendicular radial speeds in the limit of small smearing for a proton beam created by a uniform, isotropically emitting sphere of radius $a = 5 \times 10^{-4} \, \mathrm{cm}$. The numerical PDF is calculated by generating $20,000,000$ $14.7 \, \mathrm{MeV}$ particles inside the source, then propagating them a distance $r_i = 1\, \mathrm{cm}$ to a box `interaction' region, side length $l_i = 0.1 \, \mathrm{cm}$. For computational efficiency, angles are constrained so that polar angle $\theta \leq 1.1 \, a/r_i$, and then the distribution of perpendicular particle speeds is sampled in a small region at the centre of the image of radius $\delta = 5 \times 10^{-5} \, \mathrm{cm}$ (the sample contained $\sim 41,000$ particles). \label{finitesourcePDF}}
\end{figure}

We can test this calculation by creating an artificial distribution of protons produced from a finite source. This is done by simulating a large number of protons inside a sphere of radius $a$, centre $\left(0,0,-r\right)$, with exact positions inside determined by random sampling from a three-dimensional uniform distribution (see Figure \ref{PRsetup} for definition of the coordinate system). We additionally assign a direction to each proton by random sampling from a uniform distribution on the surface of a unit sphere. Then, each proton is allowed to move freely with fixed speed $V$. The distribution of perpendicular speeds of protons passing through a circular surface with centre $\left(0,0,0\right)$, normal direction parallel to the $z$-direction, and radius $\delta \ll a$ is then taken; the result should correspond directly to \eqref{perpradspeeddist}. In this test, it is computationally more efficient to constrain the possible range of initial angles to some fraction of the interaction region, then focus on some small circular surface within that. A specific example is shown in Figure \ref{finitesourcePDF}; we see good agreement between theory and simulation. 

\section{Derivation of deflection-field spectral relation \eqref{deffieldspec}} \label{DeflfieldCorr}

Section \ref{PlasMap} claims that the magnetic-energy spectrum $E_B\!\left(k\right)$ of a zero-mean stochastic magnetic field ($\bar{\mathbf{B}} = 0$, $\mathbf{B} = \delta \mathbf{B}$) can be related to the one-dimensional spectrum $E_W\!\left(k_\bot\right)$ of the perpendicular-deflection field $\mathbf{w}\!\left(\mathbf{x}_{\bot0}\right)$ by spectral relation \eqref{deffieldspec}, that is 
\begin{equation}
E_{B}\!\left(k\right) = \frac{m_p^2 c^2} {4\pi^2 l_z e^2} k E_{W}\!\left(k\right) \, .\label{deffieldspecAppend}
\end{equation}
We derive \eqref{deffieldspecAppend} in this appendix under the assumption that the correlation length $l_B$ of the stochastic magnetic field is much smaller than the path-length $l_z$ of the proton beam through the magnetic field, as well as homogeneity and isotropy of magnetic field statistics. Our approach for doing so will be to first relate the magnetic autocorrelation function $M\!\left(r\right)$ defined by \eqref{isomagcoll} in Appendix \ref{FurtherStatCharMagFieldsMagAuto} to an autocorrelation function of the perpendicular-deflection field, and then utilise the Wiener-Khinchin theorem. 

We begin by defining the deflection-field autocorrelation tensor:
\begin{equation}
C_{ll'}\!\left(\mathbf{x}_{\bot0},\tilde{\mathbf{x}}_{\bot0}\right) \equiv \left<w_{l}\!\left(\mathbf{x}_{\bot0}\right)w_{l'}\!\left(\tilde{\mathbf{x}}_{\bot0}\right)\right> \, . \label{corrtensdeflfielddef}
\end{equation}
Substituting the definition \eqref{pathintfieldSec2} of the perpendicular-deflection field into \eqref{corrtensdeflfielddef} gives
\begin{equation}
C_{ll'}\!\left(\mathbf{x}_{\bot0},\tilde{\mathbf{x}}_{\bot0}\right) = \frac{e^2}{m_p^2 c^2} \epsilon_{lmn} \epsilon_{l'm'n'} \hat{z}_m \hat{z}_{m'} \int_0^{l_z} \mathrm{d}z \int_0^{l_z} \mathrm{d}z' \left<\delta B_n\!\left[\mathbf{x}_{\bot}\!\left(z\right)\right] \delta B_{n'}\!\left[\tilde{\mathbf{x}}_{\bot}\!\left(z'\right)\right] \right> \, ,
\end{equation}
where $\mathbf{x}_{\bot}\!\left(z\right)$ and $\tilde{\mathbf{x}}_{\bot}\!\left(z\right)$ denote the trajectories of protons with initial perpendicular positions $\mathbf{x}_{\bot0}$ and $\tilde{\mathbf{x}}_{\bot0}$ respectively. We then neglect the distinction between the integrated field along the undeflected trajectories and along paths parallel to the $z$-direction, because the resulting error is $\mathcal{O}\left(\delta \alpha, \delta \theta\right)$ under assumptions of statistical homogeneity and isotropy of magnetic field statistics. The autocorrelation tensor becomes
\begin{IEEEeqnarray}{rCl}
C_{ll'}\!\left(\mathbf{x}_{\bot0},\tilde{\mathbf{x}}_{\bot0}\right) & = & \frac{e^2}{m_p^2 c^2} \epsilon_{lmn} \epsilon_{l'm'n'} \hat{z}_m \hat{z}_{m'} \int_0^{l_z} \mathrm{d}z \int_0^{l_z} \mathrm{d}z' \left<\delta B_n\!\left(\mathbf{x}_{\bot0},z\right) \delta B_{n'}\!\left(\tilde{\mathbf{x}}_{\bot0},z'\right) \right> \nonumber \\
& = & \frac{e^2}{m_p^2 c^2} \epsilon_{lmn} \epsilon_{l'm'n'} \hat{z}_m \hat{z}_{m'} \int_0^{l_z} \mathrm{d}z \int_0^{l_z} \mathrm{d}z' \, M_{nn'}\!\left(\mathbf{x}_{\bot0}+z\hat{\mathbf{z}},\tilde{\mathbf{x}}_{\bot0}+z'\hat{\mathbf{z}}\right) \, ,
\end{IEEEeqnarray}
where we have introduced the magnetic autocorrelation tensor defined by \eqref{magcorrtensdef} in \ref{FurtherStatCharMagFields}. 
Anticipating a recourse to homogeneity and isotropy of the magnetic autocorrelation tensor, change variables in the double integral by
\begin{IEEEeqnarray}{rClr}
\left(\mathbf{r}_\bot,r_z\right) & = & \left(\tilde{\mathbf{x}}_{\bot0}-\mathbf{x}_{\bot0},z'-z\right) &\,  , \\
\bar{z} & = & z & \, ,
\end{IEEEeqnarray}
which on explicitly assuming a homogeneous, isotropic magnetic autocorrelation tensor gives a homogeneous, isotropic deflection-field autocorrelation tensor
\begin{equation}
C_{ll'}\!\left(r_\bot\right) = \frac{e^2}{m_p^2 c^2} \epsilon_{lmn} \epsilon_{l'm'n'} \hat{z}_m \hat{z}_{m'} \int_0^{l_z} \mathrm{d}\bar{z} \int_{-\bar{z}}^{l_z-\bar{z}} \mathrm{d}r_z \, M_{nn'}\!\left(\sqrt{r_\bot^2+r_z^2}\right) \, .
\end{equation}
The small scale of the field $l_B$ allows for an extension of the integration limits to $\left(-\infty,\infty\right)$ via an intermediate scale $l_B \ll l \ll l_z$, producing an $\mathcal{O}\!\left(l_B/l\right)$ error from \eqref{magcorrfuntailbound} in Appendix \ref{FurtherStatCharMagFieldsCorrTail}. The outer integral can then be evaluated independently, leaving
\begin{equation}
C_{ll'}\!\left(r_\bot\right) = \frac{l_z e^2}{m_p^2 c^2} \epsilon_{lmn} \epsilon_{l'm'n'} \hat{z}_m \hat{z}_{m'} \int_{-\infty}^{\infty} \mathrm{d}r_z \, M_{nn'}\!\left(\sqrt{r_\bot^2+r_z^2}\right) \, .
\end{equation}
Taking the trace of the deflection-field autocorrelation tensor gives the deflection-field autocorrelation function $C\left(r_{\bot}\right) \equiv C_{ll}\!\left(r_\bot\right)$ in terms of the magnetic autocorrelation function:
\begin{equation}
C\!\left(r_\bot\right) = \frac{l_z e^2}{m_p^2 c^2} \int_{-\infty}^{\infty} \mathrm{d}r_z \, \left[M\!\left(\sqrt{r_\bot^2+r_z^2}\right) - M_{zz}\!\left(\sqrt{r_\bot^2+r_z^2}\right)\right]=  \frac{l_z e^2}{m_p^2 c^2} \int_0^{\infty} \mathrm{d}r_z \, M\!\left(\sqrt{r_\bot^2+r_z^2}\right) \, . \label{deflautocorr}
\end{equation}
This correlation measure has an essentially identical form to that for Faraday rotation autocorrelation function~\cite{EV03}, and so analogies can be made with results derived in that case. In particular, the correlation length of the magnetic field is related to $C\!\left(0\right)$ by
\begin{equation}
l_B = \frac{m_p^2 c^2}{l_z e^2} \frac{C\!\left(0\right)}{M\!\left(0\right)} \, ,\label{magcorlengthdefl}
\end{equation}
which in turn can be rearranged to give the RMS perpendicular-deflection field:
\begin{equation}
w_{rms} = \frac{e \left<\delta \mathbf{B}^2\right>^{1/2}}{m_p c} \sqrt{l_z l_B} \, .
\end{equation}
Deflection angle RMS \eqref{RMSdeflecangle} follows immediately.  

The relation \eqref{deffieldspecAppend} between the magnetic-energy spectrum and the perpendicular-deflection field spectrum can be found by recourse to the Wiener-Khinchin theorem, which takes the following form for an isotropic deflection-field autocorrelation function:
\begin{equation}
E_W\!\left(k_\bot\right) = 2 \pi k_\bot \hat{C}\!\left(k_\bot\right) \, , \label{WKtheoremdeflfield}
\end{equation}
where $\hat{C}\!\left(k_\bot\right)$ is the Fourier-transformed deflection-field autocorrelation function. 
Fourier-transforming \eqref{deflautocorr} in the perpendicular direction gives
\begin{equation}
\hat{C}\!\left(k_\bot\right) = \frac{\pi l_z e^2}{m_p^2 c^2} \hat{M}\!\left(k_\bot,0\right) \label{FTdeflautocorr}
\end{equation}
However, since magnetic fluctuations are assumed isotropic, we note that
\begin{equation}
\hat{M}\!\left(k_\bot,0\right) = \hat{M}\!\left(k_\bot\right) = 4 \pi k_\bot^2 E_B\!\left(k_\bot\right) \, , \label{WKtheoremmagfield}
\end{equation}
where the second equality follows from equation \eqref{magengspeccorr} in Appendix \ref{FurtherStatCharMagFieldsMagAuto}. The desired result \eqref{deffieldspecAppend} is obtained by substituting \eqref{WKtheoremdeflfield} and \eqref{WKtheoremmagfield} into \eqref{FTdeflautocorr}, and rearranging. 

In addition to the proof of \eqref{deffieldspecAppend}, we provide one comment on its validity for compact configurations. The assumption of universal homogeneity and isotropic field statistics is technically an inconsistent one for inhomogeneous fields (as is inevitable for a compact magnetic field); however, autocorrelation-type analysis can be extended to include slow variation along path length. In particular, if geometrical statistics are preserved along the path length, but $\left<\delta \mathbf{B}^2\right>^{1/2}$ slowly varies, then we can describe magnetic autocorrelation tensors of the form
\begin{equation}
M_{nn'}\!\left(r\right) = \tilde{M}_{nn'}\!\left(r\right) \left<\delta \mathbf{B}^2\right>\!\left({z}\right) \, .
\end{equation}
The the deflection-field autocorrelation function becomes
\begin{equation}
C\!\left(r_\bot\right) = \frac{e^2}{m_p^2 c^2} \int_0^{l_z} \mathrm{d}\bar{z} \left<\delta \mathbf{B}^2\right>\!\left(\bar{z}\right) \int_0^{\infty} \mathrm{d}r_z \, \tilde{M}\!\left(\sqrt{r_\bot^2+r_z^2}\right) \, . \label{deflautocorrzvar}
\end{equation}
However, such a picture is only valid if there is a sufficient separation of scales between variation of $\left<\delta \mathbf{B}^2 \right>\!\left(z\right)$ and stochastic fluctuations. In particular, if
\begin{equation}
\frac{l_z}{l_B} \sim \frac{\max_{z \in \left[0,l_z\right]}{\left\{\delta \mathbf{B}\!\left(z\right)\right\}}}{\left<\delta \mathbf{B}^2\right>^{1/2}} \, ,
\end{equation} 
then the magnitude of deflections will deviate from \eqref{deffieldspecAppend}. 

\section{Numerical illustration of scalings for the RMS proton deflection angle $\delta \theta_{rms}$ and constrast parameter $\mu$} \label{ConstrastScale}

In Section \ref{Contrast}, expressions \eqref{RMSdeflecangle} and \eqref{contrastmagfield} for the RMS proton deflection angle $\delta \theta$ and $\mu$ respectively are given, which for convenience are reproduced here:
\begin{IEEEeqnarray}{rCl}
\delta \theta_{rms} & = & \frac{e B_{rms}}{m_p c V} \sqrt{l_z l_B} \label{thetaparamAppend} \, ,\\
\mu & = & \frac{r_s r_i}{r_s+r_i} \frac{e B_{rms}}{m_p c V} \sqrt{\frac{l_z}{l_B}} \label{contrastparamsim} \, . 
\end{IEEEeqnarray}
These are then used to conclude that $\mu$ is a scale-dependent quantity which increases as the correlation length $l_B$ of the magnetic field decreases -- the opposite conclusion to the deflection angle $\delta \theta_{rms}$. In this appendix, we illustrate this result numerically, by creating a single-scale stochastic magnetic field, and then generate artificial proton images of different instantiations of this field with a range of correlation lengths $l_B$. In particular, we provide a simple visual demonstration that a small-scale field with the same magnetic field strength probability-density function as some larger scale field leads to smaller typical proton deflection angles when imaged, but larger values of $\mu$. We also show that for this example the scaling laws \eqref{thetaparamAppend} and \eqref{contrastparamsim} are indeed obtained.

\begin{figure}[htbp]
\centering
    \begin{subfigure}{.48\textwidth}
        \centering
        \includegraphics[width=\linewidth]{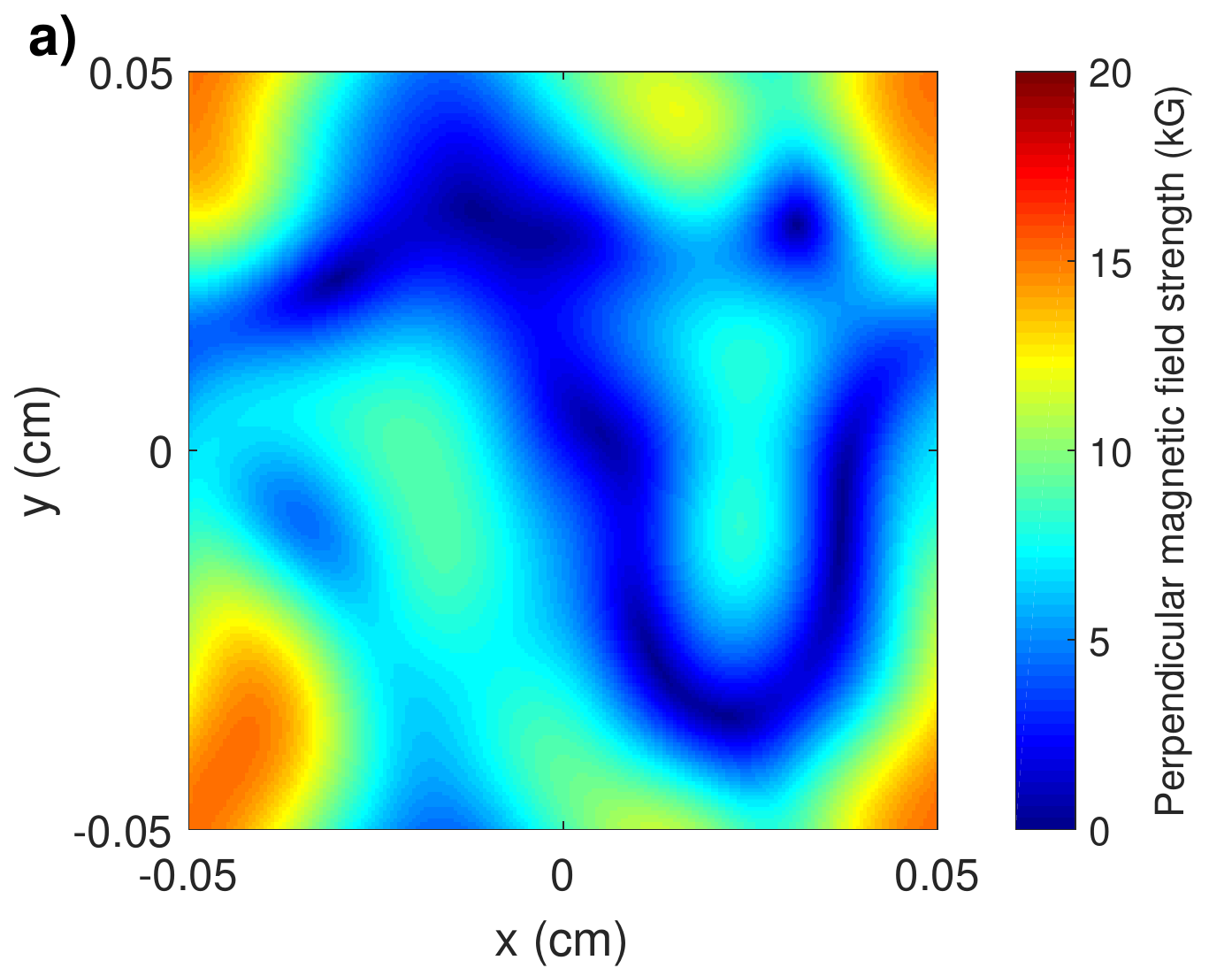}
    \end{subfigure} %
    \begin{subfigure}{.48\textwidth}
        \centering
        \includegraphics[width=\linewidth]{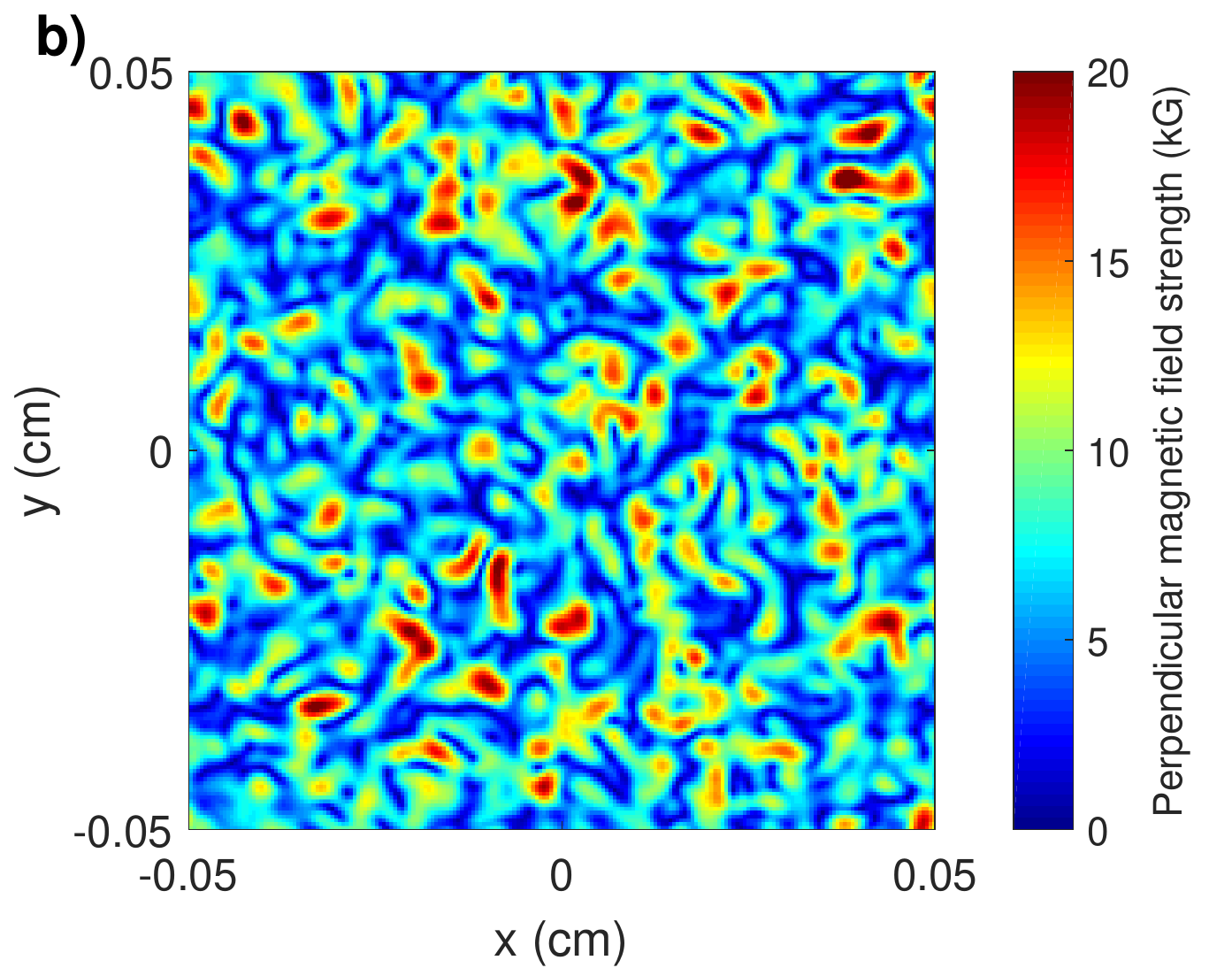}
    \end{subfigure} %
    \begin{subfigure}{.48\textwidth}
        \centering
        \includegraphics[width=\linewidth]{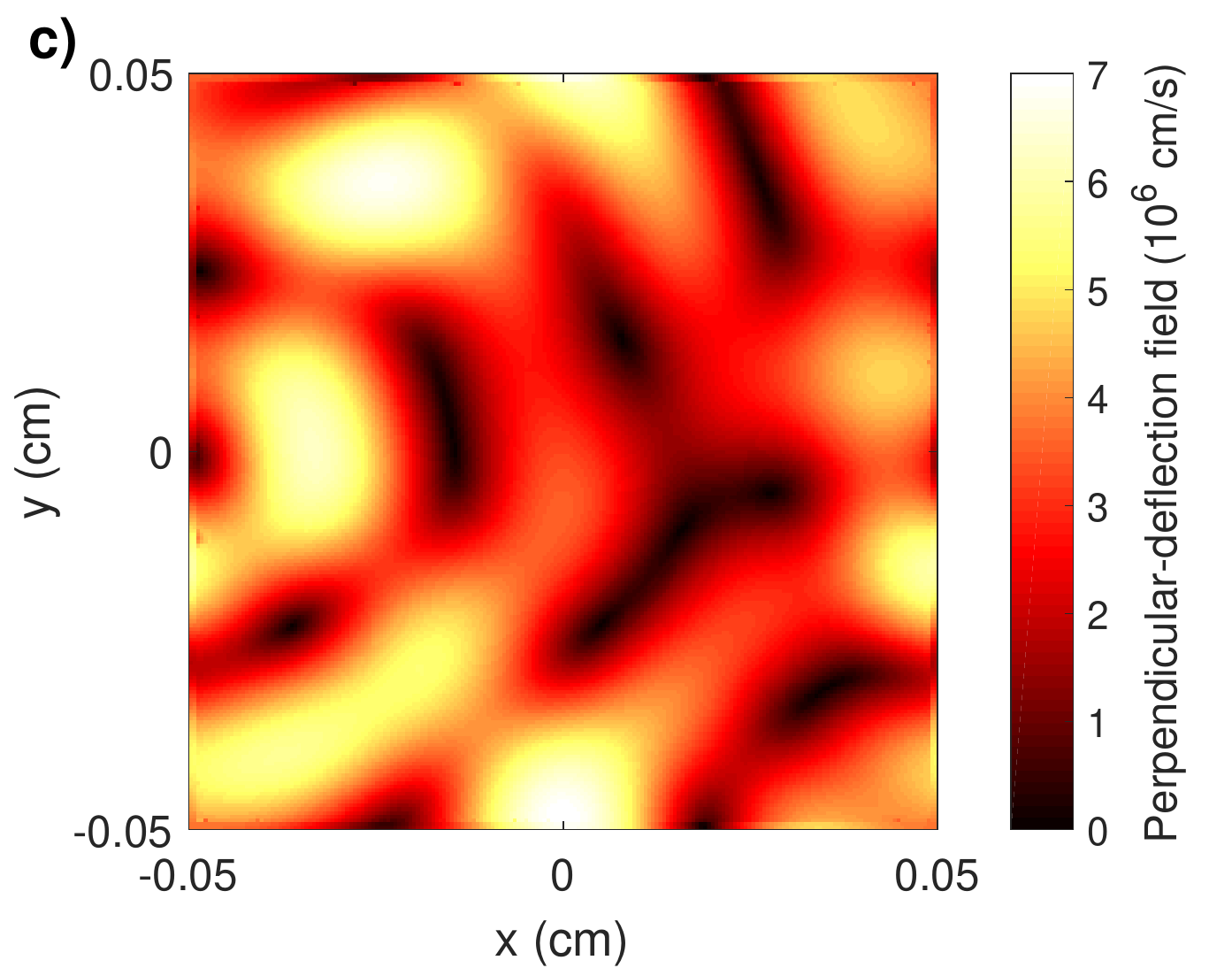}
    \end{subfigure} %
    \begin{subfigure}{.48\textwidth}
        \centering
        \includegraphics[width=\linewidth]{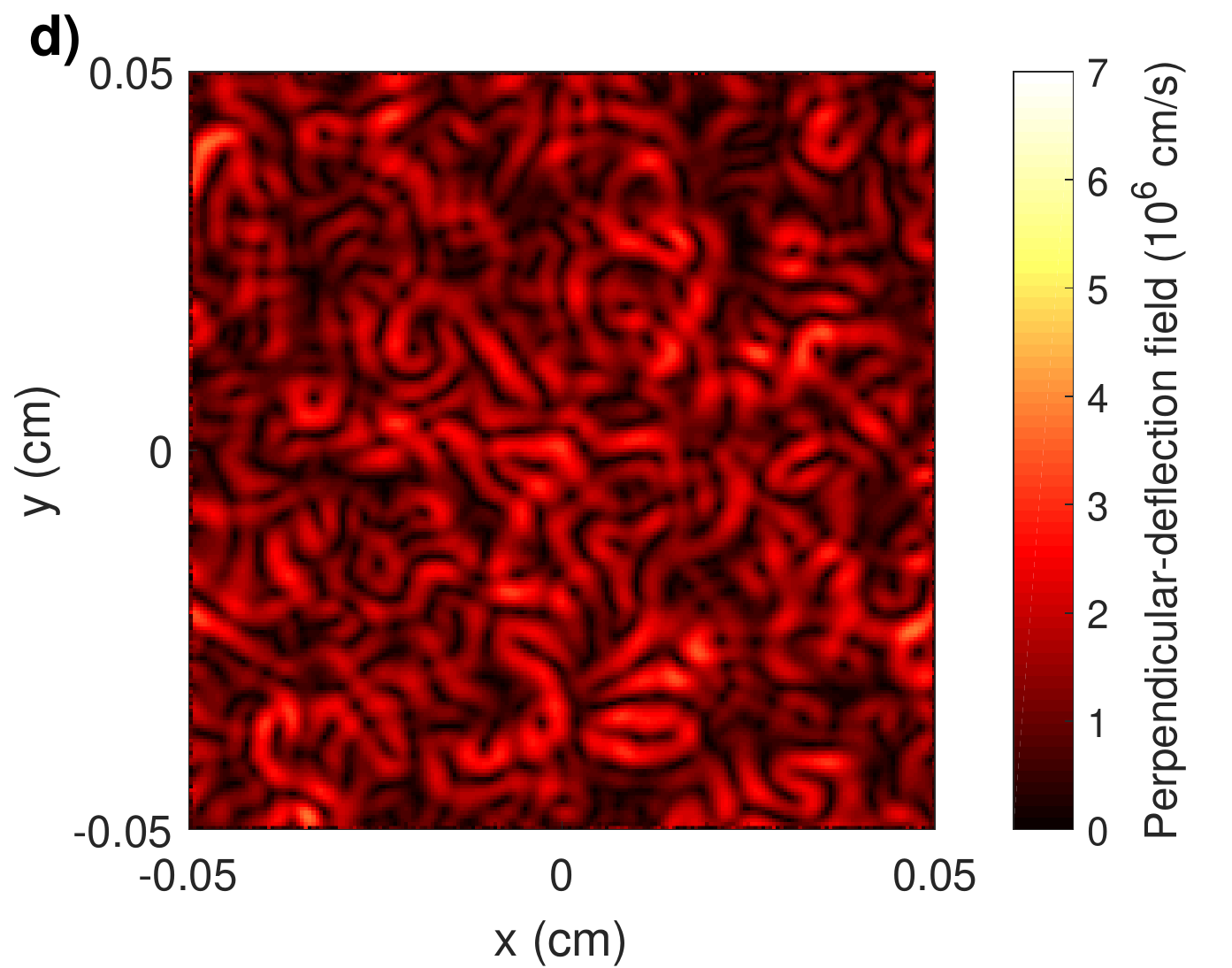}
    \end{subfigure} %
        \begin{subfigure}{.48\textwidth}
        \centering
        \includegraphics[width=\linewidth]{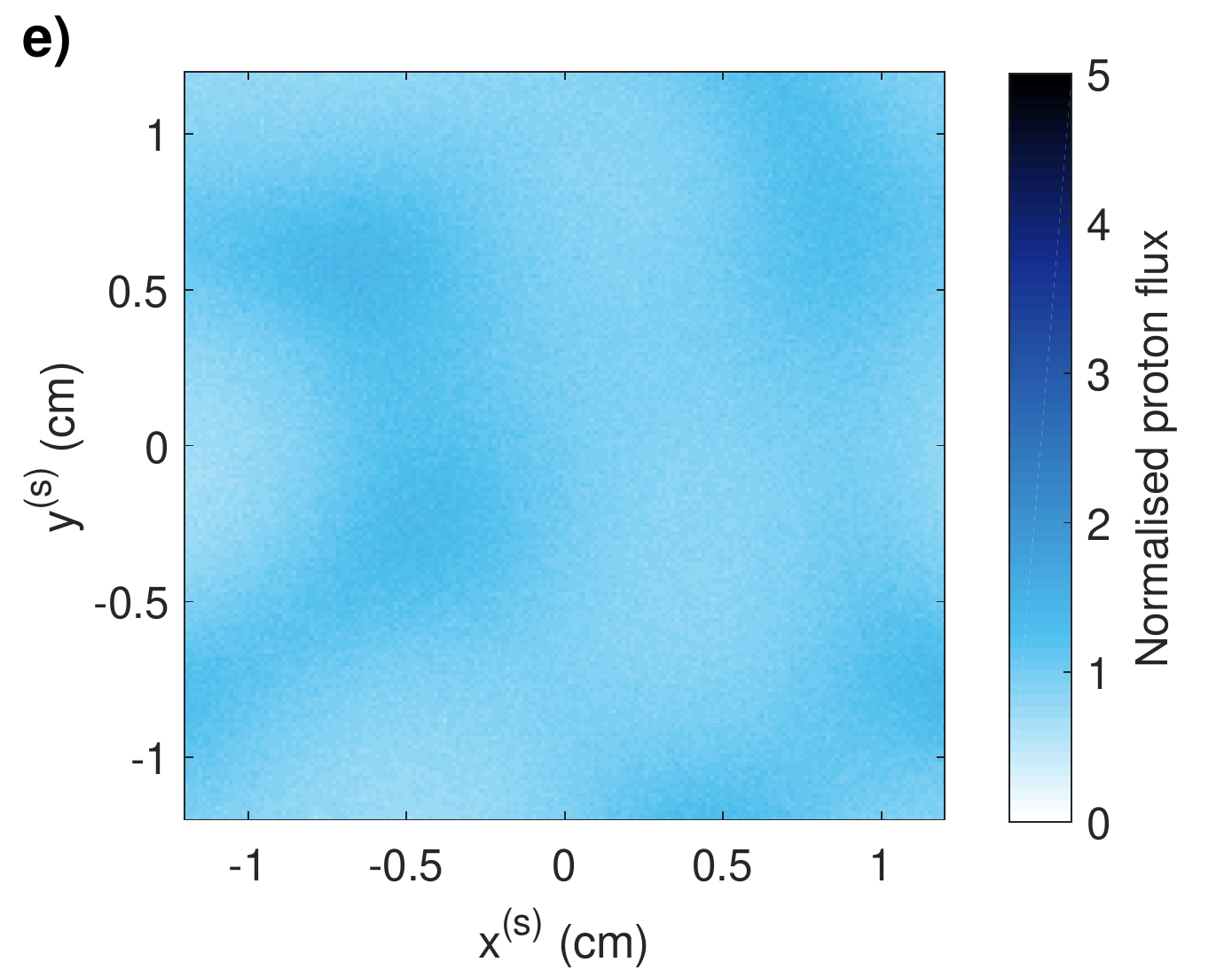}
    \end{subfigure} %
    \begin{subfigure}{.48\textwidth}
        \centering
        \includegraphics[width=\linewidth]{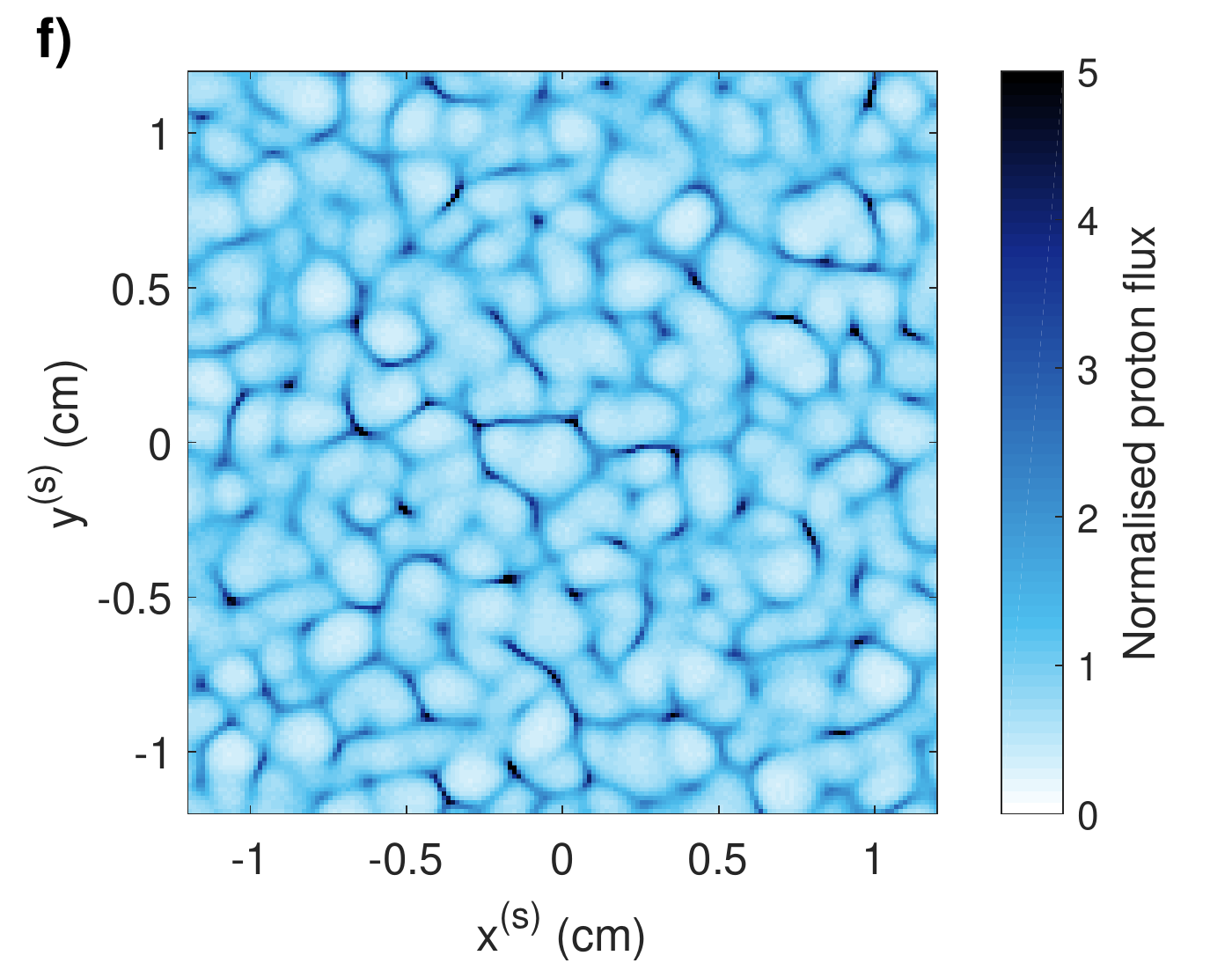}
    \end{subfigure} %
\caption{\textit{Comparison of perpendicular-deflection fields and image-flux distributions resulting from a larger- and smaller-scale stochastic magnetic field}. \textbf{a)} Central-slice plot of larger-scale stochastic magnetic field, located in interaction region side length $l_i = l_{\bot} = l_z =  0.1 \, \mathrm{cm}$, perpendicular to direction of imaging proton beam. The field is specified on a $201^3$ array (grid spacing $\delta x = l_i/201$), with magnetic-energy spectrum of the form \eqref{magcocoonspecAppendscale}, $B_{rms} = 10 \, \mathrm{kG}$, and $l_B = 170 \, \mu \mathrm{m}$. \textbf{b)} Equivalent slice plot of smaller-scale magnetic field configuration, with same $B_{rms}$, but $l_B = 30 \, \mu \mathrm{m}$. \textbf{c)} perpendicular-deflection field induced on 3.3 MeV proton beam by magnetic forces associated with larger-scale configuration as calculated using 30,000,000 test protons from a point source, located at $r_i = 1 \, \mathrm{cm}$ from the array.  \textbf{d)} Deflection field induced on 3.3 MeV proton beam by magnetic forces associated with small-scale configuration. \textbf{e)} Normalised image-flux distribution on detector ($r_s = 30 \, \mathrm{cm}$) resulting from imaging beam for larger-scale configuration. \textbf{f)} Normalised image-flux distribution for smaller-scale field.} \label{largeandsmallscalecocoon}
\end{figure}

First, we describe in greater detail the methodology of this numerical experiment. The single-scale stochastic magnetic fields considered in this appendix are Gaussian, and all have the same form of magnetic-energy spectrum: that of a isotropic random field of magnetic cocoons with a fixed size (see Appendix \ref{ToySpecLinThy} for a full mathematical characterisation~\cite{D04}). Such a field has magnetic-energy spectrum 
\begin{equation}
E_B\!\left(k\right) = \frac{B_{rms}^2 l_e}{12\sqrt{2} \pi^{3/2}} l_e^4 k^4 \exp{\left(-l_e^2 k^2/2\right)} \, , \label{magcocoonspecAppendscale}
\end{equation}
where $l_e$ is the size of a typical cocoon related to the magnetic correlation length by
\begin{equation}
l_B = \frac{\sqrt{2 \pi}}{3} l_e \, .
\end{equation}
This result is derived by substituting explicitly for the magnetic-energy spectrum in \eqref{corrlengthspec} using \eqref{magcocoonspecAppendscale}.
Particular instantiations of the cocoon magnetic field are generated inside a cube-shaped region of side length $l_i$ (that is, $l_i = l_z = l_\bot$) using the technique described in Appendix \ref{NumSimMagFieldGen}. For the visual demonstration, the `larger-scale' field (a two-dimensional slice of which is shown in Figure \ref{largeandsmallscalecocoon}a) has correlation length $l_B/l_i = \sqrt{2 \pi}/15$, while for the `smaller-scale' field (see Figure \ref{largeandsmallscalecocoon}b), $l_B/l_i = \sqrt{2 \pi}/90$. Figures \ref{largeandsmallscalecocoon}a and \ref{largeandsmallscalecocoon}b qualitatively display the same magnitude of field strength, as required for the numerical experiment.

As with the magnetic fields considered in the main text, the cocoon magnetic fields are imaged using an artificial source of protons: details of the numerical scheme used to propagate these protons through the magnetic field are also given in Appendix \ref{NumSimFluxImageGenFull}. Once the protons have passed through the region containing the magnetic field, the change in perpendicular velocity of each proton due to magnetic forces is recorded. The perpendicular-deflection field $\mathbf{w}\!\left(\mathbf{x}_{\bot0}\right)$ can then be calculated by plotting the change in perpendicular velocity as a function of the proton's initial perpendicular position before interaction with the magnetic field. The RMS proton deflection angle is proportional to $w_{rms}$, the RMS of the measured perpendicular-deflection field:
\begin{equation}
\delta \theta_{rms} \propto w_{rms} = \frac{e B_{rms}}{m_p c} \sqrt{l_z l_B}\, . \label{RMSdeflnumexp}
\end{equation}
If \eqref{thetaparamAppend} is accurate, then as the correlation length $l_B$ is changed, $\delta \theta_{rms}$ should increase linearly with $l_B^{1/2}$.

Having determined the perpendicular-deflection field, we can then use plasma-image mapping \eqref{divmappingSec2} combined with Kugland image-flux relation \eqref{screenfluxSec2} to predict the image-flux distribution $\Psi\!\left(\mathbf{x}_{\bot}^{\left(s\right)}\right)$ for any given cocoon magnetic field (see Appendix \ref{NumSimFluxImageGen} for a fuller description of this procedure). To estimate the value of $\mu$ associated with each image, we calculate the RMS of relative image-flux deviations for each predicted image-flux distribution; as described in Section \ref{LinRgme}, this scales linearly with $\mu$ provided the RMS of relative image-flux deviations is small:
\begin{equation}
\mu \propto \left<\left(\frac{\delta \Psi\left(\mathbf{x}_{\bot}^{\left(s\right)}\right)}{\Psi_0^{\left(s\right)}}\right)^2\right>^{1/2} = \frac{2 \sqrt{2 \pi}}{3} \frac{r_s r_i}{r_s+r_i} \frac{e B_{rms}}{m_p c V} \sqrt{\frac{l_z}{l_B}}\, , \label{RMSfluxnumexp}
\end{equation}
where the pre-factor can be calculated exactly for a magnetic field composed of cocoon structures (see Appendix \ref{ToySpecLinThyCocoon}). 
To agree with \eqref{contrastparamsim}, the RMS of relative image-flux deviations should be inversely proportional to $l_B^{1/2}$. 

The results of the visual demonstration test for the perpendicular-deflection field are shown in Figures \ref{largeandsmallscalecocoon}c and \ref{largeandsmallscalecocoon}d; it is qualitatively clear that the perpendicular-deflection field is significantly larger for the larger-scale cocoon field than the smaller-scale one, as anticipated according to scaling \eqref{thetaparamAppend}. 

\begin{figure}[htbp]
\centering
    \begin{subfigure}{.48\textwidth}
        \centering
        \includegraphics[width=\linewidth]{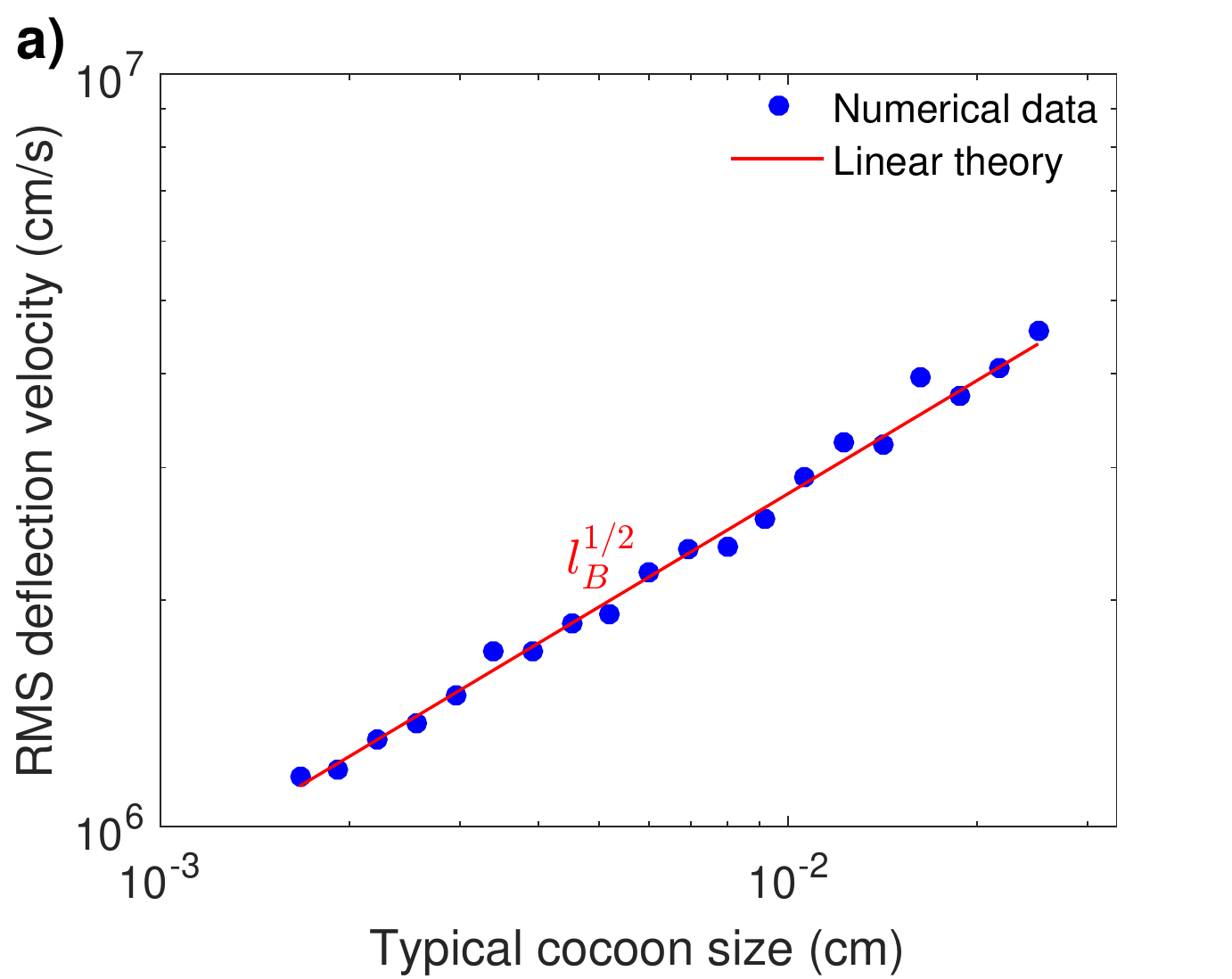}
    \end{subfigure} %
      \begin{subfigure}{.48\textwidth}
        \centering
        \includegraphics[width=\linewidth]{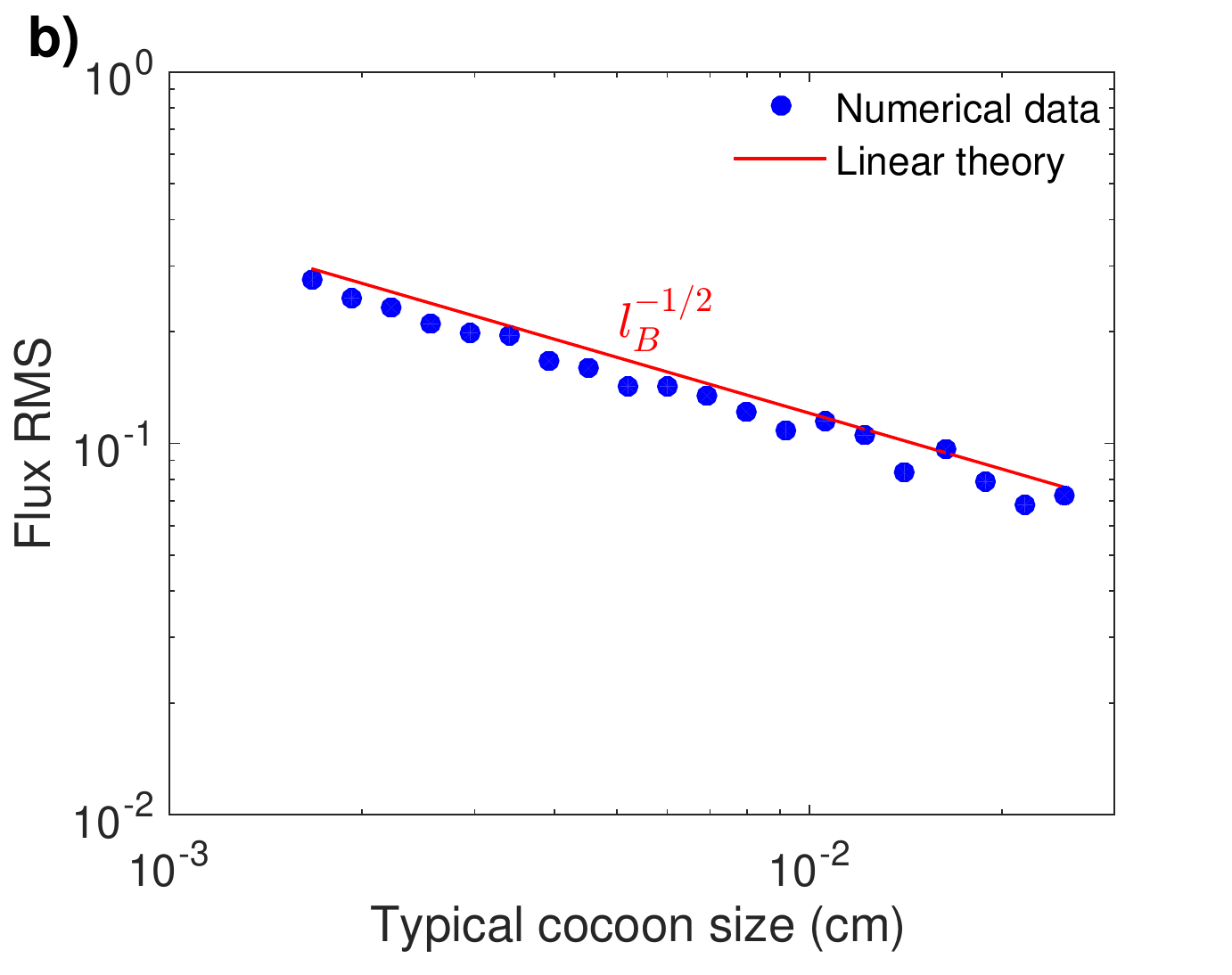}
    \end{subfigure} %
\caption{\textit{Numerical test of scaling laws for proton deflection angle $\delta \theta_{rms}$ and $\mu$}. A range of stochastic magnetic fields of the same type to those described in Figure \ref{largeandsmallscalecocoon} were generated, but with different correlation lengths $l_B$. The imaging procedure described in Figure \ref{largeandsmallscalecocoon} was then applied to calculate the perpendicular-deflection field and image-flux distribution for each image. \textbf{a)} Numerical calculation of perpendicular-deflection field RMS from tests (blue circles), plotted with prediction \eqref{RMSdeflnumexp} (red). \textbf{b)} Numerical calculation of relative image-flux deviation RMS from tests (blue), plotted with prediction \eqref{RMSfluxnumexp} (red).} \label{cocoonPDFs}
\end{figure}

The visual demonstration test for $\mu$ also gives results in agreement with the prediction from \eqref{contrastparamsim}: image-flux structures associated with the larger-scale cocoon field in Figure \ref{largeandsmallscalecocoon}e are much weaker than those for the smaller-scale field shown in Figure \ref{largeandsmallscalecocoon}f, a result suggesting the smaller-scale field has a higher $\mu$ value. The results of explicit tests of the predicted scaling laws \eqref{thetaparamAppend} for the RMS proton deflection angle and \eqref{contrastparamsim} for $\mu$ are shown in Figures \ref{cocoonPDFs}e and \ref{cocoonPDFs}f: the numerical results for the RMS of the perpendicular-deflection field and RMS of relative image-flux deviations are in strong agreement with predictions \eqref{RMSdeflnumexp} and \eqref{RMSfluxnumexp} respectively. 

\section{Derivation of linear-regime image-flux relation \eqref{linfluxscalsect3} from small-deflection Kugland image-flux relation \eqref{screenfluxSec2}} \label{LinThyDev}

In Section \ref{LinRgme}, it was stated that in the linear regime, plasma-image mapping \eqref{divmappingSec2},
and Kugland image-flux relation \eqref{screenfluxSec2} reduce to simplified expressions \eqref{linmapSec2} and \eqref{linrelfluxgenSUM}, which for convenience we reproduce here:
\begin{equation}
\mathbf{x}_{\bot}^{\left(s\right)} = \frac{r_s+r_i}{r_i}  \mathbf{x}_{\bot0} \left[1 + \mathcal{O}\!\left(\mu\right)\right] \, , \label{linmapAppend}
\end{equation}
and
\begin{equation}
\frac{\delta \Psi\!\left(\mathbf{x}_\bot^{\left(s\right)}\right)}{\Psi_0^{\left(s\right)}} =  \frac{r_s r_i}{r_s+r_i} \frac{4 \pi e}{m_p c^2 V} \int_0^{l_z} j_z\!\left(\mathbf{x}_{\bot0}\left(1+\frac{z'}{r_i}\right),z'\right) \mathrm{d}z' \, . \label{linrelfluxgenAppend}
\end{equation}
It was also claimed that the image-flux deviation  $\delta \Psi\!\left(\mathbf{x}_{\bot0}\right)$ could be related to the deflection-field potential $\varphi\!\left(\mathbf{x}_{\bot0}\right)$ defined by \eqref{deffieldpotdef} according to Poisson equation \eqref{linfluxref}:
\begin{equation}
\nabla_{\bot0}^2 \varphi\!\left(\mathbf{x}_{\bot0}\right) = - \Xi\!\left(\mathbf{x}_{\bot0}\right) \, , \label{linfluxrefAppend}
\end{equation} 
where $\Xi\!\left(\mathbf{x}_{\bot0}\right)$ is a source function defined by \eqref{linfluxref},
\begin{equation}
\Xi\!\left(\mathbf{x}_{\bot0}\right) = \frac{r_s+r_i}{r_i}\frac{V}{r_s} \frac{\delta \Psi\!\left(\mathbf{x}_\bot^{\left(s\right)}\right)}{\Psi_0^{\left(s\right)}} \, .
\end{equation}
In this appendix we derive these results using the fact that in the linear regime, $\mu \ll 1$, allowing for asymptotic expansions of plasma-image mapping \eqref{divmappingSec2} and Kugland image-flux relation \eqref{screenfluxSec2} in $\mu$. It also examines the proper magnitude of terms neglected in the asymptotic expansion of Kugland image-flux relation \eqref{screenfluxSec2} in $\mu$ -- in particular, whether it is consistent to not expand the argument of the image-flux distribution in terms of its argument, the plasma-image mapping. 
  
To derive the simplifed plasma-image mapping \eqref{linmapAppend}, we start from full mapping \eqref{divmappingSec2}, that is,
\begin{equation}
\mathbf{x}_{\bot}^{\left(s\right)}\!\left(\mathbf{x}_{\bot0}\right) = \left(
\frac{r_s+r_i}{r_i} \mathbf{x}_{\bot0} + \frac{r_s}{V} \, \mathbf{w}\!\left(\mathbf{x}_{\bot0}\right) \right)\left[1 + \mathcal{O}\!\left(\delta \alpha, \delta \theta\right)\right] \, , \label{pathintfieldAppend}
\end{equation}     
and note that
\begin{equation}
\frac{r_s \left|\mathbf{w}\!\left(\mathbf{x}_{\bot0}\right)\right|/V}{\frac{r_s+r_i}{r_i} \left|\mathbf{x}_{\bot0}\right|} \sim  \mu  \, .\label{plasmapscalgen}
\end{equation}
Thus, \eqref{pathintfieldAppend} can be re-written as the simplified linear mapping \eqref{linmapAppend} to leading order in $\mu$:
\begin{equation}
\mathbf{x}_{\bot}^{\left(s\right)} = \frac{r_s+r_i}{r_i}  \mathbf{x}_{\bot0} \left[1 + \mathcal{O}\!\left(\delta \alpha, \mu\right)\right]  \, .\label{linmap}
\end{equation}
Note that the asymptotic expansion carried out in $\delta \theta$ in order to derive \eqref{pathintfieldAppend} is superceded by the expansion in $\mu$, since by its definition $\mu$ is a larger parameter. 

Now turning to the image-flux relation, we begin with \eqref{screenfluxSec2}:
\begin{equation}
\Psi\!\left(\mathbf{x}_{\bot}^{\left(s\right)}\!\left(\mathbf{x}_{\bot0}\right)\right) = \sum_{\mathbf{x}_\bot^{\left(s\right)} = \mathbf{x}_\bot^{\left(s\right)}\!\left(\mathbf{x}_{\bot0}\right)}\frac{\Psi_{0}}{\left|\det{\nabla_{\bot0}\!\left[\mathbf{x}_{\bot}^{\left(s\right)}\!\left(\mathbf{x}_{\bot0}\right)\right]}\right|} \label{screenfluxAppend}
\end{equation}
Evaluating the determinant in the denominator of image-flux relation \eqref{screenfluxAppend} explicitly in terms of the full plasma-image mapping \eqref{pathintfieldAppend}, we find
\begin{IEEEeqnarray}{rCl}
\det{\frac{\partial \mathbf{x}_{\bot}^{\left(s\right)}}{\partial \mathbf{x}_{\bot0}}} & = & \det{\left(\frac{r_s+r_i}{r_i} \underline{\underline{\mathrm{I}}}+\frac{r_s}{V} \frac{\partial \mathbf{w}\!\left(\mathbf{x}_{\bot0}\right)}{\partial \mathbf{x}_{\bot0}}\right)} \nonumber \\
& = &  \left(\frac{r_s+r_i}{r_i}\right)^2 \left\{1+\frac{r_s r_i}{\left(r_s+r_i\right)V} \nabla_{\bot0} \cdot \mathbf{w}\!\left(\mathbf{x}_{\bot0}\right) + \left[\frac{r_s r_i}{\left(r_s+r_i\right)V}\right]^2 \det{\frac{\partial \mathbf{w}\!\left(\mathbf{x}_{\bot0}\right)}{\partial \mathbf{x}_{\bot0}}}\right\} \label{fluxdetAppend}
\end{IEEEeqnarray}
Estimating the relative size of terms on the right hand side of \eqref{fluxdetAppend}, we have
\begin{equation}
\frac{r_s r_i}{\left(r_s+r_i\right)V} \nabla_{\bot0} \cdot \mathbf{w}\!\left(\mathbf{x}_{\bot0}\right) \sim \mu \, , \qquad \left(\frac{r_s r_i}{\left(r_s+r_i\right)V}\right)^2 \det{\frac{\partial \mathbf{w}\!\left(\mathbf{x}_{\bot0}\right)}{\partial \mathbf{x}_{\bot0}}} \sim \mu^2 \, .
\end{equation}
Assuming $\mu \ll 1$, the determinant can therefore be expanded in terms of $\mu$ as
\begin{equation}
\det{\frac{\partial \mathbf{x}_{\bot}^{\left(s\right)}}{\partial \mathbf{x}_{\bot0}}} =  \left(\frac{r_s+r_i}{r_i}\right)^2 \left[1+\frac{r_s r_i}{\left(r_s+r_i\right)V} \nabla_{\bot0} \cdot \mathbf{w}\!\left(\mathbf{x}_{\bot}\right) + \mathcal{O}\!\left(\mu^2\right) \right]  \, .\label{detdivrelate}
\end{equation}
Then expanding image-flux relation \eqref{screenfluxAppend} in terms of $\mu$ leads to
\begin{equation}
\Psi\!\left(\mathbf{x}_\bot^{\left(s\right)}\right) \approx  \left(\frac{r_i}{r_s+r_i}\right)^2 \Psi_{0} \left[1-\frac{r_s r_i}{\left(r_s+r_i\right)V} \nabla_{\bot0} \cdot \mathbf{w}\!\left(\mathbf{x}_{\bot0}\right)+ \mathcal{O}\!\left(\mu^2\right)\right] \, , \label{linfluxA}
\end{equation}
which can be re-written in terms of the relative flux, normalised by the unperturbed image flux
\begin{equation}
\Psi_0^{\left(s\right)} \equiv \left(\frac{r_i}{r_s+r_i}\right)^2 \Psi_{0} \, ,
\end{equation}
to give
\begin{equation}
\frac{\delta \Psi\!\left(\mathbf{x}_\bot^{\left(s\right)}\right)}{\Psi_0^{\left(s\right)}} =  \frac{\Psi\!\left(\mathbf{x}_\bot^{\left(s\right)}\right)-\Psi_0^{\left(s\right)}}{\Psi_0^{\left(s\right)}} \approx  -\frac{r_s r_i}{\left(r_s+r_i\right)V} \nabla_{\bot0} \cdot \mathbf{w}\!\left(\mathbf{x}_{\bot0}\right) \, . \label{linrelfluxgenA}
\end{equation} 
The perpendicular-deflection field in this case is given by \eqref{deffieldundefl}, that is
\begin{equation}
\mathbf{w}\!\left(\mathbf{x}_{\bot0}\right) = \frac{e}{m_p c} \hat{\mathbf{z}} \times \int_0^{l_z} \mathrm{d}z' \; \mathbf{B}\!\left(\mathbf{x}_{\bot0}\left(1+\frac{z'}{r_i}\right),z'\right) \left[1+ \mathcal{O}\!\left(\delta \alpha, \delta \theta \, \frac{l_z}{l_B}\right)\right] \, , \label{deffieldundeflAppend}
\end{equation}
since for $\mu \ll 1$, $\delta \theta \, l_z/l_B \sim \mu \, \delta \alpha \left(r_s+r_i\right)/r_s \ll 1$. We recover the result that in the linear regime the image displays the undeflected path integrated z-component of the magnetic field curl:
\begin{IEEEeqnarray}{rCl}
\frac{\delta \Psi\!\left(\mathbf{x}_\bot^{\left(s\right)}\right)}{\Psi_0^{\left(s\right)}} & = & \frac{r_s r_i}{r_s+r_i} \frac{e}{m_p cV} \hat{\mathbf{z}} \cdot \int_0^{l_z} \nabla_{\bot0} \times \mathbf{B}\!\left(\mathbf{x}_{\bot0}\left(1+\frac{z'}{r_i}\right),z'\right) \mathrm{d}z' \left[1+ \mathcal{O}\!\left(\delta \alpha, \mu\right)\right] \nonumber \\
& = & \frac{r_s r_i}{r_s+r_i} \frac{e}{m_p cV} \hat{\mathbf{z}} \cdot \int_0^{l_z} \nabla \times \mathbf{B}\!\left(\mathbf{x}_{\bot0}\left(1+\frac{z'}{r_i}\right),z'\right) \mathrm{d}z' \left[1+ \mathcal{O}\!\left(\delta \alpha, \mu\right)\right] \, .\label{linrelfluxgenB}
\end{IEEEeqnarray}
Applying Amp\'ere's law (in the absence of the displacement current) gives the desired result \eqref{linrelfluxgenAppend}. 

To derive \eqref{linfluxref}, we start from an expression for the perpendicular-deflection field in terms of the deflection-field potential:
\begin{equation}
\mathbf{w}\!\left(\mathbf{x}_{\bot0}\right) = \frac{V}{r_s} \nabla_{\bot0} \varphi\!\left(\mathbf{x}_{\bot0}\right) \left[1+ \mathcal{O}\!\left(\delta \theta \, \frac{l_z}{l_B}\right)\right]\, , \label{lindeflfieldAppend}
\end{equation}
Substitution of \eqref{lindeflfieldAppend} into \eqref{linrelfluxgenA} gives
\begin{equation}
\frac{\delta \Psi\!\left(\mathbf{x}_\bot^{\left(s\right)}\right)}{\Psi_0^{\left(s\right)}} = -\frac{r_i}{r_s+r_i} \nabla_{\bot0}^2 \varphi\!\left(\mathbf{x}_{\bot0}\right) \, ,
\end{equation}
which on rearrangement gives the expected result  \eqref{linfluxref}. 

A potential subtlety in the derivations of \eqref{linrelfluxgenAppend} and \eqref{linfluxref} (noticed by Graziani \textit{et. al.}~\cite{GTLL16}) 
arises from the observation that the image-flux distribution is evaluated in the image-coordinate system. More specifically, we need to check whether using \eqref{pathintfieldAppend} for calculating the argument for the image-flux distribution does not result in the systematic neglect of any $\mathcal{O}\!\left(\mu\right)$ terms in the expansion of \eqref{screenfluxAppend} in $\mu$. 
We do this by first assuming sufficiently small image-flux gradients, then Taylor expanding \eqref{linfluxA}:
\begin{equation}
\Psi\!\left(\mathbf{x}_\bot^{\left(s\right)}\right) = \Psi\!\left(\frac{r_s+r_i}{r_i}  \mathbf{x}_{\bot0}\right) + \frac{r_s}{V} \mathbf{w} \cdot \nabla_{\bot}^{\left(s\right)} \Psi\!\left(\frac{r_s+r_i}{r_i}  \mathbf{x}_{\bot0}\right) \left[1+ \mathcal{O}\!\left(\frac{r_s \, \delta \theta}{l_{\Psi}}\right)\right] \, ,
\end{equation}
where $\nabla_{\bot}^{\left(s\right)} \equiv \partial/\partial \mathbf{x}_{\bot}^{\left(s\right)}$, and $l_{\Psi} \gg r_s \delta \theta $ is a typical scale on which image-flux deviations occur. If we further assume that image-flux deviations occur with the same length scale as magnetic field fluctuations including the image-magnification factor $\mathcal{M} = \left(r_s+r_i\right)/r_i$, that is $l_\Psi \sim \tilde{l}_B = \mathcal{M} l_B$, the size of the discrepancy in the image-flux distribution resulting from using \eqref{pathintfieldAppend} as the image-flux distribution function's argument is 
\begin{equation}
\Psi\!\left(\mathbf{x}_\bot^{\left(s\right)}\right) - \Psi\!\left(\frac{r_s+r_i}{r_i}  \mathbf{x}_{\bot0}\right) = \frac{r_s}{V} \mathbf{w} \cdot \nabla_{\bot}^{\left(s\right)} \Psi\!\left(\frac{r_s+r_i}{r_i}  \mathbf{x}_{\bot0}\right) \sim \mu \, \delta \Psi \ll \delta \Psi \, .
\end{equation}
a quadratic term in $\mu$. Thus we see that using \eqref{pathintfieldAppend} as the argument for the image-flux distribution is consistent to the asymptotic order of the expansion provided the perpendicular-deflection field and its derivatives vary on similar scales to the magnetic field (taking into account the image-magnification factor).

That being said, this term could become important if image-flux gradients occur on smaller scales than gradients of the magnetic field; and there are two situations when such image-flux behaviour would be expected. First, \eqref{linrelfluxgenB} shows that plasmas involving large current gradients, but small current magnitude, will inevitably result in large-image flux gradients. Second are situations involving moderately small values of $\mu$: here, as discussed in Section \ref{NonLinInjRgme}, nonlinear effects will tend to focus regions of positive relative flux, increasing gradient values of the image-flux distribution. As a consequence, Graziani \textit{et. al.} report that including this term is essential for successful implementation of linear reconstruction methods for general proton-flux images~\cite{GTLL16}.

\section{Derivation of linear-regime flux RMS relation \eqref{linfluxscalsect3} and spectral relation \eqref{linfluxspec3}} \label{RelFluxRMSLinThy}

Section \ref{LinRgme} claims that provided $\mu \ll 1$, the RMS of relative image-flux deviations can be related to the RMS magnetic field strength via equation \eqref{linfluxscalsect3}, which is 
\begin{equation}
 \left<\frac{\delta \Psi}{\Psi_0^{\left(s\right)}}^2\right>^{1/2} = \sqrt{\frac{\pi}{2}} \frac{r_i r_s}{r_s+r_i} \frac{e B_{rms}} {m c V} \sqrt{\frac{l_z}{l_{\Psi}}} = \frac{\mu}{\mu_0} \label{linfluxscalAppend} \, ,
\end{equation}
where $\mu_0$ is defined by $\mu_0 = \sqrt{2 l_\Psi/l_B \pi}$ -- $l_\Psi$ is the relative-image flux correlation length defined subsequently in equation \eqref{screenfluxcorrlength} -- and satisfies lower bound $\mu_0 \leq 2/\pi$. 

Furthermore, under the same assumption $\mu \ll 1$ the magnetic-energy spectrum $E_B\!\left(k\right)$ is related to the 2D spectrum of image-flux deviations $\hat{\eta}\!\left(k\right)$ for homogeneous and isotropic magnetic field statistics satisfying $l_B \ll l_z$ by equation \eqref{linfluxspec3}:
\begin{equation}
E_B\!\left(k\right) = \frac{1}{2\pi} \frac{m_p^2 c^2 V^2}{e^2 r_s^2 l_z} \hat{\eta}\!\left( \frac{r_i}{r_s+r_i} k\right) \, . \label{linfluxspecAppend}
\end{equation}
where $\hat{\eta}\!\left(k\right)$ is defined by
\begin{equation}
\hat{\eta}\!\left(k_\bot\right) \equiv \frac{1}{2 \pi} \int \mathrm{d}\theta \, \left(\frac{1}{\Psi_0^{\left(s\right)}}\right)^2 \left<\left|\hat{\delta \Psi}\!\left(\mathbf{k}_\bot\right)\right|^2\right> \, , \label{fluxspecdef}
\end{equation}
and $\hat{\delta \Psi}\!\left(\mathbf{k}_\bot\right)$ is the Fourier-transformed relative image-flux deviation. In this appendix we prove both results using a similar approach to that adopted in Appendix \ref{DeflfieldCorr}: we first define an autocorrelation function of the relative image-flux distribution function and then writing the magnetic autocorrelation function in terms of it. Evaluating this at the origin will give \eqref{linfluxscalAppend}. We shall then use the Wiener-Khinchin theorem to relate the respective spectra of the relative-image flux and magnetic field.

Adopting a uniform inital flux distribution $\Psi_0\!\left(\mathbf{x}_{\bot0}\right) = \Psi_0$, define the image-flux autocorrelation function by 
\begin{equation}
\eta\!\left(\mathbf{x}_{\bot}^{\left(s\right)},\tilde{\mathbf{x}}_{\bot}^{\left(s\right)}\right) = \left<\frac{\delta \Psi\!\left(\mathbf{x}_\bot^{\left(s\right)}\right)}{\Psi_0^{\left(s\right)}}\frac{\delta \Psi\!\left(\tilde{\mathbf{x}}_\bot^{\left(s\right)}\right)}{\Psi_0^{\left(s\right)}}\right> \, .
\end{equation}
This correlation function can then be related to the magnetic autocorrelation function by substituting for the relative image-flux distribution using the linear-regime image-flux relation \eqref{linrelfluxgenSUM}, which can be written as
\begin{equation}
\frac{\delta \Psi\!\left(\mathbf{x}_\bot^{\left(s\right)}\right)}{\Psi_0^{\left(s\right)}} \approx  \frac{r_s r_i}{r_s+r_i} \frac{e}{m_p cV} \hat{\mathbf{z}} \cdot \int_0^{l_z} \nabla \times \delta \mathbf{B}\!\left(\mathbf{x}_{\bot0}\left(1+\frac{z}{r_i}\right),z\right) \mathrm{d}z \left[1+ \mathcal{O}\!\left(\delta \alpha, \mu\right)\right]\, .\label{linrelfluxgenAppend}
\end{equation}
The substitution gives
\begin{IEEEeqnarray}{rCl}
\eta\!\left(\mathbf{x}_{\bot}^{\left(s\right)},\tilde{\mathbf{x}}_{\bot}^{\left(s\right)}\right) & = & \left<\frac{e^2 r_i^2 r_s^2}{m_p^2 c^2 V^2 \left(r_s+r_i\right)^2}\epsilon_{lmn} \epsilon_{l'm'n'} \hat{z}_l \hat{z}_{l'} \int_0^{l_z} \mathrm{d}z \int_0^{l_z} \mathrm{d}\tilde{z} \, \frac{\partial \, \delta B_{n}\!\left(\mathbf{x}\right) }{\partial x_{m}} \frac{\partial \, \delta B_{n'}\!\left(\tilde{\mathbf{x}}\right) }{\partial x_{m'}} \right> \nonumber \\
& = & \frac{e^2 r_i^2 r_s^2}{m_p^2 c^2 V^2 \left(r_s+r_i\right)^2}\epsilon_{lmn} \epsilon_{l'm'n'} \hat{z}_l \hat{z}_{l'} \int_0^{l_z} \mathrm{d}z \int_0^{l_z} \mathrm{d}\tilde{z} \, \frac{\partial^2 M_{nn'}\!\left(\mathbf{x}, \tilde{\mathbf{x}}\right)}{\partial x_m \partial {\tilde{x}}_{m'}}  \, , \label{lincorrfunmagcorrfun}
\end{IEEEeqnarray}
where $M_{nn'}\!\left(\mathbf{x}, \tilde{\mathbf{x}}\right)$ is again the magnetic autocorrelation tensor, and its arguments are the plasma coordinates
\begin{IEEEeqnarray}{rClr}
\mathbf{x} & = & \left[\mathbf{x}_{\bot0}\left(1+\frac{z}{r_i}\right),z\right] & \, ,\\
\tilde{\mathbf{x}} & = &  \left[\mathbf{x}_{\bot0}\left(1+\frac{\tilde{z}}{r_i}\right),\tilde{z}\right] & \, .
\end{IEEEeqnarray} 
As when calculating the deflection-field correlation function, we expand the argument of the autocorrelation function in \eqref{lincorrfunmagcorrfun} along the $z$-projection of unperturbed trajectories -- neglecting small, $\mathcal{O}\!\left(\delta \alpha, \delta \theta\right)$ terms -- to give simplified expressions for the plasma coordinates, the perpendicular components of which can then be written in terms of the perpendicular image-coordinate using linear-regime plasma-image mapping \eqref{linmapSec2}:
\begin{IEEEeqnarray}{rClr}
\mathbf{x} & = & \left[\frac{r_i}{r_i+r_s}\mathbf{x}_\bot^{\left(s\right)},z\right] & \, ,\\
\tilde{\mathbf{x}} & = & \left[\frac{r_i}{r_i+r_s}\tilde{\mathbf{x}}_\bot^{\left(s\right)},\tilde{z}\right] \, . &
\end{IEEEeqnarray}
Again we change integration variables to
\begin{IEEEeqnarray}{rClr}
\left(\mathbf{r}_\bot,r_z\right) & = & \left(\tilde{\mathbf{x}}_\bot-\mathbf{x}_\bot,\tilde{z}-z\right) & \, ,\\
\bar{\mathbf{x}} & = & \mathbf{x} & \, ,
\end{IEEEeqnarray}
which transforms partial derivatives by
\begin{IEEEeqnarray}{rClr}
\frac{\partial}{\partial \mathbf{x}} & = & \frac{\partial}{\partial \bar{\mathbf{x}}}-\frac{\partial}{\partial \mathbf{r}} & \, , \\
\frac{\partial}{\partial \tilde{\mathbf{x}}} & = & \frac{\partial}{\partial \mathbf{r}} & \, .
\end{IEEEeqnarray}
Thus,
\begin{equation}
\eta\!\left(\mathbf{x}_{\bot}^{\left(s\right)},\tilde{\mathbf{x}}_{\bot}^{\left(s\right)}\right) = \frac{e^2 r_i^2 r_s^2}{m_p^2 c^2 V^2 \left(r_s+r_i\right)^2}\epsilon_{lmn} \epsilon_{l'm'n'} \hat{z}_l \hat{z}_{l'}  \int_0^{l_z} \mathrm{d}\bar{z} \int_{-\bar{z}}^{l_z-\bar{z}} \mathrm{d}r_z \, \left(\frac{\partial^2}{\partial \bar{x}_m \partial r_{m'}}-\frac{\partial^2}{\partial r_m \partial r_{m'}}\right) M_{nn'}\!\left(\bar{\mathbf{x}}, \mathbf{r}\right) \, . \label{lincorrfunmagcorrfuniso}
\end{equation}
Under the same assumption that $l_B \ll l_z$, we extend the limits of the inner integral to $\left(-\infty,\infty\right)$, invoking at worst an $\mathcal{O}\!\left(l_B/l_z\right)$ error. If we then assume the turbulence to be locally homogeneous, the magnetic autocorrelation function is again a function of $\mathbf{r}$ alone, eliminating the first term in the integrand \eqref{lincorrfunmagcorrfuniso}. The outer integral can be evaluated independently, leaving
\begin{equation}
\eta \! \left(\mathbf{r}_{\bot}^{\left(s\right)}\right) = -\frac{e^2 r_i^2 r_s^2 l_z}{m_p^2 c^2 V^2 \left(r_s+r_i\right)^2}\epsilon_{lmn} \epsilon_{l'm'n'} \hat{z}_l \hat{z}_{l'} \int_{-\infty}^{\infty} \mathrm{d}r_z \, \frac{\partial^2 M_{nn'}\!\left(\mathbf{r} \right)}{\partial r_m \partial {r_{m'}}} \, .
\end{equation}
Alternatively writing entirely in terms of variables on the scale of the interaction region, define rescaled relative image-flux autocorrelation function
\begin{equation}
\eta^{\left(0\right)}\!\left(\mathbf{r}_\bot\right) = \eta \! \left(\mathbf{r}_{\bot}^{\left(s\right)}\right) \,  ,
\end{equation}
which gives
\begin{equation}
\eta^{\left(0\right)} \! \left(\mathbf{r}_{\bot}\right) = -\frac{e^2 r_i^2 r_s^2 l_z}{m_p^2 c^2 V^2 \left(r_s+r_i\right)^2}\epsilon_{lmn} \epsilon_{l'm'n'} \hat{z}_l \hat{z}_{l'} \int_{-\infty}^{\infty} \mathrm{d}r_z \, \frac{\partial^2 M_{nn'}\!\left(\mathbf{r} \right)}{\partial r_m \partial {r_{m'}}} \, .
\end{equation}
If we further assume isotropy, the magnetic correlation tensor takes the form given by \eqref{isomagcoll}. In this case, it can be shown (after quite a bit of tedious but elementary manipulation) that the image flux correlation function is directly related to the magnetic autocorrelation function by
\begin{equation}
\eta \! \left(r_{\bot}^{\left(s\right)}\right) = \eta^{\left(0\right)}\!\left(r_\bot\right) = - \frac{e^2 r_i^2 r_s^2 l_z}{m_p^2 c^2 V^2 \left(r_s+r_i\right)^2} \frac{1}{r_\bot} \frac{\mathrm{d}}{\mathrm{d}r_\bot}\left(r_\bot \frac{\mathrm{d}}{\mathrm{d}r_\bot}\left[\frac{1}{2}\int_{-\infty}^{\infty} \mathrm{d}r_z \, M\!\left(r\right)\right]\right) \, . \label{result1}
\end{equation}
Now, by appropriate integration of \eqref{result1} to invert the 2D Laplacian operator, we obtain an equation with exactly the same form as the deflection autocorrelation function \eqref{deflautocorr}, but substituting an integrated form of the autocorrelation function:
\begin{equation}
\zeta\!\left(r_\bot\right) \equiv \int_{r_\bot}^{\infty} \frac{\mathrm{d} {\tilde{r}}_\bot}{{\tilde{r}}_\bot} \int_{{\tilde{r}}_\bot}^{\infty} \mathrm{d} r'_\bot \,  r'_\bot \, \eta^{\left(0\right)}\!\left(r'_\bot\right) = - \frac{e^2 r_i^2 r_s^2 l_z}{m_p^2 c^2 V^2 \left(r_s+r_i\right)^2} \frac{1}{2}\int_{-\infty}^{\infty} \mathrm{d}r_z \, M\!\left(r\right) \, .\label{result2}
\end{equation} 
We can re-express $\zeta\!\left(r_\bot\right) $ as a single integral of $\eta\!\left(r_\bot\right)$ by swapping the order of integration:
\begin{equation}
\zeta\!\left(r_\bot\right) = \int_{r_\bot}^{\infty} \mathrm{d} r'_\bot \,  r'_\bot \, \eta^{\left(0\right)}\!\left(r'_\bot\right) \int_{r_\bot}^{r'_\bot} \frac{\mathrm{d} {\tilde{r}}_\bot}{{\tilde{r}}_\bot} =  \int_{r_\bot}^{\infty} \mathrm{d} r'_\bot \,  r'_\bot \, \eta^{\left(0\right)}\!\left(r'_\bot\right) \log{\frac{r'_\bot}{r_\bot}} \, . \label{result3}
\end{equation}
We can then derive an expression for the magnetic autocorrelation function in terms of $\zeta\!\left(r\right)$ by noting \eqref{result2} is an Abel integral equation which can be inverted~\cite{EV03}. The magnetic autocorrelation function is given by
\begin{equation}
M\!\left(r\right) = \frac{2}{\pi} \frac{m_p^2 c^2 V^2 \left(r_s+r_i\right)^2}{e^2 r_i^2 r_s^2 l_z} \int_{r}^{\infty} \mathrm{d}y \, \frac{\zeta'\!\left(y\right)}{\sqrt{y^2-r^2}} \, .
\end{equation}
We can rewrite this in terms of the relative image-flux autocorrelation function using
\begin{equation}
\zeta'\!\left(y\right) = -\frac{1}{y} \int_{y}^{\infty} \mathrm{d} r' \,  r' \, \eta^{\left(0\right)}\!\left(r'\right) \, ,
\end{equation}
and so
\begin{equation}
M\!\left(r\right) = -\frac{2}{\pi} \frac{m_p^2 c^2 V^2 \left(r_s+r_i\right)^2}{e^2 r_i^2 r_s^2 l_z} \int_{r}^{\infty} \frac{\mathrm{d}y }{y\sqrt{y^2-r^2}} \int_{r'}^{\infty} \mathrm{d} r' \,  r' \, \eta^{\left(0\right)}\!\left(r'\right) \, .
\end{equation}
Swapping the order of integration again, obtain
\begin{equation}
M\!\left(r\right) = -\frac{2}{\pi} \frac{m_p^2 c^2 V^2 \left(r_s+r_i\right)^2}{e^2 r_i^2 r_s^2 l_z} \int_{r}^{\infty} \mathrm{d} r' \,  r' \, \eta^{\left(0\right)}\!\left(r'\right) \int_{r}^{r'} \frac{\mathrm{d}y }{y\sqrt{y^2-r^2}}\, . \label{working}
\end{equation}
This is the desired relation for the magnetic autocorrelation function in terms of the relative image-flux autocorrelation function.

To recover \eqref{linfluxscalAppend}, we take the limit as $r \to 0$ in \eqref{working} and using the regularity condition, deducing
\begin{equation}
\frac{B_{rms}^2}{8\pi} = \frac{1}{\left(2\pi\right)^2} \frac{m_p^2 c^2 V^2 \left(r_s+r_i\right)^2}{e^2 r_i^2 r_s^2 l_z} \int_{0}^{\infty} \mathrm{d} r \, \eta^{\left(0\right)}\!\left(r\right) =\frac{1}{\left(2\pi\right)^2} \frac{m_p^2 c^2 V^2 \left(r_s+r_i\right)}{e^2 r_i r_s^2 l_z} \int_{0}^{\infty} \mathrm{d} r \, \eta\!\left(r\right) \, .\label{result6}
\end{equation}
Now, defining relative image-flux correlation length $l_\Psi$ by
\begin{equation}
l_\Psi = \int_{0}^{\infty} \mathrm{d} r \, \frac{\eta^{\left(0\right)}\!\left(r\right)}{\eta^{\left(0\right)}\!\left(0\right)} =  \left<\left(\frac{\delta \Psi}{\Psi_0^{\left(s\right)}}\right)^2\right>^{-1/2} \int_{0}^{\infty} \mathrm{d} r \, \eta^{\left(0\right)}\!\left(r\right)\, , \label{screenfluxcorrlength}
\end{equation}
we see by substituting \eqref{screenfluxcorrlength} into \eqref{result6} that the RMS of relative image-flux deviations is related to the magnetic field RMS by
\begin{equation}
 \left<\left(\frac{\delta \Psi}{\Psi_0^{\left(s\right)}}\right)^2\right>^{1/2} = \sqrt{\frac{\pi}{2}} \frac{r_i r_s}{r_s+r_i} \frac{e B_{rms}} {m_p c V} \sqrt{\frac{l_z}{l_{\Psi}}} = \frac{\mu}{\mu_0} \, , \label{linfluxscal}
\end{equation}
in agreement with \eqref{linfluxscalAppend}.

The relation \eqref{linfluxspecAppend} between the magnetic-energy spectrum and two-dimensional spectrum of the relative image-flux in the linear regime can be determined by re-writing \eqref{result1} in the following form:
\begin{equation}
\eta^{\left(0\right)}\!\left(r_\bot\right) = - \frac{e^2 r_s^2 r_i^2 l_z}{m_p^2 c^2 V^2 \left(r_s+r_i\right)^2} \nabla_\bot^2 \left[\frac{1}{2}\int_{-\infty}^{\infty} \mathrm{d}r_z \, M\!\left(r\right)\right] \, . \label{screenfluxcorrfunspecA}
\end{equation}
where $\nabla_\bot= \partial/\partial \mathbf{r}_\bot$. Now Fourier transforming with respect to $r_\bot$, \eqref{screenfluxcorrfunspecA} becomes
\begin{equation}
\hat{\eta}^{\left(0\right)}\!\left(k_\bot\right) = \frac{e^2 r_s^2 r_i^2 l_z}{2 m_p^2 c^2 V^2 \left(r_s+r_i\right)^2} k_\bot^2 \hat{M}\!\left(k_\bot,0\right) \, .
\end{equation}
Then, we note that the Fourier transform of the rescaled relative image-flux autocorrelation function is related to the Fourier transform of the relative image-flux autocorrelation function by
\begin{equation}
\hat{\eta}^{\left(0\right)}\!\left(k_\bot\right) = \hat{\eta}\!\left(k_\bot\right) \left(\frac{r_i}{r_s+r_i}\right)^2 \, . 
\end{equation}
Again invoking isotropic magnetic field statistics and the Wiener-Khinchin theorem, we deduce
\begin{equation}
\hat{\eta}\!\left(k\right) = \frac{2 \pi e^2 r_s^2 r_i^2 l_z}{m_p^2 c^2 V^2 \left(r_s+r_i\right)^2} E_B\!\left(k\right) \, .
\end{equation}
This can be rearranged to give \eqref{linfluxspecAppend} as claimed. The derived result agrees with previous work~\cite{GTLL16}. 

Finally, to demonstrate that $\mu_0 \leq 2/\pi$, we note the following result:
\begin{equation}
 \left<\left(\frac{\delta \Psi}{\Psi_0^{\left(s\right)}}\right)^2\right> = 2 \pi \int_0^{\infty} \mathrm{d}k \, k \hat{\eta}\!\left(k\right) =  \frac{r_s^2 r_i^2}{\left(r_i+r_s\right)^2} \frac{4 \pi^ 2 e^2 l_z}{m_p^2 c^2 V^2} \int_0^{\infty} \mathrm{d}k \, k E_{B}\!\left(k\right) \, . \label{intensityspecrelate}
\end{equation}
Substuting for the mean of squared relative image-flux using \eqref{linfluxscal}, this can be used to find the relative image-flux correlation length $l_\Psi$ in terms of the magnetic-energy spectrum:
\begin{equation}
l_\Psi = \frac{\int_0^{\infty} \mathrm{d}k \, E_B\!\left(k\right)}{\int_0^{\infty} \mathrm{d}k \, k E_B\!\left(k\right)} \, . \label{currentcorrlength}
\end{equation}
Recalling that the magnetic correlation length is given by
\begin{equation}
l_B = \frac{\pi}{2} \frac{\int_0^{\infty} \mathrm{d}k \, E_B\!\left(k\right)/k}{\int_0^{\infty} \mathrm{d}k \, E_B\!\left(k\right)} \, ,
\end{equation}
we conclude that
\begin{equation}
\mu_0 = \sqrt{\frac{2 l_\Psi}{\pi l_B}} = \frac{2}{\pi} \frac{\int_0^{\infty} \mathrm{d}k \, E_B\!\left(k\right)}{\left[\int_0^{\infty} \mathrm{d}k \, k E_B\!\left(k\right)\right]^{1/2} \left[\int_0^{\infty} \mathrm{d}k \, E_B\!\left(k\right)/k\right]^{1/2}} \leq \frac{2}{\pi} \,
\end{equation}
where the final step results from the Cauchy-Schwarz inequality applied to the (positive) integrands in the denominator. 

To investigate the accuracy of \eqref{linfluxscalAppend} as $\mu$ is increased, we consider a particular stochastic field configuration and then vary the effective $\mu$ of a proton-imaging set-up applied to it. In particular, we again consider the `cocoon' magnetic field introduced in Appendix \ref{ConstrastScale}, which has magnetic-energy spectrum \eqref{magcocoonspecAppendscale}, that is
\begin{equation}
E_B\!\left(k\right) = \frac{\left<\mathbf{B}^2\right> l_e}{12\sqrt{2} \pi^{3/2}} l_e^4 k^4 \exp{\left(-l_e^2 k^2/2\right)} \, ,
\end{equation}
for $l_e$ the typical size of a cocoon (see Appendix \ref{ToySpecLinThyCocoon}). It can be shown for this form of magnetic-energy spectrum that normalisation constant $\mu_0$ is given by $\mu_0 = 3/2\sqrt{2\pi} \approx 0.6$ -- derived in Appendix \ref{ToySpecLinThy}. Figure \ref{cocoonfluxRMSscalingwithmag} plots this prediction for a range of normalised $\mu$, along with the measured RMS for a particular numerical numerical instantiation of a Gaussian cocoon field. 
\begin{figure}[htbp]
\centering
\includegraphics[width=0.5\linewidth]{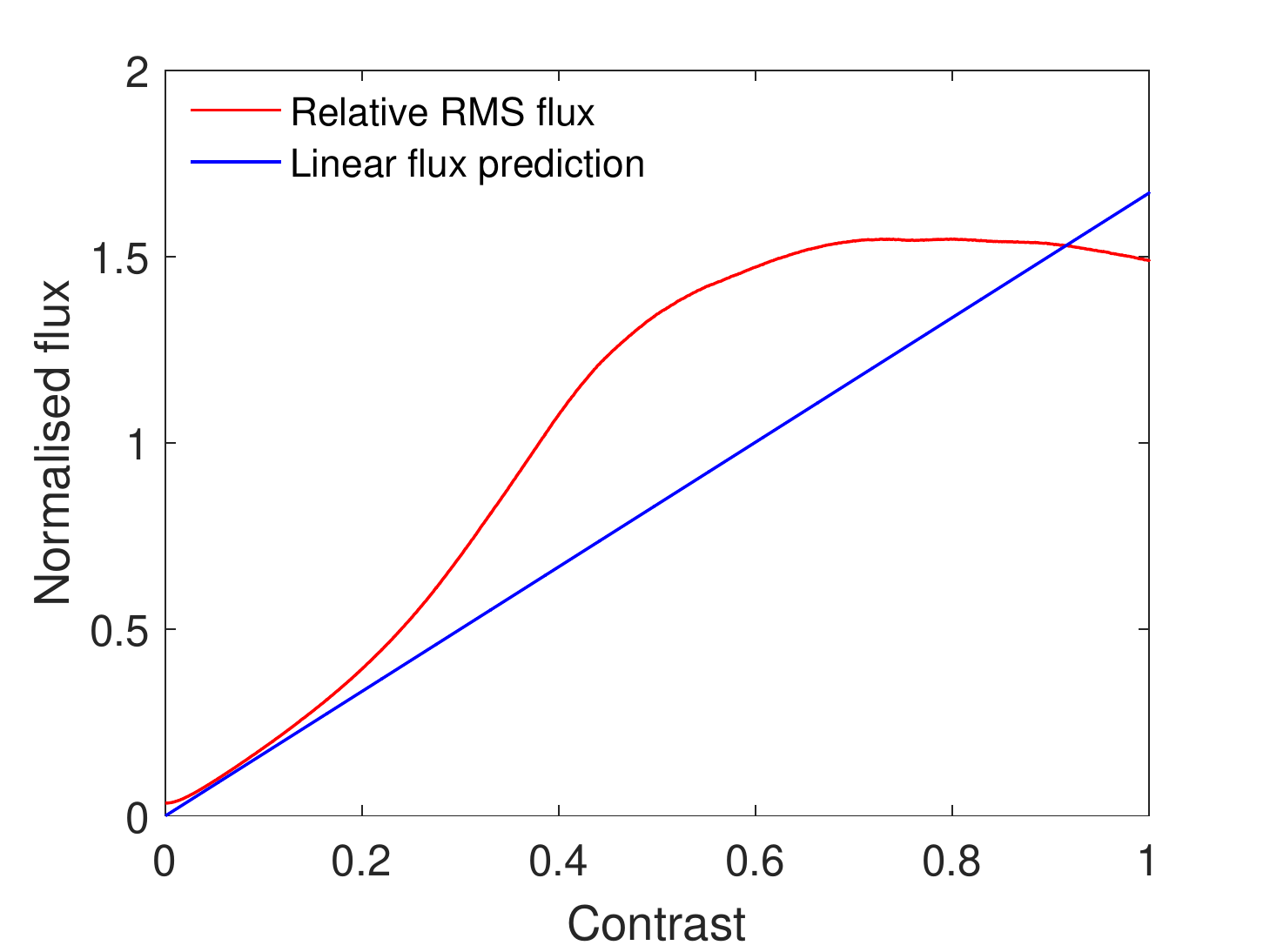}
\caption{\textit{Validity of linear-regime prediction of RMS relative image-flux variations in terms of $\mu$ for general values of $\mu$} For a particular instantiation of a cocoon configuration with correlation length $l_B = 30 \, \mu\mathrm{m}$ generated similarly to those fields described in \eqref{largeandsmallscalecocoon}, the RMS of magnetic field strength $B_{rms}$ is varied; for each different value of $B_{rms}$, the same proton imaging procedure outlined in \eqref{largeandsmallscalecocoon} is carried out, and the associated RMS of relative image-flux deviations is calculated. This is compared to linear-regime prediction \eqref{linfluxscalAppend}, with $\mu_0 = 3/2\sqrt{2 \pi} \approx 0.6$.} \label{cocoonfluxRMSscalingwithmag}
\end{figure}
We observe that \eqref{linfluxscalAppend} is very accurate for small $\mu$; however, this no longer holds as $\mu$ increases. For $0.25 > \mu > 0.9$, the RMS of relative image-flux variations is under-predicted, with 
\begin{equation}
 \left<\left(\frac{\delta \Psi}{\Psi_0^{\left(s\right)}}\right)^2\right> \sim 1 \implies \mu \approx 0.35 \approx 0.6 \mu_0 \, .
\end{equation}
For $\mu > 0.7$, the RMS of relative image-flux deviations begins to decrease with $\mu$, with the consequence that for large values of $\mu$, it becomes much larger than the RMS of relative image-flux deviations. In short, the condition of order-unity image-flux variations does not imply that $\mu$ as defined by \eqref{contrastdef} in the main text is of the same order. 

\section{Calculation of spectra/correlation scales for model power law and cocoon fields} \label{ToySpecLinThy}

In this paper, we consider three particular examples of magnetic energy spectra: the power law spectra \eqref{powerlawspecdefSec2} and \eqref{kminusonespec} specified in the main text,  and the `cocoon' spectrum (first given in \eqref{magcocoonspecAppendscale}). For each of these fields, this appendix provides a more detailed summary of the properties of magnetic fields with this spectra and associated image-flux distributions: more specifically, the magnetic correlation length $l_B$ given by \eqref{corrlengthspec}, the RMS of the perpendicular-deflection field $w_{rms}$, the relative image-flux correlation length $l_{\Psi}$ given by \eqref{screenfluxcorrlength}, the RMS of relative image-flux deviations in the linear regime, and $\mu$ normalisation constant $\mu_0$. 

\subsection{Power law spectrum \eqref{powerlawspecdefSec2}} \label{ToySpecLinThyPowerLaw}

For this paper we consider the simplest possible power law of the form (for $p \neq 1$)
\begin{equation}
E_B\!\left(k\right) = \frac{B_{rms}^2}{8 \pi} \left(p-1\right) \frac{k^{-p}}{k_l^{-p+1}-k_u^{-p+1}} \, ,
\end{equation}
where $k_l = 2 \pi/l_c$ and $k_u$ are lower and upper spectral wavenumber cutoffs. Calculating the correlation length $l_B$ with \eqref{corrlengthspec} gives
\begin{equation}
l_B = \frac{l_c}{4} \frac{p-1}{p} \frac{1-\left(k_l/k_u\right)^{p}}{1-\left(k_l/k_u\right)^{p-1}} \approx \frac{p-1}{p} \frac{l_c}{4} \, ,
\end{equation}
where the approximate expression is valid is $\left(k_l/k_u\right)^{p-1} \ll 1$. Note that the correlation length is approximately four times smaller than the wavelength associated with the lower wavenumber cutoff, even if the spectrum is very steep ($p \gg 1$). The RMS of the perpendicular deflections is then given by
\begin{equation}
w_{rms}^2 = \frac{l_c}{4} \frac{p-1}{p} \frac{1-\left(k_l/k_u\right)^{p}}{1-\left(k_l/k_u\right)^{p-1}} \frac{e^2 \left<\mathbf{B}^2\right> l_z}{m_p^2 c^2} \, .
\end{equation}
The relative image-flux correlation length -- calculated using \eqref{currentcorrlength} -- is related to the upper and lower spectral wavenumber cutoffs by
\begin{equation}
l_\Psi = \frac{2}{\pi} \frac{p\left(p-2\right)}{\left(p-1\right)^2} \frac{\left(1-\left(k_l/k_u\right)^{p-1}\right)^2}{\left(1-\left(k_l/k_u\right)^{p}\right)\left(1-\left(k_l/k_u\right)^{p-2}\right)} l_B \, .
\end{equation}
Substituting this result into RMS of relative image-flux deviations \eqref{linfluxscalsect3} gives 
\begin{equation}
 \left<\left(\frac{\delta \Psi}{\Psi_0^{\left(s\right)}}\right)^2\right> = \frac{\pi}{2} \frac{r_s^2 r_i^2}{\left(r_i+r_s\right)^2} \frac{e^2 \left<\mathbf{B}^2\right>}{m_p^2 c^2 V^2} \frac{p-1}{p-2}  \frac{k_l^{-p+2}-k_u^{-p+2}}{k_l^{-p+1}-k_u^{-p+1}} = \frac{\pi^2}{4} \frac{\left(p-1\right)^2}{p\left(p-2\right)} \frac{\left(1-\left(k_l/k_u\right)^{p}\right)\left(1-\left(k_l/k_u\right)^{p-2}\right)}{\left(1-\left(k_l/k_u\right)^{p-1}\right)^2} \mu^2 \, .
\end{equation}
In the limit of $\left(k_l/k_u\right)^{p-2} \ll 1$ (a condition which requires $p > 2$), this gives $\mu$ normalisation constant
\begin{equation}
\mu_0 \approx \frac{2}{\pi} \frac{\sqrt{p\left(p-2\right)}}{p-1} \, .
\end{equation}
For $1 < p < 2$, the $\mu$ normalisation constant becomes a function of the wavenumber range, decreasing as $k_u/k_l$ increases -- specifically
\begin{equation}
\mu_0 \approx \frac{2}{\pi} \frac{\sqrt{p\left(2-p\right)}}{p-1}  \left(\frac{k_l}{k_u}\right)^{2-p} \, .
\end{equation}
This is unsurprising, since heuristically we would expect that for spectral indices in this range the dominant wavemodes in terms of image-flux in the linear regime are at the smallest scales, on account of the image-flux distribution being closely related to projections of MHD current structure in the linear regime. 

\subsection{Power law \eqref{kminusonespec}} \label{ToySpecLinThyPowerLawB}

The special case of $p = 1$ gives power law
\begin{equation}
E_B\!\left(k\right) = \frac{\left<\mathbf{B}^2\right>}{8 \pi}  \frac{1}{k \log{k_u/k_l}} \, .
\end{equation}
The correlation length is
\begin{equation}
l_B = \frac{l_c}{4} \left(1-\frac{k_l}{k_u}\right)\frac{1}{\log{k_u/k_l}} \, ,
\end{equation}
and so
\begin{equation}
w_{rms}^2 =\frac{l_c}{4} \left(1-\frac{k_l}{k_u}\right) \frac{1}{\log{k_u/k_l}} \frac{e^2 B_{rms}^2 l_z}{m_p^2 c^2} \, .
\end{equation}
Unlike the case $p >1$, the correlation length decreases (albeit logarithmically) with the wavenumber range of the power law. The relative image-flux correlation length is
\begin{equation}
l_\Psi = \frac{l_c}{2 \pi} \frac{1}{k_u/k_l-1} \log{k_u/k_l} \, .
\end{equation}
This is inversely proportional to the wavenumber range if $k_l \ll k_u$, so is much smaller than the correlation length. The RMS of relative image-flux deviations is
\begin{equation}
 \left<\left(\frac{\delta \Psi}{\Psi_0^{\left(s\right)}}\right)^2\right> = \frac{\pi^2}{4} \frac{r_s^2 r_i^2}{\left(r_i+r_s\right)^2} \frac{e^2 B_{rms}^2}{m_p^2 c^2 V^2}  \frac{l_z}{l_B} \frac{k_u}{k_l} \frac{1}{\left[\log{\frac{k_u}{k_l}}\right]^2} \, ,
\end{equation}
which gives $\mu$ normalisation constant
\begin{equation}
\mu_0 = \frac{2}{\pi} \sqrt{\frac{k_l}{k_u}} \log{\frac{k_u}{k_l}} =  \,.
\end{equation}
Thus for a spectrum as shallow as a $k^{-1}$ power law, order-unity variations in image flux will occur at $\mu \sim \mu_0 ~ \sqrt{k_l/k_u}$. This can be interpreted as being due to the smallest structures having the highest $\mu$ for such shallow spectra -- discussed in Section \ref{TheoCom}.

\subsection{Cocoon field} \label{ToySpecLinThyCocoon}

The magnetic cocoon field is formed from spherical blobs of size $l_e$ defined precisely by
\begin{equation}
\mathbf{B} = B_0 \frac{r}{l_e} \exp{\left(-\frac{r^2}{l_e^2}\right)} \mathbf{e}_\phi \, .
\end{equation}
for azimuthal basis vector $\mathbf{e}_\phi$. As demonstrated elsewhere~\cite{K12}, when imaged in the $\mathbf{e}_z$ direction such a field can lead to either a defocusing or focusing ring of image flux, depending on its orientation; however, when imaged from the side, symmetry implies that no overall deflections can be seen. It can be shown that the spectrum of a field of randomly orientated and positioned magnetic cocoons with field energy $B_{rms}^2/8 \pi$ has magnetic-energy spectrum~\cite{D04}
\begin{equation}
E_B\!\left(k\right) = \frac{B_{rms} l_e}{12\sqrt{2} \pi^{3/2}} l_e^4 k^4 \exp{\left(-l_e^2 k^2/2\right)}
\end{equation}
The magnetic correlation length is 
\begin{equation}
l_B = \frac{\sqrt{2 \pi}}{3} l_e \, ,
\end{equation}
and the RMS deflection-field strength
\begin{equation}
w_{rms}^2 = \frac{\sqrt{2 \pi}}{3}  \frac{e^2 B_{rms}^2 l_z l_e}{m_p^2 c^2} \, .
\end{equation}
The relative image-flux correlation length becomes
\begin{equation}
l_\Psi = \frac{9}{16} l_B. 
\end{equation}
Now calculating the RMS of relative image-flux deviations for the linear regime using \eqref{linfluxscalsect3}, we find
\begin{equation}
 \left<\left(\frac{\delta \Psi}{\Psi_0^{\left(s\right)}}\right)^2\right> = \frac{8 \pi}{9} \frac{r_s^2 r_i^2}{\left(r_i+r_s\right)^2} \frac{e^2 B_{rms}^2}{m_p^2 c^2 V^2} \frac{l_z}{l_B} = \frac{8 \pi}{9} \mu^{2}
\end{equation}
This implies that for a cocoon field, the appropriate normalisation $\mu_0$ for the spectra is given by $\mu_0 = 2 \sqrt{2\pi}/3 \approx 1.65$. 

\section{Connection of Monge-Amp\`ere equation to Monge-Kantorovich problem} \label{MongeKantorovich}

In this appendix, we show that inverting the Monge-Amp\`ere equation to solve for the path-integrated field is equivalent to the $L_2$ \emph{Monge-Kantorovich problem}. Due to the coordinate perturbation itself being the argument of the image-flux distribution, it is clear that in general treating \eqref{screenfluxpotSection4} analytically is challenging. However, relating the Monge-Amp\`ere equation to the $L_2$ Monge-Kantorovich problem enables an explicit reference to be given to a proof of the existence of a solution.

We state again the image-flux relation \eqref{screenfluxnonlin121} in the case when the plasma-image mapping is injective:
\begin{equation}
\Psi\!\left(\nabla_{\bot0} \phi\!\left(\mathbf{x}_{\bot0}\right) \right) = \frac{\Psi_{0}\!\left(\mathbf{x}_{\bot0}\right)}{\det{\nabla_{\bot0}  \nabla_{\bot0}  \phi\!\left(\mathbf{x}_{\bot0}\right) }} \, . \label{screenfluxpotSection4}
\end{equation}
For convenience, we renormalise the image-coordinates by
\begin{equation}
\mathbf{x}_{\bot0}^{\left(s\right)} = \frac{r_i}{r_i+r_s} \mathbf{x}_{\bot}^{\left(s\right)} \, ,
\end{equation}
 to give modified plasma-image mapping (written in terms of the deflection-field potential)
\begin{equation}
\tilde{\phi} = \frac{1}{2}\mathbf{x}_{\bot0}^2 + \frac{r_i r_s}{r_s+r_i} \frac{\varphi\!\left(\mathbf{x}_{\bot0}\right)}{V} \, .
\end{equation}
The renormalised image-flux relation is then
\begin{equation}
\Psi\!\left(\nabla_{\bot0} \tilde{\phi}\!\left(\mathbf{x}_{\bot0}\right) \right) = \frac{\Psi_{0}\!\left(\mathbf{x}_{\bot0}\right)}{\det{\nabla_{\bot0} \nabla_{\bot0} \tilde{\phi}\!\left(\mathbf{x}_{\bot0}\right) }} \, .\label{screenfluxpotrenorm}
\end{equation}
A necessary (but not sufficient) condition for the problem of determining $\nabla_{\bot0} \tilde{\phi}$ from the image-flux distribution to be well posed is a suitable boundary condition. In the main text, we state this boundary condition in Neumann form, that is
\begin{equation}
\nabla_{\bot0} \tilde{\phi} \!\left(\Omega_I\right) =  \Omega_S \, . \label{fluxBCNeumannA}
\end{equation}
For a finite region, it can be shown that \eqref{fluxBCNeumannA} is equivalent to a global flux conservation condition of the form
\begin{equation}
\int_{\Omega_I} \Psi_0\!\left(\mathbf{x}_{\bot0}\right) \, \mathrm{d}^2 \mathbf{x}_{\bot0} = \int_{\Omega_S} \Psi\!\left(\mathbf{x}_{\bot0}^{\left(s\right)}\right) \, \mathrm{d}^2 \mathbf{x}_{\bot0}^{\left(s\right)} \, . \label{fluxseconboundcondAppend}
\end{equation}
where the mapping sends image region $\Omega_I$ to $\Omega_S$: that is $\nabla_{\bot0} \phi\!\left(\Omega_I\right) = \Omega_S$. This is known as a \emph{second boundary condition}.

Tthe $L_2$ Monge-Kantorovich problem in the context of proton flux mapping as follows: given two positive flux distributions functions $\Psi_{0}\!\left(\mathbf{x}_{\bot0}\right)$ and $\Psi\!\left(\mathbf{x}_{\bot0}^{\left(s\right)}\right)$ defined on regions $\Omega_I$ and $\Omega_S$ with equal total mass, find the mapping $\mathbf{x}_{\bot0}^{\left(s\right)} =\mathbf{x}_{\bot0}^{\left(s\right)}\!\left(\mathbf{x}_{\bot0}\right)$ which minimises the cost functional
\begin{equation}
\mathcal{C}\!\left(\mathbf{x}_{\bot0}^{\left(s\right)}\right) = \int \mathrm{d}^2 \mathbf{x}_{\bot0} \, \left|\mathbf{x}_{\bot0}^{\left(s\right)}\!\left(\mathbf{x}_{\bot0}\right)-\mathbf{x}_{\bot0}\right|^2 \, \Psi_{0} \!\left(\mathbf{x}_{\bot0}\right) \, , \label{MongeKcostfun}
\end{equation}
where global flux conservation condition \eqref{fluxseconboundcondAppend} holds~\cite{GM96,V08}. 

The equivalence with the Monge-Amp\`ere equation follows by introducting $\mathbf{x}_{\bot0}^{\left(s\right)} =\mathbf{x}_{\bot0}^{\left(s\right)}\!\left(\mathbf{x}_{\bot0}\right)$ explicitly into the right hand side of 
\eqref{fluxseconboundcondAppend}, to give local condition
\begin{equation}
\det{\frac{\partial \mathbf{x}_{\bot0}^{\left(s\right)}}{\partial \mathbf{x}_{\bot0}}} = \frac{\Psi_{0}\!\left(\mathbf{x}_{\bot0}\right)}{\Psi\!\left(\mathbf{x}_{\bot0}^{\left(s\right)}\!\left(\mathbf{x}_{\bot0}\right) \right)} \, . 
\end{equation}      
A theorem (due to Brenier~\cite{B91}) shows that the solution to the Monge-Kantorovich problem is unique, and characterised as the gradient of a convex potential mapping~\cite{S11}. Since $\nabla_{\bot0} \tilde{\phi}$ in the case of proton mapping is convex if and only if the plasma-image mapping is injective, this solution is precisely the Monge-Kantorovich potential. 

Thus, with boundary condition \eqref{fluxBCNeumannA} or \eqref{fluxseconboundcondAppend}, the inversion problem associated with \label{screenfluxpotrenorm} is well-posed. As mentioned in the main text, a variety of numerical schemes have been suggested to implement this~\cite{DG06, S11}. 

\section{Existence of caustic regime for stochastic magnetic fields} \label{CauExist}

In this appendix we provide a semi-qualitative argument that for any stochastic magnetic field, increasing $\mu$ of the imaging set-up will eventually result in caustic structures. Furthermore, we show that the critical value of $\mu$ at which this occurs is order unity i.e. $\mu_c \sim 1$. 

The existence of $\mu_{c}$ for any given magnetic field configuration is not guaranteed: indeed, any field whose associated perpendicular-deflection field has non-negative divergence everywhere will never generate caustics. However, typical stochastic fields do not reflect such a specialised case; indeed, since the divergence of the perpendicular-deflection field satisfies
\begin{equation}
\nabla_{\bot0} \cdot \mathbf{w}\!\left(\mathbf{x}_{\bot0}\right) = \frac{e}{m_p c} \hat{\mathbf{z}} \cdot \int_0^{l_z} \nabla \times \delta \mathbf{B}\!\left(\mathbf{x}_{\bot0}\left(1+\frac{z'}{r_i}\right),z'\right) \mathrm{d}z' \, , \label{deflfielddiv}
\end{equation}
for stochastic field configurations with zero mean $\nabla_{\bot0} \cdot \mathbf{w}$ will have an approximately Gaussian distribution provided $l_B \ll l_z$. The variance of this distribution is given by analogy to the result \eqref{fluxspecdef} for the relative image-flux deviation stated in Section \ref{LinRgme} :
\begin{equation}
\left<\left(\nabla_{\bot0} \cdot \mathbf{w}\right)^2\right> \approx \frac{\pi}{2} \frac{l_z e^2 B_{rms}^2}{l_\Psi m_p^2 c^2} \, .
\end{equation}
This implies the existence of negative divergences. Since for sufficent field strengths deflections will be dominated by magnetic deflections, particle trajectories will eventually be such that they cross, leading inevitably to caustics.

This argument can be further adapted to give a heuristic estimate for $\mu_{c}$. Neglecting quadratic terms in $\mu$, relation \eqref{detdivrelate} from Appendix \ref{LinThyDev} can be used to approximate the distribution of the plasma-image mapping determinant as Gaussian, with unit mean, and variance $\sigma^2$ given by
\begin{equation}
\sigma^2 \approx \frac{r_s^2 r_i^2}{V^2\left(r_i+r_s\right)^2}\left<\left(\nabla_{\bot0} \cdot \mathbf{w}\right)^2\right> \approx \frac{\pi}{2}  \frac{r_s^2 r_i^2}{\left(r_i+r_s\right)^2} \frac{l_z e^2 B_{rms}^2}{l_\Psi m_p^2 c^2 V^2} \sim \mu^2 \, .
\end{equation}
The validity of such an assumption can be tested numerically, and is usually well met in practice.

Now, a reasonable criterion for the appearance of caustics is that somewhere in a proton-flux image, the determinant mapping has vanished. Since we can approximate the flux image as consisting of $l_\bot^2/l_B^2$ `samples' of the field strength distribution, this requires that
\begin{equation}
\Pr{\left(\det{\frac{\partial \mathbf{x}_{\bot}^{\left(s\right)}}{\partial \mathbf{x}_{\bot0}}} \leq 0\right)} \sim \frac{l_B^2}{l_\bot^2} \, ,
\end{equation}
which then gives
\begin{equation}
\frac{1}{2}\left[1+\mathrm{erf}\!\left(-\frac{1}{ \sqrt{2} \mu_{c}}\right)\right] \sim \frac{l_B^2}{l_\bot^2}  \, .
\end{equation}
Since $l_B \ll l_\bot$, both sides are small, which implies that we can further approximate the error function by
\begin{equation}
\frac{1}{2}\left[1+\mathrm{erf}\!\left(-\frac{1}{ \sqrt{2} \mu_{c}}\right)\right] \approx \frac{\mu e^{-1/2\mu^2}}{\sqrt{2\pi}} \, .
\end{equation}
This then leads to
\begin{equation}
\mu_{c} \sim \frac{1}{\sqrt{\log{l_B/l_\bot}}} \, .
\end{equation}
This dependence on the correlation scale is extremely weak. Furthermore, the argument only applies in characterising the appearance of one caustic anywhere in the image, a local effect which would have a minimal impact on statistical estimates, particularly if $l_B \ll l_\bot$ (see Section 4.8.4). Global appearance of caustics would require that the probability of the mapping determinant falling below zero were moderate. Thus for most relevant situations, the condition $\mu_{c} \sim 1$ is reasonable.

\section{Ill-posedness of reconstruction for proton-flux images with caustic features} \label{IllPosNonLinRecon}

In this appendix we describe a method for constructing a simple family of magnetic fields based on a magnetic flux-rope type structure, all of which produce the same image-flux distribution. This provides an explicit example of why reconstructing the path-integrated magnetic field from a image-flux distribution in the caustic regime $\mu \geq \mu_{c}$ is not a well-posed problem. 

\subsection{Description of base field - `flux rope'}

Consider a magnetic field of the form 
\begin{equation}
\mathbf{B}\!\left(\mathbf{x}\right) = B\!\left(x,z\right) \mathbf{e}_y \, ,
\end{equation}
where
\begin{equation}
B\!\left(x,z\right) = \frac{B_0}{\pi} \, \mathrm{sech} \, \frac{x}{\delta} \, \mathrm{sech} \, \frac{z}{\delta} \, , \label{magfluxrope}
\end{equation}
for $\left|z\right| < l_i/2$ (with $\delta < l_i/2$), and a vanishing field elsewhere (here, $l_i = l_z = l_\bot$). Such a field offers an approximate model for a flux-rope structure; and although this field is not cylindrically symmetric around the axis of peak strength, for $r = \sqrt{x^2+z^2} \ll \delta$ we can write
\begin{equation}
B\!\left(x,z\right) = \frac{B_0}{\pi} \, \left(1-\frac{r^2}{\delta^2}\right) + \mathcal{O}\!\left(\frac{x^4}{\delta^4},\frac{z^4}{\delta^4}\right)
\end{equation}
giving symmetry near the origin. We use this model instead of a more conventional Gaussian flux rope~\cite{K12}, because the former has some particular analytic properties which will prove helpful for subsequent calculations. 
A plot of this field in normalised coordinates $x \mapsto x \delta$ with $\delta = l_i/10$ is shown in Figure \ref{fluxropecausticexample}a, along with a perturbation to that field constructed in a procedure described in Appendix \ref{ImFluxMagPerb} -- we show later that the perturbed field leads to exactly the same image-flux distribution. 
\begin{figure}[htbp]
\centering
    \begin{subfigure}{.48\textwidth}
        \centering
        \includegraphics[width=\linewidth]{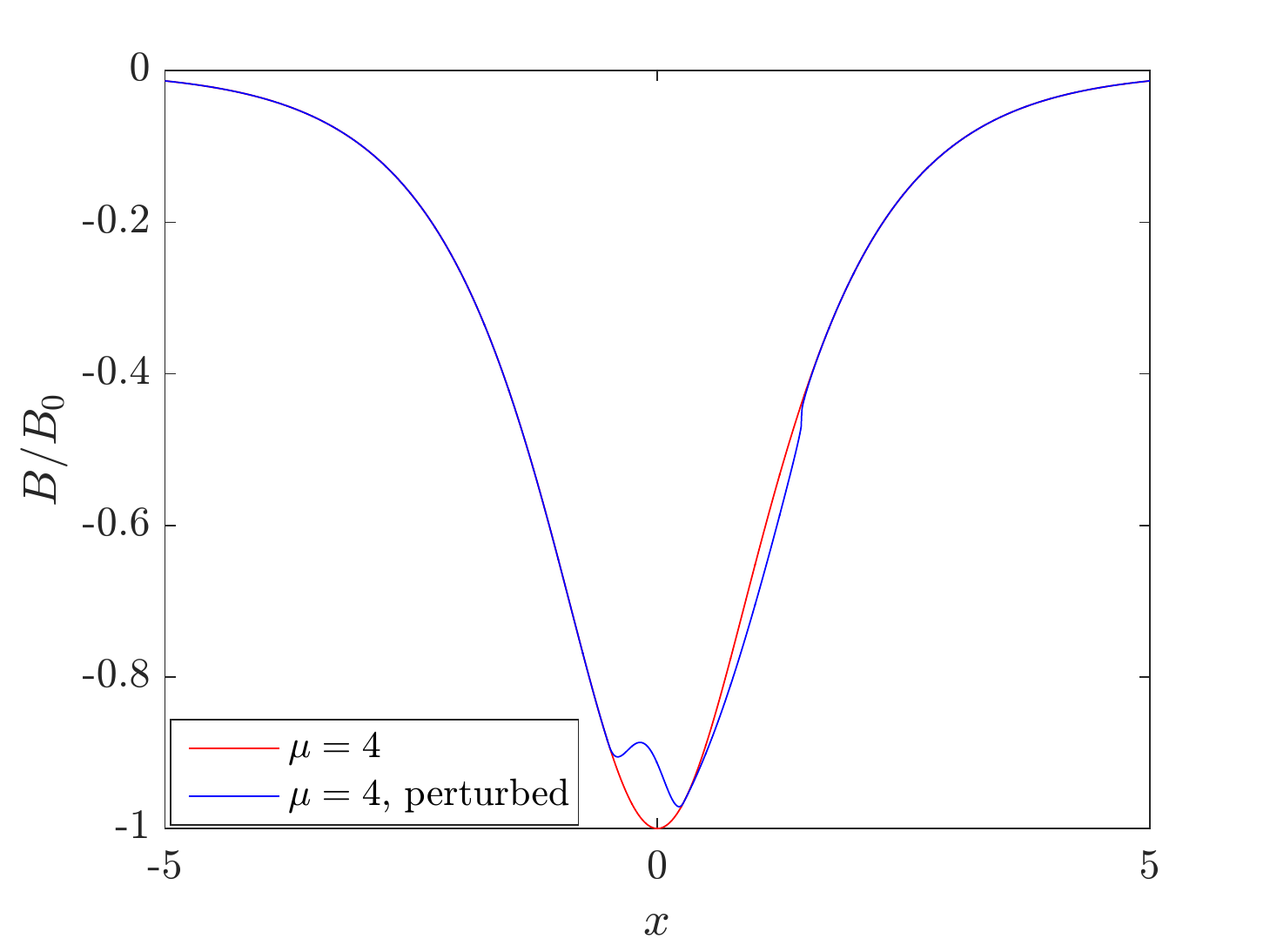}
    \end{subfigure} %
    \begin{subfigure}{.48\textwidth}
        \centering
        \includegraphics[width=\linewidth]{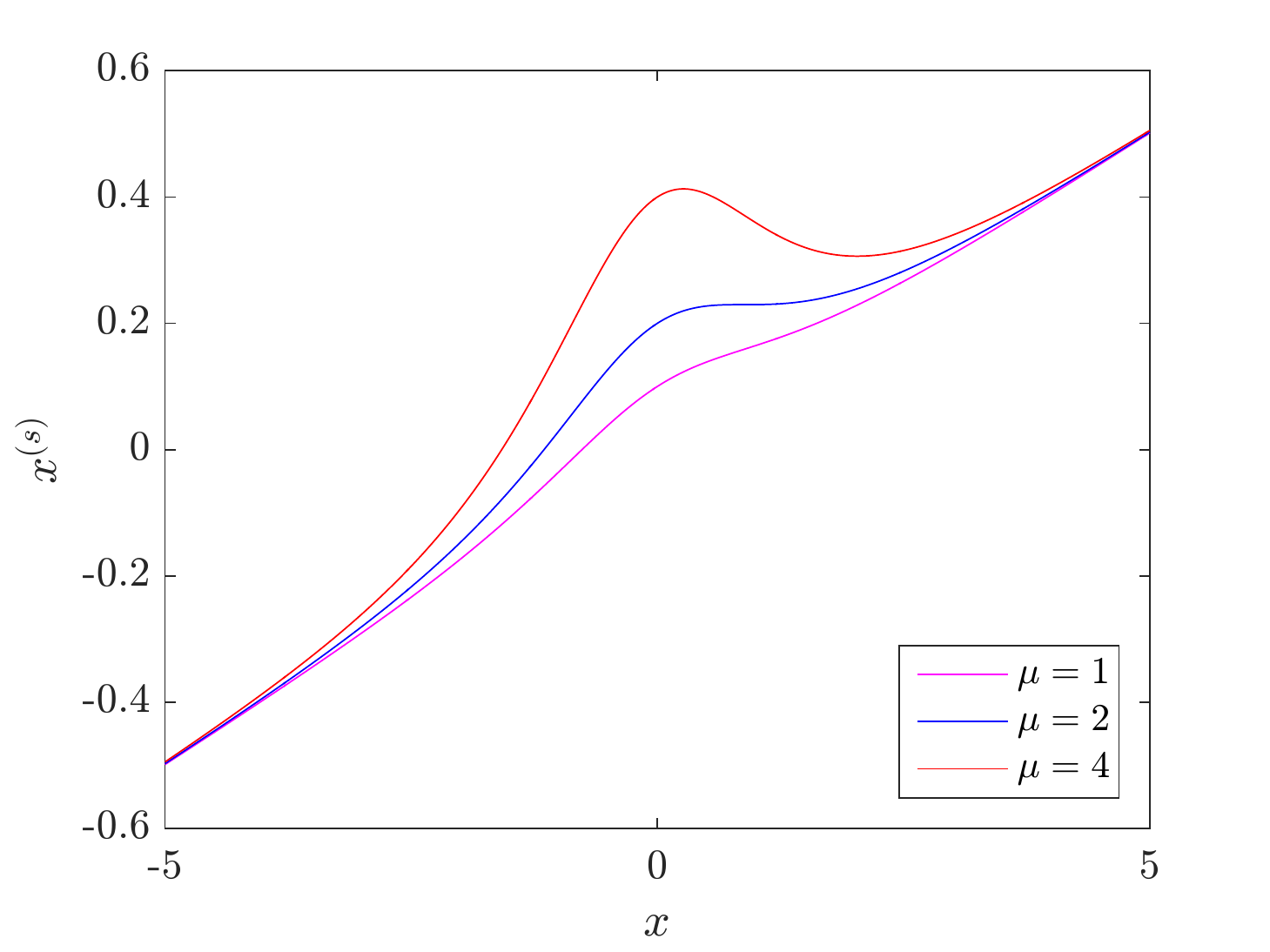}
    \end{subfigure} %
    \begin{subfigure}{.48\textwidth}
        \centering
        \includegraphics[width=\linewidth]{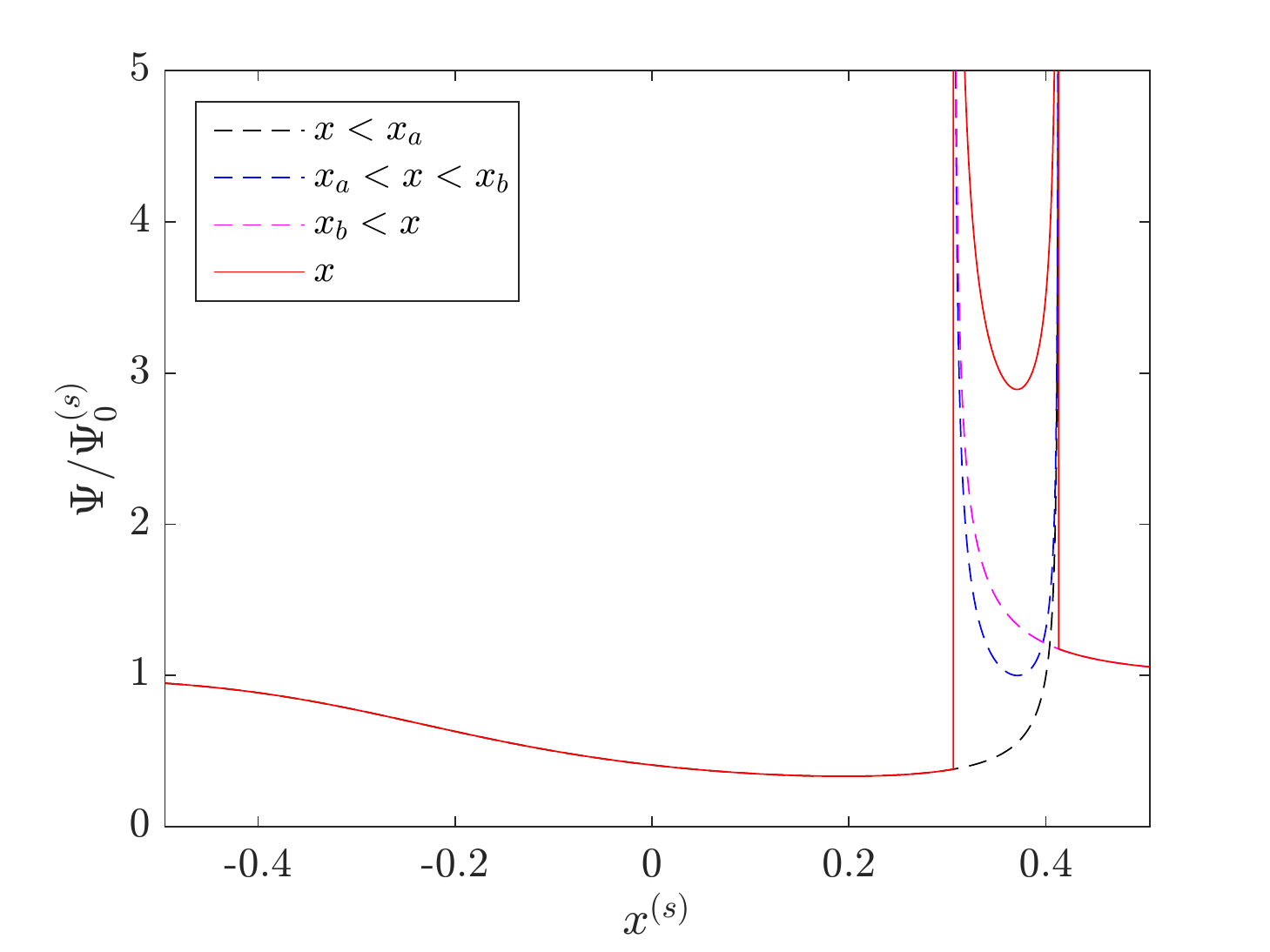}
    \end{subfigure} %
    \begin{subfigure}{.48\textwidth}
        \centering
        \includegraphics[width=\linewidth]{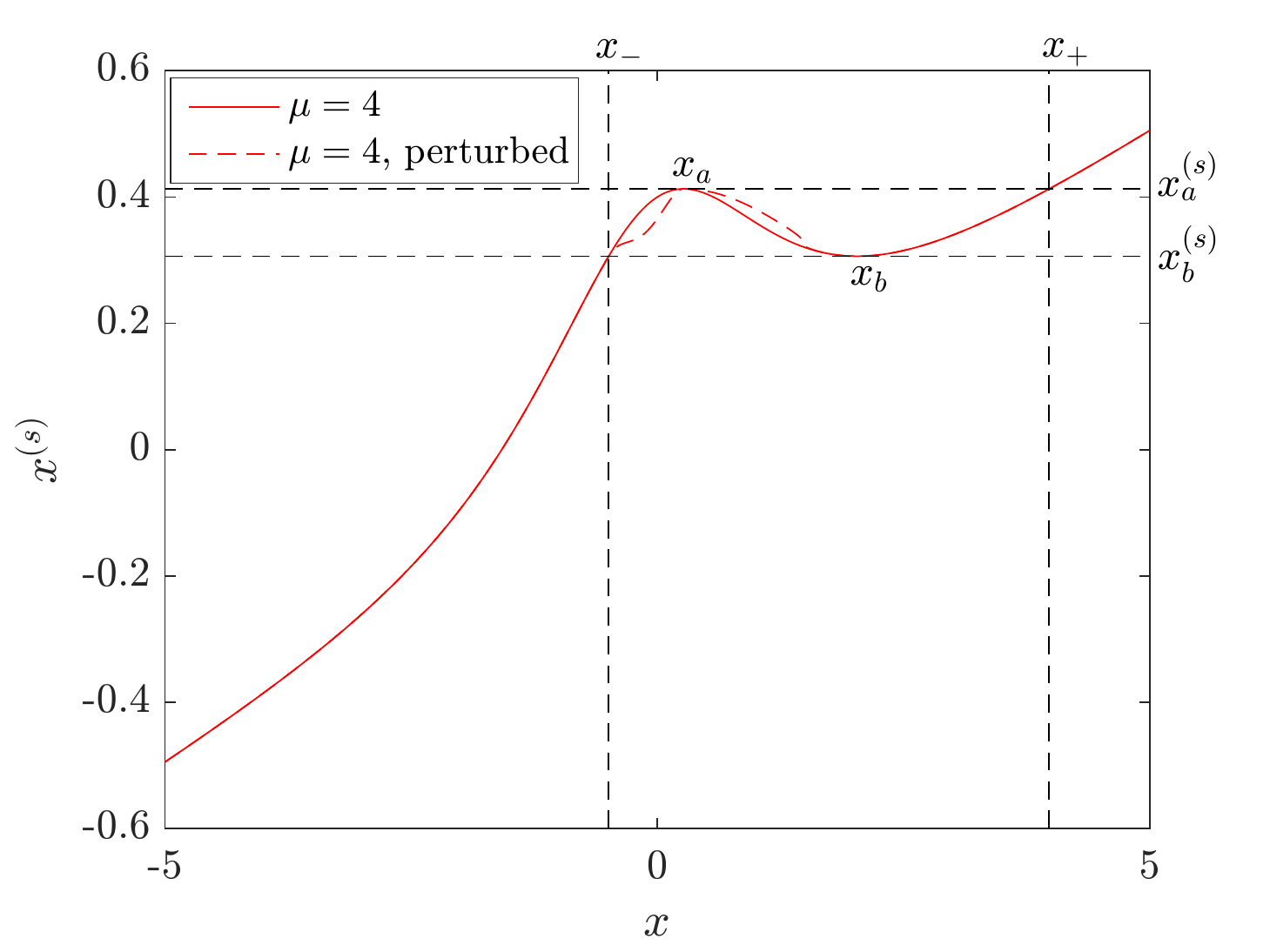}
    \end{subfigure} %
\caption{\textit{Constructing two magnetic flux-rope structures with the same image-flux distribution}. \textbf{a)} Normalised magnetic flux-rope field \eqref{magfluxrope}, plotted with perturbed magnetic flux-rope constructed via a method described in Appendix \ref{ImFluxMagPerb}; the specific perturbation is described by \eqref{mappingex}. \textbf{b)} $x$ coordinate of plasma-image mapping \eqref{fluxropeplasmap} for increasing $\mu$ [defined for the flux rope by \eqref{fluxropemu}]. \textbf{c)} One-dimensional normalised image-flux distribution \eqref{fluxropeimageflux}, for $\mu = 4$. The two flux structures with large positive flux are caustics, and the region of proton flux between them has local contributions from spatially disconnected regions of the initial proton beam. The contributions of these different regions are indicated in the legend. \textbf{d)} $x$ coordinate of plasma-image mapping for $\mu = 4$, along with mapping associated with perturbed magnetic flux-rope.} \label{fluxropecausticexample}
\end{figure}
This field is solenoidal by construction (as is the perturbed field). The base field induces deflection velocities
\begin{equation}
\mathbf{w} \approx \frac{e B_0 \delta}{m_p c}  \, \mathrm{sech} \, \frac{x}{\delta} \, \mathbf{e}_x \, ,
\end{equation}
assuming $l_i \gg \delta$ and then approximating finite limits of the $z$-integral with infinite ones. This gives mapping $\mathbf{x}_\bot^{\left(s\right)} = T\!\left(\mathbf{x}_\bot\right)$, where
\begin{equation}
T\!\left(\mathbf{x}_\bot\right) = \frac{r_i+r_s}{r_i} \mathbf{x}_\bot + r_s \, \frac{e B_0 \delta}{m_p c V}  \, \mathrm{sech} \, \frac{x}{\delta} \, \mathbf{e}_x + \mathcal{O}\!\left(\frac{l_i}{r_s}\right)
\end{equation}
between perpendicular plasma coordinates $\mathbf{x}_\bot$ and image coordinates $\mathbf{x}_\bot^{\left(s\right)}$. The $y$ coordinate mapping is linear, so for simplicity re-write the map as one-dimensional; further, introduce non-dimensionalised coordinates such that
\begin{IEEEeqnarray}{rCl}
\mathbf{x}_\bot & \mapsto & \mathbf{x}_\bot  \delta\, ,\\
\mathbf{x}_\bot^{\left(s\right)} & \mapsto &  r_s \mathbf{x}_\bot^{\left(s\right)} \, ,
\end{IEEEeqnarray}
and consider the limit $r_s \gg r_i$. Then, the mapping becomes
\begin{equation}
x^{\left(s\right)} = T\!\left(x\right) \equiv x \, \delta \alpha + \delta \theta  \; \mathrm{sech} \, x  + \mathcal{O}\!\left(\frac{l_i}{r_s}\right) \, , \label{fluxropeplasmap}
\end{equation}
where dimensionless parameters $\delta \theta$ (typical magnetic perpendicular velocity deflection) and $\delta \alpha$ (paraxial parameter) are defined by
\begin{IEEEeqnarray}{rCl}
\delta \theta & = & \frac{e B_0 \delta}{m_p c V} \, , \\
\delta \alpha & = & \frac{\delta}{r_i} \, .
\end{IEEEeqnarray}
This mapping becomes multi-valued when $T\!\left(x\right)$ is no longer monotonically increasing on its domain; since
\begin{equation}
\frac{\mathrm{d} T}{\mathrm{d} x} = \delta \alpha - \delta \theta  \; \mathrm{sech} \, x  \; \mathrm{tanh} \, x \, ,
\end{equation}
this occurs when
\begin{equation}
\mu = \frac{\delta \theta}{\delta \alpha} > 2 \, . \label{fluxropemu}
\end{equation}
Figure \ref{fluxropecausticexample} show plots of the base mapping for various $\mu$ spanning this barrier; as expected, we see that for $\mu = 4$, an interval (bounded by caustics - see later) has appeared in which multiple plasma positions map to the same position on the image. We refer to this region as a `multivalued' flux region. 

For a field varying in one direction only, the image-flux distribution (non-dimensionalised coordinates) is related to the initial flux via
\begin{equation}
\Psi\!\left(x^{\left(s\right)}\right) = \Psi_0 \frac{\delta^2}{r_s^2 \, \delta \alpha} \sum_{x: x^{\left(s\right)} = T\!\left(x\right)} \frac{1}{\left|\frac{\mathrm{d}T}{\mathrm{d}x}\!\left(x\right)\right|} = \Psi_0 \frac{r_i^2}{r_s^2}\sum_{x: x^{\left(s\right)} = T\!\left(x\right)} \frac{1}{\left| 1 - \mu  \; \mathrm{sech} \, x  \; \mathrm{tanh} \, x \right|} \, . \label{fluxropeimageflux}
\end{equation}
A plot of this flux normalised by the undeflected mean image flux (in one dimension) is shown in Figure \ref{fluxropecausticexample}c. As expected, we see two lines of extremely high intensity - the caustics. Note the contribution from the three distinct regions in the plasma to the region between the caustics. It is this ambiguity that allows a family of fields to be constructed which have the same total flux. 

The simplicity of this mapping enables an exact expression for the caustic lines to be derived in terms of $\mu$. Define $x_a$ and $x_b$ as the plasma coordinates corresponding to the stationary points of the mapping, with image points $x_a^{\left(s\right)}$ and $x_b^{\left(s\right)}$ respectively. As is clear graphically, the multi-valued region in the image is the interval $\left[x_b^{\left(s\right)}, \, x_a^{\left(s\right)}\right]$; the bounding interval corresponding to multi-valuedness in plasma coordinates is $\left[x_{-}, \, x_{+}\right]$, which encompasses $\left[x_{a}, \, x_{b}\right]$ (see Figure \ref{fluxropecausticexample}d). The mapping associated with the perturbed field is also plotted -- by construction, the mappings are identical outside the interval  $\left[x_{-}, \, x_{b}\right]$. Calculating some properties of these coordinates for the base field, it follows by definition that the stationary point values satisfy
\begin{equation}
 \mathrm{sech} \, x_a  \; \mathrm{tanh} \, x_a = \mathrm{sech} \, x_b  \; \mathrm{tanh} \, x_b = \frac{1}{\mu} \, ,
\end{equation}
from which it can be shown that
 \begin{IEEEeqnarray}{rClCrCl}
 \mathrm{sech} \, x_a & = & +\left(\frac{1+\sqrt{1-4/{\mu^2}}}{2}\right)^{1/2}  & \qquad &  \mathrm{sech} \, x_b & = & +\left(\frac{1-\sqrt{1-4/{\mu^2}}}{2}\right)^{1/2}  
\end{IEEEeqnarray}
Complicated explicit expressions for $x_a$ and $x_b$ (as well as the image points $x_a^{\left(s\right)}$ and $x_b^{\left(s\right)}$) can be found, but their form is not terribly illuminating. Qualitatively, as $\mu$ increases, $x_a$ holds a relatively constant position, while $x_b$ increases, albeit slowly. However, behaviour is reversed for the image points: $x_a^{\left(s\right)}$ grows much more rapidly with $\mu$ than $x_b^{\left(s\right)}$ does. This can be made slightly more precise by considering asymptotic forms for $\mu \gg 1$. In particular, the inner stationary point satisfies
\begin{equation}
 \mathrm{sech} \, x_a \sim 1 +  \mathcal{O}\!\left(\frac{1}{\mu}\right) \implies  x_a \sim  \mathcal{O}\!\left(\frac{1}{\mu}\right) \, ,
\end{equation} 
with (outer) image point
\begin{equation} 
 x_a^{\left(s\right)}  \sim \mu \, \delta \alpha = \delta \theta \, .
\end{equation}
Physically this can be interpreted in the following way: for fixed $\delta \alpha$, the magnetic field configuration will increasingly determine the position of the outer caustic (and this deflection will occur near the position of maximal field strength). The outer stationary point satisfies
\begin{equation}
 \mathrm{sech} \, x_b \sim \frac{1}{\mu} +  \mathcal{O}\!\left(\frac{1}{\mu^2}\right) \implies  x_b \sim \log{\mu} +\mathcal{O}\!\left(1\right) \, ,
\end{equation} 
with (inner) image point
\begin{equation} 
 x_b^{\left(s\right)}  \sim \delta \alpha \log{\mu}  = \delta \theta \,\frac{\log{\mu}}{\mu} \, .
\end{equation}
This outer feature arises where deflections have become sufficiently weak that the initial proton spread once again determines behaviour: the weak logarithmic dependence on $\mu$ is the result of exponential field decay making this switch-over point insensitive to field strength. The result is an analytic example of a general observation that caustic-like structures can hold relatively constant positions over a range of beam energies~\cite{K12}.

\subsection{Image flux-conserving magnetic field perturbations} \label{ImFluxMagPerb}

\begin{figure}[htbp]
\centering
    \begin{subfigure}{.48\textwidth}
        \centering
        \includegraphics[width=\linewidth]{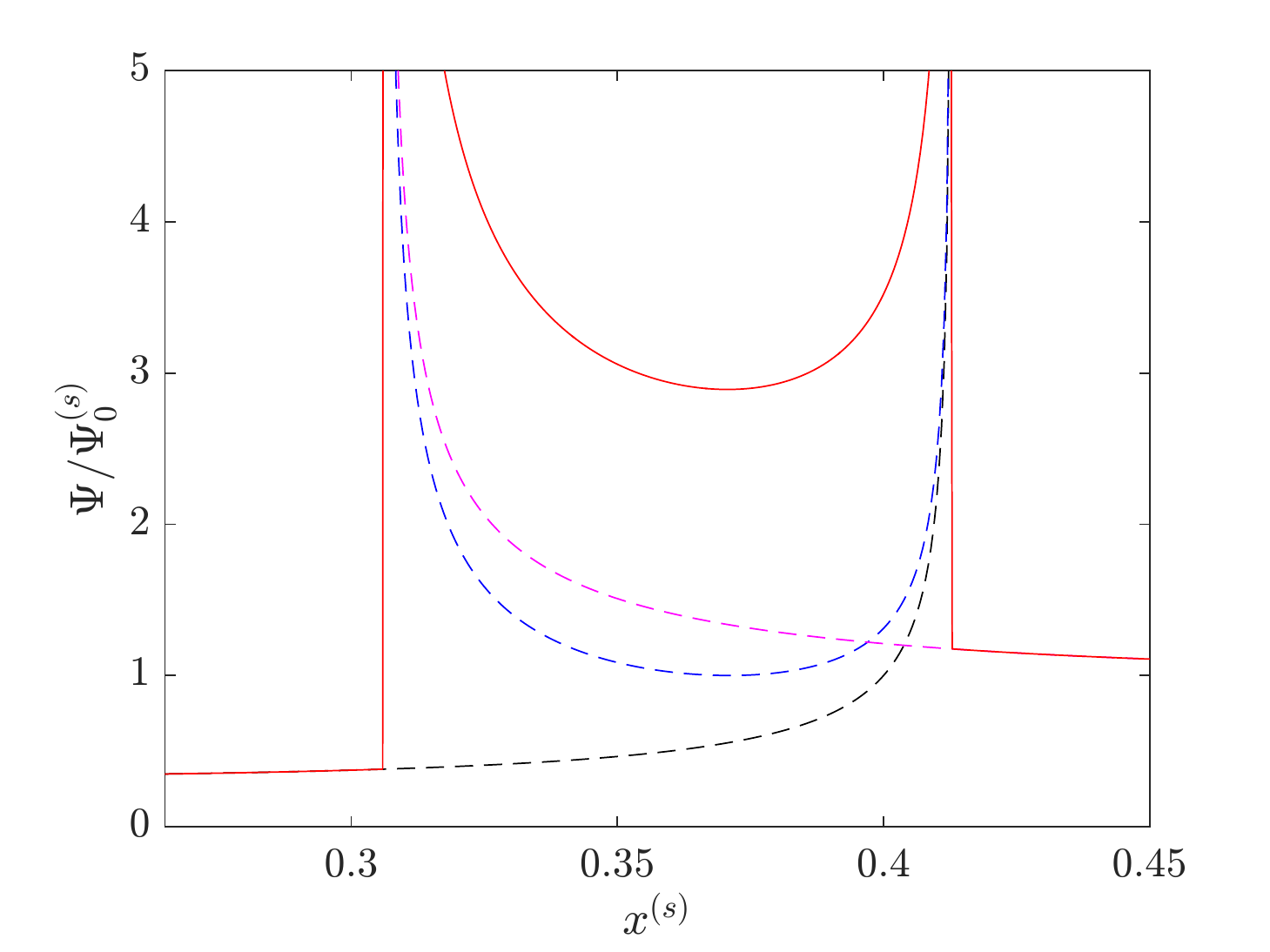}
    \end{subfigure} %
    \begin{subfigure}{.48\textwidth}
        \centering
        \includegraphics[width=\linewidth]{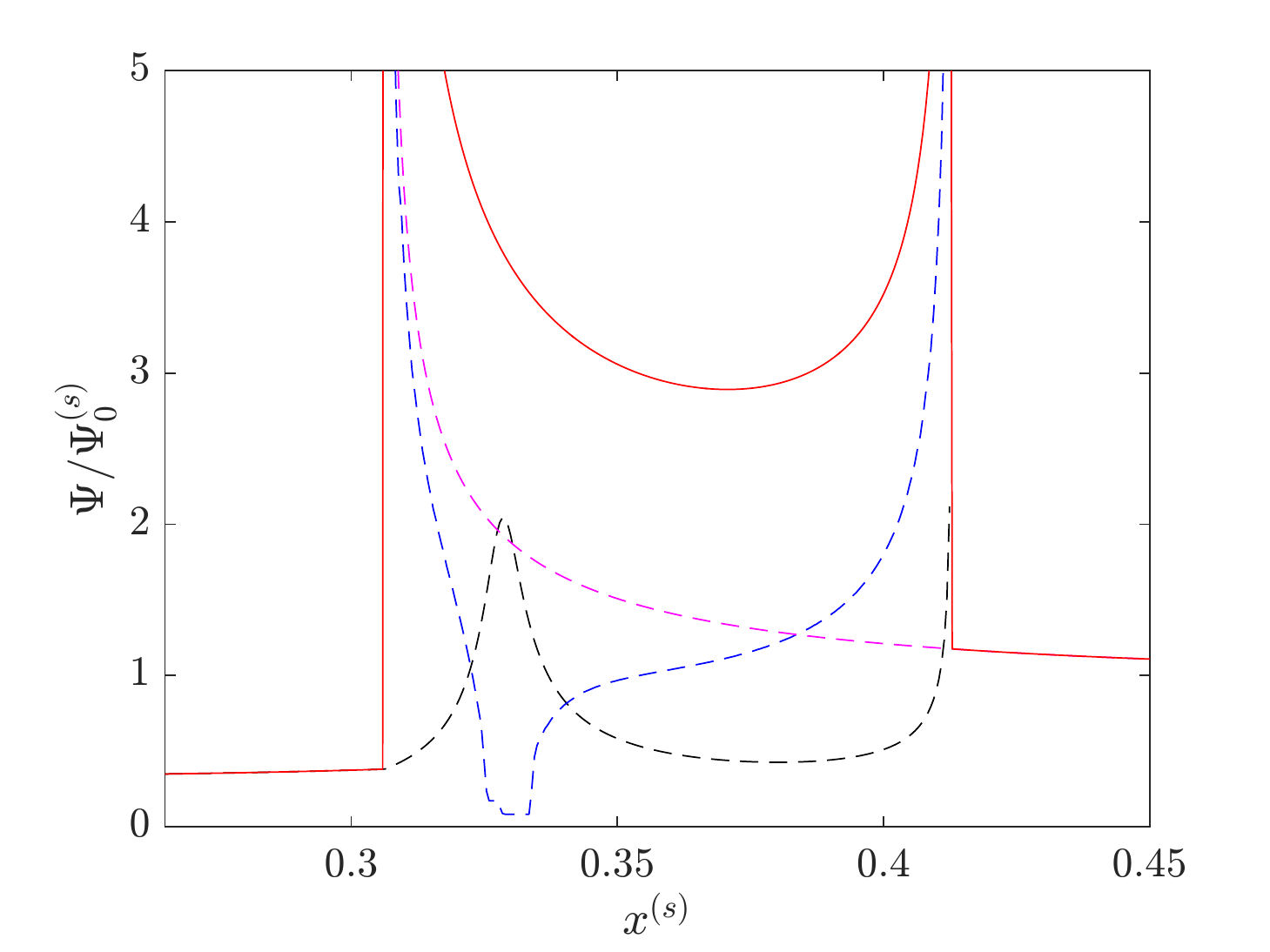}
    \end{subfigure} %
\caption{\textit{Illustration of undermination of plasma-image mapping for known image-flux distribution}. \textbf{a)} One-dimensional normalised image-flux distribution \eqref{fluxropeimageflux}, for $\mu = 4$, focused on multi-valued region. The dashed lines denote the contributions to the image-flux distribution from mappings $T_1$ (black), $T_2$ (blue), and $T_3$ (magneta) defined by \eqref{TmapA}, \eqref{TmapB}, and eqref{TmapC} respectively. \textbf{b)} One-dimensional image-flux distribution focused on multi-valued region for perturbed magnetic flux-rope field. The perturbation is defined by \eqref{mappingex}, combined with condition \eqref{S2map}. The dashed lines denotes the contributions to the image-flux distribution from $S_1$ (black), $S_2$ (blue), and $T_3$ (magenta).} \label{nonlincaustfail}
\end{figure}

To find the relevant family of mappings, first decompose $T$ in the multivalued region into three bijective functions $T_1$, $T_2$ and $T_3$, defined by
\begin{IEEEeqnarray}{lCrCl}
T_1 & : & \left[x_{-},x_a\right] & \mapsto & \left[x_b^{\left(s\right)},x_a^{\left(s\right)}\right] \, , \label{TmapA}\\ 
T_2 & : & \left[x_a,x_b\right] & \mapsto & \left[x_b^{\left(s\right)},x_a^{\left(s\right)}\right] \, , \label{TmapB}\\ 
T_3 & : & \left[x_b,x_{+}\right] & \mapsto & \left[x_b^{\left(s\right)},x_a^{\left(s\right)}\right] \, . \label{TmapC}
\end{IEEEeqnarray}
Now, in the multi-valued region we can re-write the expression for the image-flux distribution as
\begin{equation}
\Psi\!\left(x^{\left(s\right)}\right) = \tilde{\Psi}_0  \left[\frac{1}{T_1'\!\left(T_1^{-1}\!\left[x^{\left(s\right)}\right]\right)} -\frac{1}{T_2'\!\left(T_2^{-1}\!\left[x^{\left(s\right)}\right]\right)}+\frac{1}{T_3'\!\left(T_3^{-1}\!\left[x^{\left(s\right)}\right]\right)}\right] \, ,\label{fluxdef2}
\end{equation}
where $\tilde{\Psi}_0$ is the appropriate normalisation constant. 
Next, define a mapping $S_1\!: \left[x_{-},x_a\right] \mapsto \left[x_a^{\left(s\right)},x_b^{\left(s\right)}\right]$ with the following properties:
\begin{enumerate}
\item $S_1\!\left(x_{-}\right) = T_1\!\left(x_{-}\right)$, and $S_1'\!\left(x_{-}\right) = T_1'\!\left(x_{-}\right)$. 
\item $S_1\!\left(x_a\right) = T_1\!\left(x_a\right)$, and $S_1'\!\left(x_{a}\right) = T_1'\!\left(x_{a}\right) = 0$.
\item $S_1$ (and $S_2$ when constructed) remain bijective. 
\end{enumerate}
Then, define a second mapping $S_2\!: \left[x_{a},x_b\right] \mapsto \left[x_a^{\left(s\right)},x_b^{\left(s\right)}\right]$ by
\begin{equation}
S_2\!\left(x\right) \equiv S_1\!\left(x-T_2^{-1}\left[S_2\!\left(x\right)\right]+T_1^{-1}\left[S_2\!\left(x\right)\right]\right) \, . \label{S2map}
\end{equation}
This perhaps opaque definition has a more intuitive interpretation in terms of the inverse function:
\begin{equation}
S_2^{-1}\left(x^{\left(s\right)}\right) \equiv S_1^{-1}\left(x^{\left(s\right)}\right)-T_1^{-1}\left(x^{\left(s\right)}\right)+T_2^{-1}\left(x^{\left(s\right)}\right) \label{mapping1}
\end{equation}
Qualitatively, the mapping takes the $x$ coordinate associated with inverse mapping $S_1^{-1}$, adds the difference between the $x$ coordinates calculated from the base inverse mappings evaluated at the relevant image-coordinate value, and then assigns the value of $S_2$ at that point to be the image coordinate. The reason for taking this definition becomes clear on taking the derivative of \eqref{mapping1}:
\begin{equation}
\frac{1}{S_2'\!\left(S_2^{-1}\!\left[x^{\left(s\right)}\right]\right)} -\frac{1}{S_1'\!\left(S_1^{-1}\!\left[x^{\left(s\right)}\right]\right)} = \frac{1}{T_2'\!\left(T_2^{-1}\!\left[x^{\left(s\right)}\right]\right)} -\frac{1}{T_1'\!\left(T_1^{-1}\!\left[x^{\left(s\right)}\right]\right)}
\end{equation}
Substituting into \eqref{fluxdef2} gives
\begin{equation}
\Psi\!\left(x^{\left(s\right)}\right) = \tilde{\Psi}_0  \left[\frac{1}{S_1'\!\left(S_1^{-1}\!\left[x^{\left(s\right)}\right]\right)} -\frac{1}{S_2'\!\left(S_2^{-1}\!\left[x^{\left(s\right)}\right]\right)}+\frac{1}{T_3'\!\left(T_3^{-1}\!\left[x^{\left(s\right)}\right]\right)}\right] \label{flux2}
\end{equation}
which demonstrates that a complete mapping from $x$ to $x^{\left(s\right)}$ defined by
\begin{equation}
S\!\left(x\right) = 
\begin{cases} 
      T\!\left(x\right) & x\leq x_{-} \; \mathrm{or} \; x \geq x_b\\
      S_1\!\left(x\right) & x_{-} \leq x \leq x_a \\
      S_2\!\left(x\right) & x_{a} \leq x \leq x_b
\end{cases}
\end{equation}
gives exactly the same image-flux distribution (the additional conditions on $S_1$ ensure that the function and its first derivatives are continuous between sections, as well as not introducing additional caustics). The mapping can conceptually be conceived of as moderating the flux associated with the first branch of the mapping, then adapting the flux associated with the second branch so that the total flux is conserved. This is illustrated in Figures \ref{nonlincaustfail}a and \ref{nonlincaustfail}b for a particular instantiation of a perturbation field with the required properties, given below:
\begin{equation}
S_1\!\left(x\right) = T_1\!\left(x\right)-2\left(x-x_{-}\right)^2 \left(x_a-x\right)^2 \label{mappingex}
\end{equation} 
The magnetic field perturbation is proportional to the left hand term of \eqref{mappingex}.

This example shows that in a strict mathematical sense unique field reconstruction is impossible, even for a simple proton-flux image, once the mapping between plasma and image coordinates has become multi-valued. The counter-example gives a physically meaningful explanation for this: in multi-valued regions, fluxes in distinct branches of the plasma-image mapping can be rearranged without altering the total flux. For more complex fields the problem is even more acute, since once order-unity image-flux structures associated with different magnetic structures start to overlap, the mapping could become strongly multi-valued in places, leading to an increased range of possibilities for perpendicular-deflection fields for a given proton-flux image. However, it is clear that possible field configurations are somewhat strongly constrained (in magnitude, for exampl\textbf{e)}, which suggests that a statisical approach to analysing proton-flux images is possible.

\section{Derivation of lower bound \eqref{causticmaglowerbound} for magnetic field RMS by analogy to the Monge-Kantorovich problem} \label{MongeKantorovichBound}

In this appendix we demonstrate the validity of lower-bound for the RMS magnetic field strength \eqref{causticmaglowerbound} in terms of the predicted deflection-field potential resulting from the application of the field reconstruction algorithm descrived in the main text. 

First, we return to the Monge-Kantorovich formulation of the proton-imaging problem, and the observation that the reconstructed field minimised the cost functional $\mathcal{C}\!\left(\mathbf{x}_{\bot0}^{\left(s\right)}\right)$. Explicitly calculating the cost functional in terms of physical quantities for our problem, we see that for an initially uniform flux, 
\begin{equation}
\mathcal{C}\!\left(\mathbf{x}_{\bot0}^{\left(s\right)}\right) = \Psi_{0} \frac{r_s^2 r_i^2}{\left(r_s+r_i\right)^2V^2} \int \mathrm{d}^2 \mathbf{x}_{\bot0} \, \left|\mathbf{w}\!\left(\mathbf{x}_{\bot0}\right)\right|^2 \propto w_{rms}^2 \, .\label{MongeKcostfuneval}
\end{equation}
Thus, the reconstructed Monge-Kantorovich potential minimises the RMS of all possible perpendicular-deflection fields:
\begin{equation}
w_{rms}^2 \geq \left<\left(\nabla_{\bot0} \varphi\right)^2\right> \, . \label{Mongeamperelowerbounddeffield}
\end{equation}
This rigorously demonstrated lower bound can in turn be used to find a lower bound on predicted field strengths. Supposing we have a field configuration $\mathbf{B}$ in a compact volume V, with size $l_z$ in the $z$ direction. Then, it follows from the Cauchy-Schwarz inequality that
\begin{equation}
\int_V \mathbf{B}^2 \mathrm{d}V \geq \int_V \mathbf{B}_\bot^2 \mathrm{d}V \geq \frac{1}{l_z} \int \mathrm{d}^2 \mathbf{x}_\bot \, \left|\int_0^{l_z}  \mathrm{d}z \, \mathbf{B}_\bot\!\left(\mathbf{x}_\bot,z\right) \right|^2 \, .
\end{equation}
Thus, 
\begin{equation}
B_{rms}^2 \geq \frac{1}{l_z^2} \left<\left|\int_0^{l_z}  \mathrm{d}z \,  \mathbf{B}_\bot\!\left(\mathbf{x}_\bot,z\right) \right|^2\right> \, ,
\end{equation}
with equality if and only if $\mathbf{B} = \mathbf{B}_\bot$, and $\mathbf{B}_\bot = \mathbf{B}\!\left(\mathbf{x}_\bot\right)$. Since
\begin{equation}
\int_0^{l_z}   \mathrm{d}z \,  \mathbf{B}_\bot\!\left(\mathbf{x}_\bot,z\right) = -\frac{m_p c}{e} \hat{\mathbf{z}} \times \mathbf{w} \, ,
\end{equation}
it follows that
\begin{equation}
\left<\left|\int_0^{l_z}   \mathrm{d}z \,  \mathbf{B}_\bot\!\left(\mathbf{x}_\bot,z\right) \right|^2\right> = \frac{m_p^2 c^2}{e^2} w_{rms}^2 \, .
\end{equation}
Therefore, the magnetic field is bounded below by
\begin{equation}
B_{rms}^2 \geq \frac{m_p^2 c^2}{e^2 l_z^2}  w_{rms}^2 \, .
\end{equation}
The final lower bound is given by the deflection-field potential from \eqref{Mongeamperelowerbounddeffield}:
\begin{equation}
B_{rms}^2 \geq \frac{m_p^2 c^2}{e^2 l_z^2}  \left<\left(\nabla_{\bot0} \varphi\right)^2\right> \, .
\end{equation}
Operationally, we conclude that applying a field reconstruction algorithm to a particular proton-flux image to obtain the perpendicular-deflection field associated with the Monge-Kantorovich potential, then calculating the RMS deflection-field strength, provides a technique for determining a lower bound for the RMS magnetic field strength. 

\begin{figure}[htbp]
\centering
\includegraphics[width=0.5\linewidth]{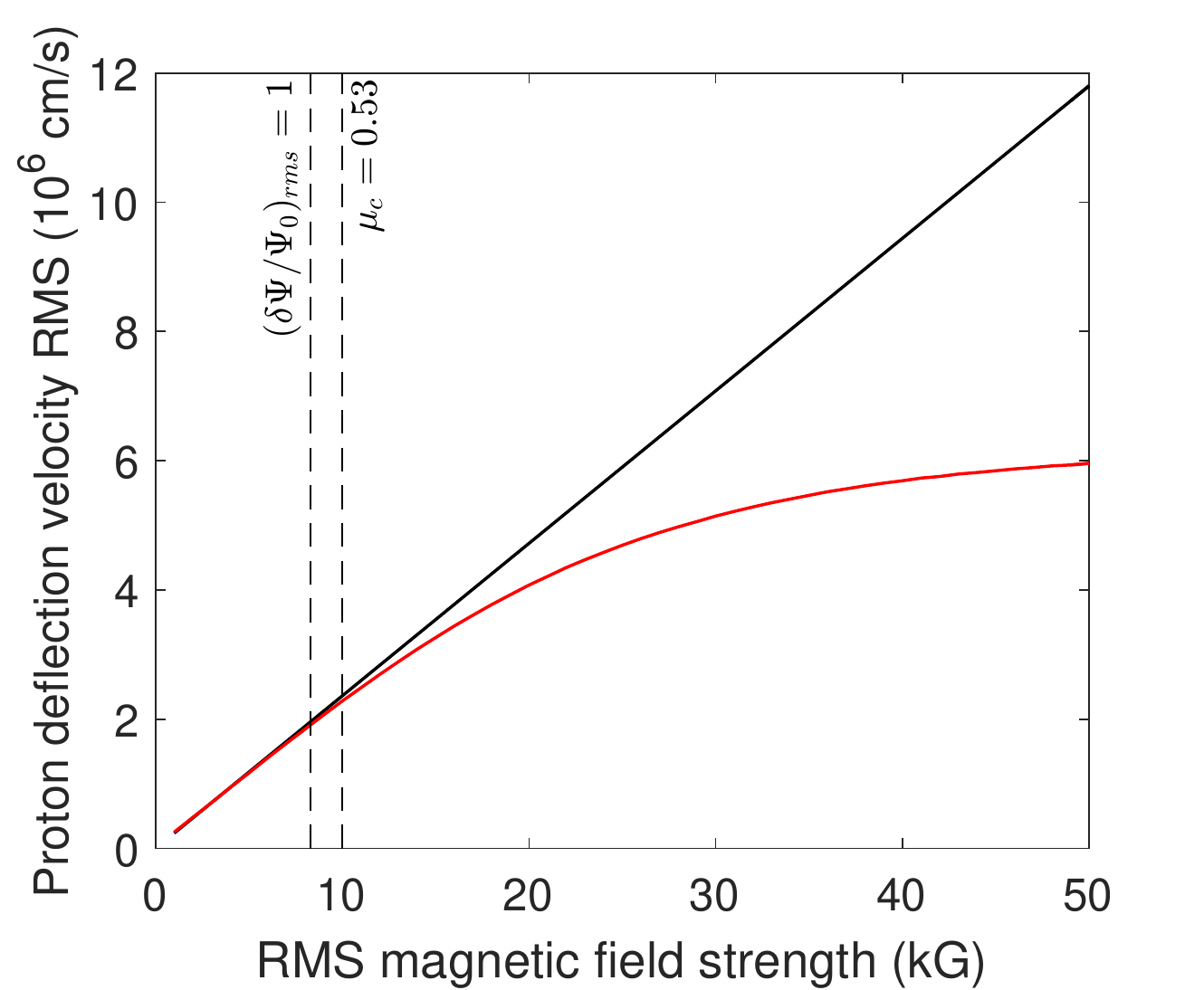}
\caption{\textit{Accuracy of lower bound \eqref{Mongeamperelowerbounddeffield} for RMS deflection-field strength over increasing RMS magnetic field strengths.} Left: actual RMS deflection-field strength (black) associated with Golitsyn magnetic field configuration defined in Figure \ref{GolitysncompactfieldSec2} for a range of RMS magnetic field strengths compared to predicted RMS deflection-field strength calculated from the perpendicular-deflection field reconstructed from numerically-generated proton-flux images (red). The imaging parameters of the proton beam were identical to those used to create proton-flux images in Figure \ref{Golitsynfluxrange}, as was the implementation.} \label{nonlinlowerbound}
\end{figure}

The accuracy of the bound is another matter; Figure \ref{nonlinlowerbound} shows the RMS deflection-field strength $w_{rms}$ for the Golitsyn field defined in Figure \ref{GolitysncompactfieldSec2} for a range of RMS magnetic field strengths. 
For values of $\mu$ not much bigger that $\mu_{c}$, the estimate of the perpendicular-deflection field RMS obtained is not that different from the true value, particularly if an image is either initially unsmeared, or de-smeared from a smeared one. However, the discrepancy grows increasingly fast if $\mu$ is raised further, diminishing the usefulness of the bound. For the predicted magnetic field strength, we have a similar story. In addition to the previously seen tends of linear flux theory weakly exaggerating field magnitude compared to the reconstructed deflection field estimate, smeared images giving reduced values, and de-smearing recovering unsmeared results, all methods give increasingly distant lower bounds as $\mu$ is increased.

\section{Diffusive model of proton imaging for $\mu \gtrsim r_s/\!\left(r_s+r_i\right)\delta \alpha \gg 1$} \label{StocMagDiffTens}

In this appendix we derive two results stated in Section \ref{HighConRgme} relating to diffusive models of proton imaging. Firstly, we present a derivation of the small-scale stochastic magnetic diffusion coefficient \eqref{difftensSection2}
\begin{equation}
D_{w} = \frac{V e^2 B_{rms}^2 l_B}{4 m_p^2 c^2} \, \label{difftensSectionAppend}
\end{equation}
in a manner consistent with asymptotic expansions done in Appendix \ref{DeflfieldCorr} when deriving the relation \eqref{deffieldspec} between the magnetic-energy spectrum and the spectrum of the perpendicular-deflection field. 
 
We then show expression \eqref{stocmagsmrdscreenflux} for the image-flux distribution, that is,
\begin{equation}
\Psi\!\left(\mathbf{x}_\bot^{\left(s\right)}\right) =  \frac{1}{\pi \delta^2} \int \mathrm{d}^2  \tilde{\mathbf{x}}_{\bot}^{\left(s\right)} \, \Psi_0^{\left(s\right)}\!\left(\tilde{\mathbf{x}}_\bot^{\left(s\right)}\right) \exp{\left[-\left(\frac{\mathbf{x}_\bot^{\left(s\right)}- \tilde{\mathbf{x}}_{\bot}^{\left(s\right)}}{\delta}\right)^2\right]} \, ,  \label{stocmagsmrdscreenfluxAppend}
\end{equation}
where
\begin{equation}
\delta = r_s \frac{\Delta w}{V} = \frac{e B_{rms} r_s}{m_p c V}\sqrt{l_z l_B} \, . \label{RWvelexactAppend}
\end{equation}
We do this by using \eqref{difftensSectionAppend} to solve for the averaged beam distribution function subsequent to interaction with the small-scale stochastic magnetic field. We then adapt the kinetic theory of proton imaging for a finite source given in Appendix \ref{ScreenDistFiniteSourceDev} to derive a image-flux point-spread function associated the effect of stochastic magnetic fields. 

Finally, we illustrate this theory with two numerical examples: diffusion of a pinhole beam through small-scale fields, and diffusion of a large-scale image-flux feature. We do this, since the sample magnetic field used for testing in the main text (described by Figure \ref{GolitysncompactfieldSec2}) does not have a sufficient separation of scales for the diffusive model \eqref{stocmagsmrdscreenfluxAppend} to manifest fully.     

\subsection{Diffusion coefficient of a fast proton beam in small-scale stochastic magnetic fields} \label{StocMagDiffTensDiffCoeff}

We calculate the diffusion coefficient \eqref{difftensSectionAppend} by deriving an evolution equation for the averaged beam distribution function. We again introduce an averaging operator $\left<\cdot\right>$, which can now be interpreted in three ways. Similarly to before, $\left<\cdot\right>$ can be taken an average over some intermediate horizontal scale, or as an ensemble average. However, in the case of very large $\mu \gg 1/\delta \alpha$, we can alternatively think of this term as naturally resulting from the presence of many protons, which originate from different initial locations and hence have experienced uncorrelated magnetic fields. This provides a local ensemble average. The first two interpretations can be applied to a greater range of $\mu$ values than the former, but only if some type of spatial averaging procedure is applied to the image-flux distribution. 

Starting with \eqref{Vlasov}, and introducing both averaged distribution function $\left<f\right>$ and difference $\delta f = f -\left<f\right>$, we can write coupled evolution equations
\begin{equation}
\frac{\partial \left<f\right>}{\partial t} + \mathbf{v} \cdot \nabla \left<f\right> + \frac{e}{m_p c} \epsilon_{ijk} v_j \left[\bar{B}_{k}\!\left(\mathbf{x}\right) \frac{\partial \left<f\right>}{\partial v_{i}}+ \left< \delta B_{k}\!\left(\mathbf{x}\right) \frac{\partial \, \delta f}{\partial v_{i}} \right> \right] = 0 \, , \label{stocavdistevolve}
\end{equation}
\begin{equation}
\frac{\partial \, \delta f}{\partial t} + \mathbf{v} \cdot \nabla \delta f + \frac{e}{m_p c} \epsilon_{ijk} v_j \left[\bar{B}_{k}\!\left(\mathbf{x}\right) \frac{\partial \, \delta f}{\partial v_{i}}+ \delta B_{k}\!\left(\mathbf{x}\right) \frac{\partial \left<f\right>}{\partial v_{i}} + \delta B_{k}\!\left(\mathbf{x}\right) \frac{\partial \, \delta f}{\partial v_{i}} - \left< \delta B_{k}\!\left(\mathbf{x}\right) \frac{\partial \, \delta f}{\partial v_{i}} \right>  \right] = 0 \, . \label{stocdiffdistevolve}
\end{equation}
Making a quasi-linear approximation, we neglect the last two terms of \eqref{stocdiffdistevolve} on the grounds that -- as deflections are small -- the difference between these two terms does not affect the evolution of the average distribution function. In particular, by construction the averaged distribution should not vary on scales $l \gg l_B$, and so its evolution should not depend explicitly on small-scale fluctuations. It should be noted here that in making this approximation, we depart from an non-averaged description of individual proton-flux images except in the case of diffusives. This is because the initial distribution has a `sharp' profile in phase space, and so small perturbations in the total velocity still lead to significant changes in the actual distribution away from the original. Large $\mu$ removes this sharpness, validating the description. 

This caveat aside, we now have
\begin{equation}
\frac{\partial \left<f\right>}{\partial t} + \mathbf{v} \cdot \nabla \left<f\right> + \frac{e}{m_p c} \epsilon_{ijk} v_j \left[\bar{B}_{k}\!\left(\mathbf{x}\right) \frac{\partial \left<f\right>}{\partial v_{i}}+ \left< \delta B_{k}\!\left(\mathbf{x}\right) \frac{\partial \, \delta f}{\partial v_{i}} \right> \right] = 0 \, , \label{stocavdistevolve}
\end{equation}
\begin{equation}
\frac{\partial \, \delta f}{\partial t} + \mathbf{v} \cdot \nabla \delta f + \frac{e}{m_p c} \epsilon_{ijk} v_j \left[\bar{B}_{k}\!\left(\mathbf{x}\right) \frac{\partial \, \delta f}{\partial v_{i}}+ \delta B_{k}\!\left(\mathbf{x}\right) \frac{\partial \left<f\right>}{\partial v_{i}}\right] = 0 \, .\label{stocdiffdistevolve}
\end{equation}
Changing frame to one moving with the initial velocity of the particles (to leading order):
\begin{IEEEeqnarray}{rCl}
\tilde{\mathbf{x}} & =  &\mathbf{x}-\mathbf{V}t \, , \\
\mathbf{w} & = & \mathbf{v}-\mathbf{V} \, ,
\end{IEEEeqnarray}
gives
\begin{equation}
\frac{\partial \left<f\right>}{\partial t} + \mathbf{w} \cdot \tilde{\nabla} \left<f\right> + \frac{e}{m_p c} \epsilon_{ijk} \left(\mathbf{V}+\mathbf{w}\right)_j \left[\bar{B}_{k}\!\left(\tilde{\mathbf{x}}+\mathbf{V}t\right) \frac{\partial \left<f\right>}{\partial w_{i}}+ \left< \delta B_{k}\!\left(\tilde{\mathbf{x}}+\mathbf{V}t\right) \frac{\partial \, \delta f}{\partial w_{i}} \right> \right]= 0 \, , \label{stocavdistevolve2}
\end{equation}
\begin{equation}
\frac{\partial \, \delta f}{\partial t} + \mathbf{w} \cdot \tilde{\nabla} \delta f + \frac{e}{m_p c} \epsilon_{ijk} \left(\mathbf{V}+\mathbf{w}\right)_j \left[\bar{B}_{k}\!\left(\tilde{\mathbf{x}}+\mathbf{V}t\right) \frac{\partial \, \delta f}{\partial w_{i}}+ \delta B_{k}\!\left(\tilde{\mathbf{x}}+\mathbf{V}t\right) \frac{\partial \left<f\right>}{\partial w_{i}} \right] = 0 \, . \label{stocdiffdistevolve2}
\end{equation}
Note that the concern about the validity of replacing the actual initial trajectories with those in the $z$ direction is not required here, since the statistical properties of either trajectory are the same up to $\mathcal{O}\!\left(\delta \alpha\right)$.

To derive a single equation for the averaged distribution, first integrate \eqref{stocdiffdistevolve2} up to some arbitrary time $t$,
\begin{equation}
\delta f\!\left(\tilde{\mathbf{x}},\mathbf{w},t\right) =  -\mathbf{w} \cdot \int_0^t \mathrm{d}t' \, \tilde{\nabla} \delta f\left(\tilde{\mathbf{x}},\mathbf{w},t'\right) - \frac{e}{m_p c} \epsilon_{npq} \left(\mathbf{V}+\mathbf{w}\right)_p \int_0^t \mathrm{d}t' \, \left[ \bar{B}_{q}\!\left(\tilde{\mathbf{x}}+\mathbf{V}t'\right) \frac{\partial \, \delta f}{\partial w_{n}} +\delta B_{q}\!\left(\tilde{\mathbf{x}}+\mathbf{V}t'\right) \frac{\partial \left<f\right>}{\partial w_{n}} \right]
\end{equation}
then evaluate the correlation with $\delta \mathbf{B}\!\left(\tilde{\mathbf{x}}+\mathbf{V}t\right)$:
\begin{IEEEeqnarray}{rCl}
\left<\delta B_k\!\left(\tilde{\mathbf{x}}+\mathbf{V}t\right)\delta f\!\left(\tilde{\mathbf{x}},\mathbf{w},t\right)\right> & = & -\mathbf{w} \cdot \int_0^t \mathrm{d}t' \left<\delta B_k\!\left(\tilde{\mathbf{x}}+\mathbf{V}t\right)\tilde{\nabla} \delta f\left(\tilde{\mathbf{x}},\mathbf{w},t'\right)\right> \nonumber \\
& & - \frac{e}{m_p c} \epsilon_{npq} \left(\mathbf{V}+\mathbf{w}\right)_p \int_0^t \mathrm{d}t' \bar{B}_{q}\!\left(\tilde{\mathbf{x}}+\mathbf{V}t'\right)\left<\delta B_k\!\left(\tilde{\mathbf{x}}+\mathbf{V}t\right) \frac{\partial \, \delta f}{\partial w_{n}}\bigg|_{\tilde{\mathbf{x}},t'} \right> \nonumber \\ 
& & - \frac{e}{m_p c} \epsilon_{npq} \left(\mathbf{V}+\mathbf{w}\right)_p \int_0^t \mathrm{d}t' \left<\delta B_k\!\left(\tilde{\mathbf{x}}+\mathbf{V}t\right) \delta B_{q}\!\left(\tilde{\mathbf{x}}+\mathbf{V}t'\right)\right> \frac{\partial \left<f\right>}{\partial w_{n}}\bigg|_{\tilde{\mathbf{x}},t'} \, .\label{stocdiffdistcorr1}
\end{IEEEeqnarray}
The first two terms on the right hand side of \eqref{stocdiffdistcorr1} involve correlations between the particle position at times $t$ and $t'$; here, we combine the fact that if $t-t' > l/V \gg l_B/V$ for some intermediate length scale $l \ll l_z$, then to $\mathcal{O}\!\left(\left(l_B/l\right)^2\right)$, the two components of each product will be uncorrelated, and so will vanish. The remaining contribution to both integrals is $\mathcal{O}\!\left(l/l_z\right)$, which is negligble compared to other terms. This ordering breaks down when $\left|\bar{\mathbf{B}}\right| \gg \left|\delta \mathbf{B}\right|$, though such a regime would be more amenably approached via a linearisation analysis. This leaves
\begin{equation}
\left<\delta B_k\!\left(\tilde{\mathbf{x}}+\mathbf{V}t\right)\delta f\!\left(\tilde{\mathbf{x}},\mathbf{w},t\right)\right> = - \frac{e}{m_p c} \epsilon_{npq} \left(\mathbf{V}+\mathbf{w}\right)_p \frac{\partial \left<f\right>}{\partial w_{n}}\bigg|_{\tilde{\mathbf{x}},t} \int_0^t \mathrm{d}t' \left<B_k\!\left(\tilde{\mathbf{x}}+\mathbf{V}t\right)B_{q}\!\left(\tilde{\mathbf{x}}+\mathbf{V}t'\right)\right> \, ,
\end{equation}
where the independence of the distribution function from $t'$ follows from the fact that $\left<f\right>$ can only vary on large scales, giving
\begin{equation}
\frac{\partial \left<f\right>}{\partial w_{n}}\bigg|_{\tilde{\mathbf{x}},t} = \frac{\partial \left<f\right>}{\partial w_{n}}\bigg|_{\tilde{\mathbf{x}},t'}\left[1+\mathcal{O}\!\left(\frac{l_B}{l}\right)\right] \, .
\end{equation}
Now substituting into \eqref{stocavdistevolve2}, we find a modified evolution equation, with an additional stochastic magnetic diffusivity:
\begin{equation}
\frac{\partial \left<f\right>}{\partial t} + \mathbf{w} \cdot \tilde{\nabla} \left<f\right> + \frac{e}{m_p c} \epsilon_{ijk} \left(\mathbf{V}+\mathbf{w}\right)_j \bar{B}_{k}\!\left(\tilde{\mathbf{x}}+\mathbf{V}t\right) \frac{\partial \left<f\right>}{\partial w_{i}} = \frac{\partial}{\partial \mathbf{w}} \cdot \left( \underline{\underline{\mathbf{\mathcal{D}}}} \cdot \frac{\partial \left<f\right>}{\partial \mathbf{w}}\right) \, . \label{stocavdistevolvegendiff}
\end{equation}
The diffusivity tensor $\underline{\underline{\mathbf{\mathcal{D}}}}$ is given by
\begin{IEEEeqnarray}{rCl}
\mathbf{\mathcal{D}}_{in} & = & \frac{V e^{2}}{m_p^2 c^2} \epsilon_{ijk} \epsilon_{npq} \left(\hat{z}_j + \frac{w_j}{V}\right) \left(\hat{z}_p + \frac{w_p}{V}\right) \int_0^z \mathrm{d}r \, M_{kq}\!\left(\tilde{\mathbf{x}}, \tilde{\mathbf{x}}+r \hat{\mathbf{z}}\right) \\
& = & \frac{V e^{2}}{m_p^2 c^2} \epsilon_{ijk} \epsilon_{npq} \hat{z}_j \hat{z}_p \int_0^\infty \mathrm{d}r \, M_{kq}\!\left(r\right)\left[1+\mathcal{O}\left(\delta \theta, \frac{l_B}{l_z}\right)\right] \, ,\label{stocdifftensA}
\end{IEEEeqnarray}
where we have assumed the fluctuating fields to be isotropic and homogeneous, and neglected $\mathcal{O}\!\left(\delta \theta\right)$ terms.  The symmetry of the diffusion term implies that we only need consider the symmetric part of $\underline{\underline{\mathbf{\mathcal{D}}}}$. Combining this with the observation from Appendix \ref{FurtherStatCharMagFieldsCorrTail} that it is asymptotically consistent to neglect the tail of the correlation function -- that is,
\begin{equation}
\int_0^z \mathrm{d}r \, M_{kq}\!\left(r\right) = \int_0^\infty \mathrm{d}r \, M_{kq}\!\left(r\right) \left[1+\mathcal{O}\!\left(\left(\frac{l_B}{l_z}\right)^2\right)\right] \, ,
\end{equation}
and then that any integral of the statistically isotropic and homogeneous $M_{kq}$ must take the form
\begin{equation}
 \int_0^\infty \mathrm{d}r \, M_{kq}\!\left(r\right) = c_1 \delta_{nq} + c_2  \hat{z}_n \hat{z}_q + c_3 \epsilon_{nql}  \hat{z}_l \, ,
\end{equation}
for constants $c_1, c_2$ and $c_3$, we conclude that the symmetric part of $\underline{\underline{\mathbf{\mathcal{D}}}}$ is both isotropic and two-dimensional, with
\begin{equation}
\mathbf{\mathcal{D}}_{in} = D_{w} \delta_{in}^{\left(2\right)} \, , \label{stocdifftensB}
\end{equation}
for some constant diffusion coefficient $D_{w}$. We can evaluate this by taking the trace of \eqref{stocdifftensA} to give
\begin{equation}
D_{w} = \frac{V e^2}{m_p^2 c^2} \int_0^\infty \mathrm{d}r \, \left[M\!\left(r\right)-M_{zz}\!\left(r\right)\right] = \frac{V e^2 B_{rms}^2 l_B}{4 m_p^2 c^2} \, ,
\end{equation}
in agreement with \eqref{difftensSectionAppend}. 

Using \eqref{stocdifftensB}, equation \eqref{stocavdistevolvegendiff} leads to diffusion equation
\begin{equation}
\frac{\partial \left<f\right>}{\partial t} + \mathbf{w} \cdot \tilde{\nabla} \left<f\right> + \frac{e}{m_p c} \epsilon_{ijk} \left(\mathbf{V}+\mathbf{w}\right)_j \bar{B}_{k}\!\left(\tilde{\mathbf{x}}+\mathbf{V}t\right) \frac{\partial \left<f\right>}{\partial w_{i}} = D_{w} \frac{\partial}{\partial \mathbf{w}_\bot} \cdot \frac{\partial \left<f\right>}{\partial \mathbf{w}_\bot} \, .
\end{equation}
In the laboratory frame, this becomes
\begin{equation}
\frac{\partial \left<f\right>}{\partial t} + \mathbf{v} \cdot \nabla \left<f\right> + \frac{e}{m_p c} \epsilon_{ijk} v_j \bar{B}_{k}\!\left(\mathbf{x}\right) \frac{\partial \left<f\right>}{\partial v_{i}} = D_{w} \frac{\partial}{\partial \mathbf{v}_\bot} \cdot \frac{\partial \left<f\right>}{\partial \mathbf{v}_\bot} \, . \label{Vlasovstocdiff}
\end{equation}
Thus we have recovered an evolution equation similar to the Vlasov equation (with large scale mean field $\bar{\mathbf{B}}$), but with an additional diffusive term. This can be thought of as a kind of collisionality due to many interactions with small-scale magnetic fields. This result agrees with the more general expression derived by Dolginov and Toptygin~\cite{DT67}.

\subsection{Derivation of image-flux distribution function \eqref{stocmagsmrdscreenflux} in presence of small-scale stochastic magnetic fields} \label{StocMagDiffTensScreenFlux}

We calculate the image-flux distribution \eqref{stocmagsmrdscreenfluxAppend} under a diffusive model in a similar manner to the method described in Appendix \ref{KinThyPrad}, except using the averaged beam distribution function $ \left<f\right>\left(\mathbf{x},\mathbf{v},t\right)$ in place of the beam distribution. $f\!\left(\mathbf{x},\mathbf{v},t\right)$. The averaged flux of particles through the detector located at $z = r_s+l_z$ at perpendicular location $\mathbf{x}_\bot^{\left(s\right)}$ and time $t$ is now given by
\begin{equation}
\psi\!\left(\mathbf{x}_\bot^{\left(s\right)},t\right) = \int \mathrm{d}^3 \mathbf{v} \, v_z \, \left<f\right>\!\left(\mathbf{x}_{\bot}^{\left(s\right)}\!,r_s+l_z,\mathbf{v},t\right) \, , \label{fluxtimediff}
\end{equation}
and the image-flux distribution as before:
\begin{equation}
\Psi\!\left(\mathbf{x}_\bot^{\left(s\right)}\right) \equiv \int_0^{\infty} \mathrm{d} t \, \psi\!\left(\mathbf{x}_\bot^{\left(s\right)},t\right) \, . \label{fluxtotdiff}
\end{equation}
Thus, we need to solve for the evolution of the averaged beam distribution function $ \left<f\right>\left(\mathbf{x},\mathbf{v},t\right)$ if we are to find the image-flux distribution.

To do this, we invoke a separation of scales between those of the small-scale magnetic fields, $l_B$, and those of the mean magnetic field (which we assume are of the same order of the global scale of the plasma $l_i \sim l_z \sim l_\bot$. Evolution equation \eqref{Vlasovstocdiff} then has solution 
\begin{equation}
\left<f\right>\!\left(\mathbf{x},\mathbf{v},t\right) = \sum_{\mathbf{x}_0, \mathbf{v}_0}\tilde{\left<f\right>}\!\left(\mathbf{x}_0\!\left(\mathbf{x},\mathbf{v},t\right)\!,\mathbf{v}_0\!\left(\mathbf{x},\mathbf{v},t\right)\right) \left|\frac{\partial\!\left(\mathbf{x}_0,\mathbf{v}_0\right)}{\partial\!\left(\mathbf{x},\mathbf{v}\right)}\right| \label{characsolB} \, ,
\end{equation}
where $\tilde{\left<f\right>}\!\left(\mathbf{x},\mathbf{v},t\right)$ locally satisfies
\begin{equation}
\frac{\partial \tilde{\left<f\right>}}{\partial t} + \mathbf{v} \cdot \nabla \tilde{\left<f\right>} = D_{w} \frac{\partial}{\partial \mathbf{v}_\bot} \cdot \frac{\partial \tilde{\left<f\right>}}{\partial \mathbf{v}_\bot} \, ,\label{stocavdistevolvenomean}
\end{equation}
and $\mathbf{x}_0 = \mathbf{x}_0\!\left(\mathbf{x},\mathbf{v},t\right)$ and $\mathbf{v}_0 = \mathbf{v}_0\!\left(\mathbf{x},\mathbf{v},t\right)$ are initial points in phase space mapping to $\left(\mathbf{x},\mathbf{v}\right)$ at time $t$ as before. Thus, to solve for the averaged beam distribution function, we first solve \eqref{stocavdistevolvenomean} locally. The global solution is then that obtained in Appendix \ref{KinThyPradBeamDist}, with characteristic rays $\left(\mathbf{x}_0,\mathbf{v}_0\right)$ determined by the mean magnetic field. 

When solving \eqref{stocavdistevolvenomean} locally, we neglect beam divergence across the the plasma (invoking an $\mathcal{O}\!\left(\delta \alpha\right)$ error), and adopt initial condition \eqref{bmdistinit}, viz.,
\begin{equation}
\tilde{\left<f_0\right>}\!\left(\mathbf{x}_{0},\mathbf{v}_{0}\right) = f_0\!\left(\mathbf{x}_{0},\mathbf{v}_{0}\right) = \Psi_{0}\!\left(\mathbf{x}_{\bot0}\right) \delta\!\left(z_0\right) \delta\!\left(\mathbf{v}_{\bot0}\right)\delta\!\left(v_{z0}-V\right) \, ,\label{bmdistinit}
\end{equation}
with horizontal spatial gradients in $\Psi_0\!\left(\mathbf{x}_{\bot0}\right)$ assumed $\mathcal{O}\!\left(l_\bot\right)$. The magnetic diffusivity term does not act on the parallel direction, so the $z$ component of the distribution function decouples, undergoing free streaming. Next, neglecting the $\mathbf{v}_\bot \cdot \nabla \tilde{\left<f\right>}$ term in \eqref{stocavdistevolvenomean} on the grounds that deflections are too small to induce $\mathcal{O}\!\left(l_\bot\right)$ density variation across the plasma, and Fourier transforming the resulting equation, we find the following solution to \eqref{stocavdistevolvenomean}:
\begin{equation}
\tilde{\left<f\right>}\!\left(\mathbf{x},\mathbf{v},t\right) =  \delta\!\left(z-Vt\right) \delta\!\left(v_{z}-V\right) \frac{\Psi_0\!\left(\mathbf{x}_\bot\right)}{\left(2\pi\right)^4}\int \mathrm{d}^2 \mathbf{k}_w \, \exp{\left[i\mathbf{k}_w \cdot \mathbf{v}_\bot - D_w k_w^2 t\right]} \, .
\end{equation}
This latter integral can be evaluated to give the smeared distribution function at the end of the interaction region
\begin{equation}
\tilde{\left<f\right>}\!\left(\mathbf{x},\mathbf{v},\frac{l_z}{V}\right) = \delta\!\left(z-l_z\right) \delta\!\left(v_{z}-V\right) \Psi_0\!\left(\mathbf{x}_\bot\right)\frac{1}{\pi \Delta w} \exp{\left[-\frac{v_\bot^2}{\Delta w^2}\right]} \, . \label{stocmagPSF}
\end{equation}
Unsurprisingly for a diffusive model, the result is a Gaussian in velocity space, with typical diffusion velocity
\begin{equation}
\Delta w = \frac{e B_{rms}}{m_p c}\sqrt{l_z l_B} \, . \label{RWvelexact}
\end{equation}
Checking that our neglected spatial advection term is indeed small, we have
\begin{equation}
\frac{\mathbf{v}_\bot \cdot \nabla \tilde{\left<f\right>}}{D_{w} \frac{\partial}{\partial \mathbf{v}_\bot} \cdot \frac{\partial \tilde{\left<f\right>}}{\partial \mathbf{v}_\bot}
} \sim \frac{t\Delta w}{l_z} \leq \frac{\Delta w}{V} \sim \delta \theta \ll 1
\end{equation}
as required. 
Equation \eqref{RWvelexact} is consistent with the relation derived in Appendix \ref{FurtherStatCharMagFieldsCorrTail} for the RMS deflection magnitude \eqref{magcorlengthdefl}, since
\begin{equation}
\left<\mathbf{v}_\bot^2\right> = \int \mathrm{d}^3 \mathbf{x} \int \mathrm{d}^3 \mathbf{v} \, \mathbf{v}_\bot^2 \tilde{\left<f\right>}\!\left(\mathbf{x},\mathbf{v},\frac{l_z}{V}\right) = \Delta w^2 \, .
\end{equation}

In short, we conclude that the effect of small-scale stochastic fields on the initial averaged distribution function is to introduce width to the distribution in velocity space. In fact, this effect is equivalent to solving for the evolution of the averaged beam distribution function in the absence of small-scale stochastic magnetic fields, but with initial averaged distribution function
\begin{equation}
\left<f\right>\!\left(\mathbf{x}_{0},\mathbf{v}_{0}\right) = \Psi_{0}\!\left(\mathbf{x}_{\bot0}\right) \, \delta\!\left(z_0\right) \delta\!\left(v_{z0}-V\right) P\!\left(\delta \mathbf{v}_{\bot0}\right) \, ,
\label{smeareddistfunc}
\end{equation}
where
\begin{equation}
P\!\left(\mathbf{v}_\bot\right) = \frac{1}{\pi \Delta w^2} \exp{\left[-\frac{v_\bot^2}{\Delta w^2}\right]}  \, . \label{stocmagPSFv2}
\end{equation}
By comparison with the general expression for the image-flux distribution \eqref{smearedscreenflux} for initial distribution functions of the form \eqref{smeareddistfunc} proven in Appendix \ref{ScreenDistFiniteSourceDev}, we deduce the desired result:
\begin{IEEEeqnarray}{rCl}
\tilde{\Psi}\!\left(\mathbf{x}_\bot^{\left(s\right)}\right) & = &\frac{V^2}{r_s^2} \int \mathrm{d}^2  \tilde{\mathbf{x}}_{\bot}^{\left(s\right)} \, \Psi\!\left(\tilde{\mathbf{x}}_\bot^{\left(s\right)}\right) P\!\left(\frac{\mathbf{x}_{\bot}^{\left(s\right)}-\tilde{\mathbf{x}}_{\bot}^{\left(s\right)}}{r_s}V\right) \, ,  \nonumber \\
 & = &\frac{1}{\pi \delta^2} \int \mathrm{d}^2  \tilde{\mathbf{x}}_{\bot}^{\left(s\right)} \, \Psi_0^{\left(s\right)}\!\left(\tilde{\mathbf{x}}_\bot^{\left(s\right)}\right) \exp{\left[-\left(\frac{\mathbf{x}_\bot^{\left(s\right)}- \tilde{\mathbf{x}}_{\bot}^{\left(s\right)}}{\delta}\right)^2\right]} \, ,  \label{smearedscreenfluxstocmag} 
\end{IEEEeqnarray}
where $\delta = r_s \Delta w/V$, and the $\Psi\!\left(\tilde{\mathbf{x}}_\bot^{\left(s\right)}\right) = \Psi_0^{\left(s\right)}\!\left(\tilde{\mathbf{x}}_\bot^{\left(s\right)}\right)$ in the absence of large-scale magnetic fields. 

For the purposes of deriving the image-flux distribution including the effect of small-scale magnetic fields, \eqref{stocmagPSF} is adequate. However, for completeness, we note that an exact solution for $\left<f\right>$ can be derived exactly from \eqref{stocavdistevolvenomean}. To undertake this calculation, we consider an initial distribution function in the form of a delta function beam:
\begin{equation}
\tilde{\left<f_0\right>}\!\left(\mathbf{x},\mathbf{v},t\right) = f_0\!\left(\mathbf{x},\mathbf{v},t\right) = \delta\!\left(\mathbf{x}_{0\bot}\right) \delta\!\left(z_0\right) \delta\!\left(\mathbf{v}_{0\bot}\right)  \delta\!\left(v_{0z}-V\right) \, . 
\end{equation}
As with the previous case, the $z$-component of this solution separates out. We can then Fourier transform the perpendicular components of \eqref{stocavdistevolvenomean} in phase space to give a first order differential equation in time for $\left<f\right>$. Solving yields
\begin{equation}
\tilde{\left<f\right>} = \delta\!\left(z-Vt\right) \delta\!\left(v_{z}-V\right)  \frac{\Psi_0\!\left(\mathbf{x}_\bot\right)}{\left(2\pi\right)^4} \int \mathrm{d}^2 \mathbf{k}_{\bot} \, \mathrm{d}^2 \mathbf{k}_w \, \exp{\left[i\mathbf{k}_\bot \cdot\left(\mathbf{x}_\bot-\mathbf{v}_{\bot}t\right)+i \mathbf{k}_w \cdot \mathbf{w} - D_w \left(k_w^2 t - \mathbf{k}_\bot \cdot\mathbf{k}_w t^2 + \frac{t^3}{3} k_\bot^2\right)\right]} \, .
\end{equation}
The integral can again be evaluated to give
\begin{equation}
\tilde{\left<f\right>}\!\left(\mathbf{x},\mathbf{v},\frac{l_z}{V}\right) = \delta\!\left(z-l_z\right) \delta\!\left(v_{z}-V\right)\frac{1}{\pi \Delta w^2} \exp{\left[-\frac{v_\bot^2}{\Delta w^2}\right]}\frac{1}{\pi \Delta x^2} \exp{\left[-\frac{\left(\mathbf{x}_\bot-\mathbf{v}_\bot t/2\right)^2}{\Delta x^2}\right]} \, ,
\end{equation}
where in addition to velocity spread \eqref{RWvelexact} we now have spatial diffusion determined by
\begin{equation}
\Delta x = \frac{1}{3}\frac{e B_{rms}}{m_p c V}l_z\sqrt{l_z l_B} \, .\label{RWspace}
\end{equation}
In the context of proton imaging, the spatial diffusion of the beam in the plasma will always be dominated by free-streaming of the velocity-diffused beam, so this effect is unlikely to be seen. 

\subsection{Numerical illustration of diffusive model \eqref{stocmagsmrdscreenflux} for image-flux distribution} \label{StocMagDiffTensNumSim}

To illustrate the validity of a diffusive model of proton imaging -- in particular, image-flux distribution expression \eqref{stocmagsmrdscreenflux} -- we now consider two numerical examples. 

First, we consider the evolution of a pinhole beam of protons through a small-scale stochastic magnetic field. We find the numerically determined image-flux distribution by the following procedure. First, the proton beam (size $d_{pin} = 0.02 \, \mathrm{cm}$) is generated from a point source, which is passed through a cube-shaped region of side length $l_i$ (that is, $l_i = l_z = l_\bot$) containing the magnetic field. The propagation is carried out using the ray-tracing code described in Appendix \ref{NumSimFluxImageGenFull}, while the magnetic field is of the same form as the fields of magnetic cocoons described in Appendix \ref{ToySpecLinThyCocoon}. Having left the magnetic field configuration, the protons propagate to the artificial `detector', where their locations are binned into pixels, giving the numerical image-flux distribution. This is compared to analytic prediction \eqref{stocmagsmrdscreenflux}, which is calculated by applying a Gaussian filter of the correct magnitude (determined by \eqref{RWvelexactAppend}) to the initial flux distribution. 

\begin{figure}[htbp]
\centering
    \begin{subfigure}{.48\textwidth}
        \centering
        \includegraphics[width=\linewidth]{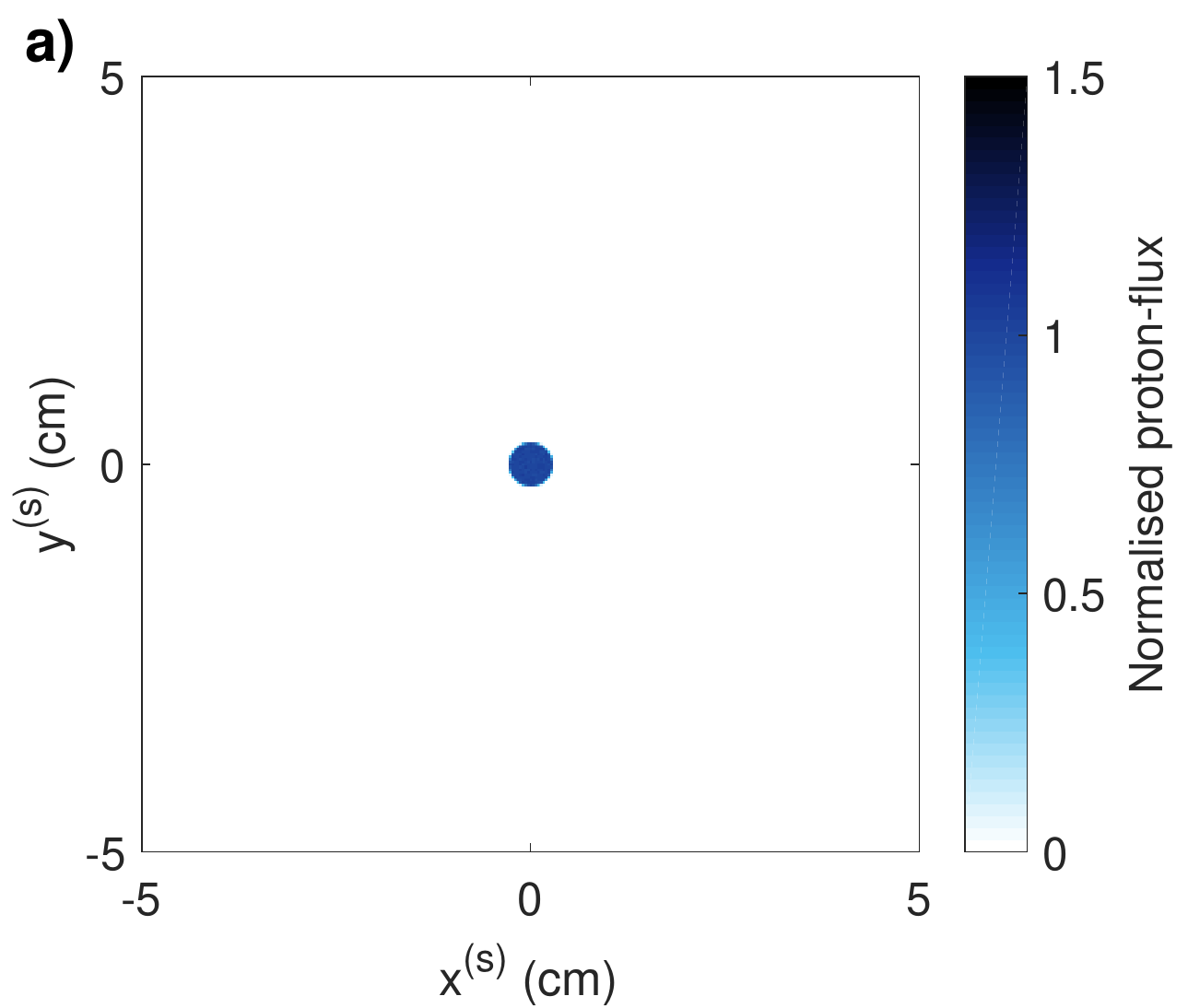}
    \end{subfigure} %
    \begin{subfigure}{.48\textwidth}
        \centering
        \includegraphics[width=\linewidth]{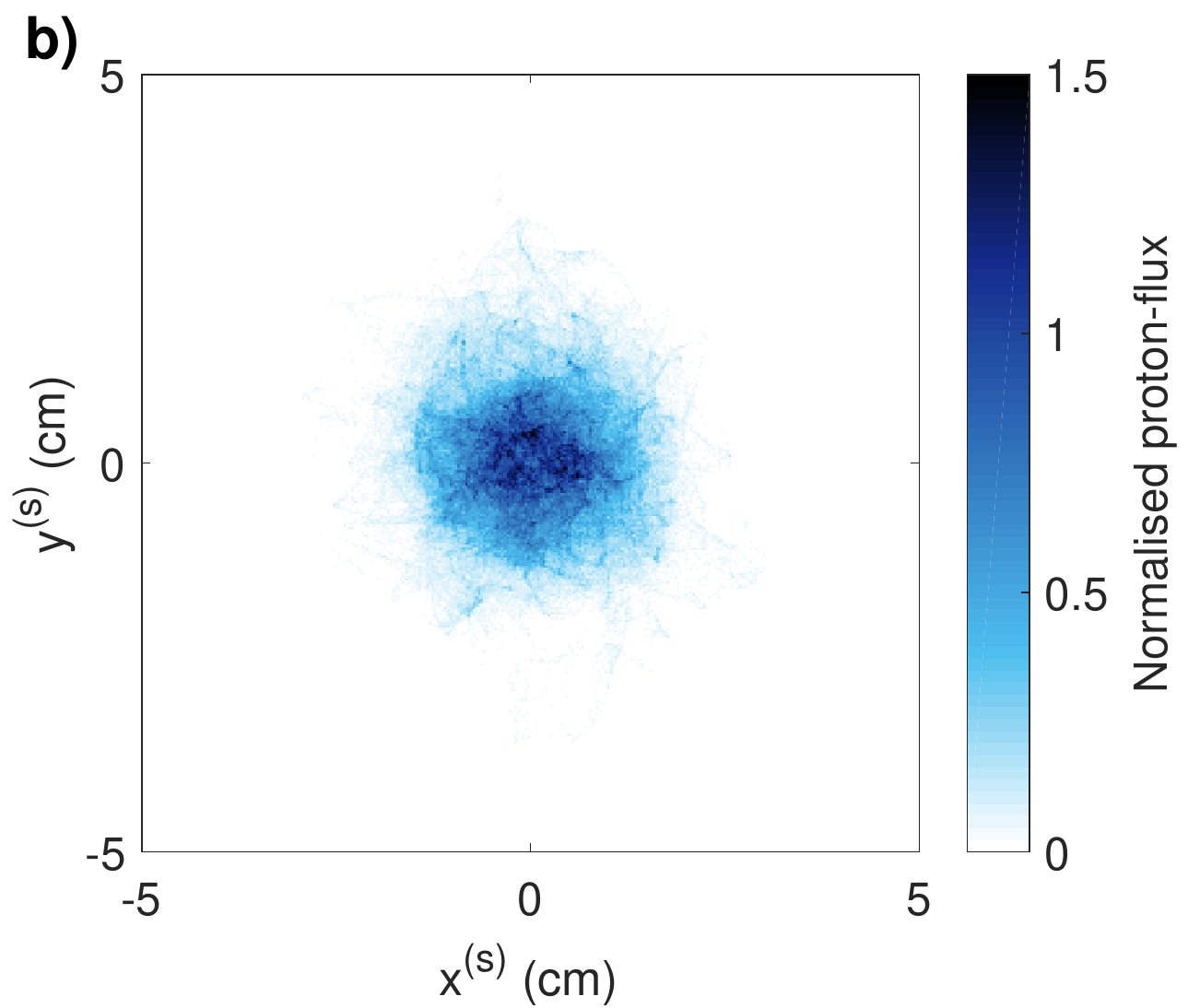}
    \end{subfigure} %
        \begin{subfigure}{.48\textwidth}
        \centering
        \includegraphics[width=\linewidth]{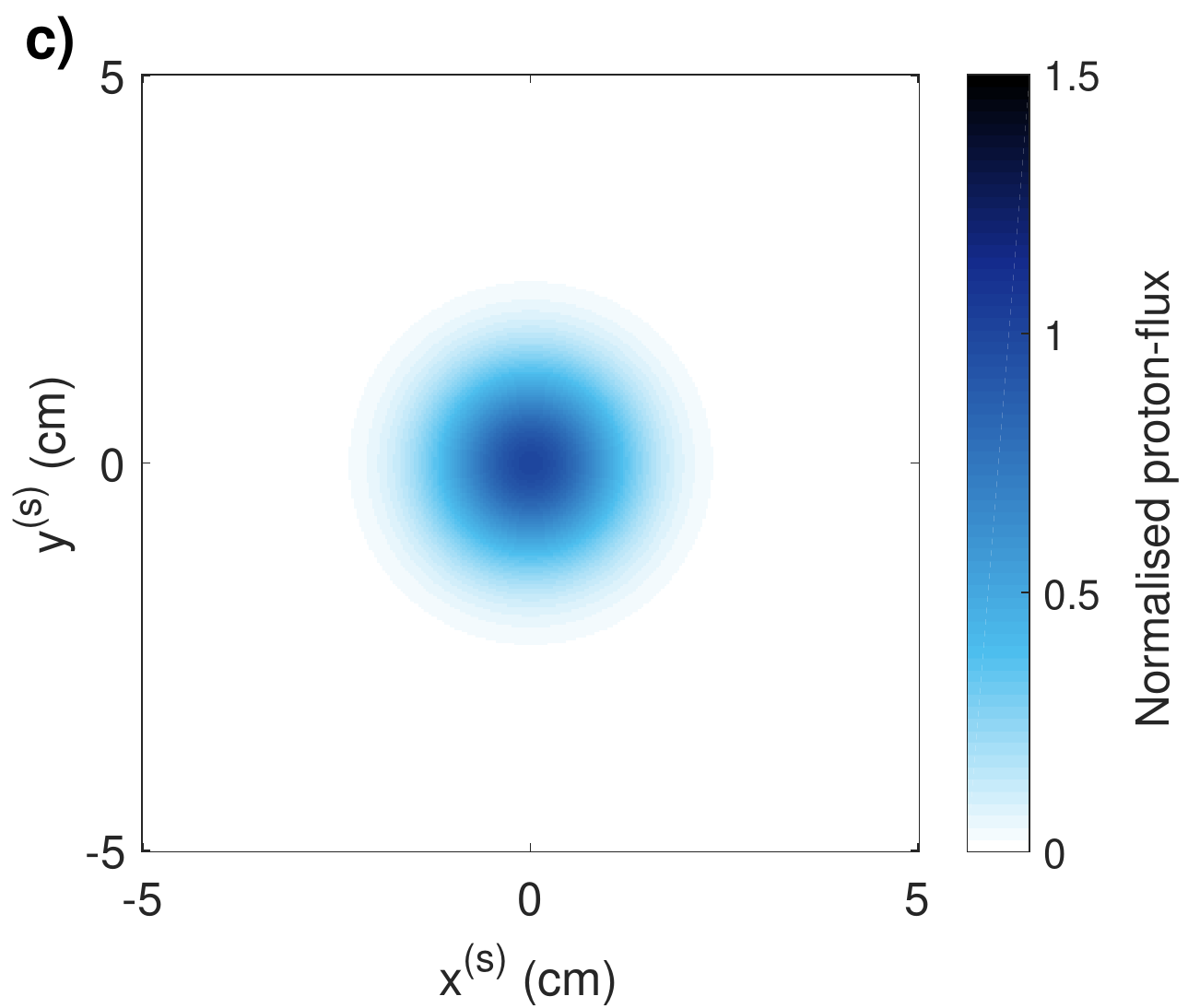}
    \end{subfigure} %
    \begin{subfigure}{.48\textwidth}
        \centering
        \includegraphics[width=\linewidth]{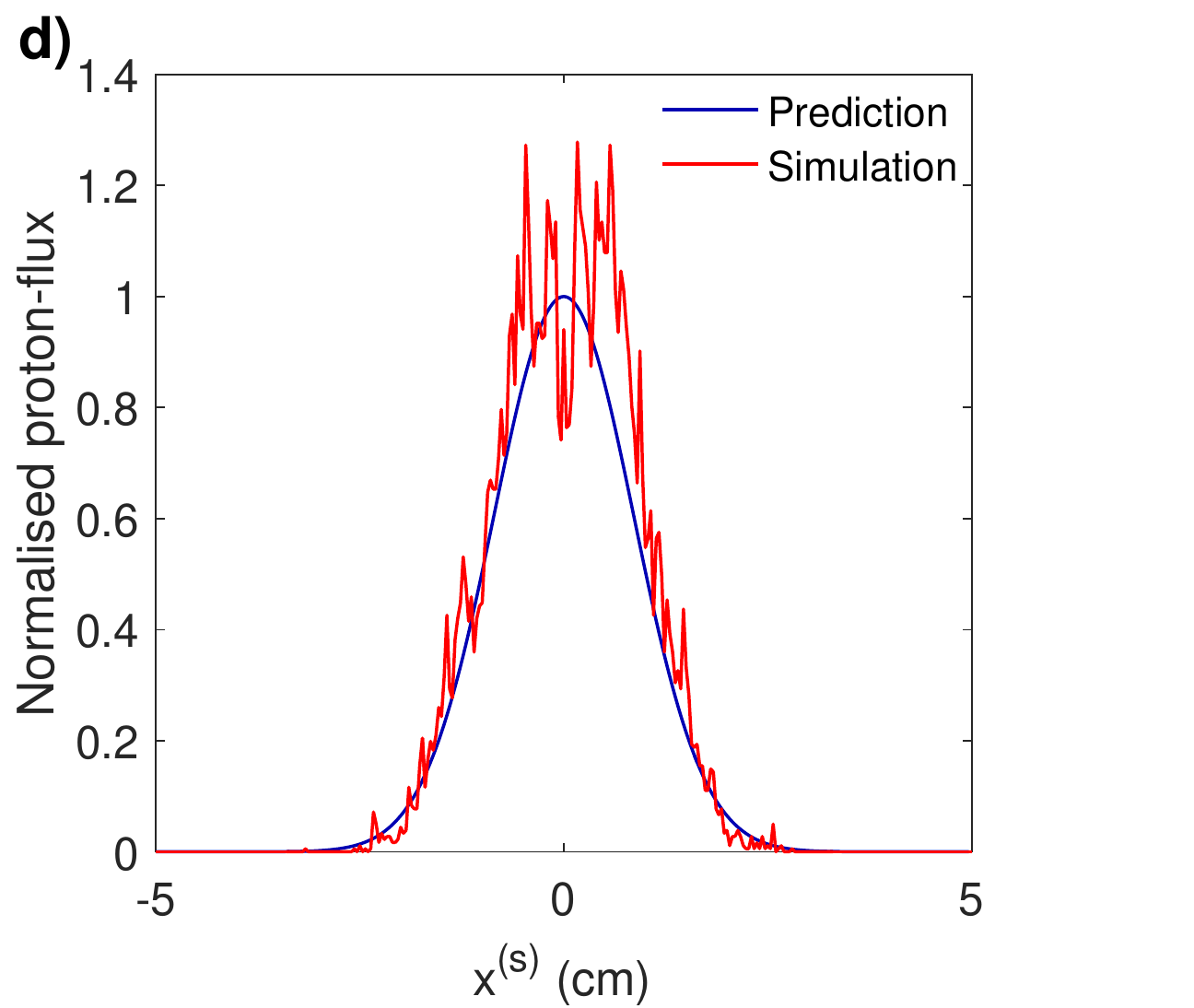}
    \end{subfigure} %
\caption{\textit{Diffusion of a pinhole beam through a strong, small-scale stochastic magnetic field.} \textbf{a)} image-flux distribution created by pinhole beam of 3.3 MeV protons in the absence of magnetic fields, normalised by the maximum image-flux value. The pinhole beam contains 800,000 3.3 MeV protons, released from a point source at distance $r_i = 1 \, \mathrm{cm}$ from the $301^3$, $l_i = l_z = l_\bot = 0.1 \, \mathrm{cm}$ interaction region, and constrained to fall within the pinhole diameter $d_{pin} = 0.02 \, \mathrm{cm}$. The detector plane is located at $r_s = 30 \, \mathrm{cm}$ on the opposing side from the source. \textbf{b)} Normalised image-flux distribution in the case of the beam passing through a magnetic field consisting of a random collection of `cocoons', with $l_B = 5.6 \mu \mathrm{m}$, and $B_{rms} = 2 \, \mathrm{MG}$. The field was generated using the technique described in Appendix \ref{NumSimMagFieldGen}, while the artificial imaging procedure was carried out using the procedure given in Appendix \ref{NumSimFluxImageGenFull}. The normalisation factor in this case was the maximum value of the analytically predicted image-flux value. \textbf{c)} Analytic prediction \eqref{stocmagsmrdscreenflux} of normalised image-flux distribution using isotropic diffusion model, with perpendicular-deflection field RMS matching numerical field. \textbf{d)} Lineouts along screen $x$-axis of numerically determined and analytically predicted image-flux distribution.} \label{pinholebeamdiffA}
\end{figure}

The results of this numerical experiment are given in Figure \ref{pinholebeamdiffA}. Comparing the numerically determined image-flux distribution in the absence of magnetic fields (Figure \ref{pinholebeamdiffA}a) and in their presence (Figure \ref{pinholebeamdiffA}b) shows that small-scale fields of sufficent strength do indeed scatter a pinhole beam of fast protons; furthermore, the analytic prediction for the image-flux distribution qualitatively matches the numerical experiment (Figures \ref{pinholebeamdiffA}c). 

The extent of agreement between analytic and numerical image-flux distributions can be seen more clearly in a lineout of both (Figure \ref{pinholebeamdiffA}d): though the width of both distributions is similiar, Figure \ref{pinholebeamdiffA}d shows that a diffusive model does not account for fluctuations associated with departures from the quasi-linear approximation. In short, the diffusive model for the image-flux distribution \eqref{stocmagsmrdscreenflux} is a reasonable (though not perfect) description in this case.

\begin{figure}[htbp]
\centering
    \begin{subfigure}{.48\textwidth}
        \centering
        \includegraphics[width=\linewidth]{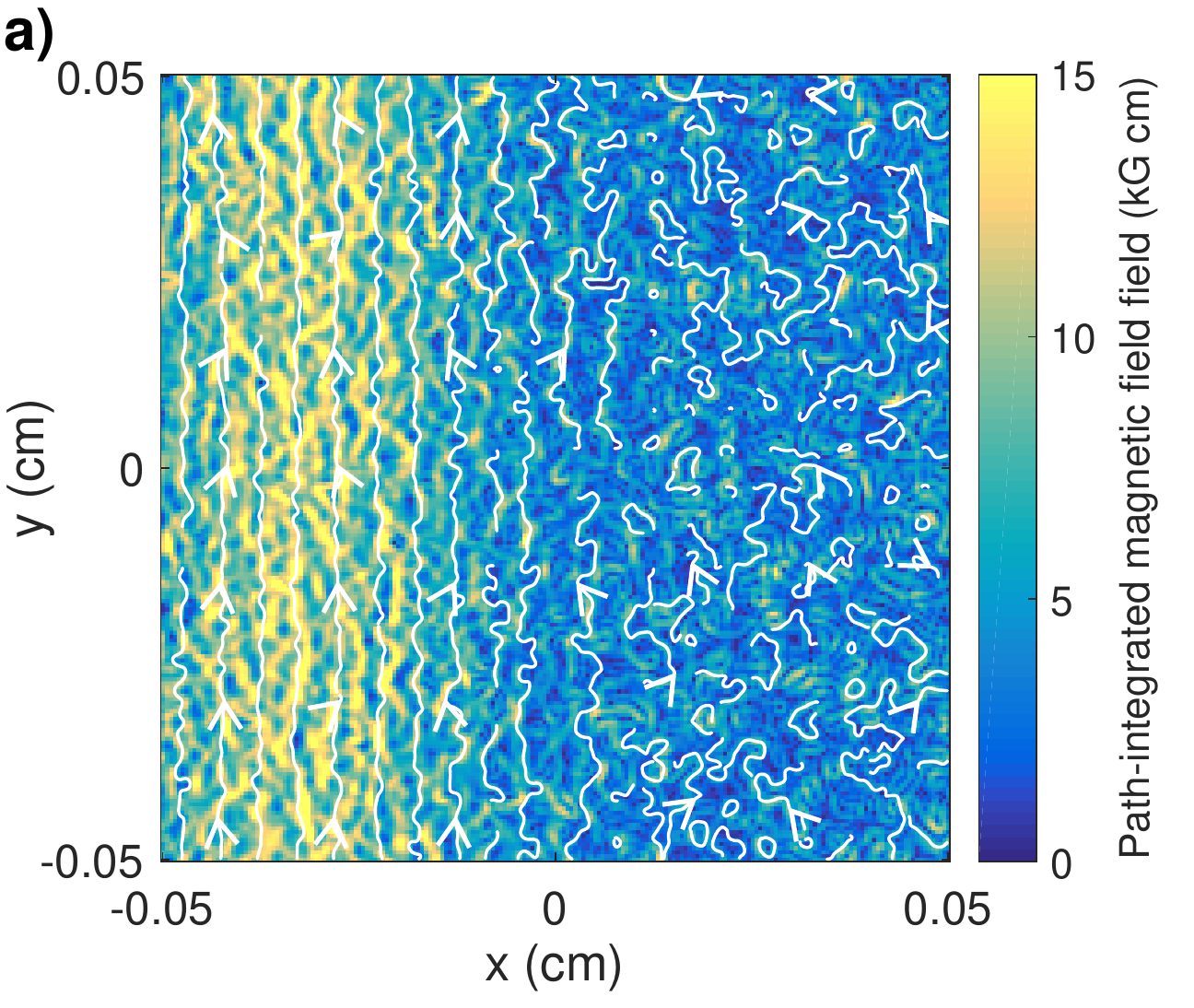}
    \end{subfigure} %
    \begin{subfigure}{.48\textwidth}
        \centering
        \includegraphics[width=\linewidth]{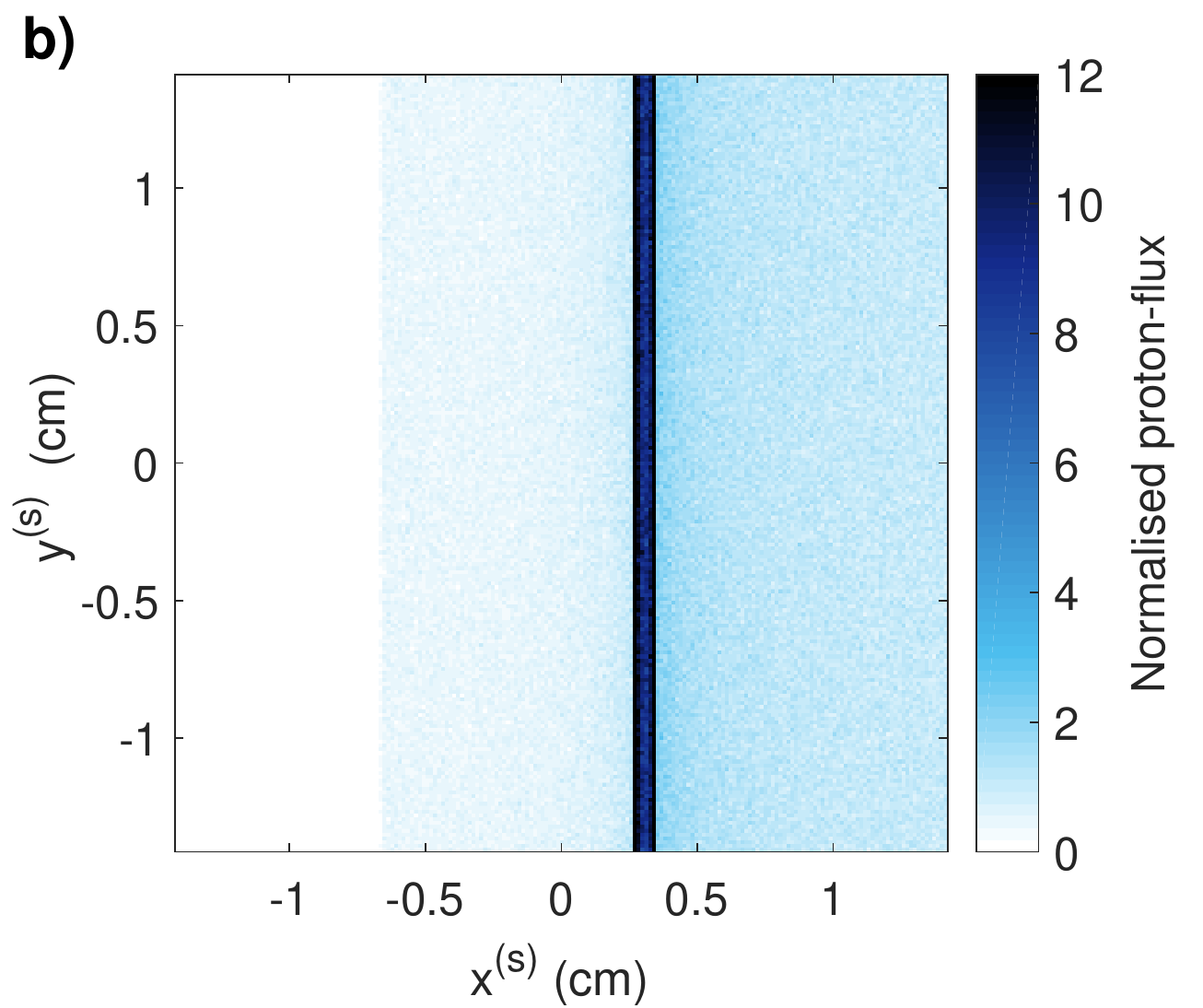}
    \end{subfigure} %
    \begin{subfigure}{.48\textwidth}
        \centering
        \includegraphics[width=\linewidth]{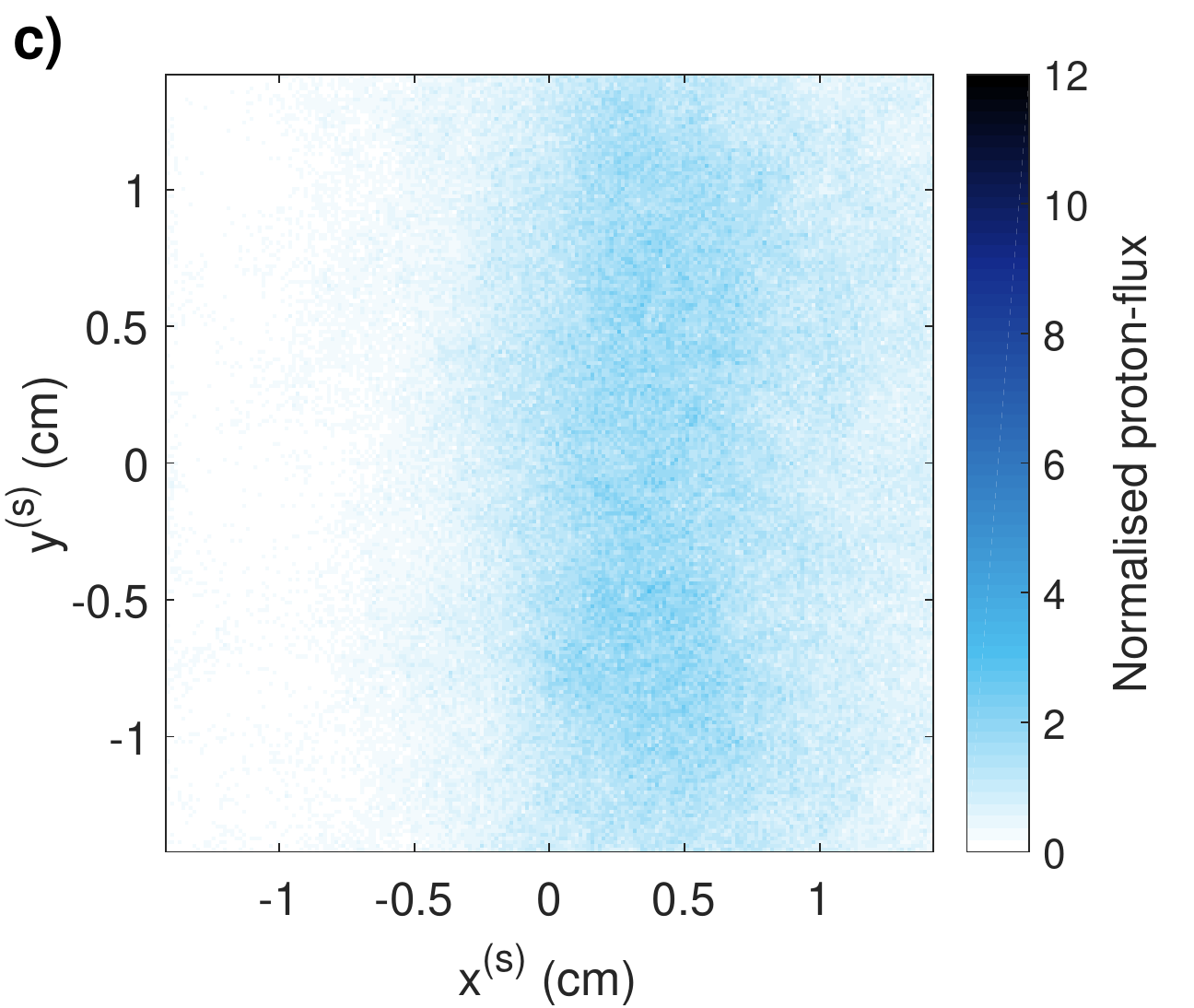}
    \end{subfigure} %
    \begin{subfigure}{.48\textwidth}
        \centering
        \includegraphics[width=\linewidth]{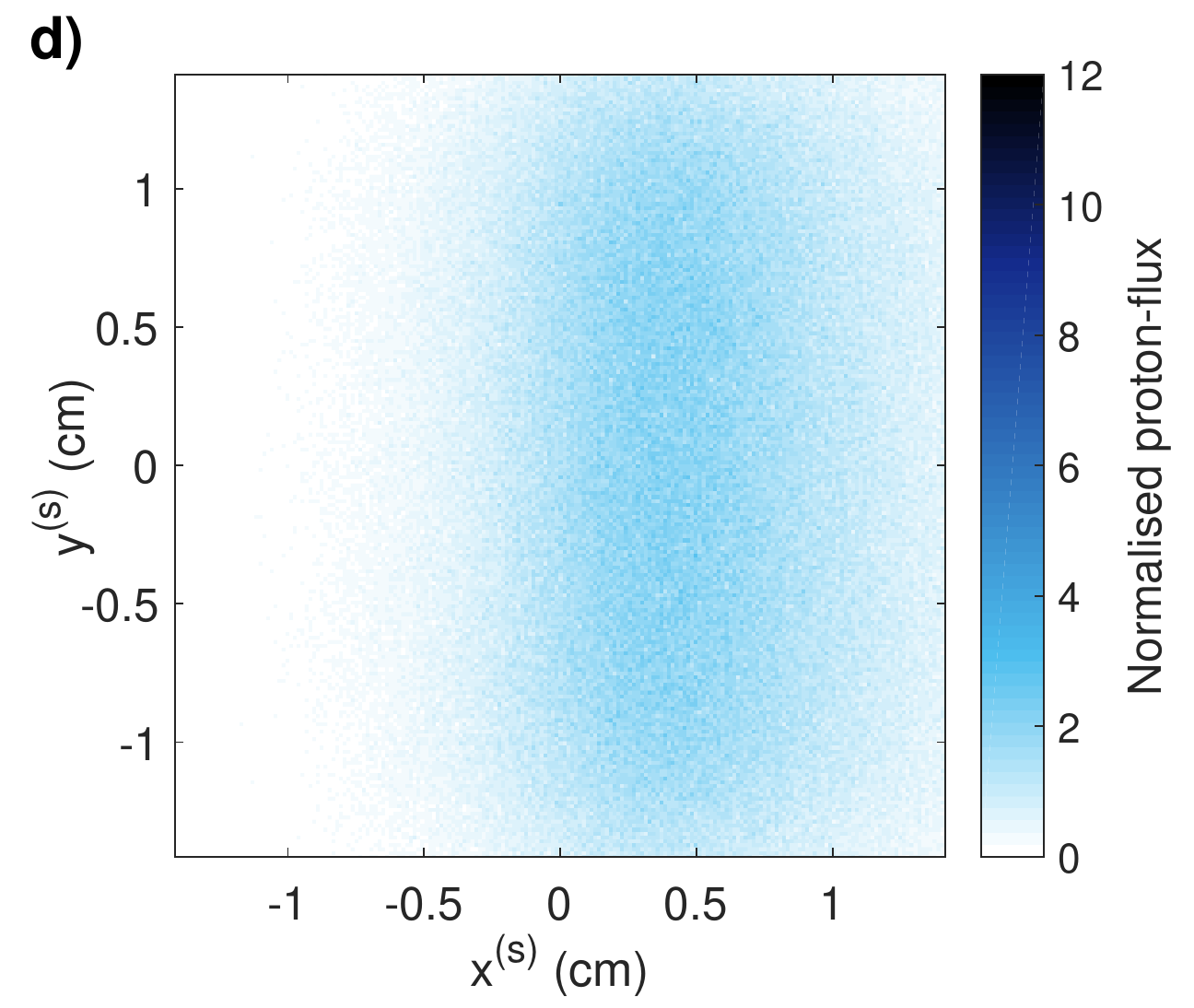}
    \end{subfigure} %
\caption{\textit{Diffusion of image-flux distribution associated with magnetic flux-rope structure due to strong small-scale fields.} \textbf{a)} Path-integrated magnetic field experienced by imaging proton beam (parameters identical to those used for the numerical experiment described in Figure \ref{smearedscreenfluxstocmaginhom}) resulting from combination of a magnetic flux-rope field, maximum field strength $B_{max} = 110 \, \mathrm{kG}$, with width $\delta = l_i/6$ and centre at $\left(x_0,z_0\right) = \left(-l_i/3,0\right)$, with statistically homogeneous small-scale field component consisting of magnetic cocoons with correlation length $l_B = 8.3 \, \mu \mathrm{m}$, and $B_{rms} = 500 \, \mathrm{kG}$. \textbf{b)} 3.3-MeV normalised image-flux distribution in absence of small-scale fields obtained using numerical imaging scheme described in Appendix \ref{NumSimFluxImageGenFull}. \textbf{c)} 3.3-MeV numerically determined normalised image-flux distribution including small-scale fields. \textbf{d)}  3.3 MeV normalised image-flux distribution predicted by analytic relation \eqref{smearedscreenfluxstocmaginhom}.} \label{fluxropesmlsclediff}
\end{figure}

The second example tests numerically the effect of a small-scale magnetic field on a larger-scale, non-stochastic magnetic structure, and compares the result to the analytical prediction arising from \eqref{smearedscreenfluxstocmag}:
\begin{equation}
\tilde{\Psi}\!\left(\mathbf{x}_\bot^{\left(s\right)}\right) = \frac{1}{\pi \delta^2} \int \mathrm{d}^2  \tilde{\mathbf{x}}_{\bot}^{\left(s\right)} \, \Psi^{\left(s\right)}\!\left(\tilde{\mathbf{x}}_\bot^{\left(s\right)}\right) \exp{\left[-\left(\frac{\mathbf{x}_\bot^{\left(s\right)}- \tilde{\mathbf{x}}_{\bot}^{\left(s\right)}}{\delta}\right)^2\right]} \, ,  \label{smearedscreenfluxstocmaginhom} 
\end{equation}
where $\delta = r_s \Delta w/V$ as before, and $\Psi^{\left(s\right)}\!\left(\mathbf{x}_\bot^{\left(s\right)}\right)$ is the image-flux distribution associated with the non-stochastic magnetic field. For this test, the non-stochastic magnetic field is chosen to be a magnetic flux-rope system of the form 
\begin{equation}
\mathbf{B}\!\left(\mathbf{x}\right) = \frac{B_0}{\pi} \, \mathrm{sech} \, \frac{x-x_0}{d} \, \mathrm{sech} \, \frac{z-z_0}{d} \mathbf{e}_y \, , \label{magfluxropestoc}
\end{equation}
contained inside a cube of side length $l_i = l_z = l_\bot$ (a more detailed consideration of this magnetic field configuration is given in Appendix \ref{IllPosNonLinRecon}). Here, $B_0/\pi$ is the peak field strength of the flux-rope, $d$ its typical size, and $\left(x_0,z_0\right)$ the $x$- and $z$-coordinate of the central axis of the flux-rope. This is combined with a similar small-scale stochastic magnetic field to that utilised for the previous numerical experiment (Figure \ref{pinholebeamdiffA}). The numerical image-flux distribution is obtained by implementing an artificial imaging set-up, with the same imaging parameters as the previous numerical test, except no longer constraining the proton beam to the pinhole. The resulting path-integrated magnetic field experienced by the beam is shown in Figure \ref{fluxropesmlsclediff}a. 

Results of the test are shown in Figure \ref{fluxropesmlsclediff}. Figure \ref{fluxropesmlsclediff}b shows the image-flux distribution $\Psi^{\left(s\right)}\!\left(\mathbf{x}_\bot^{\left(s\right)}\right)$ resulting from the magnetic-flux rope structure in the absence of the small-scale fields; the presence of two caustic structures (a defocusing caustic pair~\cite{K12}) indicate that with the specified parameters, the imaging of the non-stochastic magnetic field falls into the caustic regime. Adding in the small-scale stochastic field results in the numerical image-flux distribution shown  Figure \ref{fluxropesmlsclediff}c, while the analytic prediction using \eqref{smearedscreenfluxstocmaginhom} leads to Figure \ref{fluxropesmlsclediff}d: there is qualitative similarity. We conclude that the diffusive model \eqref{smearedscreenfluxstocmaginhom} provides an reasonable description for the image-flux distribution in the presence of small-scale fields, with the same caveat of local variations in image flux which occur in the numerical image-flux distribution being absent in the analytical prediction. It is clear that the effect of small-scale magnetic fields on larger scale image-flux structures (including caustics) is to smear them out. 

\section{Numerical algorithms} \label{NumSim}

\subsection{Generating proton-flux images numerically}\label{NumSimFluxImageGenFull}

In the main text and other appendices, we carry out various numerical experiments, for which artificial proton-flux images of magnetic fields are generated. In this appendix, we briefly explain how such images are created. 

For a proton point-source, the following procedure is followed. An artificial proton is generated at a position with Cartesian coordinate $\left(0,0,-r_i\right)$, and is assigned a random velocity, with speed fixed at $V$. The direction vector is chosen from a uniform distribution on a constrained surface region of the unit sphere; more specifically, defining a spherical polar coordinate system around the $z$-axis, the polar angle $\theta$ is restricted to the interval $\theta \in \left[0,\arctan{\left(l_\bot/r_i\right)}\right]$. The proton is then mapped to the plane $z = 0$, assuming its motion is free. If the initial perpendicular coordinate $\mathbf{x}_{\bot0}$ of the proton on intersection with the $z = 0$ plane has either $\left|x_{\bot0}\right| > l_\bot/2$ or $\left|y_{\bot0}\right| > l_\bot/2$ -- that is, it does not pass through the cuboid containing the simulated magnetic field (`interaction region') -- then the artificial proton is discarded. The procedure is repeated until the specified number of imaging protons is reached.

Once inside the interaction region, the proton's position and velocity are evolved using a Boris algorithm combined with magnetic field interpolation~\cite{BL91,W04}. When a given proton leaves the interaction region, it is then mapped to the plane $z = r_s+l_z$ (if its $z$-velocity is negative, it is discarded), again assuming free motion. To create synthetic proton-flux images, protons with perpendicular image-coordinates $\mathbf{x}_{\bot}^{\left(s\right)}$ satisfying $\left|x_{\bot}^{\left(s\right)}\right| < \mathcal{M} l_\bot/2$ or $\left|y_{\bot}^{\left(s\right)}\right| < \mathcal{M} l_\bot/2$ -- for $\mathcal{M}$ the image-magnification factor -- are binned into pixels. 

For a proton source with finite spatial extent -- a sphere, radius $a$ -- emitting protons isotropically, a similar process is carried out to the point source, with one modification. Each artificial proton is assigned a random position drawn from uniform distribution defined on the unit ball, which is then scaled by the radius $a$. 

\subsection{Generating stochastic Gaussian fields} \label{NumSimMagFieldGen}

The creation of artificial stochastic magnetic fields, and then undertaking numerical experiments simulating proton-flux images, is a useful way of testing the various analytic theories of proton imaging derived in this paper. Of the many possibilities, Gaussian stochastic magnetic fields are particularly convenient, since they are entirely characterised statistically by their magnetic-energy spectrum~\cite{A81}.
As mentioned in the main text, it is well known that stochastic magnetic fields in many situations of interest are not Gaussian~\cite{SCTM04};
nevertheless, for the purposes of testing spectral extraction methods, assuming Gaussian statistics is perfectly adequate.

The technique used in this paper to generate stochastic Gaussian fields is a spectral method based on an approach due to Yamazaki and Shinozuka~\cite{YS88}. 
For a scalar stochastic field, it consists of the following steps:
\begin{enumerate}
\item Create an array of random uncorrelated Gaussian noise with zero mean, and transform into Fourier space with a fast Fourier transform (FFT).
\item Calculate the moduli of the Fourier wavemodes, and `colour' these with a term proportional to $\left[E\!\left(k_{nm}\right)\right]^{1/2}$, where $E\!\left(k\right)$ is the desired spectrum, and $k_{nm}$ is a centralised array of wavevectors. This must be done in a way to preserve the symmetries of the discrete Fourier transform associated with a real field.
\item Applying the inverse FFT gives a Gaussian, zero-mean field with that spectrum.
\end{enumerate}
Generating random field with other types of statistics can be done in a similar manner~\cite{SD96}.

For a stochastic magnetic field, we need three such components, but also have the further requirement that $\nabla \cdot \mathbf{B} = 0$. We enforce this solenoidality condition by generating three uncorrelated components of a vector potential $\mathbf{A}$, then calculating $\mathbf{B} = \nabla \times \mathbf{A}$. To obtain the desired spectrum, we note that for a vector potential with Fourier-transformed autocorrelation tensor
\begin{equation}
\hat{M}_{ij}^{\left(A\right)}\!\left(\mathbf{k}\right) = \frac{\hat{M}^{\left(A\right)}\!\left(k\right)}{2} \delta_{ij} \, ,
\end{equation}
the relation $\hat{B}_{n} = i \epsilon_{lmn} k_m \hat{A}_n$ implies that
\begin{equation}
\hat{M}_{ij}^{\left(B\right)}\!\left(\mathbf{k}\right) = \frac{1}{V} \left<\hat{B}_i\!\left(\mathbf{k}\right) \hat{B}_j^{*}\!\left(\mathbf{k}\right)\right> =  \frac{k^2 \hat{M}^{\left(A\right)}\!\left(k\right)}{2} \left(\delta_{ij}-\frac{k_i k_j}{k^2}\right) \, .
\end{equation}
Since this is proportional to the general form of an isotropic Fourier-transformed autocorrelation tensor for a solenoidal vector field~\cite{EV03},
we conclude that taking
\begin{equation}
 \hat{M}^{\left(A\right)}\!\left(k\right) = \frac{\hat{M}^{\left(B\right)}\!\left(k\right)}{k^2}
\end{equation} 
will give a stochastic magnetic field with the desired properties. When calculating the curl of the vector potential $\mathbf{A}$ for this procedure numerically, it is best done spectrally, since naive application of a discrete finite-difference operator in real space without respecting periodic boundary conditions will result in significant spectral distortion under the curl operation. The desired field-strength normalisation for a generated magnetic field can be found simply by re-scaling. Similar approaches can be undertaken to generate two-dimensional perpendicular-deflection fields from random 2D Gaussian noise slices in terms of an assumed spectral form for the Monge-Kantorovich potential. 

In many situations, a periodic cube (side length $l_i = l_\bot = l_z$) of homogenenous stochastic magnetic field is not sufficient: including some general variation in the RMS of a magnetic field
gives for a more realistic configuration. This can be achieved by multiplying the vector potential used to generate the magnetic field by the desired smooth envelope function $f\!\left(\mathbf{x}\right)$ which only varies over $\mathcal{O}\!\left(l_i\right)$ length scales:
\begin{equation}
\tilde{\mathbf{A}}\!\left(\mathbf{x}\right) = f\!\left(\mathbf{x}\right) \mathbf{A}\!\left(\mathbf{x}\right) \, .
\end{equation}
Then, 
\begin{equation}
\tilde{\mathbf{B}}\!\left(\mathbf{x}\right) = \nabla \times \tilde{\mathbf{A}}\!\left(\mathbf{x}\right) = f\!\left(\mathbf{x}\right) \nabla \times \mathbf{A}\!\left(\mathbf{x}\right) + \nabla f \times \mathbf{A}\!\left(\mathbf{x}\right) \approx f\!\left(\mathbf{x}\right) \mathbf{B}\!\left(\mathbf{x}\right) \, ,
\end{equation}
where the second term is smaller than the first for $\mathbf{x} < l_i$. Over larger scales, for appropriately decaying $f$ the second term can often also be ignored for scales $\mathbf{x} \sim l_i$: for example, with a Gaussian envelope
\begin{equation}
f\!\left(\mathbf{x}\right) = \exp{\left(-\frac{4 \sigma \mathbf{x}^2}{l_i^2}\right)} \, , \label{Gaussianwindowappendix}
\end{equation}
it follows that even for $\mathbf{x} \sim l_i$
\begin{equation}
\frac{\nabla f \times \mathbf{A}\!\left(\mathbf{x}\right) }{ f\!\left(\mathbf{x}\right) \nabla \times \mathbf{A}\!\left(\mathbf{x}\right)} \sim \frac{l_B}{l_i} \ll 1 \, .
\end{equation} 
Since the strength of the field is much reduced at the edges anyway (where field strengths are weak), the approximation is a good one. In terms of affecting the spectrum, the impact of multiplying by a envelope function in real space is to apply a convolution in Fourier space, with the subsequent result of slight resolution loss. For $l_i \gg l_B$ this effect on spectral shape is very small; however, the overall normalisation of the field is altered. In particular, the RMS field strengths of the two fields are related by
\begin{equation}
\left<\tilde{\mathbf{B}}^2\right> = \frac{1}{V} \int \mathrm{d}V  f\!\left(\mathbf{x}\right)^2 \left|\mathbf{B}\right|^2 \, .
\end{equation}
Again invoking separation of scales, we find $\left<\tilde{\mathbf{B}}^2\right>  \approx \left<\mathbf{B}^2\right> \left<f^2\right>$. For a Gaussian envelope of the form \eqref{Gaussianwindowappendix}, this gives 
\begin{equation}
\frac{\left<\tilde{\mathbf{B}}^2\right>}{B_{rms}^2} \approx \frac{1}{16} \sqrt{\frac{\pi^3}{2 \sigma^3}} \left[\mathrm{erf}\!\left(\sqrt{2\sigma}\right)\right]^3 \, .
\end{equation}
If the RMS magnetic field strength of the enveloped field $\tilde{\mathbf{B}}$ is renormalised to match that of its parent field $\mathbf{B}$, this gives a `maximum' RMS $\left<\mathbf{B}^2\right>_{max}$ (localised at the centre of the envelope) with value
\begin{equation}
\left<\mathbf{B}^2\right>_{max} = B_{rms}^2 \left\{\frac{1}{16} \sqrt{\frac{\pi^3}{2 \sigma^3}} \left[\mathrm{erf}\!\left(\sqrt{2\sigma}\right)\right]\right\}^{-1} \, .
\end{equation}
However, from the perspective of an imaging proton beam, this is not the effective increase in the RMS magnetic field strength observed -- since the proton beam experiences a path-integrated magnetic field along the $z$-coordinate direction. Denoting the effective RMS magnetic field strength along the perperpendicular origin $\mathbf{x}_{\bot0} = \left(0,0\right)$ resulting from a enveloping procedure combined with renormalisation by $B_{rms,0}$, we find
\begin{equation}
B_{rms,0} = B_{rms} \left[\frac{1}{l_i^2} \int \mathrm{d}^2 \mathbf{x}_{\bot0} \, f\!\left(\mathbf{x}_{\bot0}\right)^2 \right]^{-1/2} = B_{rms} \left[\frac{\pi}{8 \sigma} \mathrm{erf}\!\left(\sqrt{2 \sigma}\right)\right]^{-1/2} \, , \label{Brms0modAppend}
\end{equation}
which is equation \eqref{Brms0mod} in the main text. Substituting $\sigma = 3$ gives 
\begin{equation}
B_{rms,0} \approx 2.7 B_{rms} \, .
\end{equation}
The validity of such estimates can be checked numerically with simulated fields, and estimate \eqref{Brms0modAppend} is usually accurate for $\sigma > 1$.

Applying an envelope to numerically generated stochastic fields is also a convenient way of avoiding edges effects when performing analysis on flux samples in testing. One such example is the loss of periodicity of flux samples generated from a periodic magnetic field sample as $\mu$ increases from small to moderate; another comes when applying a field reconstruction algorithm to a sample whose edges do not strictly satisfy the required boundary conditions due to loss of proton flux from the detector. The latter results in a loss of accuracy in the field reconstruction algorithm in an edge region of size similar to screen displacement magnitude; the former can lead to global spectral distortion if sufficiently strong. For the purposes of analysis of actual samples (where non-periodic boundaries cannot be avoided), additional techniques have to be applied, such as sample windowing or $\Delta$-variance methods~\cite{A12}. 

\subsection{Generating proton-flux images from perpendicular-deflection fields} \label{NumSimFluxImageGen}

For general EM particle configurations, synthetic proton-flux images can be created as just described in Appendix \ref{NumSimFluxImageGenFull}.
However, for three-dimensional magnetic fields defined on a refined grid, this can quickly become quite computationally expensive -- for example, the (non-optimised) MATLAB ray-tracing code used to create synthetic images for this paper on a 32 processor parallelised system took 8 hours to propagate two million particles through a $201^3$ grid. This can be improved by a more efficient implementation; however, various alternatives based on the analytic theory of proton imaging derived in Appendix \ref{KinThyPrad} can be used to achieve order of magnitude improvements. 

As a first step, perpendicular-deflection fields can be used numerically to create proton images. This process involves assigning a random collection of proton positions on the imaging side of the array containing the desired magnetic field configuration, and then calculating the perpendicular-deflection field for that configuration. This latter procedure can be carried out without making any asymptotic approximations at all by sending a selection of test protons though the field configuration, then using a scattered interpolation algorithm to determine the perpendicular-deflection field. Provided the correlation length of the perpendicular-deflection field is much greater than the pixel size, this numerically determined perpendicular-deflection field will accurately represent the actual perpendicular deflection experienced by protons. Once this perpendicular-deflection field is obtained, the full collection of protons positioned on the initial coordinate grid can be allocated perpendicular velocities according to the perpendicular-deflection field, and then the resulting proton-flux image created. Since only $\mathcal{O}\!\left(N^2\right)$ particles need to propagated through a $N^3$ array to give a reasonable description of the perpendicular-deflection field, and the process of random particle position allocation is typically much less demanding, this enables the same number of particle (indeed, far more) to be used in a fraction $\mathcal{O}\!\left(1/N\right)$ of the time for a full ray-tracing set-up. This whole process removes a redundancy implicit in a full ray-tracing implementation associated with randomly generated protons travelling along similar trajectories. 

\begin{figure}[htbp]
\centering
\includegraphics[width=0.80\textwidth]{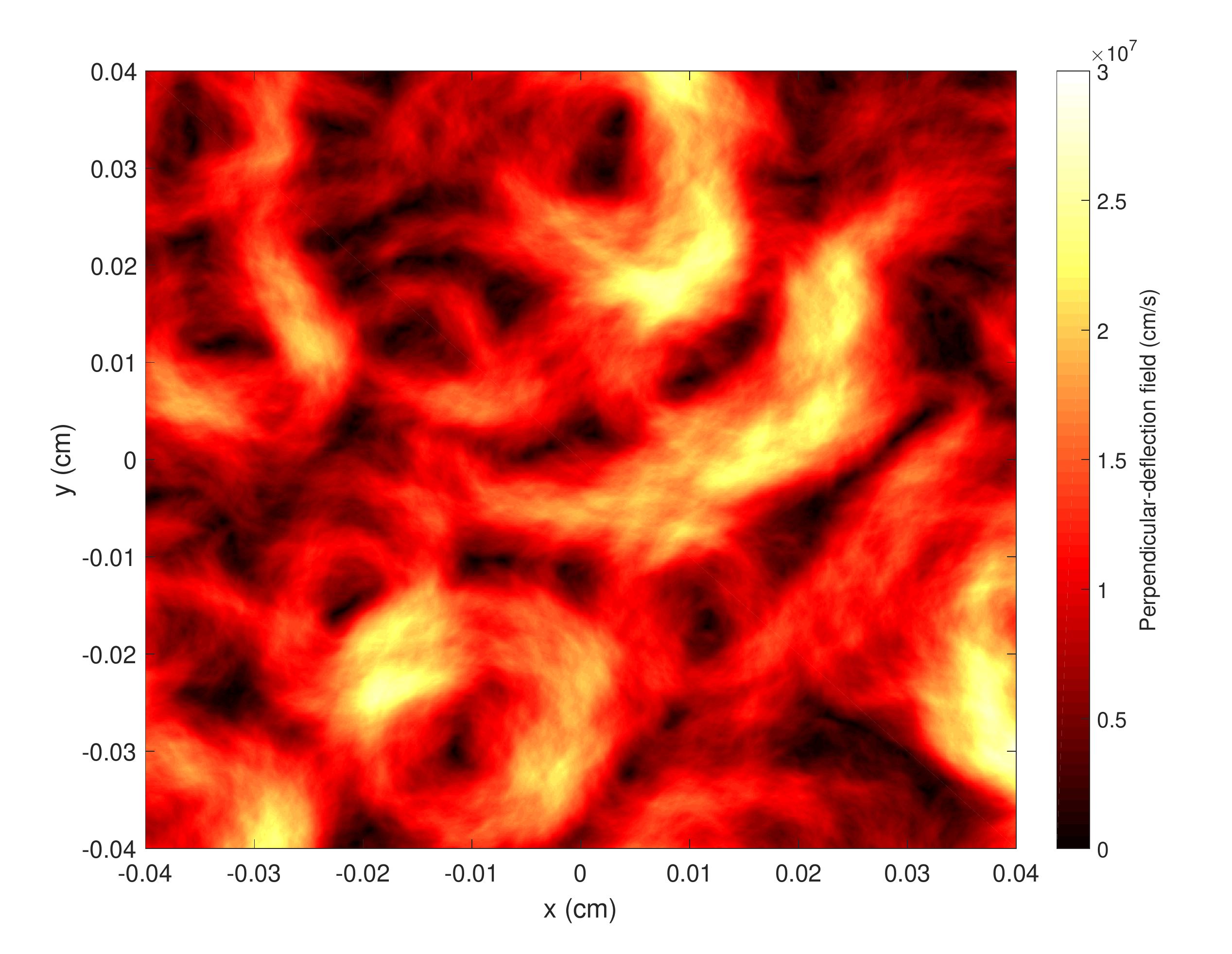}
\includegraphics[width=0.80\textwidth]{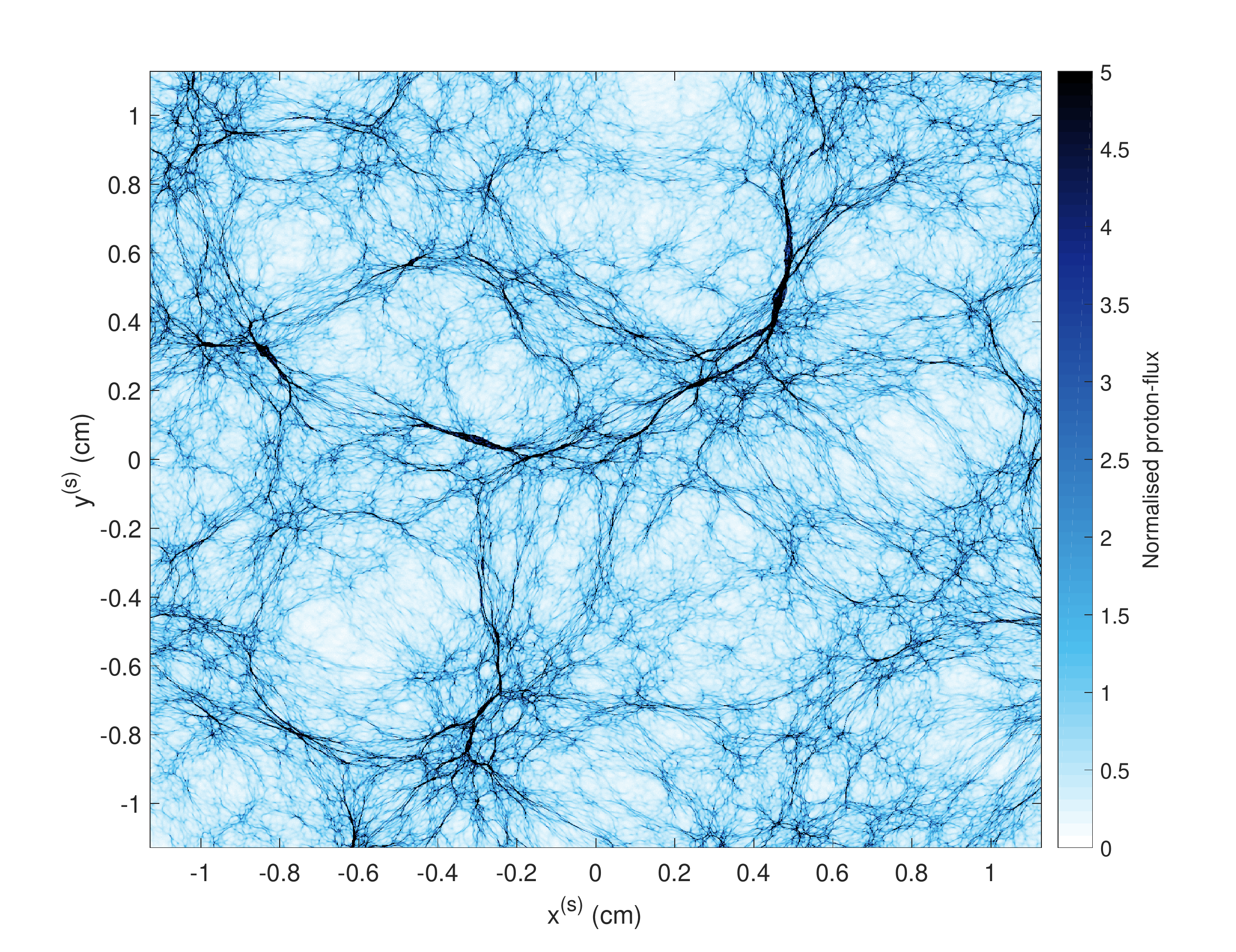}
\caption{\textit{Example of high-resolution perpendicular-deflection fields and proton-flux image created using analytic theory of proton imaging.} Top: $4001^2$ perpendicular-deflection field predicted analytically using \eqref{deffieldspec} for a stochastic magnetic field with power law spectrum $E_B\!\left(k\right) \propto k^{-2}$, magnetic field strength RMS $B_{rms} = 50 \, \mathrm{kG}$, and spectral cutoffs $k_l = 4 \pi/l_i$, $k_u = 4000 \pi/l_i$. Bottom: $4001^2$ proton-flux image generated by assigning 6.4 billion synthetic 15.0 MeV protons with positions on the far side of the interaction region, and hence velocities as determined by the perpendicular-deflection field. \label{highdefradiographs}}
\end{figure}

A further extention of this technique allows for the generation of extremely high resolution proton-flux images. Noting that the spectrum of the perpendicular-deflection field has the simple relation \eqref{deffieldspec} to the magnetic-energy spectrum -- and by the central limit theorem, is likely to be Gaussian at small-scales -- a good approximation to an actual perpendicular-deflection field can be obtained simply by generating a Gaussian perpendicular-deflection field with an appropriate spectral curve. Since this latter quantity is two-dimensional, the computational difficulty of the problem is greatly reduced. For large grids ($N \geq 1000$), this enables proton-flux images to be created on a single processor which otherwise could only be conceivably attempted with hundreds. An example is given in Figure \ref{highdefradiographs}: a perpendicular-deflection field with a shallow $k^{-1}$ spectrum is generated in $4001^2$ grid, and then 6.4 billion particles allocated deflection velocities. The resulting $4001^2$ proton-flux image was created in under ten minutes with 32 processors, and gives a resolution far exceeding current experimental possibilities.  

The only exception to this rule arises when proton trajectories cross inside the plasma -- which occurs if and only if the diffusive regime is reached. In this case, a full ray-tracing scheme is needed to obtain a truly accurate proton-flux image.

\subsection{Nonlinear field reconstruction algorithm for perpendicular-deflection field} \label{NumSimNonLinRecon}

There are various approaches for solving the Monge-Amp\`ere equation numerically, using finite-element methods, or converting the problem to its Monge-Kantorovich equivalent and implementing an optimatisation~\cite{DG06}. Indeed, the latter approach has recently been used in the very context of proton imaging~\cite{K16}. However, for this paper, we use a particularly simple alternative based on the treating the deflection-field potential as the steady-state solution of the logarithmic parabolic Monge-Amp\`ere equation:
\begin{equation}
\frac{\partial \Phi}{\partial t} = \log{\frac{\Psi\!\left[\nabla_{\bot0} \Phi\!\left(\mathbf{x}_\bot\right)\right] \det{\nabla_{\bot0} \nabla_{\bot0} \Phi\!\left(\mathbf{x}_{\bot0}\right)}}{\Psi_0\!\left(\mathbf{x}_{\bot0}\right)}} \, .
\end{equation}
This in turn is solved on a square grid using a finite-difference scheme (first order in time, second order in space) combined with interpolation methods. The imposed boundary conditions are the same Neumann condition we wish to impose for the Monge-Amp\`ere equation:
\begin{equation}
\nabla_{\bot0} \Phi\!\left(\mathbf{x}_{\bot0}\right) \cdot \hat{\mathbf{n}} = \mathbf{x}_\bot \cdot \hat{\mathbf{n}} \, .
\end{equation}
A more detailed outline of this field reconstruction algorithm is given by Sulman \textit{et. al.}~\cite{S11}, where existence and uniqueness of the solution of the parabolic Monge-Amp\`ere equation (along with its convergence to the solution to the Monge-Amp\`ere equation) is proven. 

The field reconstruction algorithm seems to be faster than previous approaches -- a single processor can usually reconstruct a perpendicular-deflection field for a $201^2$ proton-flux image in under an hour~\cite{K16}. 

Further modifications can be made to deal with non-rectangular boundaries. First, any given image-flux distribution can be embedded in a image-flux distribution defined on a larger rectangular region, with the exterior region filled with a small `shadow' image-flux. The field reconstruction algorithm can then be applied using the initial flux distribution embedded in the same larger rectangular region, but also with the exterior region filled with a  shadow initial flux whose sum is equal to that of the shadow image-flux. The calculated perpendicular-deflection field in the interior region should be a close match to the true perpendicular-deflection field, by conservation of particle number in the interior and exterior regions respectively. Alternatively, the image-flux distribution can be extended to a boundary, and then a window function used. Since the field reconstruction algorithm becomes linear (and local) for small image-flux deviations, the external flux region will not distort calculated internal perpendicular-deflection fields significantly. In both cases, the result can be tested by forward-propagating particles with the desired perpendicular-deflection field. It is acknowledged that both approaches are approximate, and more rigorous techniques could be implemented with alternative schemes. 

\tableofcontents

\end{document}